\documentclass[12pt,preprint]{aastex}

\lefthead{Antoniou et al.}  \righthead{The \chandra Survey of the SMC ``Bar'': II. Optical counterparts of X-ray sources}

\def\ergcm2s{~erg cm$^{-2}$ s$^{-1}$ } 
\def\funit{~erg cm$^{-2}$ s$^{-1}$ } 
\def\ergs{~erg s$^{-1}$}             
\def\lunit{~erg s$^{-1}$}

\def\etal{et al.~}              

\def\msun{~M$_{\odot}$\,}

\def\Zsun{Z$_{\odot}$}

\def\deg{$^{\circ}$}

\def\chandra{{\it Chandra~}}
\def\chandrasimple{{\it Chandra}}
\def\x2{$\chi^{2}$}

\def\Ha{H$\alpha$~}

\def\E(B-V){{\it E(B-V)}}
\def\Rv{{\it $R_{V}$}}
\def\Av{$A_{V}$}
\def\Ab{$A_{B}$}
\def\Au{$A_{U}$}

\def\mM{$(m-M)_{V}$} 

\def\BVo{$(B-V)_{\rm{o}}$}



\begin{document}

\title{The \chandra Survey of the SMC ``Bar'': II. Optical counterparts of X-ray sources}
\author{V. Antoniou\altaffilmark{1,2}, A. Zezas\altaffilmark{1},
  D. Hatzidimitriou\altaffilmark{2}, J. C. McDowell\altaffilmark{1}}
\altaffiltext{1}{Harvard-Smithsonian Center for Astrophysics, 60 Garden Street, Cambridge, MA 02138; vantoniou@head.cfa.harvard.edu}
\altaffiltext{2}{Physics Department, University of Crete, P.O. Box 2208, GR-710 03, Heraklion, Crete, Greece}

\begin{abstract}
We present the most likely optical counterparts of 113 X-ray sources detected in our \chandra survey of the central region of the Small Magellanic Cloud (SMC) based on the OGLE-II and MCPS catalogs. We estimate that the foreground contamination and chance coincidence probability are minimal for the bright optical counterparts (corresponding to OB type stars; 35 in total). We propose here for the first time 13 High-Mass X-ray Binaries (HMXBs), of which 4 are Be X-ray binaries (Be-XRBs), and we confirm the  previous classification of 18 Be-XRBs. We estimate that the new candidate Be-XRBs have an age of $\sim15-85$ Myr, consistent with the age of Be stars. We also examine the ``overabundance'' of Be-XRBs in the  SMC fields covered by \chandrasimple, in comparison with the Galaxy. In luminosities down to $\sim10^{34}$\ergs\,, we find that SMC Be-XRBs are $\sim1.5$ times more common when compared to the Milky Way even after taking into account the difference in the formation rates of OB stars. This residual excess can be attributed to the lower metallicity of the SMC. Finally, we find that the mixing of Be-XRBs with other than their natal stellar population is not an issue in our comparisons of Be-XRBs and stellar populations in the SMC. Instead, we find indication for variation of the SMC XRB populations on kiloparsec scales, related to local variations of the formation rate of OB stars and slight variation of their age, which results in different relative numbers of Be stars and therefore XRBs.
\end{abstract}


\section{Introduction}\label{introduction}

X-ray binaries (XRBs) are
stellar systems consisting of a compact object (neutron star, black
hole, white dwarf) accreting
material from a close companion star. They are the end
points of stellar evolution, and thus by studying them we can set
constraints on stellar evolution and compact object formation models. They constitute a numerous class of
X-ray bright objects, with typical luminosities of
$\sim10^{36}-10^{38}$\lunit\, when in outburst. They are divided into two classes
depending on the mass of the donor star, with M $\leq1$\msun for the
Low-Mass XRBs (LMXBs), and M $\geq10$\msun for the High-Mass XRBs (HMXBs). 

The companion star in HMXBs is of O or B spectral type, with optical
bolometric luminosity usually exceeding that of the accretion disk (e.g. van Paradijs \& McClintock 1995). HMXBs can further be divided into two groups, the supergiant X-ray
binaries (hereafter SG-XRBs), and the Be/X-ray binaries (hereafter
Be-XRBs). In the SG-XRBs the primary is a
supergiant of spectral type earlier than B2, or an Of star, and it has evolved away from the main sequence (MS), while in the Be-XRBs, the primary is an Oe or Be star lying close to the
MS. The optical spectra of Be-XRBs are characterized by emission lines
(mostly of the Balmer series of hydrogen; for
a review of HMXBs and their properties see for example van Paradijs \& McClintock 1995).

Different mechanisms are believed to be responsible for the mass transfer in these two
groups of HMXBs. The most luminous SG-XRBs usually 
have an accretion disk fueled via Roche-lobe overflow, while the less
luminous systems may be fed by a supersonic stellar wind.
In the Be-XRBs, Be stars are characterized by a low velocity, high density equatorial
wind, resulting in periodic accretion episodes during the passage of the compact object through the decretion disk of the donor.

Be-XRBs show pulsations and have hard 1-10 keV spectra (i.e. with a power-law energy index of
$\Gamma\sim0-1$; e.g. White, Nagase \& Parmar 1995, Yokogawa \etal 2003), which are signatures of accretion onto strongly magnetized neutron stars. In contrast, SG-XRBs have generally softer spectra, and do not always have a pulsar as the compact object.

Be-XRBs are the most numerous sub-class of HMXBs. In the Milky Way as well as in the Large Magellanic Cloud (LMC), they constitute 60-70\%       
of all HMXBs (Sasaki, Pietsch \& Haberl 2003), while in the Small Magellanic Cloud (SMC) only one out of 92 known
or probable HMXBs (Liu, van Paradijs \& van den Heuvel 2005) is a supergiant system.

 The SMC is an excellent laboratory to study the Be-XRB
populations. It is the second nearest star-forming galaxy, and it has a
large number of Be-XRBs seen through moderate Galactic foreground
absorption ($N_{H}\simeq6\times10^{20} cm^{-2}$; Dickey \& Lockman
1990), and well mapped extinction (Zaritsky \etal 2002). In addition,
its well measured distance (60 kpc; e.g. van den Bergh 2000, Hilditch, Howarth \& Harries 2005), small line-of-sight depth of the young populations at its main body ($<$10 kpc; e.g. Crowl \etal 2001; Harries, Hilditch \& Howarth 2003), well studied recent star-formation (SF) history (Harris \& Zaritsky 2004), and generally uniform metallicity of the young populations facilitate the interpretation of the results. Therefore, one can study in great detail and in an homogeneous way the faint end of the XRB populations.

Several studies of the optical characteristics of the XRB
systems in the SMC, have been published in the last few years.
Haberl \& Sasaki (2000), based on the ROSAT surveys of the SMC,
identified 25 X-ray sources as new Be-XRBs. Haberl \& Pietsch (2004) extended this work presenting 65 SMC HMXBs,
of which 45 are associated with an emission-line object indicating that they are
Be-XRBs. In the latest census of HMXBs by Liu \etal (2005) 62
Be-XRBs are listed (for 92 confirmed or proposed HMXBs). In addition out of the 38 currently known X-ray pulsars in
the SMC, 34 have identified optical counterparts (Coe
\etal 2005a).

In this paper we study in a systematic way the young SMC XRB populations. The SMC ROSAT survey reached a non-uniform detection limit of $\sim5\times10^{34}-10^{35}$\ergs\, (e.g. Kahabka \& Pietsch 1996, Haberl \etal 2000, Sasaki \etal 2000). Here we are using data down to $\sim4\times10^{33}$\ergs\, from the \chandra survey of the SMC (Zezas \etal 2003) that allow to investigate the faintest of the HMXB populations.

\subsection{The \chandra survey of the SMC}\label{Chandrasurvey}

\begin{figure}
\centering
\includegraphics[height=12.5cm]{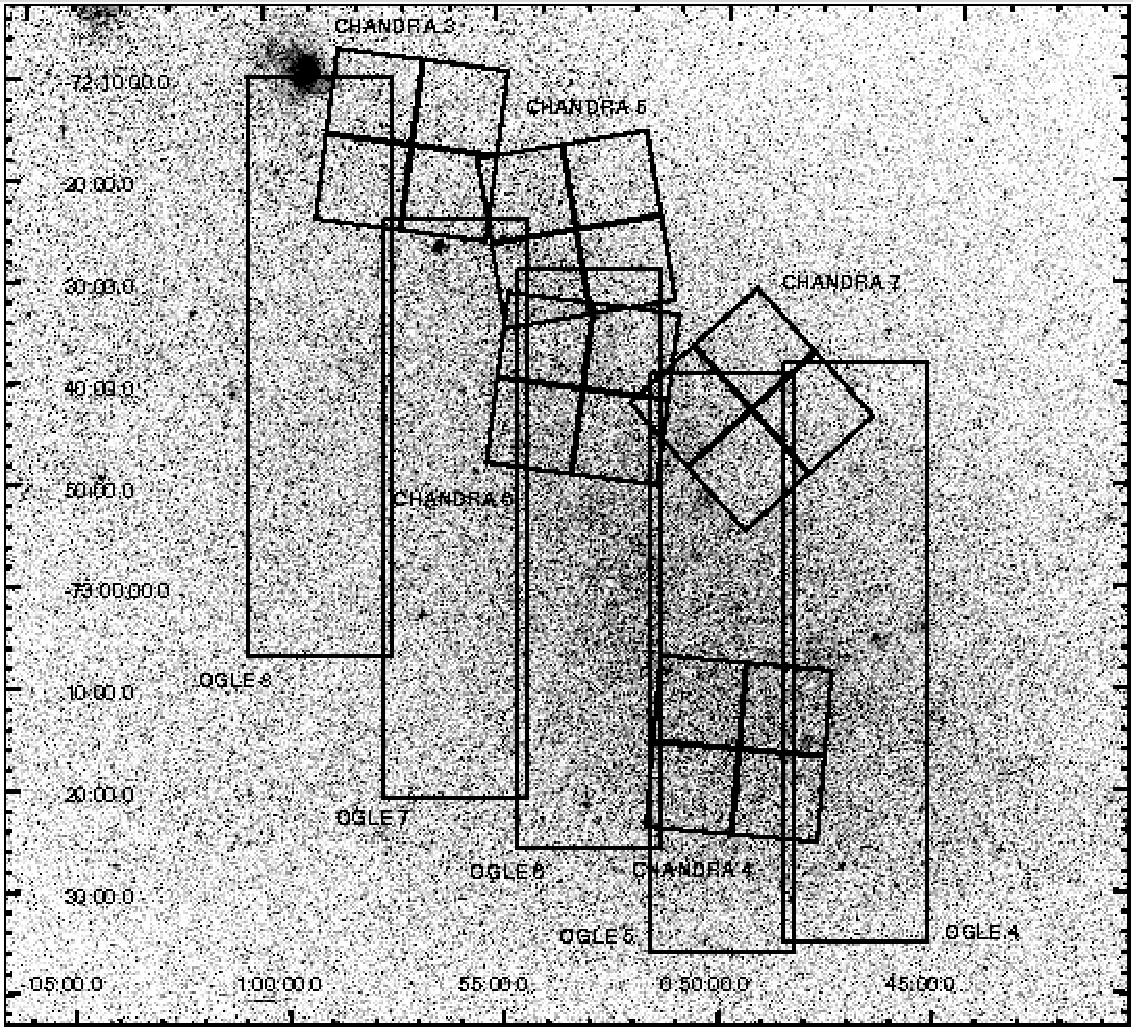}
\caption{A DSS optical image of the center of the SMC (outline of the 5 observed \chandra fields ($16\arcmin\times16\arcmin$) and the 5 overlapping OGLE-II fields).}\label{DSS}
\end{figure}

The SMC as a prime target to study the
faint end of the XRB populations (with typical ${\rm L_{x}}<10^{34}$\lunit), overcoming in this way the inherent problems of
observations in the Milky Way (e.g. distance determination, and
obscuration). For this reason we initiated a \chandra survey, consisting of five observations of the central part of the SMC performed
between May and October 2002 with the \chandra ACIS-I detector (Advanced CCD
Imaging Spectrometer; Garmire \etal 2003). The survey covers
an area of $1280\,arcmin^{2}$ along the central, most actively star forming, region
of the SMC (also referred to as the SMC ``Bar'', although it bears no
relation to a dynamical bar; van den Bergh, 2000).  In Figure
\ref{DSS} we plot the footprints of the \chandra and OGLE-II fields
(for the latter see \S \ref{opticalcat}) overlaid on a Digitized Sky
Survey (DSS) optical image of the main body of the SMC.

 The survey yielded a total of 158 sources, detected on the ACIS-I CCDs, down to a limiting
luminosity of $\sim4\times10^{33}$\lunit, which is within the
luminosity range of quiescent HMXBs (typical ${\rm L_{x}}\sim10^{32}-10^{34}$\lunit\,; van Paradijs \& McClintock 1995). Each of the 5 fields contained
between 8 and 16 sources with fluxes at least of $3\sigma$
significance (assuming Gehrels statistics; Gehrels 1986) above their local background. A brief description of the
survey and the first results are presented in Zezas \etal (2003). The
final source-list (including the X-ray luminosity functions) and the
spectral properties of the sources are
presented in Zezas \etal (2008, in prep.). Because of the large
off-axis angle (and therefore large positional uncertainty) of sources
on the ACIS-S3 and ACIS-S4 CCDs we do not include them in the
present work.

The absolute astrometric accuracy of on-axis sources is dominated by
the boresight error of \chandra ($<0.6\arcsec$ at the 90\% confidence level\footnotemark)\footnotetext{http://cxc.harvard.edu/cal/docs/cal\_present\_status.html\#abs\_spat\_pos}.
However, at larger off-axis angles the Point Spread Function (PSF) becomes
broader and asymmetric, resulting in larger positional
uncertainties (see \chandra POG, 2005). To estimate the positional errors of the off-axis sources, we used the empirical formula of Kim \etal (2004), which gives positional uncertainties for sources with 20
and 100 counts as a function of their off-axis angle (up to a maximum of $10\arcmin$). For sources with 
net number of counts within the range we linearly interpolated between these two estimates,
while for sources with less than 20 or more than 100 counts we used the appropriate
branch of the equation.

Five sources out of the 158, which were at large off-axis angles
($\rm{D_{off-axis}>10}\arcmin$), yielded high positional errors and
were discarded from further consideration in this paper. The \chandra
field IDs, and the coordinates of their centers are listed in Table \ref{tablesurvey} (Columns (1)-(3)). In Column (4)
we present the number of all detected X-ray sources in each \chandra
field, while in parenthesis we give the number of sources with fluxes at $3\sigma$ or greater level above their
local background. In Column (5) we give the number of \chandra sources
for which we searched for counterparts (i.e. detected in $\rm{D_{off-axis}<10}\arcmin$).

 The identification and classification of the optical counterpart of
an X-ray source allow us to identify the interlopers, while it is the
only way to confirm the different types of XRBs (e.g. Be-XRBs, SG-XRBs, LMXBs). Although the classification of the counterparts is only secure via
optical spectroscopic observations, color-magnitude and/or color-color diagrams (hereafter CMDs and 2-CDs, respectively) can be used to obtain rough classifications
for the optical counterparts, and at least identify HMXBs. From the appropriate isochrones and the stellar
tracks we can also obtain a feel for the mass and the age of the donor
star, if it dominates the optical emission of the X-ray
sources.

In this paper we have used the \chandra survey of the SMC (Zezas \etal 2003), and searched for optical counterparts of the X-ray sources included in this survey. In particular, we confirm or suggest new counterparts of the above X-ray sources, based on their optical properties, and when possible we give a tentative classification. In Section \ref{opticalcat} we present the optical catalogs used in this study, and in Section \ref{cc} the
counterparts of the X-ray sources. We also present their finding
charts (Section \ref{charts}), and we examine the photometry of the optical data (Section \ref{photometry}). In Section \ref{CMDs} we describe the data used in the construction of the $V$, $B-V$ CMD for the classification of the sources, and we estimate the expected contamination by
foreground stars, while in Section \ref{chance} we estimate the chance coincidence
probability. In the discussion (Section \ref{discussion}) we present the criteria used to classify the X-ray sources, and distinguish among multiple matches the most likely counterpart. We also discuss the properties of the optical counterparts, and the relative numbers of Be-XRBs in the SMC, LMC and Milky Way. In the Appendix we present notes on individual sources.

\section{Optical data}\label{opticalcat}

The SMC has been surveyed extensively in the optical and in the near
infrared. In the past decade, several photometric catalogs have become
available, with varying photometric and astrometric accuracy, limiting
magnitudes, completeness and spatial coverage. The most extensive of
these are the US Naval Observatory USNO-B1.0 (Monet \etal 2003), the USNO CCD Astrograph Catalog (UCAC2, 2nd release;
Zacharias \etal 2004), the Massive Compact Halo
Objects database (MACHO, Alcock \etal 1997), the
Optical Gravitational Lensing Experiment survey (OGLE-II;
Udalski \etal 1998a)\footnotemark,\footnotetext{http://ogle.astrouw.edu.pl/ogle2/smc\_maps.html}
the $UBVR$ CCD Survey by Massey (2002), and the Magellanic Clouds
Photometric Survey (MCPS; Zaritsky \etal 2002). We also mention, for completeness, the Two Micron All Sky Survey (2MASS; Skrutskie \etal 2006), which is sensitive to red stars, and thus it is not used in the present work (see \S \ref{multiplediscussion}).

Taking into consideration the photometric and astrometric accuracy of
each of these surveys, and their spatial coverage of the \chandra fields, we
selected the OGLE-II and MCPS surveys, as the main optical catalogs for the present work. We also
looked for additional possible counterparts of the X-ray sources
within the SIMBAD Astronomical Database\footnotemark,\footnotetext{http://simbad.u-strasbg.fr/Simbad/} and for
previously identified optical counterparts of known X-ray sources
within the NASA's Astrophysics Data System\footnotemark
(ADS)\footnotetext{http://www.adsabs.harvard.edu} (for more details see \S \ref{otherbiblio}).

\subsection{OGLE-II catalog}\label{OGLE}

 The second release of the OGLE survey (hereafter OGLE-II) is a {\it BVI}\, survey of the SMC, carried out with the 1.3m Warsaw telescope at Las Campanas Observatory (Udalski \etal 1998a). The survey provides photometric (in {\it B}, {\it V} and {\it I} band) and astrometric data for about 2.2 million stars in the
dense central regions of the SMC ``Bar'', down to $B\sim20.0\,$ mag,
$V\sim20.5\,$ mag, and $I\sim20.0\,$ mag, with typical completeness
$\sim$ $75\%$, $80\%$ and $85\%$, respectively, down to these limits. In
the densest fields (e.g. corresponding to \chandra field 6), the
completeness is reduced almost by half. The vast majority of observations was performed in 
the {\it I}\, band (because of the microlensing observing strategy of
this survey). However meticulous photometry was also
performed in both {\it B}\, and {\it V}\, bands
(Udalski et al., 1998a). For these reasons, and because of its small astrometric
error ($\sim0.7\arcsec$), matching well the positional uncertainty
of the \chandra detections (see \S \ref{Chandrasurvey}), the small pixel size
($0.417\arcsec$), and the good average seeing (as good as
$0.8\arcsec$ with typical median value of $1.25\arcsec$) during the
survey, we opted to use the OGLE-II catalog in order to identify
optical counterparts for the X-ray sources detected in our \chandra
fields. However, due to incomplete coverage of our \chandra fields by the OGLE-II survey ($\sim72\%$ of the area of our \chandra survey), we were able to search for optical counterparts for only 102 ($\sim66.7\%$) of our sources. Four of our fields are partially covered by OGLE-II, and only field 4 is fully covered. However, due to the distribution of the X-ray sources in our \chandra fields, we were able to search for
counterparts for all our sources also in
field 6. For the remaining fields, we searched for counterparts in the OGLE-II catalog for $\sim22\%$, $\sim28\%$ and $\sim86\%$ of the detected sources in \chandra fields 3, 5 and 7, respectively.

\subsection{MCPS catalog}\label{MCPS}

 Because of the incomplete coverage of the \chandra fields by the
OGLE-II survey, we supplemented the optical data with the MCPS catalog
(Zaritsky \etal 2002). The MCPS survey is based on drift-scan
images of the SMC in the Johnson {\it U}\,, {\it B}\, and {\it V}\, and Gunn
{\it I}\, filters obtained with the Las Campanas Swope telescope
(1m). The MCPS catalog contains stellar photometry for more than 5 million stars in
the central 18 deg$^{2}$ area of the SMC, which fully covers our
\chandra survey. Only for one of our X-ray sources we were not able to look for counterparts in the
MCPS catalog, because it falls in the gap between the different subscans of this survey. The final MCPS catalog includes only
stars with both {\it B}\, and {\it V}\, detections,
while there are not always measurements in the {\it U}\, band. The
incompleteness is significant below $V\sim20$ mag, while the typical seeing for this survey is
$\sim1.5\arcsec$, and the pixel size $0.7\arcsec$.

In general the two catalogs have consistent photometry: Evans \etal
(2004) reported a difference of $\delta$$B_{\rm{(MCPS - OGLEII)}} = +
0.02$ mag for stars in their 2dF spectroscopic sample. Zaritsky \etal
(2002), in their comparison between MCPS and OGLE-II magnitudes, found
mean photometric offsets of $\delta$$B = 0.011$, $\delta$$V =
0.038$, and $\delta$$I = 0.002$
mag. However, in crowded fields, actual photometric (as well astrometric)
uncertainties can be larger, due to source confusion (see also
Zaritsky \etal 2002), which we consider to be more severe in the MCPS catalog, due to the larger pixel size, and worse overall seeing. In addition, observations were not repeated in the
MCPS survey, which makes it more likely to contain a higher percentage
of stars with less reliable photometry, especially in the faint end, and crowded regions.

\section{Cross-correlation analysis}\label{cc}

\subsection{OGLE-II and MCPS}\label{OGLEMCPS}

In order to identify the optical counterparts of the \chandra sources we have first cross-correlated their coordinates with the OGLE-II, and the MCPS catalogs. The search radius for each X-ray source was calculated from the
combination, in quadrature, of the astrometric uncertainty of the corresponding optical catalog, and the positional uncertainty for
each X-ray source (including the absolute astrometric uncertainty of
{\it Chandra}). When the resulting search radius was less than
$1.5\arcsec$ we enforced a conservative minimum of
$1.5\arcsec$ ($\sim48\%$ of our sources).  
 The maximum resulting radius was $5.29\arcsec$ with the average
search radius being $1.99\arcsec$.

The optical counterparts of the X-ray
 sources are presented in Tables \ref{field3}-\ref{field7} (one for each \chandra field). In Columns (1) and (2) we give the \chandra source ID and the search
radius (in arcseconds), respectively. The X-ray sources
(Column (1)) are named as F\_NN, where F is the \chandra field
number, and NN is the source ID in this field (from Zezas \etal 2008,
 in prep.). 
In Column (3) we give the proposed counterparts, following a similar
notation: OGLE-II sources are named as O\_F\_NNNNNN, where F and NNNNNN
are the field and optical source number, respectively (from
Udalski \etal 1998a), while MCPS sources are named as Z\_NNNNNN where NNNNNN is the line number of the source in Table 1 of Zaritsky \etal (2002). In Columns (4) and (5) we present the right ascension and declination (J2000.0) of the
counterparts from their respective catalogs. The distance (in
arcseconds) of the counterpart to the \chandra source is given in
Column (6). The photometric data of the sources are presented in Columns (7) to
(12) (these data are taken directly from the original catalogs without
applying any reddening or zero-point correction): apparent magnitude
in the {\it
V}\, band (Column (7)), the {\it B - V}\, and {\it U - B}\, colors (Columns (9) and (11)), and their
errors (Columns (8), (10) and (12), respectively). Non detections in a band are indicated by 99.99, while three dots indicate that an
entry was not available at the original catalog. In Column (13) we
give notes on individual sources: ``n'' indicates a new candidate counterpart
for the X-ray source, ``c'' denotes that there are additional
comments for this source in Table \ref{tableotherinfo}, while ``u'' indicates sources for which we were not able to uniquely identify an optical counterpart.

Matched sources between the OGLE-II and MCPS catalogs are grouped
together, with different source groups separated by blank
lines, while single horizontal lines separate the counterparts of different X-ray sources (see \S \ref{Chandrasurvey}). When we were not
able to uniquely identify a MCPS source with an OGLE-II 
source, we list all possible matches (because of
source confusion, in several cases, two or more
OGLE-II sources are detected as a single brighter source in MCPS). In cases of unclear matches between the two catalogs, we show the MCPS
sources in parenthesis. The matched pairs are based on positional
coincidence, visual inspection of the OGLE-II {\it I}\, band images,
and comparison of the magnitudes and colors of the sources. MCPS
sources that may be associated with OGLE-II sources but lie just outside the search radius of each X-ray source are marked with an ``o'' (e.g. Z\_2132222 for X-ray source 4\_36).

We indicate in bold face the most likely counterpart. In all but one
case (\chandra source 3\_19, which is discussed in \S \ref{properties}), this is the brightest OGLE-II match, or, when
there is no OGLE-II match the brightest MCPS source.
 For sources with more than one stars with the same or very similar
magnitudes we indicate all of them in bold face. There are 3 such
cases (\chandra sources 4\_6, 7\_16, and 7\_17), for which
even if we apply additional criteria (see \S \ref{multiplediscussion})
we cannot identify the true
counterpart, and classify the sources.

From the total of 102 X-ray sources covered by the
OGLE-II fields, we found that 34\% have a unique candidate optical counterpart, 23\% have two, and 29\% have three or more matches. In total, we identified 229 optical matches for 87 X-ray sources.

Since the MCPS survey covers the entire area of the \chandra survey, we were able to search for counterparts for all X-ray sources apart from source 7\_12, which falls on the gap between two 
scans of the MCPS survey (we
identified a unique counterpart for this source from the OGLE-II
survey). We found that 45\% of the 152 X-ray sources have a unique
candidate optical counterpart, 13\% have two and 6\% have three or more, resulting in a
total of 138 matches. Fifty five X-ray
sources do not have any match in the MCPS
catalog; from them, 22 have match(es) in the deeper OGLE-II catalog.

In Table \ref{tableresults} we summarize the above results. In Column
(1) we give the field ID, while in Column (2) we indicate the complete
or partial
coverage of these fields by the OGLE-II survey. In Column (3) we specify the
optical catalog used for the cross correlation (O for OGLE-II, Z for MCPS and ``combined" for the results of the matched lists of OGLE-II and MCPS sources presented in Tables \ref{field3}-\ref{field7}). In Columns (4), (5), (6) and (7) we give the number of sources with no counterparts, and with 1, 2 or more matches, respectively. Only in four cases (sources 4\_23, 4\_36, 5\_34 and 7\_5) we found two separate OGLE-II and MCPS matches within the search radius (derived as described in \S \ref{OGLEMCPS}). The coordinates and photometry of these stars indicate that they are distinct objects, and thus we consider them as two individual sources.  Because, in many cases, the MCPS sources are resolved into two or more OGLE-II stars, and because of the cases of non-matched associations just mentioned, the number of counterparts in the ``combined'' entry is always smaller than the numbers corresponding to individual catalogs.
 
In total, out of the 153 \chandra sources, 
$34\%$ have unique candidate optical counterparts, $22\%$ have two, and $23\%$ have three or more
optical matches. There are also 33 X-ray sources with no matches.

\subsection{Other bibliographic sources}\label{otherbiblio}

For all detected sources we performed an extensive (but by no means complete) literature search for previously published identification and classifications. In Table \ref{tableotherinfo} we present (when available) information
on the X-ray sources and/or the optical sources detected within
$1.5\arcsec$, based on a search of the ADS and SIMBAD Astronomical Databases (see also \S \ref{opticalcat}). In particular, in Column (1) we give the X-ray source ID (same as in Tables \ref{field3}-\ref{field7}). In columns (2) and (4) we give any reported classification for the
X-ray source\footnotemark \footnotetext{Comparison with
previously published X-ray catalogs is based on positional coincidence
and the spectral and timing properties of the sources (Zezas \etal
2008, in prep.)} and previously identified optical counterparts (the
relevant references are given in Columns (3) and (5)). In the last 2 columns we present any other sources
encompassed by the minimum search radius of $1.5\arcsec$ around each
X-ray source. In Column (6) we present the source ID, and in Column (7) its reference.

We note that we have made no attempt to match the sources given in
Columns (2), (4), and (6) since they are produced from several
different surveys, with different areal coverage and positional accuracy. However, for completeness we present them all.

Summarizing the above results, within the search radius, we find 2 AGN (Dobrzycki \etal
2003a, 2003b; [DMS03] and [DSM03], respectively), 1 foreground star
(Sasaki \etal 2000; [SHP00]), 16 confirmed pulsars (e.g. Edge \etal 2004, [EHI04]; Haberl \etal 2004,
[HPS04]), 1 eclipsing binary (Wyrzykowski
\etal 2004; [WUK04]), 1 planetary nebula (Murphy \& Bessell
2000; [MB00]), 4 variable stars from the OGLE-II survey (one of which
is also a candidate HMXB; Ita \etal 2004, Zebrun \etal 2001), 5 stars
with spectral classification (using the 2dF spectrograph on AAT; 3 by
Evans \etal 2004, and 2 by Antoniou \etal 2008), 1 B[e] star (Massey \& Duffy 2001; reported here for the first time as single counterpart of an XRB), 11 optical matches within OB stellar associations (Oey \etal 2004), and 10 candidate Be stars (Mennickent \etal 2002; see \S \ref{properties} for more details).

\section{Finding charts}\label{charts}

\begin{figure} 
\begin{center}
\includegraphics[width=.45\textwidth]{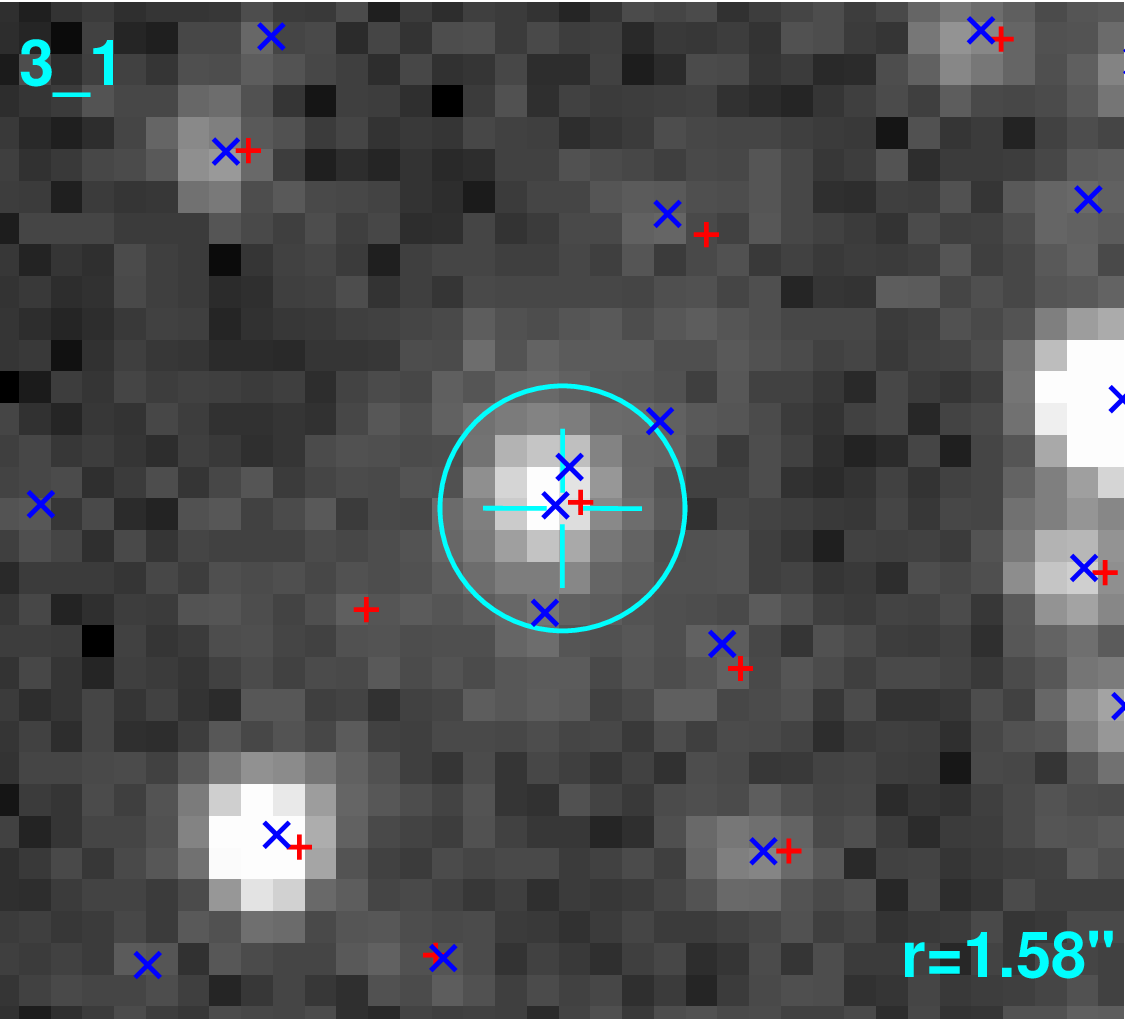} 
\includegraphics[width=.45\textwidth]{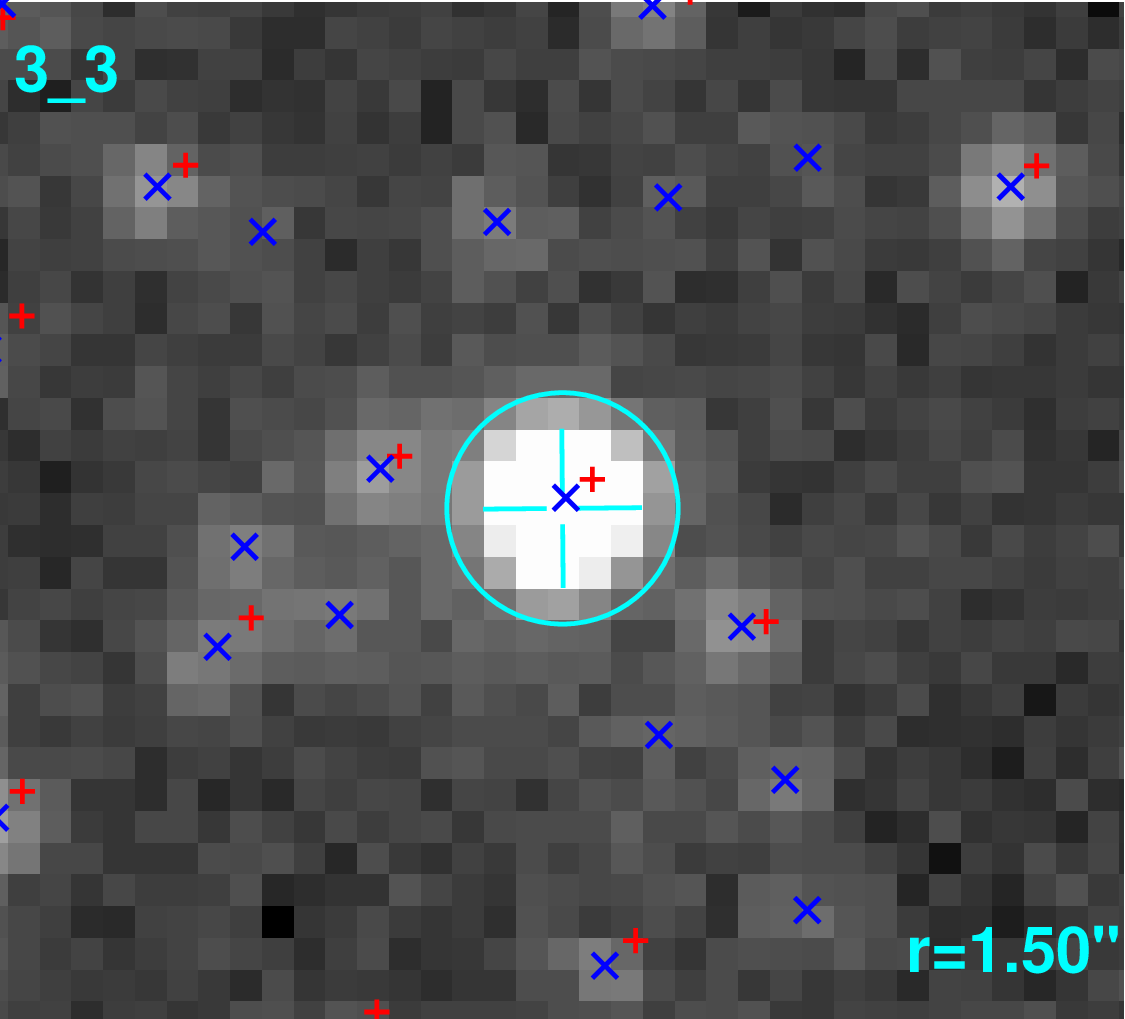} 
\includegraphics[width=.45\textwidth]{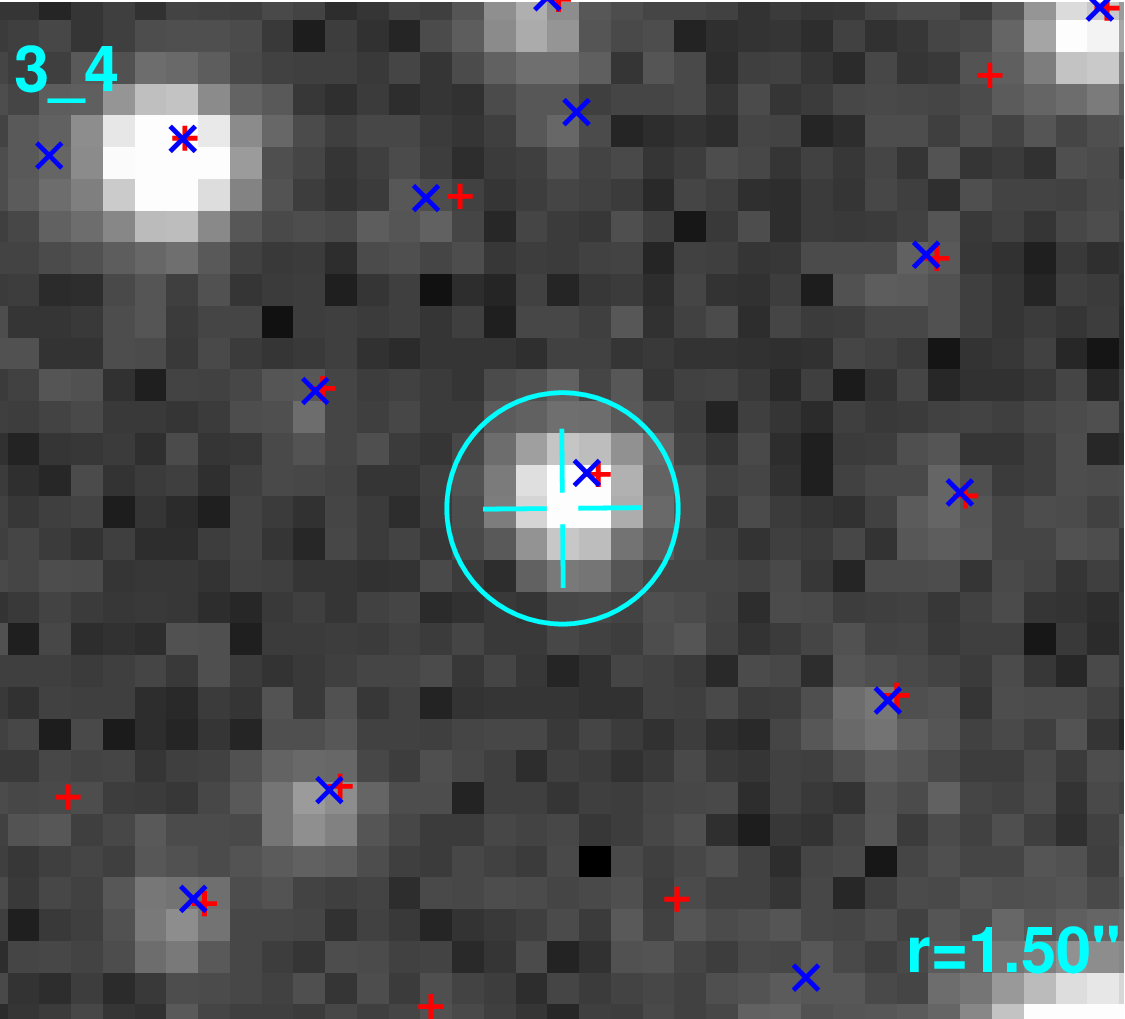} 
\includegraphics[width=.45\textwidth]{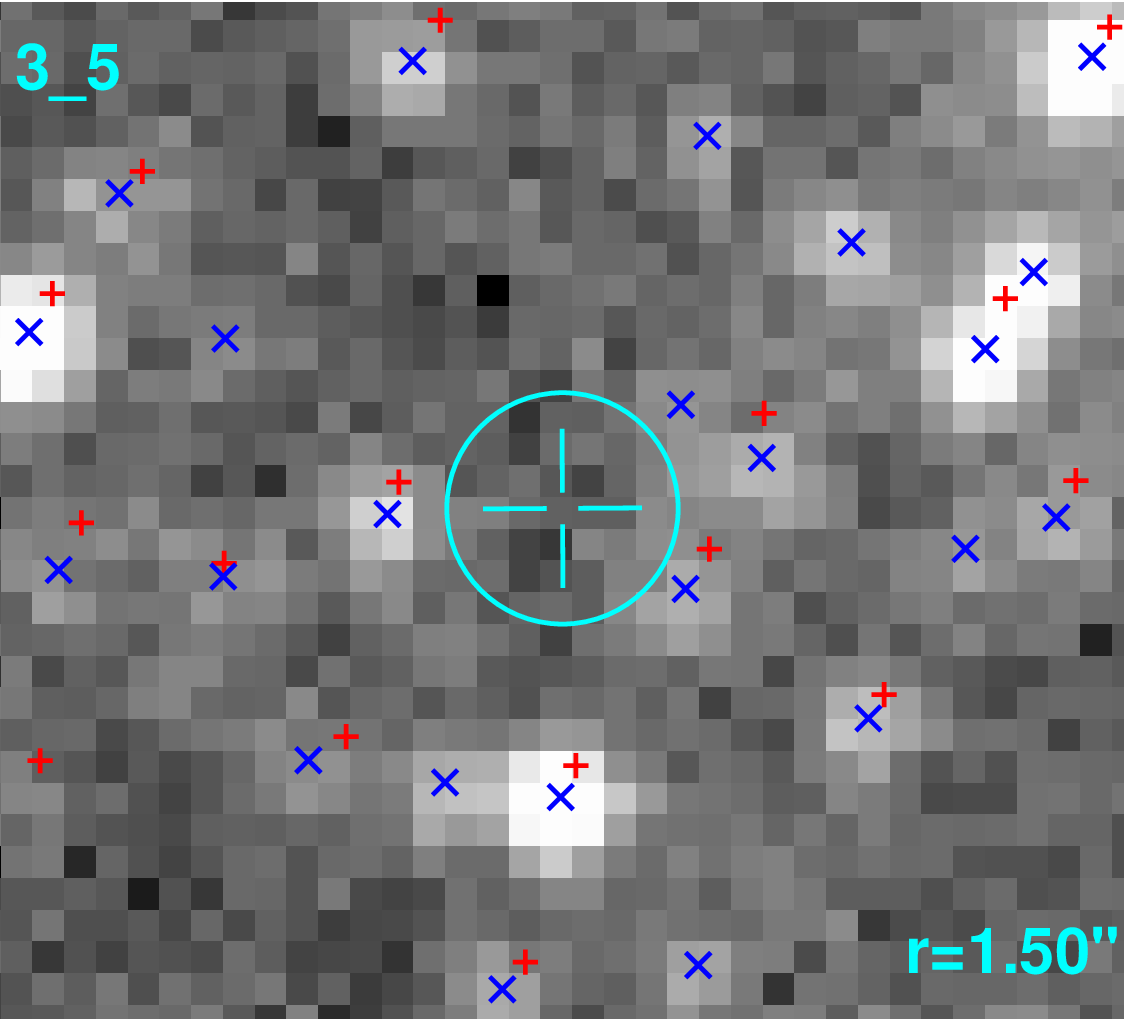}
\end{center}
\caption{OGLE-II {\it I}-band images for Chandra sources in field
  3. North is up and East to the right. $\times$ symbols are used for OGLE-II
  sources and $+$ for MCPS sources.\label{fcfield3} {\it This is an electronic Figure.}}
\end{figure}

\begin{figure} 
\centering
\includegraphics[width=.45\textwidth]{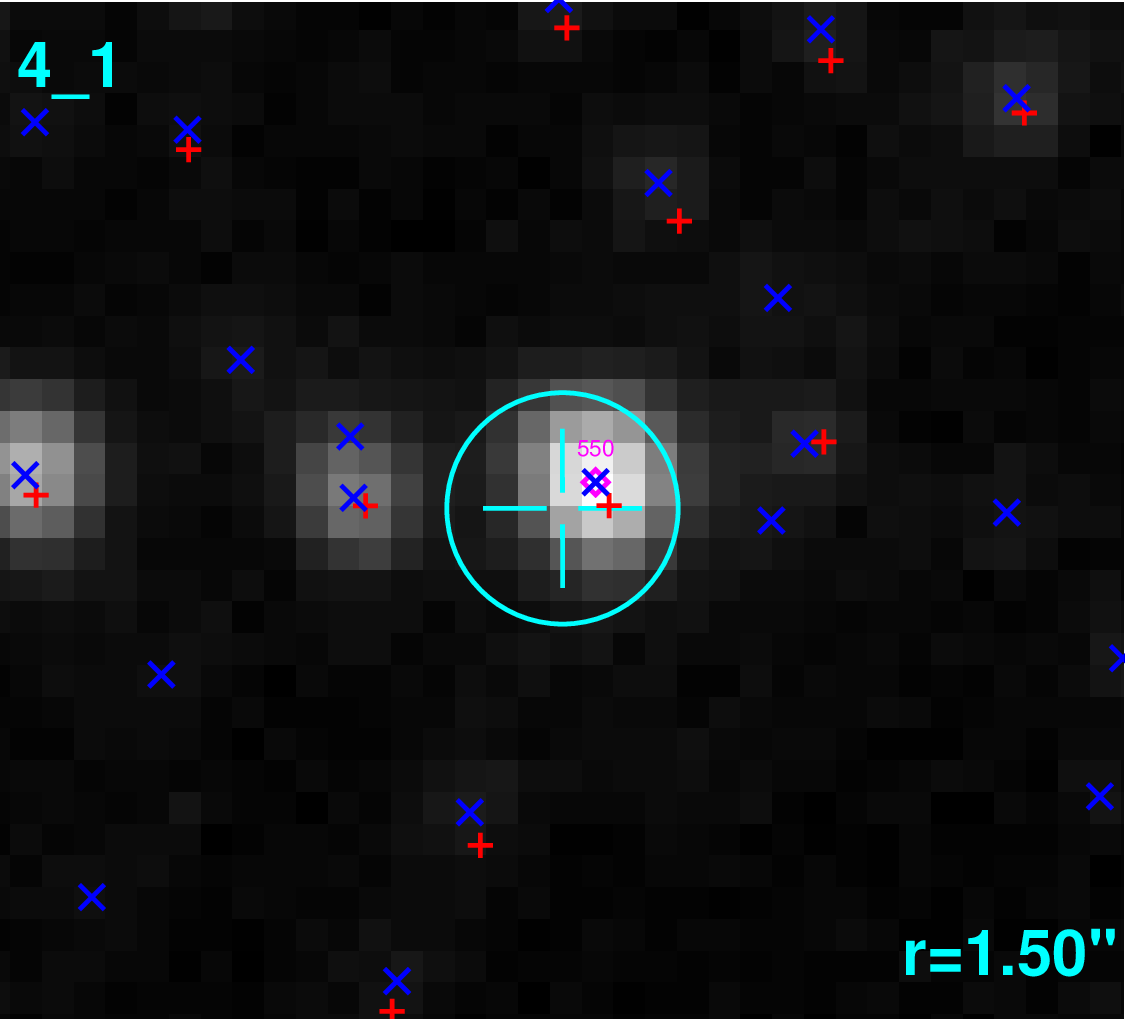} 
\includegraphics[width=.45\textwidth]{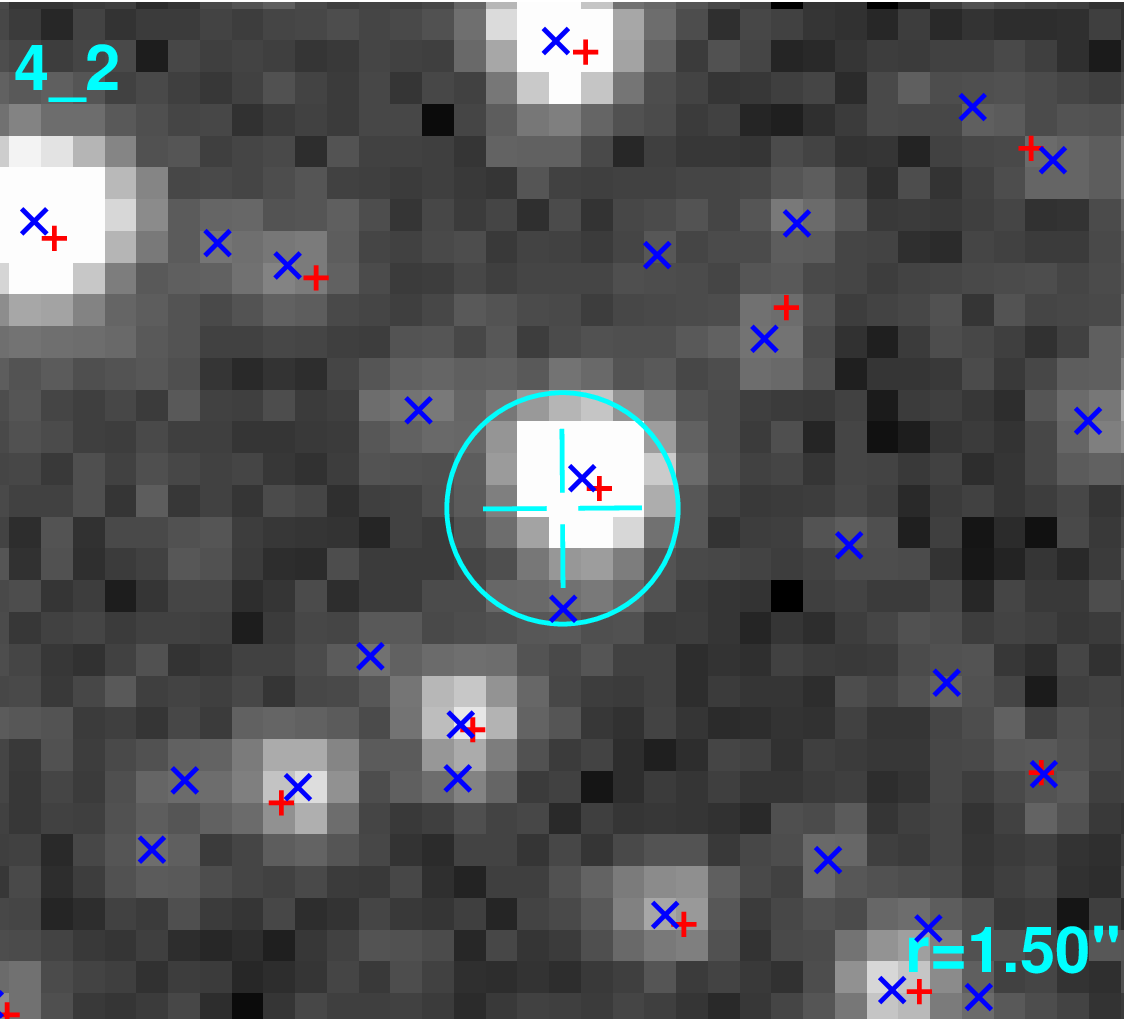} 
\includegraphics[width=.45\textwidth]{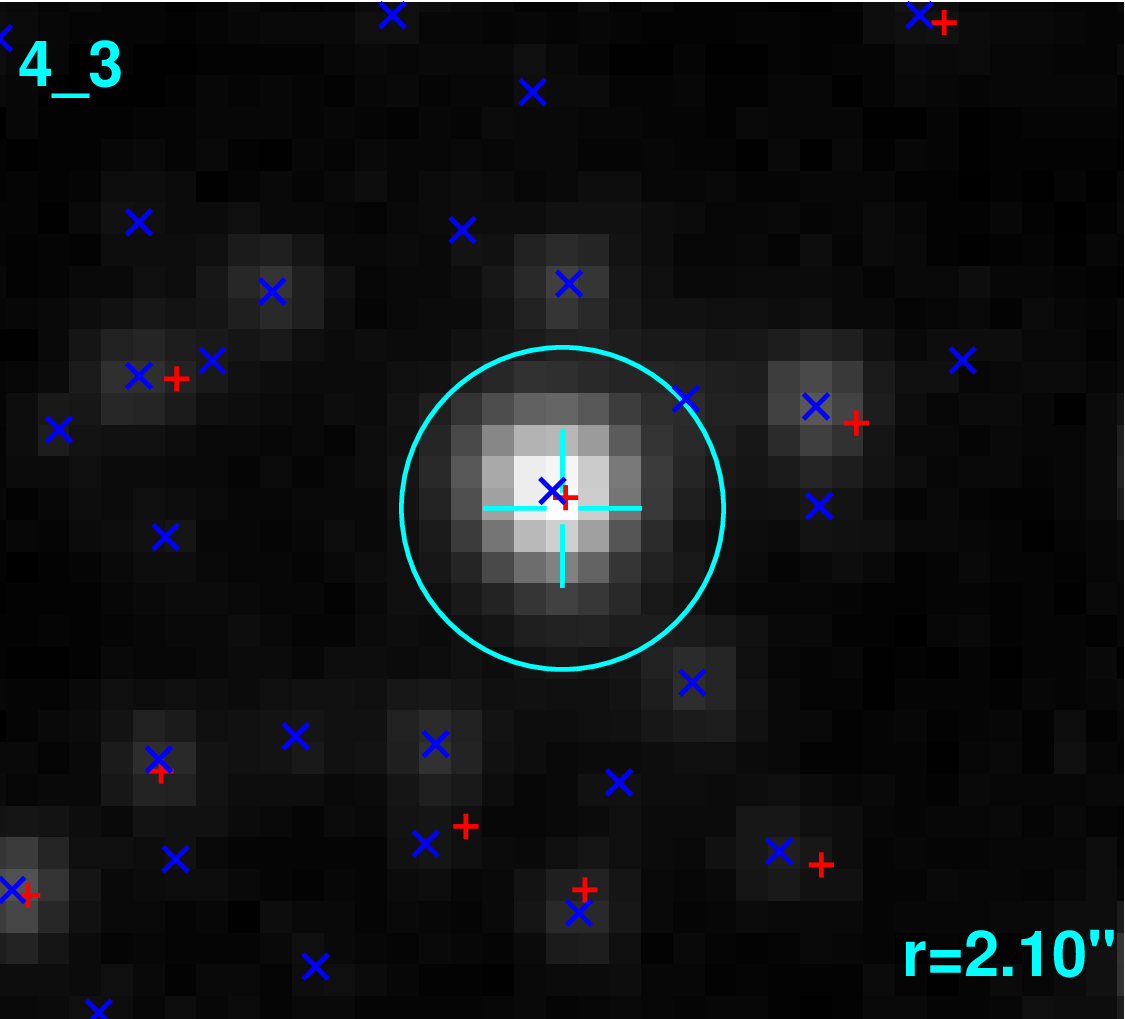} 
\includegraphics[width=.45\textwidth]{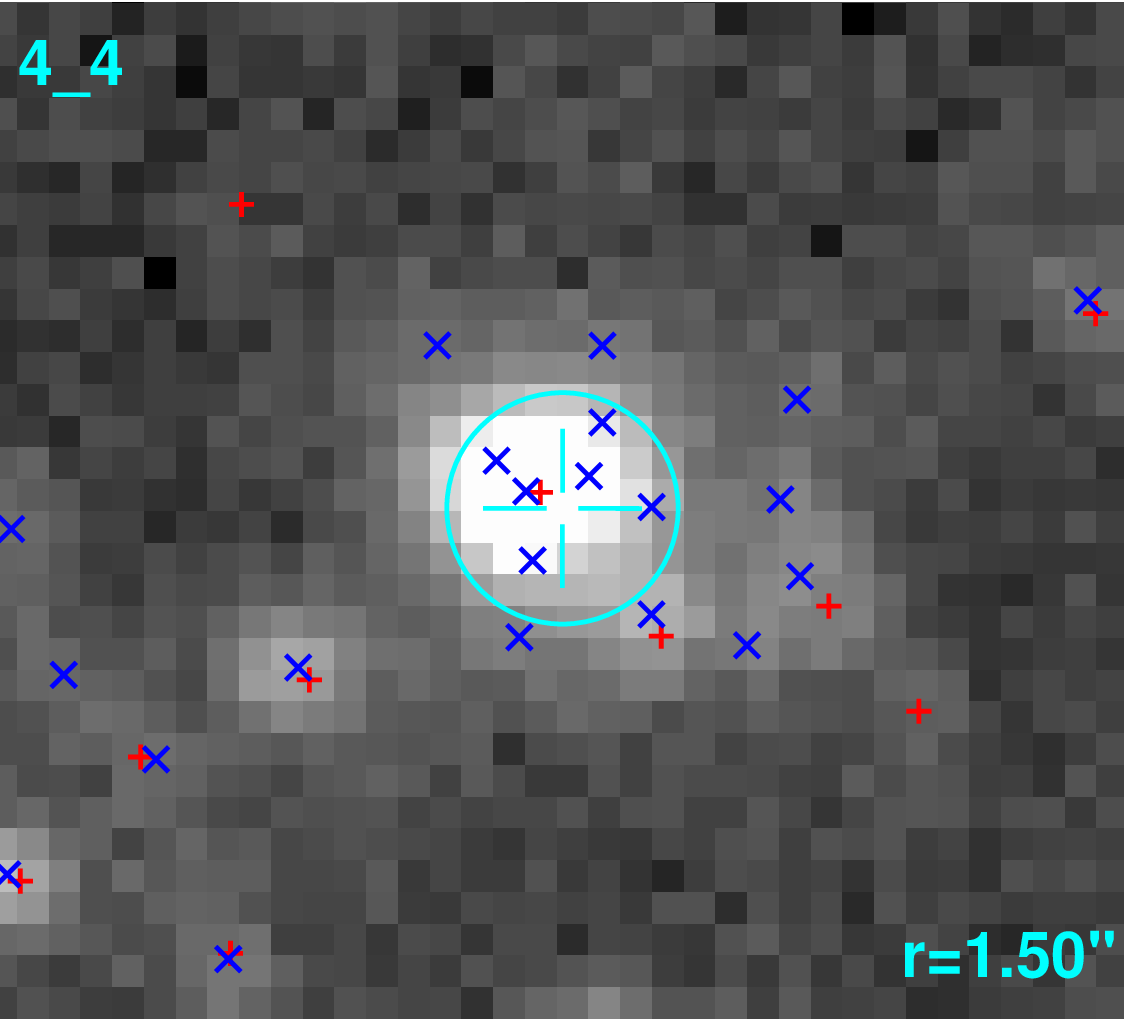} 
\caption{OGLE-II {\it I}-band images for Chandra sources in field
  4. North is up and East to the right. $\times$ symbols are used for OGLE-II
  sources and $+$ for MCPS sources.\label{fcfield4} {\it This is an electronic Figure.}}
\end{figure}

\begin{figure} 
\centering
\includegraphics[width=.45\textwidth]{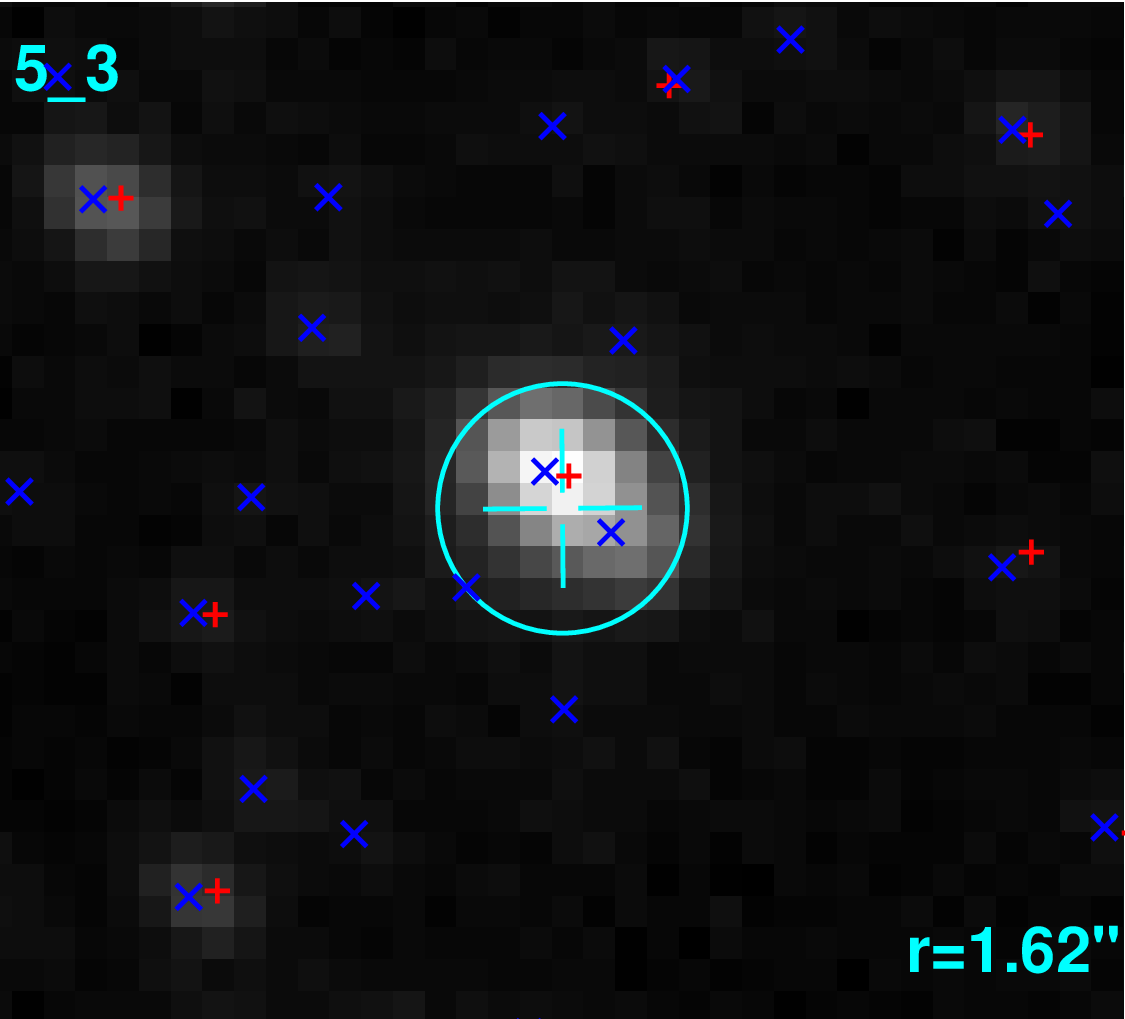} 
\includegraphics[width=.45\textwidth]{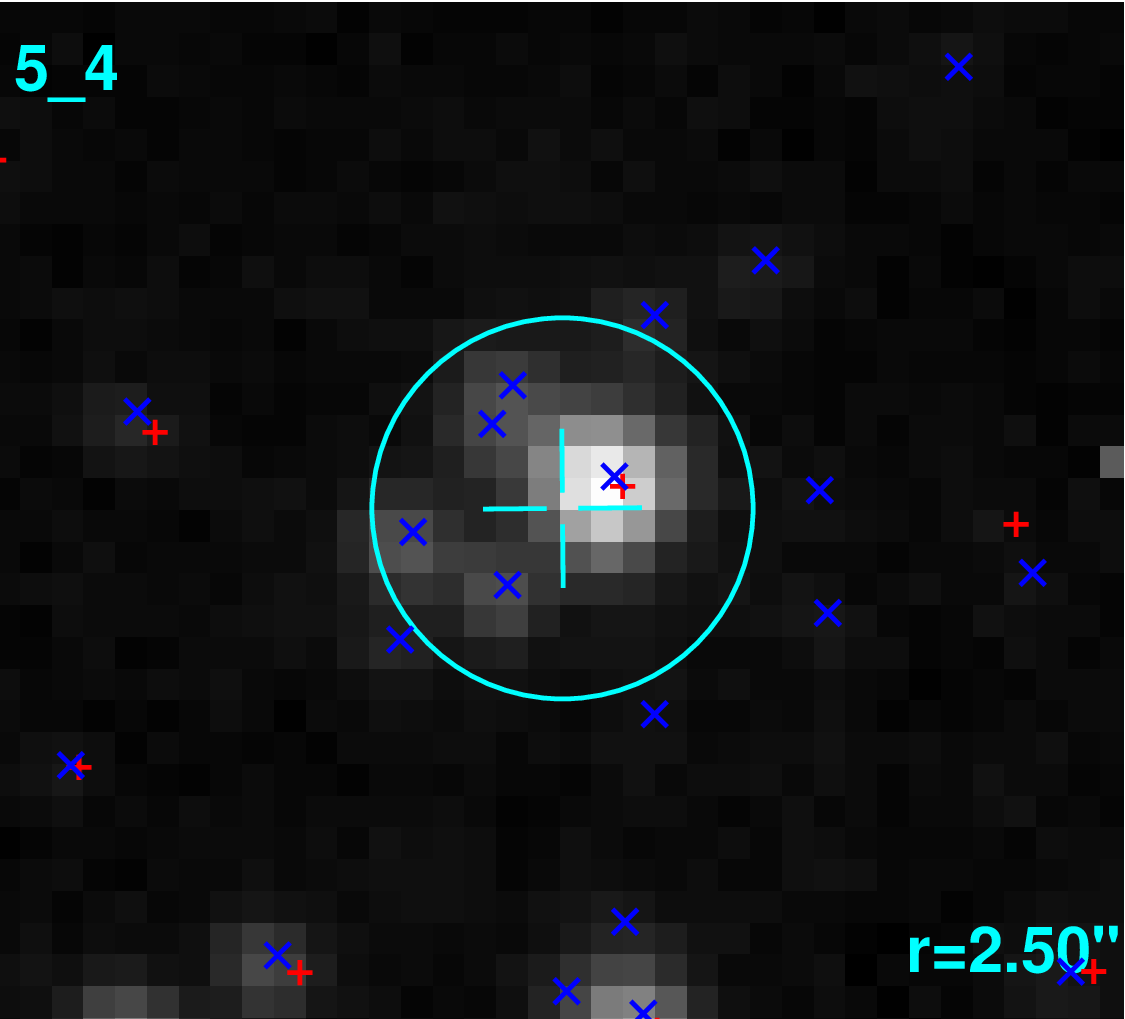} 
\includegraphics[width=.45\textwidth]{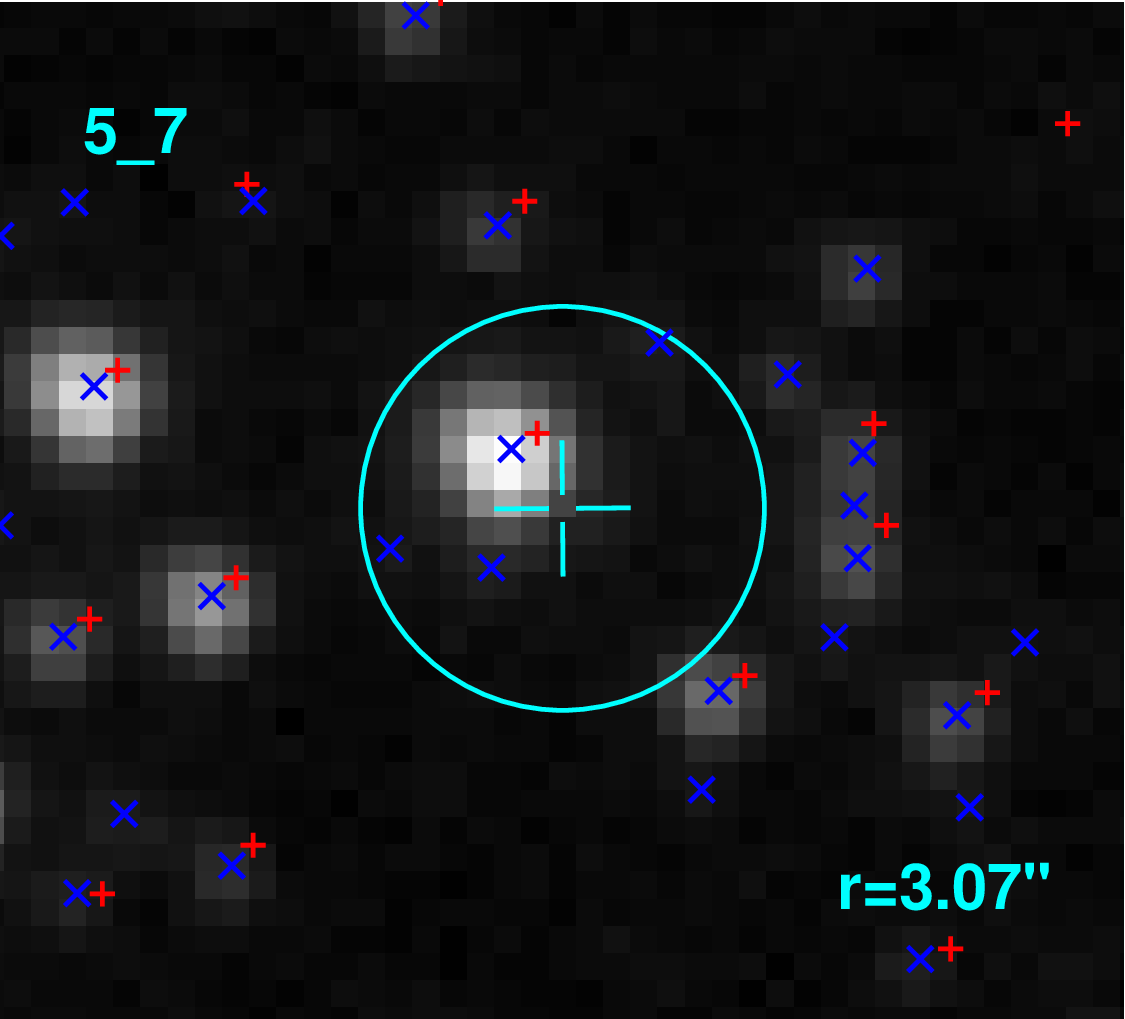} 
\includegraphics[width=.45\textwidth]{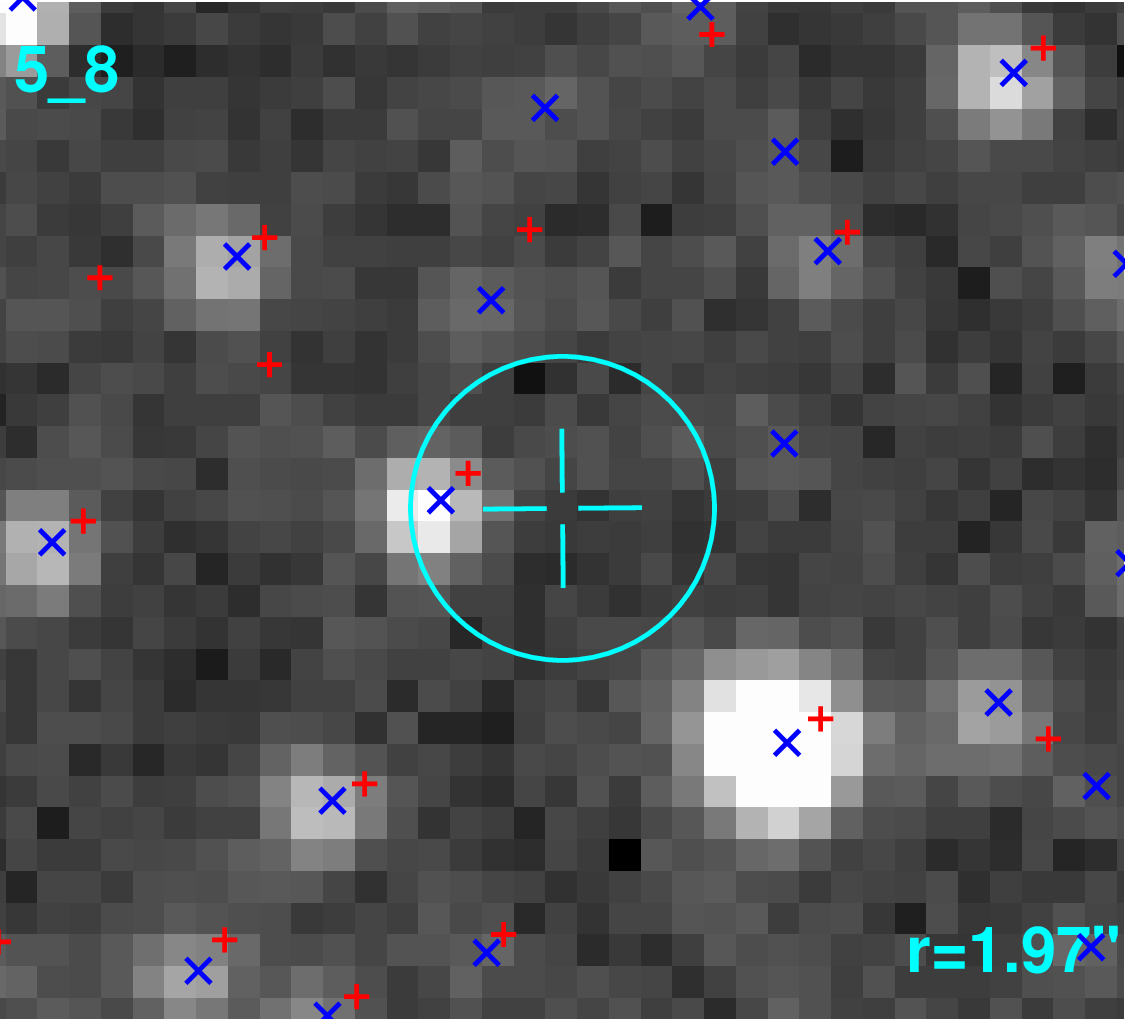} 
\caption{OGLE-II {\it I}-band images for Chandra sources in field
  5. North is up and East to the right. $\times$ symbols are used for OGLE-II
  sources and $+$ for MCPS sources.\label{fcfield5} {\it This is an electronic Figure.}} 
\end{figure}

\begin{figure} 
\centering
\includegraphics[width=.45\textwidth]{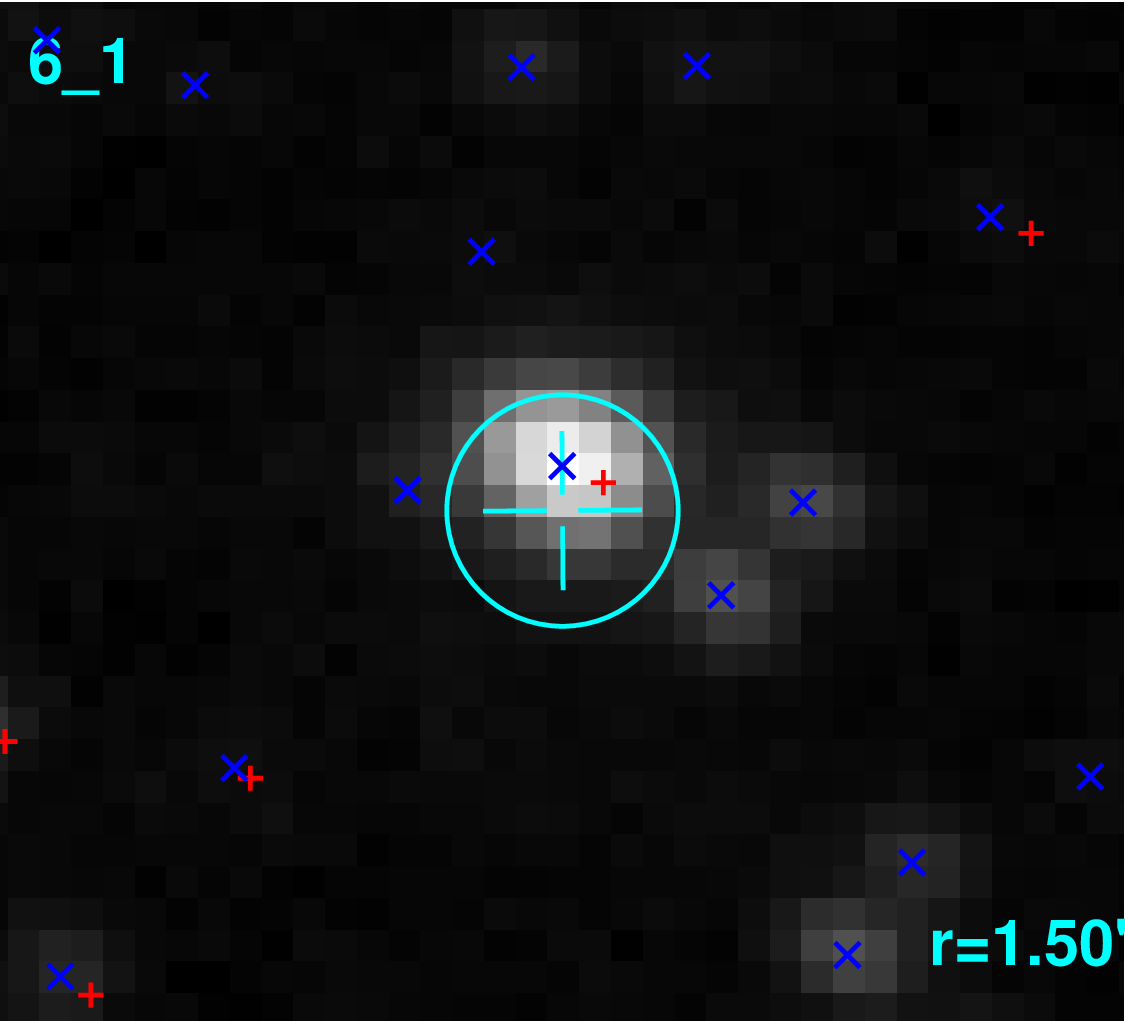} 
\includegraphics[width=.45\textwidth]{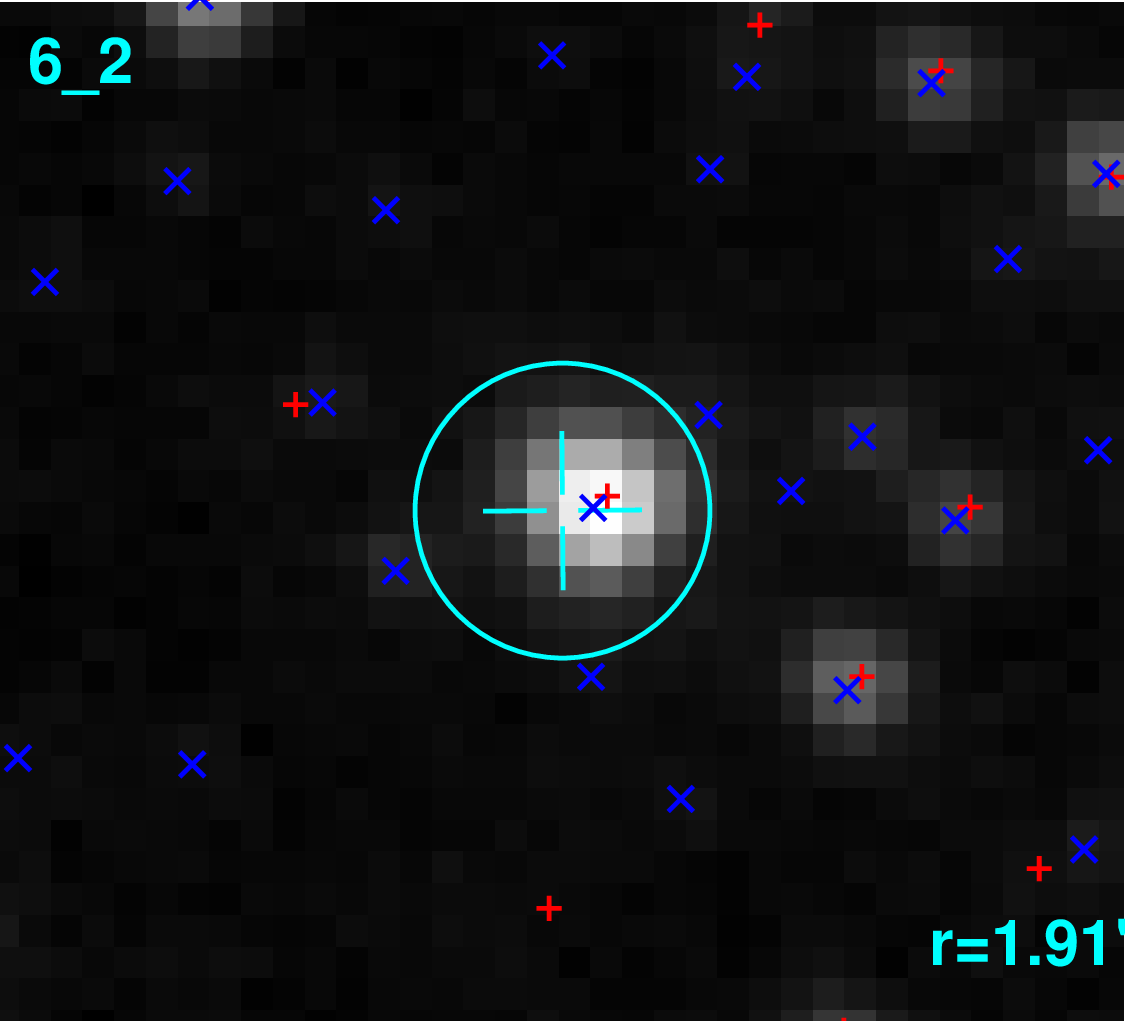} 
\includegraphics[width=.45\textwidth]{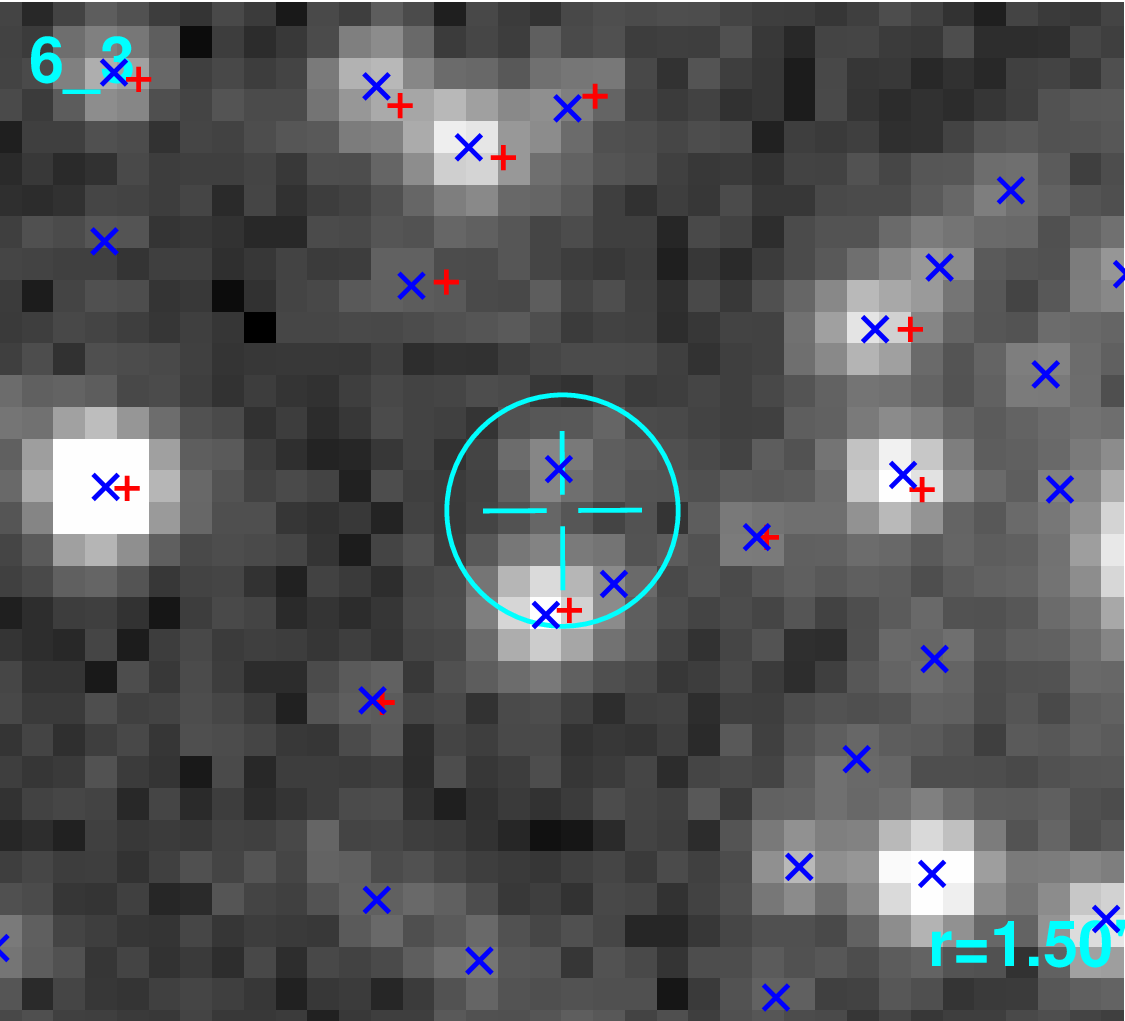} 
\includegraphics[width=.45\textwidth]{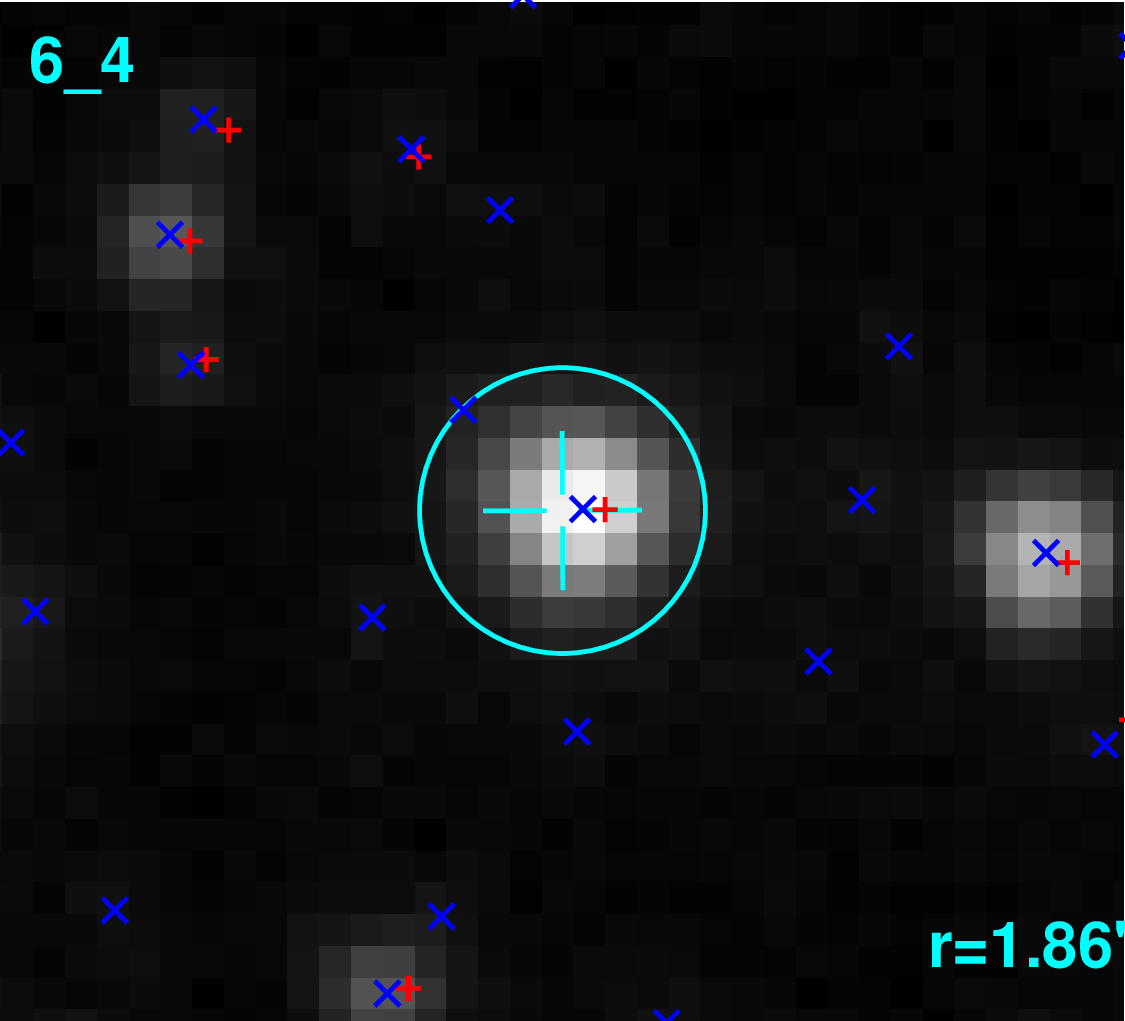} 
\caption{OGLE-II {\it I}-band images for Chandra sources in field
  6. North is up and East to the right. $\times$ symbols are used for OGLE-II
  sources and $+$ for MCPS sources.\label{fcfield6} {\it This is an electronic Figure.}} 
\end{figure}

\begin{figure} 
\centering
\includegraphics[width=.45\textwidth]{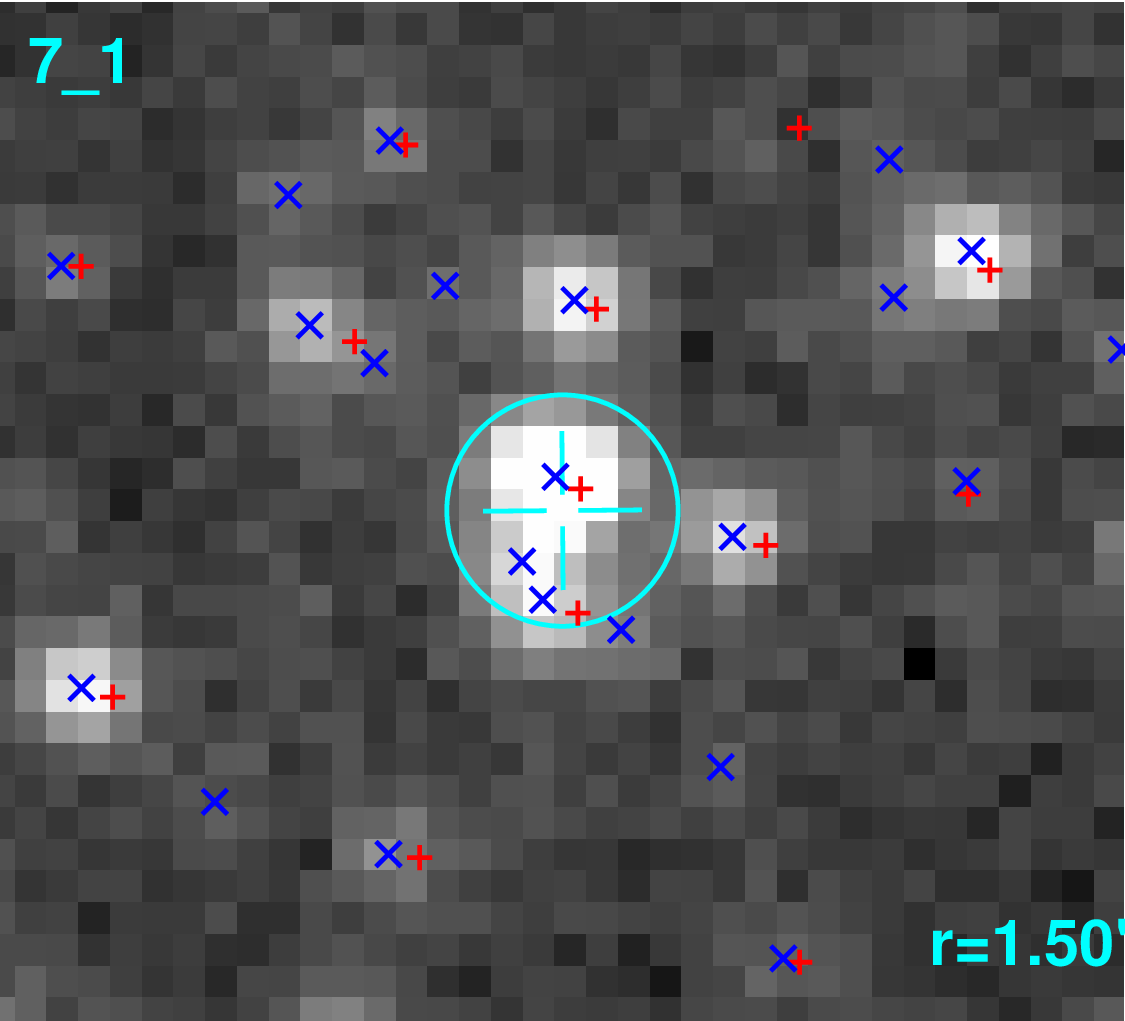} 
\includegraphics[width=.45\textwidth]{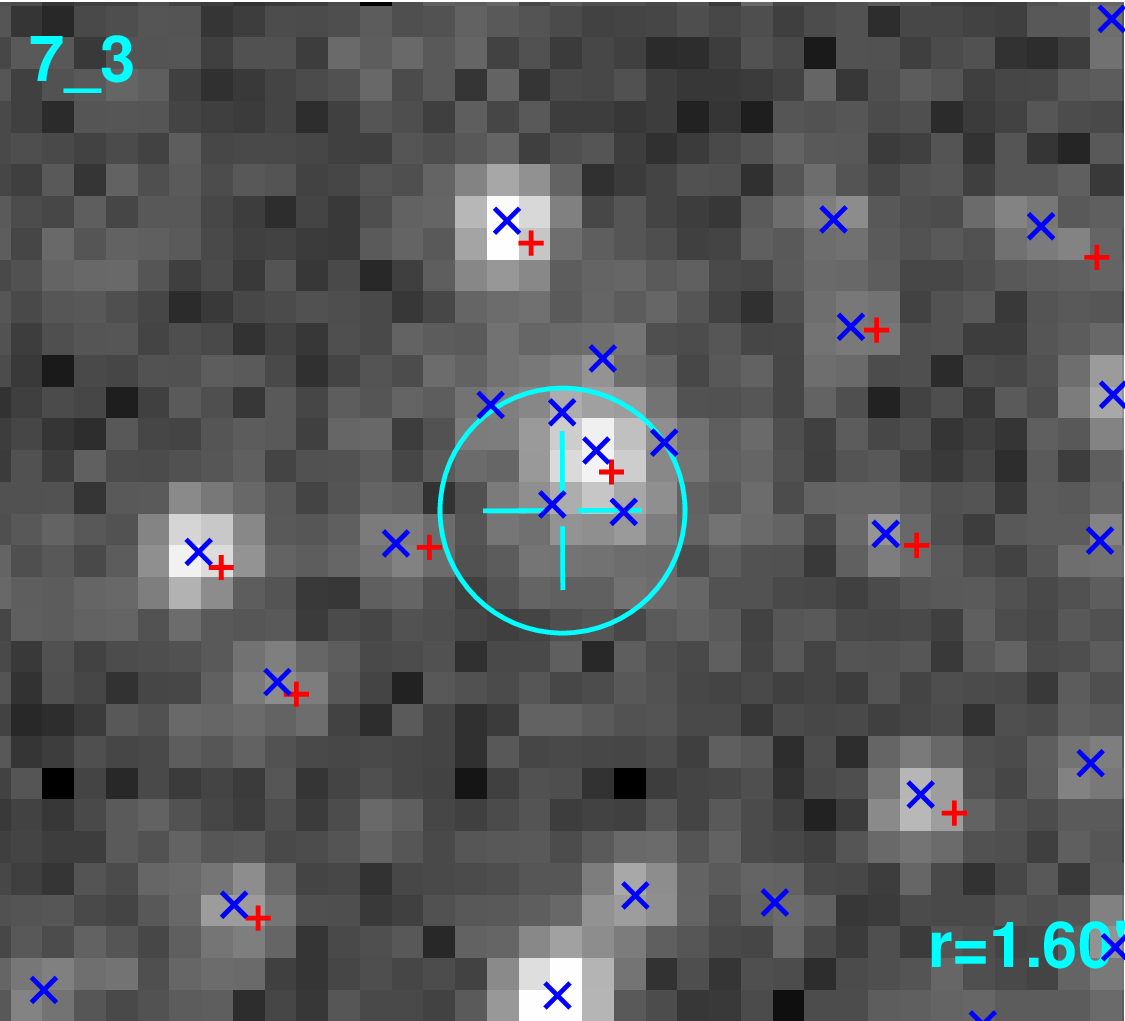} 
\includegraphics[width=.45\textwidth]{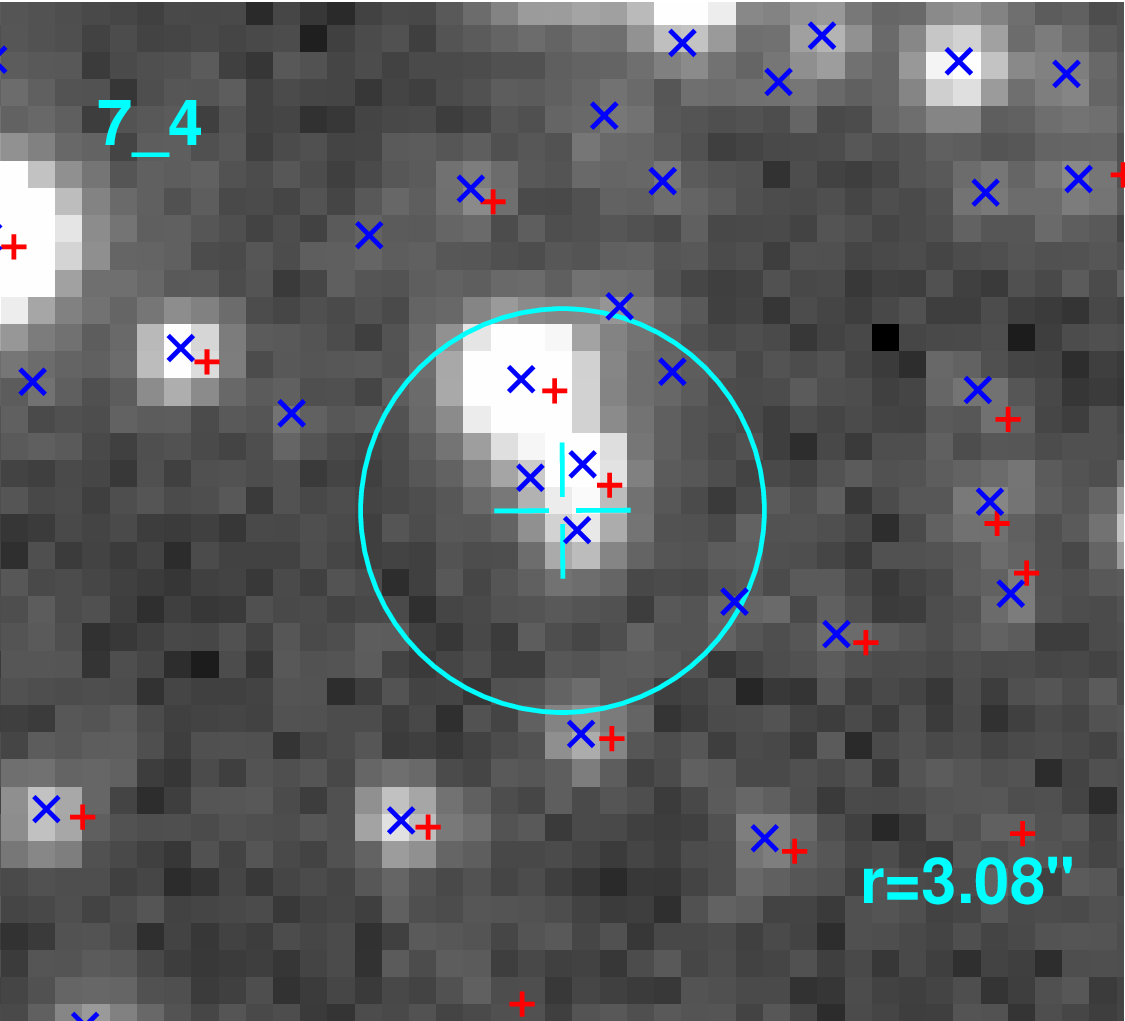} 
\includegraphics[width=.45\textwidth]{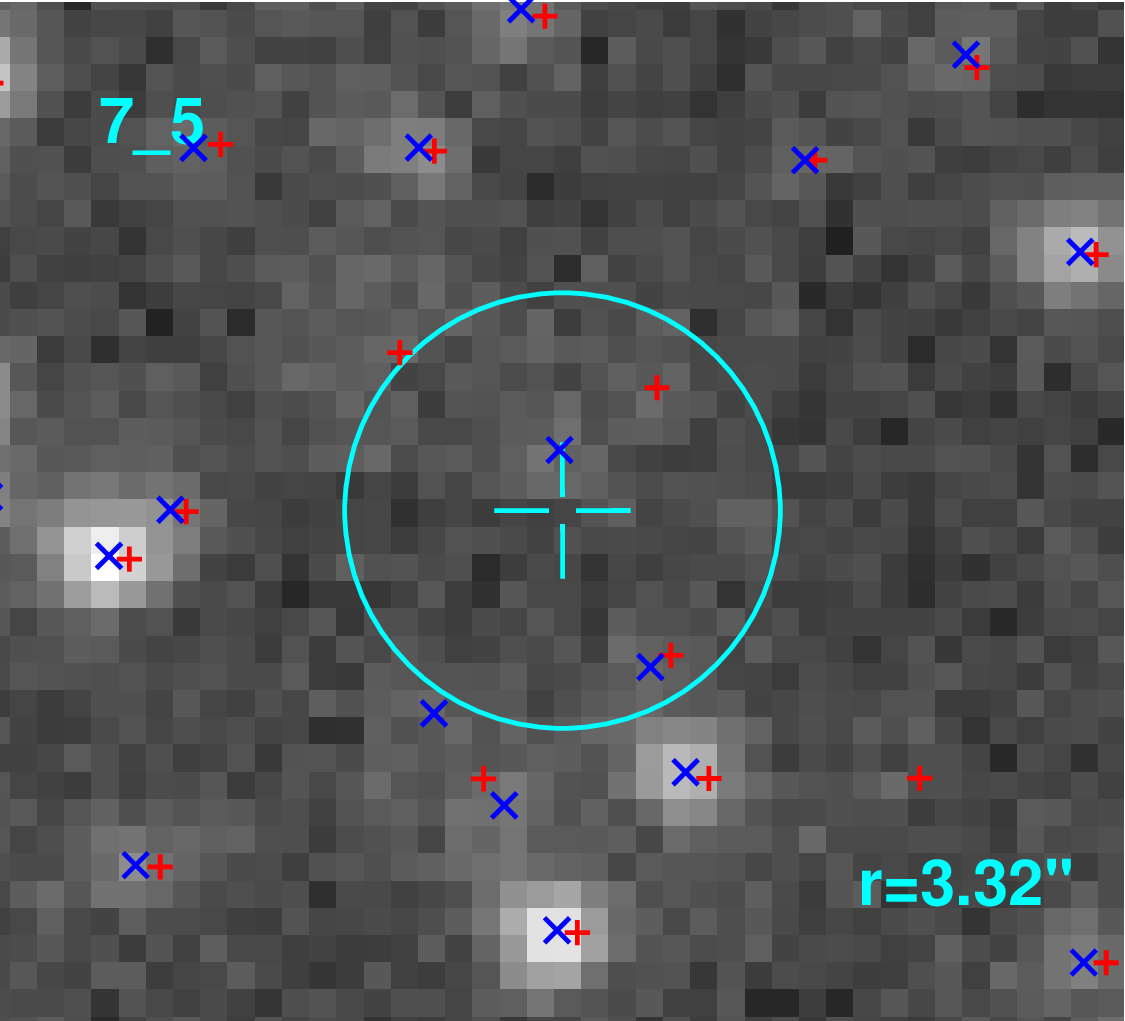} 
\caption{OGLE-II {\it I}-band images for Chandra sources in field
  7. North is up and East to the right. $\times$ symbols are used for OGLE-II
  sources and $+$ for MCPS sources.\label{fcfield7} {\it This is an electronic Figure.}} 
\end{figure}

For the \chandra sources which are covered by the OGLE-II
survey, we created optical finding charts, presented in
the on-line Figures \ref{fcfield3} - \ref{fcfield7}. These charts are created from the {\it I}\,
band OGLE-II fields (Udalski \etal 1998a), they are $\sim15\arcsec\times13\arcsec$ in size and they are centered on the position of the X-ray source, which
is indicated by a cross-hair (each bar of the cross-hair 
is $0.8\arcsec$ long). The positional uncertainty of the finding charts is $\sim0.7\arcsec$, based on the astrometry of the OGLE-II images. The search radius used
in the cross-correlation is shown by a circle centered on the X-ray source, and is also given (in arcseconds) at the bottom right corner
of each chart. All optical sources from the OGLE-II survey 
are indicated with an X symbol, while those from the MCPS survey are
indicated with a cross. In addition, we present with a diamond the eclipsing binary star
identified for \chandra source 4\_1 (source No.550 in Udalski \etal 1998b). For the X-ray sources which fall on two neighboring OGLE-II fields we only show the finding chart with the largest coverage. We note here that we
present finding charts only for those sources with OGLE-II coverage,
as the MCPS images are not publicly available.

\section{Photometric data}\label{photometry}

The photometric parameters of the optical matches of the X-ray sources in the OGLE-II and MCPS catalogs are from the second OGLE
release (Udalski \etal 1998a), and the catalog of
Zaritsky \etal (2002), respectively. These parameters are presented in
Columns (7) to (12) of Tables \ref{field3}-\ref{field7} (see \S \ref{cc}).

Although there are no known significant photometric offsets between the OGLE-II and MCPS catalogs (see \S \ref{MCPS}), we look for photometric discrepancies specific to the objects found in our cross correlation by comparing the {\it V}\, and {\it B}\,
band magnitudes, and {\it B-V}\, colors of the 18 stars identified as unique counterparts in both OGLE-II and
MCPS. We find $\delta$$V_{\rm{(OGLEII - MCPS)}} = 0.00\pm0.30$ mag,
$\delta$$B_{\rm{(OGLEII - MCPS)}} = - 0.06\pm0.30$ mag and
$\delta$$BV_{\rm{(OGLEII - MCPS)}} = -0.06\pm0.27$ mag (error is the standard deviation). The zero-point differences are comparable to the corresponding errors quoted in the two catalogs. The relatively large scatter (0.30) is mainly caused by the large photometric errors of faint sources, close to the detection limit (particularly of the MCPS catalog), and partly by actual source variability.

\section{Color-Magnitude Diagram}\label{CMDs}

CMDs and 2-CDs are a standard
tool for the characterization of stars (e.g. Johnson 1966). In this study, we use the $V$, $B-V$ CMD in order to classify the X-ray sources. An important parameter in CMD studies is the extinction correction. Several different values have been proposed for the SMC reddening. For this work, we use reddening
values based on the red clump stars (Udalski \etal 1999) in regions as close as possible to the \chandra
fields. The mean color excess is $E(B-V)=0.09\pm0.02$ mag. Similar results ($E(B-V)\sim0.09\pm0.07$) have been obtained by
the spectroscopic study of Massey \etal (1995). Larsen \etal (2000)
derived somewhat higher values of ($E(B-V)\sim0.21\pm0.10$ intrinsic
to the SMC and $\sim0.07\pm0.02$ due to the foreground dust), although
generally compatible with the previously mentioned values, within the
combined errors. In conclusion, we adopt the value of $E(B-V)=0.09$ for the entire area covered by the \chandra fields.

To derive the reddening correction for $V$, $A_{V}$ ($=R_{V}E(B-V)$),
we use the standard Galactic value of \Rv\, $=3.24$ (Schlegel \etal 1998), which was also used in previous
published photometric catalogs of the SMC. A lower value of $R_{V}=2.74\pm0.13$, which has been recently proposed for the SMC (Gordon \etal 2003), would result in a negligible difference of 0.04 mag in $A_{V}$. The adopted \Rv\,= 3.24, gives a mean reddening value for the 5
OGLE-II fields of \Av = 0.29 mag. Based on the
extinction curve of Cardelli \etal (1989), we estimate
\Ab\, and \Au\, of 0.39 mag and 0.46 mag, respectively. We correct the original data presented in
Tables \ref{field3}-\ref{field7} for the extinction to the SMC ($\E(B-V)=0.09$ and \Av=0.29)
and we use these corrected numbers for all the plots described below. 

Throughout this study we adopt a distance modulus of
\mM$=18.89\pm0.11$
mag (derived from eclipsing O, B-type binaries in the SMC; Harries
\etal 2003). The small depth of the young populations ($<$10 kpc; e.g. Crowl \etal 2001, Harries \etal 2003)
results in a difference in the above distance modulus of 
$<0.2$ mag, while the corresponding difference in the X-ray luminosity is $\sim5.9\times10^{34}$\ergs.

\begin{figure}
\centering
\rotatebox{0}{\includegraphics[height=12.5cm]{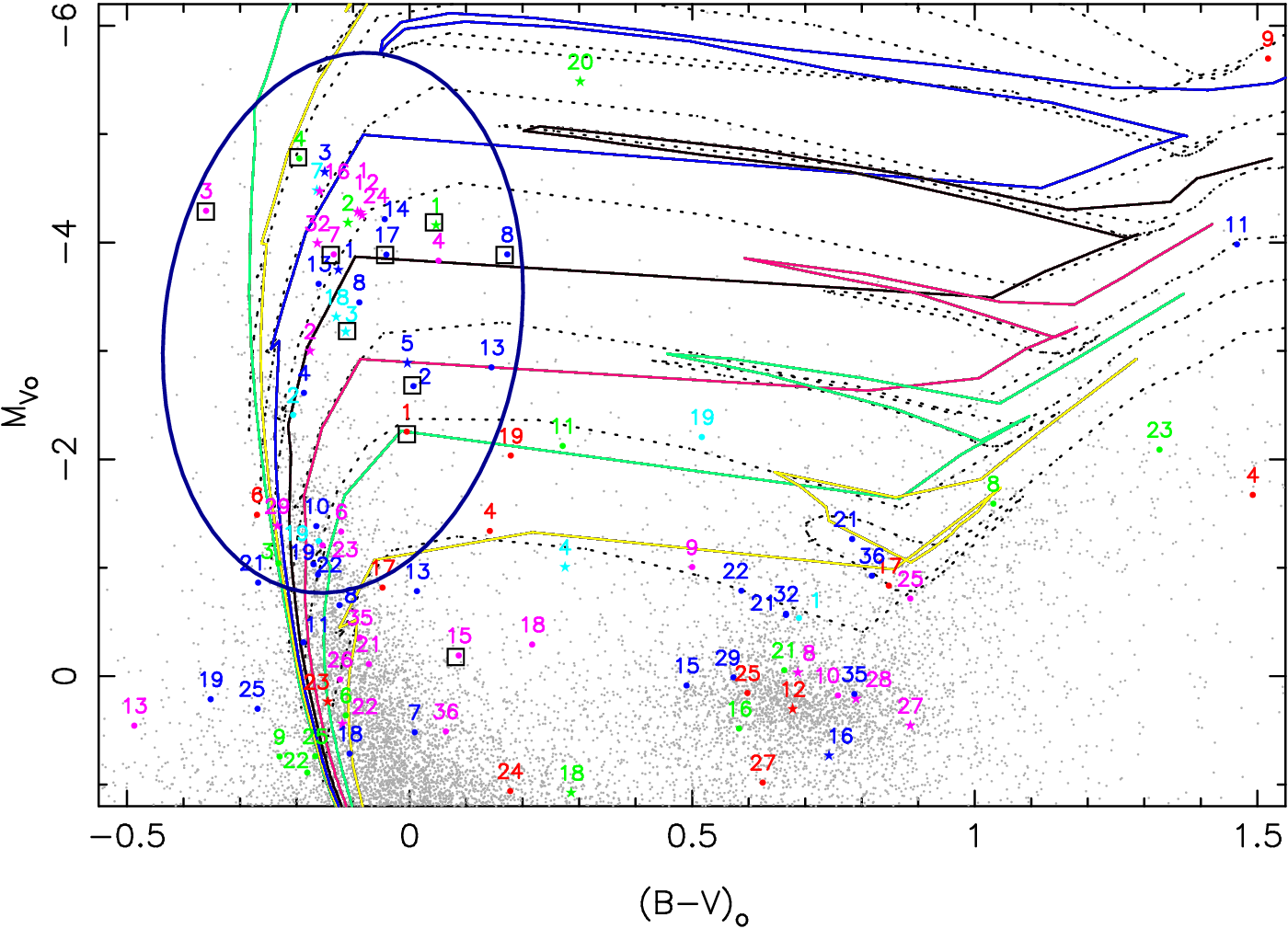}}
\caption{CMD of all single and the brightest of multiple matches of
 our \chandra sources. We present the $M_{{V}_{o}}$ vs. $(B-V)_{o}$ CMD of all single matches (asterisk) and the brightest source of those with multiple matches (open circle) of our \chandra sources (which are indicated in bold face in Tables \ref{field3}-\ref{field7}). The optical sources are color coded as, \chandra field 3: cyan; 4: 
 blue; 5: magenta; 6: green; 7: red; while the OGLE-II stars that lie
in our \chandra field 4 are indicated in gray small dots. Part of the MS, the
red giant branch and the red clump loci are clearly visible. With black
squares we present the 10 candidate Be stars of Mennickent \etal
(2002) that we have identified as optical counterparts. Overlaid are the isochrones (solid
lines) and
stellar evolutionary tracks (dotted lines) from Geneva database
(Lejeune \& Schaerer 2001) for ages of 8.7 Myr, 15.5 Myr, 27.5 Myr, 49.0
Myr, 87.1 Myr, 154.9 Myr and 275.4 Myr and
initial stellar masses of 12\msun, 9\msun, 7\msun, 5\msun, 4\msun\, and
3\msun\, stars (from top to bottom).}\label{VBV}
\end{figure}

In Figure \ref{VBV} we present the extinction corrected $M_{{V}_{o}}$ vs. $(B-V)_{o}$ CMD
diagram (absolute $V$ magnitude vs. $B-V$ color) for all single and the brightest of multiple matches of
our sources (which are indicated in bold face in Tables
\ref{field3}-\ref{field7}). Whenever both OGLE-II and MCPS photometric data
exist we use the OGLE-II data. The identifier of each point is the X-ray source number (see \S \ref{OGLEMCPS}), while we plot in the
same color sources from the same \chandra field. On the same diagram
we overlay with small dots the stars from the OGLE-II catalog in the area
common with the \chandra field 4 (after having corrected the magnitudes and colors for reddening and
distance as described above).

 In order to identify on the CMD the locus of stars of different
spectral types, we use data from the 2dF spectroscopic survey of SMC
stars (the most extended such catalog available; Evans \etal 2004). This survey provides
spectroscopic classification for 4161 stars in the SMC\footnotemark,
\footnotetext{In this paper we only utilized photometric catalogs, as
we found only 3 matches of our \chandra sources with the 2dF catalog;
see below Table \ref{tableotherinfo}.} covering
spectral types from O to FG, and sampling the MS to
$\sim$mid-B. However, the Be stars are generally redder than B-type stars
(e.g. McSwain \& Gies 2005). Mennickent \etal (2002), studying the light curves of the OGLE-II data
for the SMC, identified $\sim1000$ candidate Be stars. We have identified
10 of our proposed counterparts with their type-4 stars (objects with light curves
similar to Galactic Be stars) from this catalog (shown in Figure \ref{VBV} as black squares). For the classification of the \chandra sources we used the
combined locus (blue curve) of early-type stars (O and B, including the
Be stars) identified in
the 2dF spectroscopic survey of Evans \etal (2004) and the
candidate Be stars of Mennickent \etal (2002). This choice is in
agreement with the $B-V$ color range of O and B spectral-type
stars from Massey (2002), and is confirmed by follow-up spectroscopy
of selected targets from our survey (Antoniou \etal 2008). 
Even a difference of $\sim$0.2 mag in the distance modulus (due to the
depth of the young populations discussed above) does not affect the
classification of a bright source. On the other hand, faint
counterparts do not
``contaminate'' the locus of OB stars even if we adopted the lower value
of the distance
modulus ($\sim18.7$ mag). 

For stars brighter than $M_{{V}_{o}}\sim-0.25$ the majority of the
candidate optical counterparts (either single or brightest of the
multiple) lie within the loci of OB and Be stars. Fainter
objects are mainly found in regions of higher spatial density, and
they populate the high density regions of the CMD (lower MS and red clump), clearly indicating high incidence of chance
coincidences in these dense regions. This issue is discussed in detail
in \S \ref{chance}.

Isochrones and stellar evolutionary tracks can give further insight on the properties of the companion star. In Figure \ref{VBV}, we overplot the isochrones from the Geneva database (Lejeune \&
Schaerer 2001)
for the metallicity of the young SMC stars ($\rm{[Fe/H]}=-0.68\pm0.13$ dex,
equivalent to $\rm{Z}=0.004$\footnotemark; \footnotetext{For the conversion, we use the relation
$\rm{[Fe/H]}\equiv \log(\rm{Z}/$\Zsun$)$, with \Zsun=$0.02$ for the
solar metallicity of $\rm{[Fe/H]}=0$; Russell \& Dopita (1992).}Luck \etal 1998), for
ages of 8.7 Myr to 275.4 Myr and initial stellar masses from 12\msun to
3\msun, respectively. The age of the OB stars, as they are defined
based on their locus in the CMD, is estimated to range from $\sim15.5$ to $\sim85$ Myr.

Given the large area covered by the \chandra survey, we expect a
significant contamination by Galactic foreground stars. Based on the tables of Ratnatunga \& Bahcall (1985), the fraction of Galactic stars at the locus of early-type stars is very small ($< 3.6\%$), and thus, 
the foreground contamination can be safely considered
negligible. Exception to the above is the region of the CMD between
$-6 < M_{{V}_{o}} < -4$ and $(B-V)_{o} > 0.75$, in which the contamination is more significant
($\sim$11\%). However, there are no counterparts identified in this
range (see Figure \ref{VBV}), except for
\chandra source 7\_9 that has been previously identified as a foreground star (Sasaki \etal 2000). These estimates are consistent with those of Massey (2002), although direct comparison is difficult because of the different detection limits of the two surveys.

\section{Chance coincidence probability}\label{chance}

To estimate the number of possible chance associations between the X-ray
sources and the stars in the OGLE-II and MCPS catalogs, we followed
a Monte-Carlo procedure similar to that in Zezas \etal (2002). We
simulated 1000 random samples of X-ray sources by applying to the
position of each source a random offset in R.A. and Dec.. The offsets were drawn from a uniform distribution, taking care that the new position is outside the search radius of each source but
within the boundaries of the OGLE-II fields (the latter constraint does
not apply to the MCPS catalog since in that case the overlap is complete). However the maximum offset was restricted so a source does not fall in a region of different stellar density. Each of these samples was cross-correlated with the optical catalogs in the same way as the observed data.

We calculated the chance coincidence probability for detecting {\it
  (a)} unique, and {\it (b)} any number of multiple optical matches
(one or more) for various search radii. We note that our minimum
  search radius is $1.5\arcsec$ (see \S \ref{OGLEMCPS}), while only 20\% of the \chandra sources have search radii larger than $2.5\arcsec$. In addition, we estimated the expected chance coincidence probability for matches with optical sources of different spectral types (O, B, and later, following Binney \& Merrifield 1998) depending on their position on the CMD: 

(i) MS and post-MS O-type stars (\BVo\,$\leq-0.31$ and \BVo\,$>-0.31$, respectively, for $M_{{V}_{o}}\leq-4.5$)

(ii) MS and post-MS B-type stars (\BVo\,$\leq-0.11$ and \BVo\,$>-0.11$, respectively, for $-4.5<M_{{V}_{o}}\leq-0.25$)

(iii) later type stars ($M_{{V}_{o}}>-0.25$, with no restriction in the color range), 

where $M_{{V}_{o}}$ denotes the absolute {\it V} magnitude. The photometric data used here (both in $B$ and $V$ band) are corrected for extinction (see \S \ref{CMDs}).

\begin{figure}
\centering
\rotatebox{270}{\includegraphics[height=14.0cm]{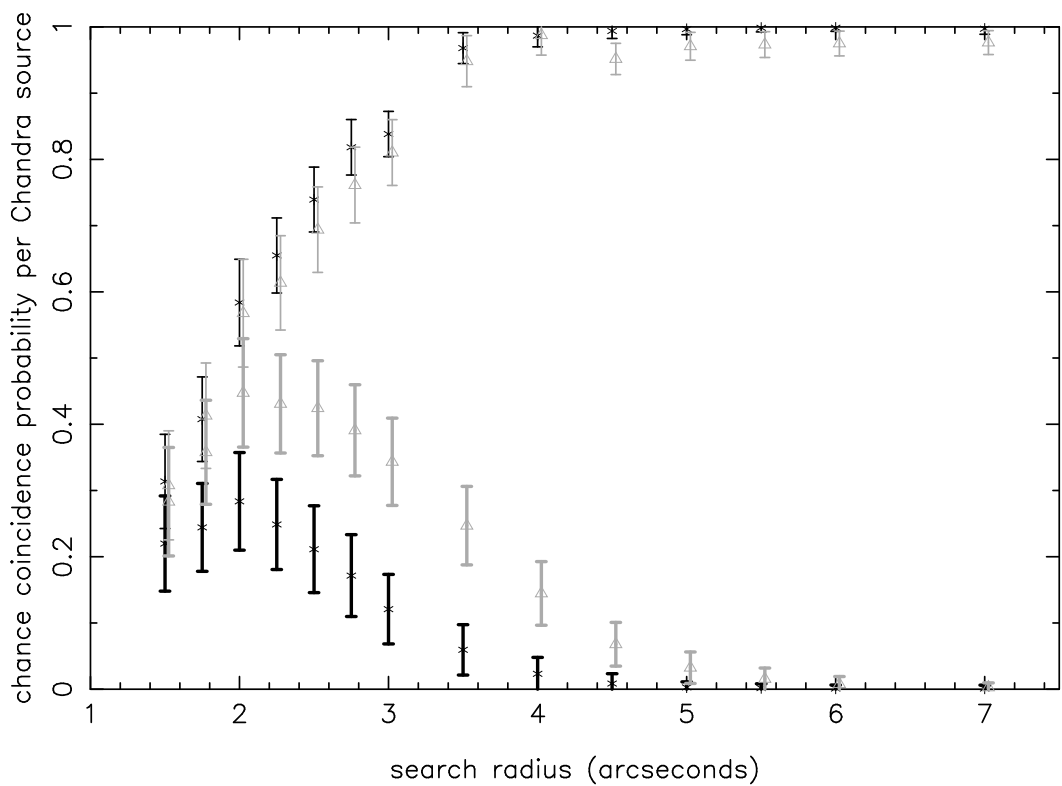}}
\caption{Normalized chance coincidence probability (matches per source) as a function of the search radius around the \chandra sources. The estimated chance coincidence probability for each
  X-ray source is shown in black (asterisks) for the OGLE-II data
  and in gray (open triangles) for data from the MCPS catalog (for
  clarity a small offset in the MCPS data has been applied). The thick lines represent the estimated probability of detecting by chance one source for a single match, while the thin lines indicate the probability of detecting one or more spurious matches for a source with one or more associations (i.e. total number of matches).}\label{ccradius}
\end{figure}

In Figure \ref{ccradius}, we plot the expected chance coincidence
probability per X-ray source (i.e. the normalized probability and not
the additive) as a function of the search radius for the cases of single, and multiple matches. The errors are estimated from the variance of the combined simulations of the 5 fields, and reflect field to field stellar density variations. These values are the average for all the \chandra fields (weighted by the number of sources in each field; a quantile-quantile plot of the distribution of the simulated chance associations shows that they follow approximately a Gaussian distribution). It is clear from Figure \ref{ccradius} that for the same search radius the OGLE-II data result in fewer spurious matches for single associations than the data from the MCPS catalog. This is mainly due to the smaller pixel size and better overall seeing of the OGLE-II catalog resulting in better positions, especially in crowded fields. On the other hand, we find no difference between the two catalogs for the case of one or more (multiple) matches.

The (normalized) chance coincidence probability for OGLE-II and MCPS sources of different spectral types is presented in Tables \ref{chanceOGLE} and \ref{chanceMCPS}, respectively. In Column (1) we give the magnitude range ($M_{{V}_{o}}$) of the stars in the optical catalog, and in Column (2) their color range ($(B-V)_{o}$). In Columns (3) and (4) we present the estimated chance coincidence probability for the single counterparts for $1.5\arcsec$ and $2.5\arcsec$ search radius, respectively, while in Columns (5) and (6)
we present the corresponding results for sources with one or more optical matches.

 These results indicate that $\leq19\%$ of
the bright blue ($M_{{V}_{o}}\leq-0.25$ and $(B-V)_{o}\leq-0.11$)
counterparts can be considered spurious matches. Therefore for 
sources with multiple matches, the brightest blue object is the most
likely counterpart. On the other hand, the high chance coincidence probability for the fainter optical sources does not allow us to securely identify an optical counterpart.

\section{DISCUSSION}\label{discussion}

In the previous sections we described the results from the
cross-correlation of the sources detected in our \chandra survey of
the SMC with the OGLE-II and MCPS optical catalogs. We find 52 sources
with a single counterpart, 68 sources with two or more matches,
while for 33 sources we do not find any matches, within an average
search radius of $1.99\arcsec$ (see \S \ref{OGLEMCPS}). We also find that the 
chance coincidence probability and foreground contamination are minimal for early-type (O, B) stars. Here, we will use the optical photometric data
(discussed in \S \ref{photometry} and \S \ref{CMDs}), and the X-ray
properties of the \chandra sources with early-type optical
counterparts (from Zezas \etal 2008, in prep.) in order
to set constraints on their nature. We do not attempt to classify
X-ray sources with faint optical matches ($M_{{V}_{o}}> -0.25$), because of their large photometric errors and high chance coincidence probability.

\subsection{Classification criteria}\label{criteria}

In order to classify the X-ray sources we use a scheme involving the position of the counterparts on the CMD and their spectral and timing X-ray properties:

(i) The broad spectral types of the optical counterparts of the X-ray
sources are determined on the basis of their position on the CMD. In
this work we focus on OB stars (see \S \ref{CMDs}). It should be noted
that the contribution of the accretion disk in the optical band is not
expected to be significant for early-type stars, while it becomes
dominant for the later spectral types (e.g. van Paradijs \& McClintock
1995). On the other hand, it is expected a substantial contribution from the
decretion disk of any Be star, which gets redder in B-V than the Be
star itself (e.g. Janot-Pacheco, Motch \& Mouchet 1987).
 
(ii) A hard ($\Gamma < 1.6$) X-ray spectrum or hardness ratio is
indicative of a pulsar binary (e.g. Yokogawa \etal 2003). Sources
with softer spectra could be either background AGN, black-hole
binaries or neutron stars with weak magnetic fields. Although a subset
of AGN have hard X-ray spectra (Compton thick AGN; e.g. Matt \etal
2000), the identification of a hard X-ray source with an early-type
star would strongly suggest that it is a HMXB pulsar.

(iii) The X-ray to optical ``color index", $\xi=B_{o}+2.5log(F_{X})$
(where $B_{o}$ is the reddening corrected apparent {\it B}\,
magnitude, and $F_{X}$ is the 2.0-10.0 keV X-ray flux in $\mu$Jy)
introduced by van Paradijs \& McClintock (1995), is another standard
means of classifying X-ray sources, as HMXBs have lower values of
$\xi$ than the LMXBs. Since many of our optical counterparts do not
have {\it B}\, band photometry, we define $\xi$ in terms of the
extinction corrected {\it V}\, band magnitude, and we calibrate it for
different types of XRBs based on the HMXBs and LMXBs catalogs of Liu
\etal (2005 for the SMC and LMC HMXBs, 2006 for the Galactic HMXBs,
and 2007 for the Galactic and LMC LMXBs). From these catalogs we only used data for sources which have X-ray fluxes in the 2.0-10.0 keV band, {\it V}\, band photometry, and reported $E(B-V)$
reddening correction (used to calculate $V_{o}$ assuming
$R_{V}=3.24$). However,
for the Magellanic Clouds (MCs) HMXBs we included even sources without reported
$E(B-V)$, since we opted to correct all of them for extinction of $A_{V}=0.29$ mag
and $A_{V}\sim0.55$ mag (mean value for the hot SMC and LMC
populations, respectively; Zaritsky
\etal 2002, 2004). So far, no LMXBs have been detected in the
SMC, while for the LMC there is only one source listed in Liu \etal
(2007). 

\begin{figure}
\centering
\rotatebox{270}{\includegraphics[height=18cm]{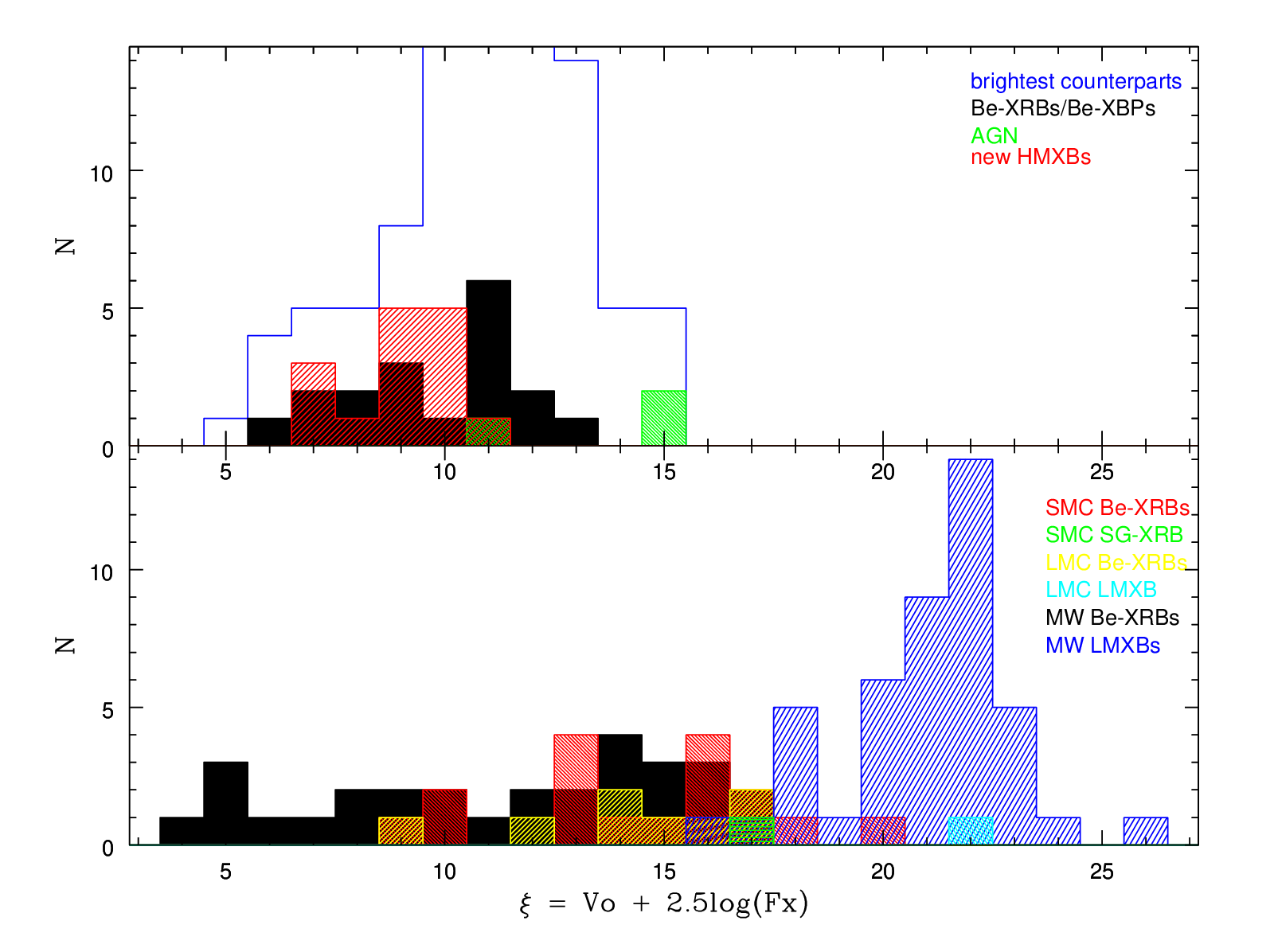}}
\caption{Histogram of $\xi$ parameter values. {\bf(a)} {\it Lower
  panel}: MCs Be-XRBs (SMC in red; LMC in yellow) and SMC SG-XRBs (in
  green) are taken from Liu \etal (2005). Milky Way Be-XRBs are taken
  from Liu \etal (2006; black), while the data for LMXBs (Galactic in
  blue; LMC in cyan) are taken from Liu \etal (2007). {\bf(b)} {\it Upper
  panel}: The $\xi$ values for the brightest optical counterparts of
  all \chandra sources (indicated in bold face in Tables
  \ref{field3}-\ref{field7}) are shown with an open histogram (in
  blue). Identified AGN (shown in green) and known Be-XRBs and
  Be-XBPs (shown in black) are also included. New HMXBs (including new candidate
  Be-XRBs) from this work are over-plotted (shown in red). More details are given in \S \ref{criteria}.}\label{xiplot}
\end{figure}

The histogram of the $\xi$
parameter values for previously identified HMXBs and LMXBs (using data from the above 
catalogs), and for HMXBs and AGN using data from
this study, is presented in Figure \ref{xiplot} (bottom and top panel,
respectively). In particular, in the bottom panel we present
the $\xi$ values for 17 SMC Be-XRBs (shown in red), the only SMC
SG-XRB (in green), 8 LMC Be-XRBs (in yellow), 1 LMC LMXB (in cyan), 29
Galactic Be-XRBs (in black), and 44 Galactic LMXBs (in blue), using
data from the catalogs of Liu \etal (2005, 2006, 2007). In the
top panel, we present the $\xi$ values for the
brightest counterparts of all \chandra sources (indicated in bold face
in Tables \ref{field3}-\ref{field7} and shown with an open blue
histogram). Previously known Be-XRBs and XRB pulsars (hereafter XBPs),
which have been detected in our \chandra survey, are
shown in black. We note that the majority of
these sources is included in the SMC Be-XRB sample of Liu \etal
(2005), thus in the bottom panel. However, for the $\xi$
parameter shown in the top panel we used their fluxes from the \chandra survey
of the SMC (Zezas \etal 2008, in prep.; also presented below in Table \ref{propertiestable}). Two \chandra sources previously identified as AGN are also included (shown in
 green; for \chandra source 3\_1 we show both counterparts). Finally, newly identified HMXBs from this study are shown in
 red. The flux limits of the Be-XRBs samples from the work of Liu \etal used
here are $1.2\times10^{-13}$\funit\,
for the SMC, $2.2\times10^{-13}$\funit\, for the LMC, and
$1.2\times10^{-10}$\funit\, for the Milky
Way (in the 2.0-10.0 keV energy band)\footnotemark.\footnotetext{Using a flat power-law ($\Gamma$=1), and
  ${\rm N_{H}=6.0\times10^{21}cm^{-2}}$ and $6.0\times10^{20}{\rm cm^{-2}}$ for the SMC and the LMC, respectively. For the Milky Way
  sample, ${\rm N_{H}}$ appropriate for each
  source has been used (using 
  http://heasarc.gsfc.nasa.gov/cgi-bin/Tools/w3nh/w3nh.pl).} The
identified AGN with $\xi\sim11$ (top panel), which
coincides with the locus of known Be-XRBs and Be-XBPs, is associated to \chandra
source 5\_15 and is discussed in detail in \S \ref{uncertainnature}.

HMXBs and LMXBs are well
separated in the $\xi$ parameter space, with only exception the value range
between $\sim16-17$, for which there is overlap among the two types
of XRBs and small number of sources, as it is shown in
Figure \ref{xiplot}. Regarding the Galactic Be-XRBs, we find two
populations that could be considered as the ``outbursting'' and ``in
quiescence'' sources with  $\xi$ parameter values $>$11 and $<$11,
respectively. Sources with $\xi$
values around 11 most probably represent the low tail of active
Be-XRBs, and can be considered as HMXBs being in an intermediate state, e.g. soon after outburst. From Figure \ref{xiplot}, it is also confirmed the
classification of HMXBs derived in this work. Using this plot as an
interpretation tool, the classification of the above sources as
candidate new Be-XRBs is suggested, given that they all fall well
within the locus of known Be-XRBs. In addition, all but one of
the identified HMXBs and candidate Be-XRBs from this work have $\xi$
values $<$11, thus suggesting a quiescent nature, or at least that
there is small probability these detected sources to be in outburst. Thus, although for the classification of an X-ray source, the $\xi$ parameter is not as sensitive as the position of the optical counterparts on the CMD and the X-ray
spectral properties, nevertheless, this criterion can be used in order to confirm the classification derived using the first two criteria and as a way to distinguish between outbursting and less active HMXBs.

 Given the large area covered by the Chandra survey, we expect a
 significant number of background AGN. However, we cannot use the
 $\rm{F_{x}/F_{opt}}$ ratio in order to identify them, since its range
 ($0.1\leq\rm{F_{x}/F_{opt}}\leq10$; e.g. Silverman \etal 2005) also
 covers the bona fide Be-XRBs. Therefore, we do not attempt to
 identify any AGN, but we note that the majority of the unclassified
 \chandra sources (see below Table \ref{propertiestable}) could be
 background AGN. In addition, this work is focused on the early-type
 counterparts of the \chandra sources, for which the chance
 coincidence probability and the photometric errors are small, thus we do not attempt to
 classify any source as a LMXB. On the other hand, this work is not
 sensitive on identifying black-hole XRBs
 (BH-XRBs; with either an early or a later type counterpart). A tell-tale
 signature of BH-XRBs is the detection of periodic modulations
 in their X-ray lightcurves, which then requires their detailed optical spectroscopic
 follow-up in order to identify them as such.  

A more detailed presentation of these classification criteria will be presented in a forthcoming paper.

\subsection{Sources with multiple optical matches}\label{multiplediscussion}

For 61 X-ray sources in our survey with more than one optical
match (out of the 68 in total), we
tentatively identify them with the brightest optical source in the region of low
chance coincidence within their error circle (\S \ref{chance}). Whenever possible, we confirm this identification with
the spectral properties of the X-ray source: a hard ($\Gamma < 1.6$)
spectrum and/or detection of pulsations are tell-tale signatures of  
pulsar binaries. For example, we suggest that the counterpart of \chandra
source 5\_7 is the early-type star O\_7\_70829 from the OGLE-II catalog,
and not one of the three fainter stars within its error circle, which do not lie on the locus of OB stars, in agreement with the same identification by Sasaki \etal
(2003), and also supported by its optical variability (Zebrun \etal
2001). We also classify this X-ray source as Be-XRB pulsar (in
agreement with Haberl \& Sasaki 2000), due to its
possibly hard spectrum (for details see below \S \ref{properties}),
and its detected pulsations (Marshall \etal 1998). For the few (4 out
of the 68) X-ray sources which do not have optical
counterparts in the region of low chance coincidence, we cannot
propose a likely counterpart (discussed in \S \ref{properties}).

\subsection{Properties of the optical counterparts and implications for the nature of the X-ray sources}\label{properties}

Following the above approach, out of the 120 X-ray sources with
optical matches, we were able to identify 35 with early-type
counterparts. The locus of O and B spectral-type sources (as defined in
\S \ref{CMDs}) consists of objects with $-5.75 \leq
M_{{V}_{o}}\leq-0.75$ and $-0.45\leq (B-V)_{o}\leq0.2$.

In Table \ref{propertiestable} we summarize the optical and X-ray
properties of the \chandra sources with optical counterparts brighter
than $M_{{V}_{o}}\leq-0.25$, which have low chance coincidence
probability (as discussed in \S \ref{chance}). We also present sources with
counterparts fainter than $M_{{V}_{o}}>-0.25$, only in cases of
reliable public classification (\chandra sources 3\_1, 3\_2, and
5\_15). In total there are 54 \chandra sources for which we present their
optical and X-ray properties in
Table \ref{propertiestable}. In Column
(1) we give the \chandra source ID, in Column (2) the optical
counterpart ID, and in Columns (3), (5) and (7) their $V$ magnitude,
$B-V$ and $U-B$ colors, respectively (along with their errors in
Columns (4), (6), and (8)). In Column (9) is given the offset from the X-ray source (in arcseconds).
 In Column (10) we
indicate if the X-ray source has a hard ($\Gamma < 1.6$) or soft
($\Gamma > 1.6$) spectrum, and whether this is based on spectral
fits (S) (Zezas \etal 2008, in prep.) or X-ray colors (HR; calculated
using the Bayesian method of Park \etal 2006). The
X-ray color derived parameters are based on color-color diagrams
involving X-ray colors defined as Col1=log(S/M) and
Col2=log(M/H), where S, M and H correspond to the
soft (0.5-1.0 keV), medium (1.0-2.5 keV), and hard (2.5-7.0 keV) band,
respectively. A ``possibly'' hard or soft X-ray spectrum is used in
order to indicate some uncertainty due to the small number of source
counts, thus large errors in the X-ray colors. In Column (11) we indicate if the X-ray source has detected pulsations
or not and its reference, while in Column (12) we give its observed
flux in the 2.0-10.0 keV energy band\footnotemark \footnotetext{Using
 the unabsorbed flux values in the 0.5-7.0 keV band (Zezas \etal 2008,
 in prep.), and
 assuming $\Gamma=1.7$ and $\rm{N_{H}=5.94\times10^{20}cm^{-2}}$,
 $4.69\times10^{20}\rm{cm^{-2}}$, $6.33\times10^{20}\rm{cm^{-2}}$,
 $6.33\times10^{20}\rm{cm^{-2}}$, $6.19\times10^{20}\rm{cm^{-2}}$ for \chandra
 fields 3, 4, 5, 6, and 7, respectively.}. In Column (13) we give the $\xi$
parameter value, and in Column (14) the identified emission-line object from
the catalog of Meyssonnier \& Azzopardi (1993; hereafter [MA93]) and
its distance from the X-ray source (given in parenthesis). Finally, in Column (15) we present a tentative
classification for the X-ray sources derived in this study. New HMXBs are
proposed on the basis of early-type counterparts and hard (or
possibly hard) X-ray spectrum and/or pulsations (listed
as ``new HMXBs'' in Table \ref{propertiestable}), while as new candidate HMXBs (listed as ``new
HMXBs ?'') we consider those sources 
with early-type counterparts but no X-ray spectral information available
(due to their small number of source counts). Sources with a soft X-ray
spectrum, unavailable X-ray spectral information and/or sources without an early-type
optical counterpart cannot be classified without further information
(e.g. optical spectroscopy, X-ray timing or spectroscopic
analysis). We note that in Column (15) we list as unclassified sources, some few previously classified for which we did not identify an OB
counterpart in this work and they do not have hard X-ray spectra or
pulsations. However, we mention that we agree with the previous
classification in all such cases. In Column (16) we present any previous classification of
\chandra sources (including those with fainter than
$M_{{V}_{o}}\sim-0.25$ counterparts). We note here that since all but
one known SMC HMXBs are Be-XRBs (there is only one spectroscopically
confirmed SG-XRB), a
classification of a source as a ``new HMXB'' may well suggest that
this source could be a new candidate Be-XRB. This is also in agreement
with the $\xi$ parameter values of the newly identified HMXBs, which
coincide with those of previously known Be-XRBs (as discussed in
detail in \S \ref{criteria}).

We mention here that in the above table we included \chandra sources
4\_19 and 5\_35 for which there is a bright optical match in the $2\sigma$
search radius (also discussed in \S \ref{brighter}, and \S
  \ref{SevenNew}, respectively). Source 5\_35 does not have any counterpart in the
$1\sigma$ radius, while source 4\_19 has a single faint match in the $1\sigma$ radius.

In this paper we present classifications for 34 sources of which 13
are previously unclassified. \chandra source 4\_3, which has been
previously listed as a candidate Be-XRB (Haberl \& Pietsch 2004), is
further classified at the present work as a new candidate Be-XBP, for
which it remains to be confirmed the X-ray pulse period (found by
Laycock, Zezas \& Hong 2008). Our results are consistent with previous
classifications in all cases of overlap (21 sources in total all of
which are Be-XRBs). There are 20 additional sources with bright
optical counterparts which remain
unclassified: 4 with OB spectral type
counterparts and soft X-ray spectra (discussed
below), and 16 sources with bright optical counterparts
($M_{{V}_{o}}\leq-0.25$), but not within the locus of OB
stars. Nine out of these 16 sources do not have X-ray spectral
information, 6 have soft spectra and 1 has a possibly hard
spectrum (\chandra source 4\_11, discussed below). For 2 sources (out of the 54 in
total presented in Table \ref{propertiestable}), we cannot
confirm the previously published classification based on the data we
used (1 AGN, and 1 Quasar), thus we list them as unclassified in this
work. The 13 sources with new classification (excluding \chandra
source 4\_3) have a mean
value of $\xi=8.90\pm1.15$, which is in the range of quiescent HMXBs
(as discussed in \S \ref{criteria}). We mention here that for the
$\xi$ value of \chandra
sources 4\_8 and 4\_13, for which we find 2 optical counterparts
within the locus of OB stars, we used the $V$ magnitude of the
brightest of the two objects. In addition,
these 13 X-ray sources have low luminosities
(${\rm L_{x}}\sim5\times10^{33}$-$10^{35}$\ergs), and all but 2 (for which we
cannot derive spectral information) have hard or possibly hard 
X-ray spectra, which together with the intermediate value of the
$\xi$ parameter are indicative of residual accretion. Therefore, these 13 sources
can be considered as HMXBs being in an intermediate state, e.g. soon after outburst.

Four \chandra sources (sources 4\_10, 5\_6, 5\_32, and 6\_3) have optical matches within the locus of OB stars, but
have a soft or possibly soft X-ray spectrum. These sources could be
XBPs with a weak accretion component (e.g. XBPs close to quiescence),
possibly as a result of the propeller effect (Illarionov \& Sunyaev
1975). This interpretation is consistent with their typical low X-ray
luminosity (${\rm L_{x}}\sim5\times10^{33}$-$10^{34}$\ergs), and mean value of the $\xi$ parameter ($9.79\pm2.29$) characteristic of intermediate state XRBs.   

\chandra source 4\_11 is not classified in the current work, although it has a hard
X-ray spectrum and a bright single optical counterpart. The latter is located just below
the locus of OB stars in the CMD (see also \S \ref{OGLEvariables} for
more details). Thus, optical spectroscopy is needed in order to unambiguously
identify it as a new Be-XRB. The same holds for the optical
counterpart of \chandra source 5\_35 (MCPS source Z\_2438540), for
which there is no optical match in the $1\sigma$ search radius and no
X-ray spectral information. On the other hand, \chandra source 7\_9 is
classified as a foreground star (in agreement with previous studies),
because its brightest optical counterpart (source O\_5\_316703) has
$M_{{V}_{o}}\sim-6$ and $(B-V)_{o}\sim1.5$ (following the discussion
in \S \ref{CMDs}). Furthermore, we classify \chandra source 6\_11 as a
new HMXB. Its brightest counterpart (MCPS source Z\_2657679) does not have an OGLE-II match within the $1\sigma$ search radius. However, it is associated with OGLE-II source O\_7\_57270 ($\equiv$O\_6\_320020) just outside this radius, which is an
OB-type star (with $V=17.14$, and $B-V=0.25$). The early-type counterpart coupled with the possibly hard X-ray spectrum of this source, make \chandra source 6\_11 a new HMXB. 

Moreover, there are 8 sources with more than one optical match for
which it is not possible to select the most likely counterpart. The
optical matches of \chandra sources 3\_1, 3\_2, 4\_6 and 7\_16 are
very faint, and because of their large photometric errors and high chance
coincidence probability, we cannot select one of the two counterparts
(\chandra source 3\_1 is discussed in \S \ref{OGLEvariables}). This is also the case for \chandra source 4\_23, for which we find two optical matches, one in each of the studied optical catalogs. The O\_5\_101577 source has only been detected in the $I$ band (i.e. no $B$, $V$ photometric data available for it), while the Z\_2029266 source is also faint ($V>19$ mag). Likewise,
for \chandra sources 7\_4 and 7\_17 we cannot choose the most likely counterpart
without optical spectroscopic or variability information, as the 2
brightest sources within the search radius have similar $V$
magnitudes. In particular, for \chandra source 7\_4 we do not have
X-ray spectral information, while source 7\_17 has possibly a soft spectrum
(also discussed in the Appendix, \S \ref{uncertainnature}). However,
the blue object of the bright identified counterparts is the best choice
since it results in smaller chance coincidence probability. In the case of \chandra source 3\_19 the brightest counterpart (source
Z\_3050592) is not an OB star, but it lies in the area between the red giant branch and the MS. Instead, we find that a fainter source
(Z\_3049033) is located in the OB star locus. For completeness, we
present the optical properties of both sources in Table
\ref{propertiestable}. However, the low X-ray luminosity of source
3\_19 indicates that the compact object does not accrete via
Roche-lobe overflow as would be expected in the case of an evolved
donor (most probably this is a wind-fed
system). Furthermore, the hard X-ray spectrum of this source, that is
indicative of a pulsar, suggests a younger system, possibly a
Be-XRB. Thus, we propose the OB star candidate Z\_3049033 as the most likely
counterpart of \chandra source 3\_19.

\chandra sources 5\_4 and 5\_12 have early-type counterparts and hard
(or possibly hard) X-ray spectra. Following the above classification
scheme, they would be classified here as new HMXBs. However, optical spectra for these sources have been recently published (Antoniou \etal 2008), revealing their Be nature,
thus in Table
\ref{propertiestable} they are listed as new Be-XRBs. We mention that \chandra source 5\_4 is identified with the
emission-line object [MA93]798, while \chandra source 5\_12 has no match in the latter
catalog (within a search radius of 5$\arcsec$).  Although it is
difficult to estimate the incompleteness of the [MA93] catalog, the
fact that there are Be-XRBs confirmed spectroscopically for which
there is not a match in this catalog, implies that if
a source does not have a [MA93] counterpart it does not
necessarily mean that the source does not exhibit any H$\alpha$
emission (typical of Be-XRBs). Actually, from the
cross-correlation of MCPS stars within our \chandra fields with the
[MA93] catalog, we find that the vast majority of the
matches have $V<16.6$ mag. This suggests that sources with
early (O, B) type counterparts but fainter magnitudes may have
H$\alpha$ emission, but not a [MA93] counterpart. \chandra source 5\_12 has a bright
optical counterpart ($V\sim14.9$ mag), thus revealing that even brighter
sources may have remained undetected in the [MA93] work. Other sources
which are previously known Be-XBPs and do not have [MA93]
counterparts\footnotemark \footnotetext{This problem is not limited to the present work. Previous
works on Be-XRBs (e.g. by Haberl \& Pietsch 2004), were not successful in identifying a [MA93] counterpart even for
confirmed Be-XRBs.}
 include \chandra sources 4\_2, 4\_8, 5\_16, 6\_1, and 7\_1. 

 In total, we identify 35 sources with early-type counterparts, 27 of
 which have hard X-ray spectra, strongly suggesting that they are
 XBPs. All but one pulsars (i.e. 14) that have been identified as
 X-ray sources within our \chandra fields have hard spectra and
 early-type counterparts (for completeness in our \chandra fields lie
 22 confirmed and 11 candidate pulsars\footnotemark, however only 16
 have been detected in our survey).\footnotetext{The confirmed
 systems are taken from http://www.astro.soton.ac.uk/$\sim$mjc/ (as of
 08/22/2008), while the candidate XBPs are from the work of Laycock
 \etal (2008).} The only exceptions are \chandra sources 3\_18 and 5\_24: although their optical counterparts are OB spectral-type star, their small number of counts does not provide us the necessary spectral information.

Ten of the candidate Be stars of Mennickent \etal (2002) have been
identified as counterparts to our \chandra sources. One of those is
proposed as counterpart to \chandra source 4\_17 for
the first time, while for the remaining 9 X-ray sources (3\_3, 4\_2, 4\_8, 5\_3,
5\_7, 5\_15, 6\_1, 6\_4, and 7\_1) there is a known counterpart in the literature (see Table \ref{tableotherinfo} for more details). An extensive
presentation of these 10 sources is given in \S
\ref{Becandidatecounterparts}. This supports our classification of the
aforementioned source as new Be-XRB, given also its possibly hard
X-ray spectrum and the emission-line object [MA93]396 within its
search radius.

From the finding charts we see that a few sources have brighter
counterparts that fall outside the $1\sigma$ search radius. Thus, we
extended the search radius to twice the one used in the original
search and we cross-correlated again the coordinates of the \chandra
sources with the OGLE-II and MCPS catalogs. We find that out of the 33
X-ray sources without an optical counterpart identified in the
$1\sigma$ search radius, there are 7 X-ray sources with bright optical
matches ($\rm{M_{V_{o}}}\leq-0.25$ mag) between the $1\sigma$ and
$2\sigma$ search radii. None of these matches lie in the locus of OB
stars, while we ignore fainter sources. In addition there are 33 \chandra sources with optical matches between the $1\sigma$ and
$2\sigma$ search radii, which are brighter than those found within the
$1\sigma$ search radius. In particular, for 10 of these \chandra
sources their optical matches lie in the locus of OB stars, and 3 of
them have hard or possibly hard X-ray spectrum (sources 3\_2, 4\_13,
4\_19). \chandra source 3\_2 is a known pulsar, possibly Be-XRB (Edge
\etal 2004), while sources 4\_13 and 4\_19 have not been previously
identified. The hard spectrum of these sources and the early-type
counterpart make them candidate XBPs (see  \S \ref{brighter} in the
Appendix). We note that although the chance coincidence probability
for the bright optical matches within the $2\sigma$ search radius is
very high ($\sim$70\%; see Tables \ref{chanceOGLE} and
\ref{chanceMCPS} for more details), the identification of a hard X-ray
source with an optical source within the locus of OB stars makes it a
HMXB (these additional 3 sources are included in
Table \ref{propertiestable}  and in our census of confirmed and
candidate Be-XRBs in the SMC). For the remaining sources we do not propose any of their optical matches as a counterpart, and we only present them for completeness (Tables \ref{BrightNor} and \ref{properties2r} in the Appendix, \S \ref{SevenNew}).

\subsection{Comparison with the Galactic and LMC Be-XRBs}\label{comparisonMCsMW}

In order to understand the HMXB population of the SMC, several studies have compared the number of Be-XRBs in the MCs and
the Milky Way (e.g. Haberl \& Sasaki 2000, Majid, Lamb \& Macomb 2004, Coe \etal
2005a, Haberl \& 
Pietsch 2004; hereafter [HP04]). One approach is to compare the ratios of Be-XRBs to
normal (i.e. non-emission line) OB stars. By studying the Be-XRB population with respect to
their related stellar populations, we minimize age effects or
variations due to SF rate differences for populations of different ages, allowing us to probe for intrinsic differences in the XRB formation efficiency. 
For the same reason, we do not compare the ratio of Be-XRBs to
SG-XRBs, since this is sensitive to the age of these systems, and
therefore to variations of the SF rate over these time scales. We note that because of the transient nature of the Be-XRBs, their numbers can be considered only as lower limits. This has been demonstrated by the increasing number of Be-XRBs detected in long term monitoring surveys of the SMC (e.g. Laycock \etal 2005). So far, various studies have confirmed 21 Be-XRBs
in the SMC ([HP04], Raguzova \& Popov
2005; hereafter [RP05]) and only 1 supergiant, which is located in the
SMC Wing (SMC X-1, Webster \etal 1972). In the compilation of
[HP04], 17 (out of the 21) Be-XRBs
are X-ray pulsars, while for another 8 pulsars the identification
with a Be star is suggested (see [HP04] and references therein). There
are also 23 proposed Be-XRBs without detected pulsations, and one
suggested Be-XRB with uncertain pulse period ([HP04]). Since the
detection of pulsations requires high S/N X-ray data, in our study, we
include all Be-XRBs, and not only those with detected X-ray
pulsations. In total, there are 44 confirmed
and candidate Be-XRBs in the 5 \chandra fields: 19 listed in [HP04] (with 4 of those not
detected in the \chandra survey), 12 listed in more recent works (such
as by Liu \etal 2005, and Antoniou \etal 2008), 2 newly identified
Be-XRBs (\chandra sources 4\_17 and 7\_19), and 11 HMXBs from this
work, which can be considered as new candidate Be-XRBs (see also the discussion in \S \ref{properties}).

 At this point, we revisit the relation between Be-XRBs and OB stars, taking into account the latest census of XRBs, and comparing it with the corresponding ratios for the LMC and Milky Way. The numbers of Be-XRBs in the 3 galaxies are based on the compilations of Liu \etal (2005, 2006). Since these compilations consist of different samples of sources detected in various surveys with different sensitivities, we use a subset of Be-XRBs down to the same detection limit
of $10^{34}$\ergs\, (in the 2.0-10.0 keV band; equivalent to $\sim1.1\times10^{34}$\ergs\, in the 0.7-10.0 keV band). This cut-off X-ray
luminosity was chosen in order to assemble uniform samples of sources
detected in surveys of the SMC and the Milky Way, given also that this is
the detection limit of the \chandra survey used in this work
(Zezas \etal 2008, in prep.). From those catalogs
we used all sources with available spectral types for their
counterparts. We also converted their flux densities from the various
energy bands to the 2.0-10.0 keV band, assuming a flat power-law
spectrum of $\Gamma=1$.

In the catalog of Liu \etal (2006) we find 31 Galactic Be-XRBs located within 10 kpc
from the Sun (based on distances from [RP05]\footnotemark)\footnotetext{http://xray.sai.msu.ru/$\sim$raguzova/BeXcat/ as of 08/25/2006 for the
Milky Way.} with ${\rm L_{x}}\geq
10^{34}$\ergs\, (Liu \etal 2006) and 17 confirmed Be-XRBs in the SMC area covered 
by our \chandra survey (Liu \etal 2005). Including 7 Be-XRBs\footnotemark\footnotetext{These systems are: 3 Be-XRBs not listed
in Liu \etal (2005; \chandra sources
3\_2, 5\_7, and 6\_1), one newly identified system (\chandra source 5\_4),
and 3 HMXBs from this work (\chandra sources
3\_7, 4\_4, and 7\_4 which are candidate Be-XRBs).} from the
present work with an X-ray luminosity of $\sim10^{34}$\ergs, results in a
total of 24 SMC Be-XRBs within our \chandra fields. In this comparison
(down to ${\rm L_{x}}\geq
10^{34}$\ergs) we do not consider LMC sources since the LMC has not been surveyed as extensively and in the same depth as the Milky Way or the SMC.

We then estimate the number of OB stars using their locus on the CMD
defined in \S \ref{CMDs}. Based
on the MCPS catalog (Zaritsky \etal 2002) we find $\sim13720$ OB stars
within our \chandra fields ($\sim$2220, $\sim$4060, $\sim$2730, $\sim$3040, and $\sim$1670 OB stars for \chandra field 3, 4, 5, 6, and 7, respectively). The photometric data of these sources have
been reddening corrected as described in \S \ref{CMDs}. In the case of
the Milky Way, Reed (2001) estimated that there are $\sim25800$ OB
stars within 10 kpc of the Sun. Based on the above numbers we find that the Be-XRBs with
${\rm L_{x}}\geq10^{34}$\ergs\, are $\sim1.5$ times more common in the SMC
when compared to the Milky Way, with respect to its populations of
young stars, thus confirming the notion
that the SMC has a large number of Be-XRBs (e.g. Liu \etal 2005; Coe \etal 2005b).

For completeness, we extend this comparison to the LMC,
considering only sources with ${\rm L_{x}}\geq10^{36}$\ergs, which is the
average completeness level of the ROSAT observations of LMC fields. In this case we find 6
Be-XRBs in the LMC (Liu \etal 2005), 10 in the SMC, and 29 in the Milky Way (in
the case of the LMC we find $\sim42200$ OB stars, in the area covered by
the ROSAT fields where the 6 Be-XRBs were detected,
based on the MCPS catalog of Zaritsky \etal (2004) and assuming
$A_{V}\sim0.55$ mag as in \S \ref{criteria}). Hence, the
Be-XRBs are $\sim5.5$ times more common in the SMC, with respect to
the LMC, while when compared to the Milky Way we find almost equal
ratios of Be-XRBs to OB stars. However, one caveat in the case of the
LMC, is that its small number of Be-XRBs may be due to the fact that
it has not been monitored as extensively and in the same depth as the SMC or the Milky Way.

The high number of Be-XRBs in the SMC could simply be the result of enhanced star forming activity in the SMC $\sim30$ Myr ago (e.g. Majid \etal 2004). In particular, the formation and lifetimes of
wind-fed HMXBs are driven by the stellar evolution of the donor. Be
stars, which are the most common type of donor in wind-fed XRBs,
develop their decretion disks at ages of $25-80$ Myr (McSwain \& Gies
2005), with a peak at $\sim35$ Myr. Indeed, using the isochrones
presented in \S \ref{CMDs}, we determine that bright counterparts of
the X-ray sources in our fields (of OB spectral type) have ages
ranging from $\sim15.5$ to $\sim85$ Myr. Based on the SF history of
the SMC (Harris \& Zaritsky 2004),and for Z=0.008, we find that the most
recent major burst occurred $\sim42$ Myr ago. The only exceptions are
\chandra field 6, where a similarly strong burst occurred somewhat
earlier ($\sim27$ Myr ago), and \chandra field 4 which shows an
additional burst only $\sim7$ Myr ago. The duration of the major
bursts is $\sim40$ Myr. In addition, there were older SF episodes
($\sim0.4$ Gyr ago) with lower intensity but longer duration. However,
we do not expect a population of Be-XRBs associated with either the
relatively old ($\sim0.4$ Gyr) widespread star forming activity, or
the most recent episode ($\sim7$ Myr) at the current time. Therefore, the large number of Be-XRBs in the SMC is most likely due to its enhanced star-forming activity $\sim42$ Myr ago. OB stars formed during this episode are expected to reach now the maximum rate of decretion disk formation. Even in smaller spatial scales the excess of Be-XRBs in fields 4 and 5 is consistent with their higher SF rate $\sim42$ Myr ago, in comparison to the other fields, and with the fact that the age of their stellar populations is in the range of maximum Be-star formation.

However, the comparison of the ratios of Be-XRBs over OB stars in the
SMC, and the Milky Way indicates that there is still a residual excess
of a factor of $\sim1.5$ that cannot be accounted for by SF and age
differences. This residual excess can be attributed to the different
metallicity of the two galaxies. Population synthesis models predict a
factor of $\sim3$ higher numbers of HMXBs in galaxies with
metallicities similar to that of the SMC, when compared to the Milky
Way (Dray 2006). In addition, there
is observational evidence for higher proportion of Be stars in lower
metallicity environments (at least in the case of younger
systems). Wisniewski \& Bjorkman (2006) found a
ratio of $\sim2$ for the numbers of Be stars in SMC and Galactic
metallicities, while Martayan \etal (2007) found that the B and Be
stars rotate faster in the SMC than in the LMC and faster in the LMC
than in the Milky Way. This could also explain the residual excess of
binaries in the SMC, since only a fraction of the B stars that reach
the zero-age MS with a sufficiently high initial rotational velocity can
become Be stars (Martayan \etal 2006). Moreover, by noticing the lack of massive Be
stars in the Milky Way at ages for which Be stars are found in the MCs,
Martayan \etal (2007) suggested that the Be star phase can last longer in low metallicity
environments, such as in the MCs, than in the Milky Way.

\subsection{Local variations of the Be-XRB populations}

\begin{figure}
\centering
\rotatebox{270}{\includegraphics[height=13.0cm]{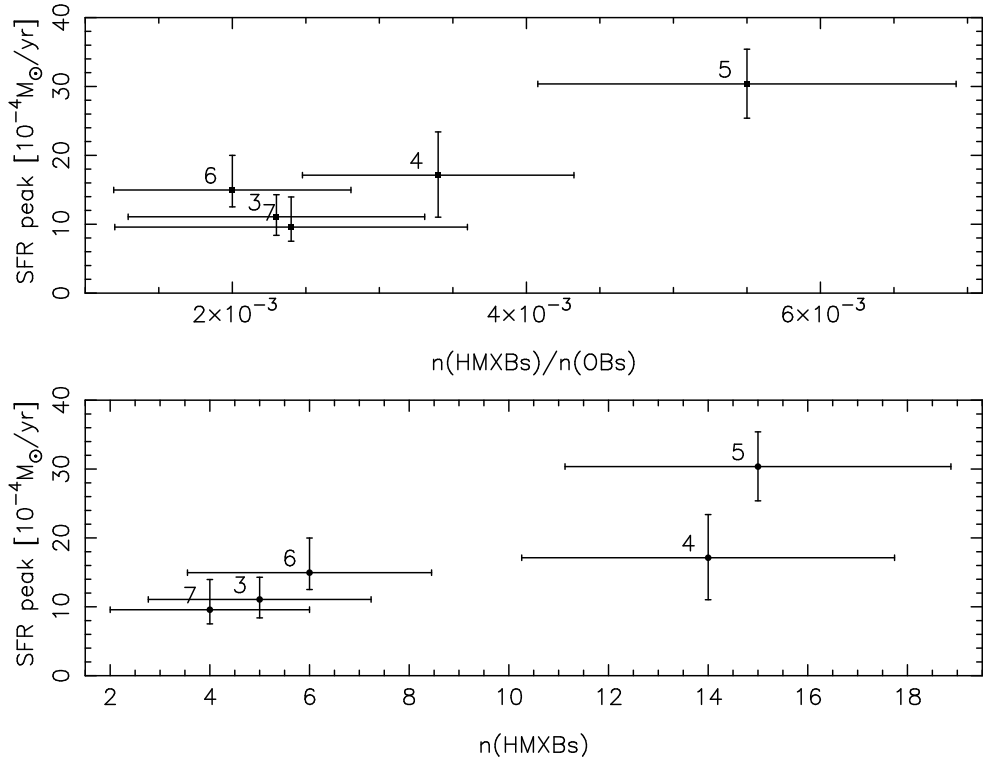}}
\caption{{\bf(a)} {\it Lower panel}: The star-formation rate at the age of $\sim42$ Myr versus the number of HMXBs in each \chandra field. {\bf(b)} {\it Upper panel}:  The star-formation rate at the age of $\sim42$ Myr versus the ratio of the number of HMXBs to the number of OB spectral type stars in each field. The error bars in the x-axis were derived assuming Poisson statistics. In the y-axis error bars indicate lower and upper limits of the mean SF rate in each \chandra field.}
\label{peak}
\end{figure}

When we compare the number of SMC HMXBs (or equivalently the number of
Be-XRBs, since all but one of the confirmed HMXBs is a SG-XRB) in the different \chandra fields
we find an excess of these objects in fields 4 and 5. In
particular, these two fields have 14 and 15 HMXBs, respectively, versus an average of 5 HMXBs in each of the other 3 fields. Field 5
also has the largest number of identified pulsars (8; fields 3, 4, 6,
and 7 have 3, 6, 3, and 2 confirmed pulsars, respectively). Although given the
small numbers of sources this excess is only marginally statistically
significant, it is intriguing that fields 4 and 5 had the highest SF
rate at the age of maximum Be-star production ($\sim42$ Myr ago). In
Figure \ref{peak} (lower panel), we present the height of the SF rate
at the age of $\sim42$ Myr versus the number of HMXBs in each
\chandra field. In the upper panel of Figure \ref{peak} we present the
SF rate versus the ratio of the number of HMXBs to the number of OB stars in each field. In this plot we include all HMXBs detected in our
\chandra fields (based on our survey and the census of Liu \etal 2005;
44 in total), and the OB stars in each field (selected as described in \S
\ref{comparisonMCsMW}) from the MCPS catalog. Predominantly, field 5 which has the most intense SF peak at
the peak age of Be-star formation  ($\sim2$ times more intense when
compared to that of the other fields) appears to have almost double
ratio of HMXBs to OB stars than the other fields.

The direct comparison between the XRB populations and the SF history of different regions, may be complicated by kicks imparted on the compact object
during the supernova explosion (e.g. van den
Heuvel \etal 2000). The result of these kicks is that
the XRBs may be spread over a larger volume with respect to their parent
stellar population, and thus complicating the study of their
connections. In the case of the SMC, and given a typical projected
runaway velocity of $15\pm6$ km s$^{-1}$ (van den Heuvel \etal 2000), we estimate that a
Be-XRB would have traveled a maximum
distance of $\sim640$ pc from the time it was formed (i.e. $\sim42$
Myr ago, given the SF history of the SMC). This distance is an upper limit, since the age of 42 Myr is only based on the formation time-scale of Be binaries, and it does not include the formation time-scale of the pulsar, which is a few Myr. For a distance of 60 kpc this
translates to a projected angular distance of $\sim36.7\arcmin$, roughly twice the dimension of the \chandra
fields. However, the scales of the star-forming regions at the age
ranges of interest ($<$20 Myr, $20-60$ Myr,
$>$60 Myr) are much larger than the size of the \chandra
fields. Therefore, our fields across the SMC ``Bar'' provide a representative picture of the Be-XRB population in its main body. Moreover, the excess of Be-XRBs at fields 4 and 5 which have
the highest SF rates in ages of $\sim42$ Myr (which corresponds to
the age of maximum production of Be stars) gives further confidence
that the kick velocities are small enough not to smear the correlation
between X-ray sources and their parent stellar populations.

\section{Conclusions}

   In this paper we report the results for the optical counterparts of 153 X-ray
sources detected in the \chandra survey of the central region of the
SMC. For 120 sources we find optical matches in the OGLE-II and/or MCPS catalogs. Using these photometric data, we propose the most likely
optical counterpart for 113 \chandra sources, while for 7 sources the
candidate counterparts are equally likely. By combining their optical
and X-ray properties (i.e. spectral
and timing properties), we classify 34 \chandra sources of which 13
were previously unclassified. In particular:

\begin{enumerate}

\item{We find that 52 X-ray sources have a single counterpart, 68
have two or more matches, while 33 sources do not have
counterparts in either OGLE-II or MCPS catalog within the $1\sigma$
search radius (7 out of these 33 sources have bright optical matches
($M_{{V}_{o}}\leq-0.25$) between the $1\sigma$ and $2\sigma$ search
radii, but none of them is associated with an OB star).}

\item{Early-type counterparts (of OB spectral type) have been
identified for 35 X-ray sources (with a chance coincidence probability
$\leq19\%$). Based on spectroscopic observations of stars in the SMC
we define the photometric locus of early-type (OB) stars in the
$M_{{V}_{o}}$ vs. $(B-V)_{o}$ CMD. Based on the position on the CMD
and their X-ray spectral properties, we propose: (a) 4 new Be-XRBs,
and (b) 7 new HMXBs, and 2 new candidate HMXBs. Two of the new HMXBs
have hard X-ray spectra and early-type (OB) counterparts within their $2\sigma$
search radius.}

\item{Based on the isochrones of Lejeune \& Schaerer (2001), we
estimate the age of the 13 new (confirmed and candidate Be-XRBs and HMXBs) to be $\sim15-85$
Myr, in agreement with the age of Be stars.}

\item{Twenty X-ray sources with bright counterparts
($M_{{V}_{o}}\leq-0.25$) could not be classified, because we do not
have X-ray spectral information for them and/or they have other than
OB spectral-type counterparts.}

\item{We find that the mixing of Be-XRBs with other than their natal
stellar population is not an issue in
our comparisons of Be-XRBs and stellar populations in the SMC, because
the SF activity across the SMC ``Bar'' is generally uniform
in scales larger than the size of the \chandra fields. Instead we find
indication for variation in the XRB populations across the ``Bar'' and
in scales of $\sim1$ kpc: fields with higher SF rate at ages which are
more prone to produce Be stars show increased number of Be-XRBs.}

\item{Using the catalogs of Liu \etal (2005, 2006) we find that for
luminosities down to ${\rm L_{x}}\sim10^{34}$\ergs\, (chosen as a moderate
minimum cut-off between different SMC and Galactic X-ray surveys) the
\chandra SMC fields contain $\sim1.5$ times more Be-XRBs in comparison
to the Milky Way (including the new confirmed and candidate Be-XRBs from this study), even
after taking into account the different OB star formation rates in the
two galaxies. This residual excess can be explained when we account for the
different metallicity of the SMC and the Milky Way. This is in good
agreement with theoretical predictions derived from population synthesis
studies (Dray 2006), and previous observational (both photometric and
spectroscopic) works (e.g. Wisniewski \& Bjorkman 2006; Martayan \etal 2006, 2007).}

\end{enumerate}

\acknowledgments

We would like to thank the anonymous referee for helpful comments
which have improved this paper. We would also like to thank Dr.\,A.\,Udalski for providing unpublished light curves of the OGLE-II stars used in this paper. This research has made use of the SIMBAD database,
operated at CDS, Strasbourg, France.
This work was supported by NASA LTSA grant NAG5-13056, and NASA grant G02-3117X.

\appendix

\section{Notes on individual sources}

\subsection{Sources with uncertain nature}\label{uncertainnature}

Most of the sources that remain unclassified in this work have soft
(or possibly soft) X-ray spectra or no X-ray spectral information
(because of their small number of source counts in the 0.5-7.0 keV energy band; Zezas \etal
2008, in prep.).

{\bf \chandra source 3\_1:} One of the two optical counterparts of
this source has been identified as a variable star. More details of
source 3\_1 are presented below in \S \ref{OGLEvariables}.

{\bf \chandra source 3\_4:} A single optical match was identified in
both OGLE-II and MCPS catalogs. However, its $V=18.19$ magnitude and
$B-V=0.37$ color are not indicative of the nature of the companion
star. Its soft X-ray spectrum (Zezas \etal 2008, in prep.) possibly suggests an AGN. 

{\bf \chandra sources 5\_9 \& 6\_8:} These sources have possibly a soft X-ray
spectrum and multiple optical matches within their search radius. The
brightest matches of 5\_9 and 6\_8 are not O or B-type stars ($B-V$ color $\sim0.59$
and $\sim1.12$ mag, respectively). Recently, Laycock \etal (2008) detected source 5\_9 as a candidate quiescent
X-ray pulsar (SXP7.83).

{\bf \chandra source 5\_15:} Within the search radius of this source
we identified 5 matches, all of them fainter than V$\sim$19 mag. Its
possibly soft X-ray spectrum suggests an
AGN. However, Laycock \etal (2008) detected this source as a candidate
quiescent X-ray pulsar, while the brightest of its matches has been
identified as a Be candidate star (Mennickent \etal 2002). See also
the discussion in \S \ref{Becandidatecounterparts}.

{\bf \chandra source 5\_23:} We identify MCPS source Z\_2288276 (no
OGLE-II overlap) as the optical counterpart of X-ray source 5\_23. Based
on its V magnitude and B-V color, this is an O or B-type
star. Thus, we classify this source as a new candidate HMXB. The small
number of net counts of source 5\_23 ($\sim10$ counts in the
0.5-7.0 keV energy band; Zezas \etal
2008, in prep.) does not allow us to derive its X-ray spectral
information, thus prohibiting us to classify it further as a new
candidate Be-XRB.   
 
{\bf \chandra sources 4\_10, 5\_6, 5\_32, 6\_3, 7\_17:} The soft
X-ray spectrum of these sources (HR method, more details in Table
\ref{propertiestable} and its description in \S \ref{properties}) and
their early-type counterparts could be indicative of a candidate
black-hole X-ray binary (BH-XRB) nature. However, only X-ray
spectral fits and optical spectroscopy of the companion stars could
possibly identify their true nature.  
 
{\bf \chandra source 6\_20:} The small number of net counts of this source
($<10$ counts in the
0.5-7.0 keV energy band; Zezas \etal 2008, in prep.) does not allow us to
derive spectral information for it. Its $V=13.71$ magnitude and
$B-V=0.39$ color are not indicative of the nature of the companion
star, thus this source remains unclassified in this work. More details
on this source are presented to \S \ref{appendix2dF}.  
 
{\bf \chandra source 7\_17:} This source has possibly a soft
X-ray spectrum. The large
search radius of it ($5.29\arcsec$, maximum search radius
used in this study because of the large off-axis angle)
resulted in the identification of multiple matches. However, only 2
sources are bright ($V\sim18.4$ mag, thus resulting in smaller chance
coincidence probability). One of
the two sources lies on the red giant branch, while the other one (O\_5\_140992) falls just
outside the locus of OB spectral type stars. Source 7\_17 remains
unclassified in this work.

{\bf \chandra source 7\_19:} Due to the small number of counts ($<10$ counts in the
0.5-7.0 keV energy band; Zezas \etal 2008, in prep.) we do
not have X-ray spectral information for this source. However, there is
an emission-line object within $\sim2\sigma$ of the search radius (source
[MA93]316; Meyssonier \& Azzopardi 1993). Actually this source is
located in the star cluster SMC-N32 (Bica \& Schmitt 1995), a crowded region surrounded
by a diffuse emission (HII region). The detection algorithm in
this case resulted in many spurious sources that they were not
deblended (see also the on-line finding chart for this source). We note
here that the brightest of the 5 matches within the search radius of
this source is located just to the right border of the ellipse of OB
stars in the V, B-V CMD. Its large error in B-V color could
place it within the ellipse, thus suggesting that it could be another
new HMXB.

{\bf \chandra sources 4\_32, 4\_36, 5\_18, 5\_25, 6\_23, 7\_9:} All
but two of the brightest counterparts of these sources lie on the red
giant branch in the V, B-V CMD (Fig. \ref{VBV}), while that of
source 5\_18 lies on the right of the MS (most probably in the giant
branch). The counterpart of source 7\_9 has $M_{{V}_{o}}=-5.7$ and
$(B-V)_{o}>0.75$, thus it lies on the locus of foreground stars (for
more details see \S \ref{CMDs}). In addition, these sources do not have X-ray spectral
information (because of their small number of counts), except from
source 7\_9 which has possibly a hard X-ray spectrum.

\subsection{Point sources associated with known SNRs}\label{SNRs}
  In \chandra field 4 there are three known supernova remnants
  (SNRs). Below are given details for the point sources associated
  with two of them. The third SNR (\chandra source 4\_38) is located
  in a large off-axis angle ({$\rm{D_{off-axis}>10}\arcmin$}),
  thus not considered in this study.

{\bf \chandra source 4\_16:} This is a point source associated with
the previously known SNR 0047-735 (Mathewson \etal 1984). For this source it has
been found one single optical match (O\_5\_27768), while it
has been recently classified as a new candidate quiescent X-ray pulsar (CXO J004905.1-731411$\equiv$SXP8.94; Laycock \etal 2008).

{\bf \chandra source 4\_20:} This point source is associated with the previously known
SNR 0049-73.6 (Mathewson \etal 1984). We identify source O\_5\_262679
as the optical counterpart of source 4\_2. CXO J005059.0-732055 (SXP9.39; Laycock \etal 2008)
has been classified as a candidate quiescent X-ray pulsar.

{\bf \chandra source 4\_21:} This point source is also associated with the 
SNR 0049-73.6 (Mathewson \etal 1984; Sasaki \etal 2000). There are 3 bright counterparts
($M_{{V}_{o}}<-0.25$) within its search radius, while one of those (O\_5\_256826) is an early-type
star (O or B).

\subsection{Candidate Be nature of identified counterparts}\label{Becandidatecounterparts}

All sources that are presented below have been classified as type-4 stars (light curves similar to Galactic Be stars) in the compilation of Be candidate stars by Mennickent \etal 2002 (hereafter [MPG02]).

{\bf \chandra source 3\_3:} The identified counterpart O\_8\_49531 (single match) of X-ray source 3\_3 is a Be candidate star ([MPG02]No.494), in agreement with the identification of Sasaki \etal (2003).

{\bf \chandra source 4\_2:} We find two matches for X-ray source
4\_2. Although none of them is located within the locus of OB stars
(criterion (i)), we identify source O\_5\_111490 as the counterpart
for 4\_2, because of its Be nature ([MPG02]No.206), while it was also
found to be a variable star (Ita \etal 2004). This is in agreement with the identification of Coe \etal (2005a).

{\bf \chandra source 4\_17:} Source O\_5\_180008 is identified as
the optical counterpart of X-ray source 4\_17, while it is also
identified as a Be candidate star ([MPG02]No.265). This
source coincides with the emission-line object [MA93]396. In combination with the possibly
hard X-ray spectrum, we classify this source as a new Be-XRB. 

{\bf \chandra source 5\_3:} The identified optical counterpart
O\_6\_85614 (in agreement with Schmidtke \& Cowley 2006) is a Be
candidate star [MPG02]No.321. It is of B0-5 spectral type
(source [2dF]No.0828; see below \S \ref{appendix2dF}), and it has been found to coincide with the emission-line object [MA93]No.506 (in
agreement with Haberl \& Sasaki 2000).

{\bf \chandra source 5\_7:} The identified optical counterpart of
X-ray source 5\_7 is source O\_7\_70829. It is identified as a Be
candidate star ([MPG02]No.426), while Sasaki \etal (2003) presented the
same counterpart, identified also as the emission-line object [MA93]No.810.  

{\bf \chandra source 5\_15:} We identify source O\_7\_71429 as the
optical counterpart of X-ray source 5\_15 (see also \S \ref{uncertainnature}). This is also reported as a
Be candidate star ([MPG02]No.423). However, Dobrzycki \etal (2003a)
identified this object as a quasar ($z=1.79$) located behind the SMC (based on a spectroscopic study).
 
{\bf \chandra source 6\_1:} The identified counterpart O\_6\_77228 of
source 6\_1 is a candidate Be star ([MPG02]No.341; see also \S \ref{appendix2dF}).  

{\bf \chandra source 6\_4:} Source O\_6\_147662 is identified as the optical counterpart of X-ray source 6\_4. This optical source is also
reported as a candidate Be star ([MPG02]No.364), and it is located
$\sim0.57\arcsec$ away from the emission-line object [MA93]618 (in
agreement with Haberl \& Sasaki 2000).

{\bf \chandra source 7\_1:} We identify source O\_5\_65517 as the
optical counterpart of X-ray source 7\_1. This is star 1 of Stevens
\etal (1999), and it is also in agreement with the identification of Coe \& Orosz (2000). This OGLE source is identified as a Be candidate star ([MPG02]No.199).

\subsection{Sources identified with OGLE-II variable stars}\label{OGLEvariables}

The following OGLE-II optical counterparts have been identified as variable
stars in the catalogs of Zebrun \etal (2001) or Ita \etal (2004).

{\bf \chandra source 3\_1:} We identify source O\_7\_267163 as the
optical counterpart of X-ray source 3\_1. From visual inspection of
the finding chart, from comparison of the $V$ magnitude for sources
O\_7\_267163 and O\_7\_267132, and taking into account the $\sim0.7\arcsec$ astrometric accuracy of
OGLE, we believe these stars are a single source. Although none of
the identified matches is an early-type star, and the
parameter $\xi$ cannot distinguish in this case the right
counterpart, source O\_7\_267163 is identified as a variable star
(OGLE00571981-72253375) in the study of Zebrun \etal (2001). Sasaki \etal (2003) identified the same object, while they also associated it with star USNOA2.0
0150-00625436. The soft X-ray spectrum of 3\_1 and the identification
of the optical counterpart as a background quasar (Dobrzycki et
al. 2003b), suggest that this is possibly an AGN (in agreement with Sasaki \etal 2003).   

{\bf \chandra source 4\_2}: Source O\_5\_111490 is also identified as
a variable star in the analysis of Ita \etal (2004; see also \S \ref{Becandidatecounterparts}.

{\bf \chandra source 4\_5:} The optical counterpart of X-ray source
4\_5 is the variable star O\_5\_111500 (single optical match; $P_{pulse,
  opt}=89.835$ days, Ita \etal 2004). Haberl, Eger \& Pietsch (2008)
have recently confirmed the Be-XRB nature of this source, while no
significant pulsations have been found in their study. In a later
work, Laycock \etal (2008) identified \chandra source 4\_5 as the
X-ray pulsar SXP892.

{\bf \chandra source 4\_11:} Source O\_4\_164855 is the single match
of \chandra source 4\_11 within the search radius with $V$ magnitude
and $B-V$ color which place it just below the locus of OB stars in the
CMD. We mention here that in $\sim2\sigma$ from the X-ray source position there is
a bright variable star (O\_4\_163513$\equiv$OGLE 004802.55-731659.8;
Ita \etal 2004). However,
the B-V color of this optical source is not indicative of an O or
B-type star, thus even if its X-ray spectrum is possibly hard this
source remains unclassified.

\subsection{Sources with available spectral classification for their counterparts}\label{appendix2dF}

The sources presented here (with the exception of \chandra source 5\_16) have been identified spectroscopically in the 2dF survey of the SMC by Evans \etal (2004; hereafter [2dF]) and/or are reported in the emission-line object compilation of Meyssonnier \& Azzopardi (1993; hereafter [MA93]). The luminosity classes shown in parenthesis indicate an hybrid photometric/spectroscopic approach and are not true, MK-process, morphological types (for more details see Evans \etal 2004).

{\bf \chandra source 5\_1:} The identified optical counterpart in this case is MCPS source Z\_2311496, in agreement with Coe \etal (2005). This source is also defined as a B0-5(II)e star in the 2dF spectroscopic catalog of Evans \etal (2004; source [2dF]No.0839).

{\bf \chandra source 5\_3:} The optical counterpart O\_6\_85614 of
source 5\_3 is identified as a B0-5(II) spectral-type source ([2dF]No.0828). 

{\bf \chandra source 5\_16:} A single match (source Z\_2573354) is identified as the optical counterpart of this source in the MCPS catalog (no OGLE-II overlap), in agreement with Coe \etal (2005a). Buckley \etal (2001) identified source 5\_16 as a transient pulsar
 and spectroscopically classified the suggested counterpart (star A or
 B mentioned in their study) as a B1-B2III-Ve star. On the other hand,
 the variable MACHO source 207.16202.30 lies $>14\arcsec$ away of
 X-ray source 5\_16 (unlikely the correct counterpart; Coe \etal 2005a).

{\bf \chandra source 6\_1:} We identify source O\_6\_77228 as the
optical counterpart of X-ray source 6\_1 (in agreement with Coe \etal
2005a). This is source [2dF]No.5054 classified as a B1-5(II)e star
(Evans \etal 2004). It also appears as a candidate Be star in the
compilation of Mennickent \etal (2002; [MPG02]No.341). For more
details see \S \ref{Becandidatecounterparts}.  

{\bf \chandra source 6\_20:} Source O\_6\_311169 (single match) is
proposed here as the optical counterpart of X-ray source 6\_20. It is
the brightest counterpart identified in this study, while it has been
identified as the emission-line object [MA93]No.739 (peculiar \Ha
emission-line star with FeII and [FeII] emission; $\sim0.17\arcsec$
from the OGLE-II source, and $\sim0.8\arcsec$ from the X-ray
source). The latter source has been identified as a B[e] star by
Massey \& Duffy (2001). We only propose the above OGLE-II source as
the counterpart of \chandra source 6\_20, and we note that further
analysis of this peculiar system is required so as to confirm its
nature (see also \ref{uncertainnature}).

\subsection{\chandra sources with bright optical counterparts
($M_{V_{o}}\leq-0.25$) only between the $1\sigma$ and $2\sigma$ search
radii}\label{SevenNew}

   Out of the 33 X-ray sources without any OGLE-II or MCPS match in the
$1\sigma$ search radius, there are 7 \chandra sources with
counterparts in the $2\sigma$ search radius. These sources are
presented in Table \ref{BrightNor}. The description of this table is
similar to that of Tables \ref{field3}-\ref{field7} (given in \S \ref{OGLEMCPS}), with the only difference that there is only one table
for all the sources found in the different \chandra fields (separated
by double horizontal lines). In Table
\ref{properties2r} we also present the optical and X-ray properties of
these sources (see also \S \ref{properties}).

\subsection{\chandra sources with optical counterparts in the locus of
OB stars and hard or possibly hard spectrum between the $1\sigma$ and
$2\sigma$ search radii}\label{brighter}

{\bf \chandra source 3\_2:} This source is a pulsar (Edge \etal
2004). However, both of its optical matches in the MCPS catalog
presented in Table \ref{field3} (no overlap with OGLE-II) are faint stars. Due to the high 
chance coincidence probability of faint sources, we do not propose
any of them as counterpart of X-ray source 3\_2. Instead, looking for
counterparts between the $1\sigma$ and $2\sigma$ search radii, we find
MCPS source Z\_2806702 (in agreement with
Edge \etal 2004). This is a bright star ($V=16.78$,
$B-V=-0.12$), which is located $4.27\arcsec$ away from the X-ray
position. Although no spectroscopic data for these 3
sources are available, the early-type nature of MCPS source Z\_2806702, make us believe this is the most likely
counterpart. Recently, Haberl \& Eger (2008; ATel\#1529) suggested
that \chandra source 3\_2 is a background AGN and not a Be-XRB pulsar,
with an optical counterpart one of the two faint stars with V$=$20.3
mag found in the $1\sigma$ search radius.

{\bf \chandra source 4\_13:} We have identified 
source O\_5\_90858 as the optical counterpart of 4\_13. However, when we look
for counterparts between the $1\sigma$ and $2\sigma$ search radii, we find 2 
sources brighter than O\_5\_90858. These sources, O\_5\_90493 and
O\_5\_90535, ($5.47\arcsec$ and $4.10\arcsec$ away from
the X-ray source, respectively) are both of OB spectral type ($V=15.57$,
$B-V=-0.07$ and $V=16.34$, $B-V=0.23$, respectively). The possibly
hard X-ray spectrum of 4\_13 suggests that this is a new candidate
Be-XRB. However, only optical spectroscopy could determine the exact
type of these bright stars and the true counterpart of
\chandra source 4\_13.

{\bf \chandra source 4\_19:} The optical source O\_5\_261182 is found
$2.27\arcsec$ away from \chandra source 4\_19 (between the $1\sigma$
and $2\sigma$ search radius), and it is an OB spectral-type star
($V=18.16$, $B-V=-0.08$). In combination with its possibly hard X-ray
spectrum, we propose this is a new HMXB. Optical spectroscopy will definitely 
determine if this source is also a new Be-XRB.

{}

\clearpage

\makeatletter
\def\jnl@aj{AJ}
\ifx\revtex@jnl\jnl@aj\let\tablebreak=\nl\fi
\makeatother
\begin{deluxetable}{ccccc}
\tabletypesize{\scriptsize}
\tablecolumns{5}
\tablewidth{0pt}
\tablecaption{\chandra source data statistics\label{tablesurvey}}
\tablehead{\colhead{\chandra} & \colhead{Right Ascension (J2000.0)} & \colhead{Declination (J2000.0)} & \colhead{Detected} & \colhead{Sources with} \\
\colhead{field} & \colhead{(h m s)} & \colhead{(\deg\, \arcmin\, \arcsec\,)} & \colhead{sources} & \colhead{$\rm{D_{off-axis}<10}\arcmin$} \\[0.2cm]
\colhead{[1]} & \colhead{[2]} & \colhead{[3]} & \colhead{[4]} & \colhead{[5]}}
\startdata
3 & 00 56 46.14  & -72 18 10.78 & 28 (11) & 27 (10) \\
4 & 00 49 30.74  & -73 16 52.34 & 38 (16) & 36 (16) \\
5 & 00 53 11.45  & -72 26 29.91 & 37 (11) & 36 (11) \\
6 & 00 53 04.40  & -72 42 18.22 & 28 (8) & 28 (8) \\
7 & 00 49 25.09  & -72 44 22.80 & 27 (11) & 26 (11) \\
\enddata
\end{deluxetable}

\makeatletter
\def\jnl@aj{AJ}
\ifx\revtex@jnl\jnl@aj\let\tablebreak=\nl\fi
\makeatother
\begin{deluxetable}{ccccccccccccc}
\tablewidth{0pt}
\tabletypesize{\scriptsize}
\tablecolumns{13}
\tablecaption{Optical counterparts of X-ray sources in \chandra field 3\label{field3}}
\tablehead{\multicolumn{2}{c}{X-ray} & \multicolumn{10}{c}{Optical counterpart} & \colhead{Notes} \\
\colhead{Src} & \colhead{Rad.} & \colhead{Src ID} & \colhead{R.A.} & \colhead{Decl.}  & \colhead{Offset} & \colhead{{\it V}} & \colhead{$\rm{\delta}V$} & \colhead{{\it B-V}} & \colhead{$\rm{\delta}\rm{(}B-V\rm{)}$} &  \colhead{{\it U-B}} & \colhead{$\rm{\delta}\rm{(}U-B\rm{)}$}  & \colhead{} \\
\colhead{} & \colhead{($\arcsec$)}  & \colhead{} &\multicolumn{2}{c}{(J2000.0)} & \colhead{($\arcsec$)} & \colhead{} & \colhead{} & \colhead{} & \colhead{} & \colhead{} & \colhead{} & \colhead{} \\[0.2cm]
\colhead{[1]} & \colhead{[2]} & \colhead{[3]} & \colhead{[4]} & \colhead{[5]} & \colhead{[6]} & \colhead{[7]} & \colhead{[8]} & \colhead{[9]} & \colhead{[10]} & \colhead{[11]}  & \colhead{[12]}  & \colhead{[13]} }
\startdata
3\_1   &  1.58  &   {\bf O\_7\_267132}   &  00 57 19.80   &  -72 25 34.0   &  0.13  & 19.76 & 0.30 & 99.99 & 99.99 &  \nodata   &  \nodata     & u, c \\  
    &    &   {\bf O\_7\_267163}  &  00 57 19.84   &  -72 25 33.5   &  0.54  & 19.54 & 0.33 & 99.99 & 99.99 &  \nodata   &  \nodata   & \\ 
    &    &   (Z\_3066294)    &  00 57 19.87   &  -72 25 34.0   &  0.21  & 18.66 & 0.06 & 0.78 & 0.08 & -0.63 & 0.09   & \\[0.1cm]
 
    &    &   O\_7\_269933   &  00 57 19.77   &  -72 25 35.4   &  1.39  & 21.51 & 0.55 & 99.99 & 99.99 &  \nodata   &  \nodata   & \\ 
 
\hline
 
3\_2   &  2.18  &   {\bf Z\_2804383}     &  00 55 27.52   &  -72 11 00.5   &  2.01  & 20.30 & 0.07 & -0.38 & 0.14 & -0.33 & 0.16 & u, c \\[0.1cm] 
 
    &    &   {\bf Z\_2804939}     &  00 55 27.75   &  -72 10 58.7   &  0.28  & 20.30 & 0.08 & 1.17 & 0.20 & -1.85 & 0.22   & \\ 
 
\hline 
3\_3   &  1.50  &   {\bf O\_8\_49531}    &  00 57 36.01   &  -72 19 33.8   &  0.14  & 16.01 & 0.02 & -0.02 & 0.04 &  \nodata  &  \nodata   & c  \\ 
 
    &    &   Z\_3103982     &  00 57 36.09   &  -72 19 33.6   &  0.53  & 15.99 & 0.03 & 0.01 & 0.17 & -1.05 & 0.17  &  \\
\enddata
\tablenotetext{\ast}{Sources which have OGLE-II coverage (thus available on-line finding charts) but no counterpart(s) (either in OGLE-II or MCPS catalogs).}
\tablecomments{Sources 3\_6, 3\_8, 3\_10, 3\_11, 3\_13, 3\_16, 3\_20, 3\_21, 3\_23 to 3\_25, 3\_27, and 3\_28 do not have OGLE-II coverage (thus no finding charts), and any MCPS counterpart(s).}
\end{deluxetable}

\makeatletter
\def\jnl@aj{AJ}
\ifx\revtex@jnl\jnl@aj\let\tablebreak=\nl\fi
\makeatother
\begin{deluxetable}{ccccccccccccc}
\tablewidth{0pt}
\tabletypesize{\scriptsize}
\tablecolumns{13}
\tablecaption{Optical counterparts of X-ray sources in \chandra field 4\label{field4}}
\tablehead{\multicolumn{2}{c}{X-ray} & \multicolumn{10}{c}{Optical counterpart} & \colhead{Notes} \\
\colhead{Src} & \colhead{Rad.} & \colhead{Src ID} & \colhead{R.A.} & \colhead{Decl.}  & \colhead{Offset} & \colhead{{\it V}} & \colhead{$\rm{\delta}V$} & \colhead{{\it B-V}} & \colhead{$\rm{\delta}\rm{(}B-V\rm{)}$} &  \colhead{{\it U-B}} & \colhead{$\rm{\delta}\rm{(}U-B\rm{)}$}  & \colhead{} \\
\colhead{} & \colhead{($\arcsec$)}  & \colhead{} &\multicolumn{2}{c}{(J2000.0)} & \colhead{($\arcsec$)} & \colhead{} & \colhead{} & \colhead{} & \colhead{} & \colhead{} & \colhead{} & \colhead{} \\[0.2cm]
\colhead{[1]} & \colhead{[2]} & \colhead{[3]} & \colhead{[4]} & \colhead{[5]} & \colhead{[6]} & \colhead{[7]} & \colhead{[8]} & \colhead{[9]} & \colhead{[10]} & \colhead{[11]}  & \colhead{[12]}  & \colhead{[13]} }
\startdata
4\_1   & 1.50   &   {\bf O\_5\_180026}  &  00 50 44.71  &  -73 16 05.0  &  0.55  & 15.44 & 0.04 & -0.04 & 0.05 &  \nodata   &  \nodata  & c \\ 
 
    &    &   Z\_2131651    &  00 50 44.75  &  -73 16 05.3  &  0.60  & 15.48 & 0.04 & -0.11 & 0.05 & -0.91 & 0.04 & \\ 

\hline 
4\_2   & 1.50   &   {\bf O\_5\_111490}  &  00 49 13.63  &  -73 11 37.4  &  0.47  & 16.52 & 0.02 & 0.10 & 0.04 &  \nodata     &  \nodata  & c \\ 
    &    &   Z\_1938257    &  00 49 13.68  &  -73 11 37.5  &  0.54  & 16.44 & 0.04 & 0.19 & 0.05 & -0.92 & 0.04 &   \\[0.1cm] 

    &    &   O\_5\_115042  &  00 49 13.57  &  -73 11 39.1  &  1.31  & 21.53 & 0.48 & 99.99 & 99.99 &  \nodata      &  \nodata  &   \\ 
  
\hline 
4\_3   & 2.10   &   {\bf O\_5\_271074}  &  00 50 57.12  &  -73 10 07.7  &  0.28  & 14.54 & 0.01 & -0.06 & 0.01 &  \nodata     &  \nodata  & c \\ 
 
    &    &   Z\_2159045    &  00 50 57.16  &  -73 10 07.8  &  0.14  & 14.35 & 0.05 & 0.08 & 0.06 & -0.88 & 0.04 & \\
 
\hline 
4\_4   & 1.50   &   {\bf O\_4\_159896}  &  00 48 18.67  &  -73 20 59.7  &  0.54  & 16.58 & 0.02 & -0.10 & 0.03 &  \nodata    &  \nodata  & n, c \\
\enddata
\tablenotetext{\dag}{Sources detected in two neighboring OGLE-II fields.}
\tablenotetext{o}{Just outside the error circle of the X-ray source there is another OGLE-II source (not listed in these tables), that it is most likely related to the detected MCPS source within the search radius.}
\tablenotetext{d}{OGLE-II detected two sources instead of one star (suggested by the visual inspection of the finding chart).}
\tablenotetext{\ast}{Sources which have OGLE-II coverage (thus available on-line finding charts) but no counterpart(s) (either in OGLE-II or MCPS catalogs).}
\tablecomments{Sources 4\_9 and 4\_38 do not have OGLE-II coverage (thus no finding charts), and any MCPS counterpart(s).}
\end{deluxetable}

\makeatletter
\def\jnl@aj{AJ}
\ifx\revtex@jnl\jnl@aj\let\tablebreak=\nl\fi
\makeatother
\begin{deluxetable}{ccccccccccccc}
\tablewidth{0pt}
\tabletypesize{\scriptsize}
\tablecolumns{13}
\tablecaption{Optical counterparts of X-ray sources in \chandra field 5\label{field5}}
\tablehead{\multicolumn{2}{c}{X-ray} & \multicolumn{10}{c}{Optical counterpart} & \colhead{Notes} \\
\colhead{Src} & \colhead{Rad.} & \colhead{Src ID} & \colhead{R.A.} & \colhead{Decl.}  & \colhead{Offset} & \colhead{{\it V}} & \colhead{$\rm{\delta}V$} & \colhead{{\it B-V}} & \colhead{$\rm{\delta}\rm{(}B-V\rm{)}$} &  \colhead{{\it U-B}} & \colhead{$\rm{\delta}\rm{(}U-B\rm{)}$}  & \colhead{} \\
\colhead{} & \colhead{($\arcsec$)}  & \colhead{} &\multicolumn{2}{c}{(J2000.0)} & \colhead{($\arcsec$)} & \colhead{} & \colhead{} & \colhead{} & \colhead{} & \colhead{} & \colhead{} & \colhead{} \\[0.2cm]
\colhead{[1]} & \colhead{[2]} & \colhead{[3]} & \colhead{[4]} & \colhead{[5]} & \colhead{[6]} & \colhead{[7]} & \colhead{[8]} & \colhead{[9]} & \colhead{[10]} & \colhead{[11]}  & \colhead{[12]}  & \colhead{[13]} }
\startdata
5\_1\tablenotemark{\ast\ast}   & 1.50    &   {\bf Z\_2311496}    &  00 52 05.69  &  -72 26 04.0   &  0.55  & 14.91 & 0.02 & 0.00 & 0.03 & -0.97 & 0.04 & c \\ 
 
\hline 
5\_2\tablenotemark{\ast\ast}   & 1.50    &   {\bf Z\_2498173}    &  00 53 23.90  &  -72 27 15.4   &  0.24  & 16.19 & 0.12 & -0.09 & 0.12 & -1.05 & 0.04 & c \\ 
 
\hline 
5\_3   & 1.62    &   {\bf O\_6\_85614}   &  00 51 53.11   &  -72 31 48.3  &  0.54  & 14.90 & 0.12 & -0.27 & 0.13 &  \nodata  &  \nodata  & c \\ 
    &     & Z\_2282823 &  00 51 53.18  &  -72 31 48.4   &  0.43  & 14.38 & 0.02 & 0.41 & 0.03 & -1.13 & 0.11 &  \\[0.1cm] 
  
    &     &   O\_6\_85986   &  00 51 53.30   &  -72 31 49.1  &  0.69  & 99.99 & 99.99 & 99.99 & 99.99 &  \nodata  &  \nodata  &  \\ 
 
\hline 
5\_4   & 2.50    &   {\bf O\_7\_70843}   &  00 54 46.37   &  -72 25 22.6  &  0.79  & 15.58 & 0.02 & 99.99 & 99.99 &  \nodata  &  \nodata  & n, c \\  
    &     &   Z\_2707354    &  00 54 46.39  &  -72 25 22.7   &  0.83  & 15.36 & 0.05 & 0.14 & 0.06 & -1.11 & 0.13 &  \\[0.1cm]  
  
    &     &   O\_7\_71473   &  00 54 46.06   &  -72 25 24.0  &  1.23  & 19.57 & 0.13 & 99.99 & 99.99 &  \nodata  &  \nodata  &  \\
\enddata
\tablenotetext{\dag}{Sources detected in two neighboring OGLE-II fields.}
\tablenotetext{o}{Just outside the error circle of the X-ray source there is another OGLE-II source (not listed in these tables), that it is most likely related to the detected MCPS source within the search radius.}
\tablenotetext{\ast}{Source with OGLE-II coverage (thus available on-line finding chart) but no counterpart(s) (either in OGLE-II or MCPS catalogs).}
\tablenotetext{\ast\ast}{Sources which do not have OGLE-II coverage (thus no available on-line finding charts) but they have MCPS counterparts within their search radius.}
\tablecomments{Sources 5\_5, 5\_11, 5\_17, 5\_19, 5\_20, and 5\_35 do not have OGLE-II coverage (thus no finding charts), and any MCPS counterpart(s).}
\end{deluxetable}

\makeatletter
\def\jnl@aj{AJ}
\ifx\revtex@jnl\jnl@aj\let\tablebreak=\nl\fi
\makeatother
\begin{deluxetable}{ccccccccccccc}
\tablewidth{0pt}
\tabletypesize{\scriptsize}
\tablecolumns{13}
\tablecaption{Optical counterparts of X-ray sources in \chandra field 6\label{field6}}
\tablehead{\multicolumn{2}{c}{X-ray} & \multicolumn{10}{c}{Optical counterpart} & \colhead{Notes} \\
\colhead{Src} & \colhead{Rad.} & \colhead{Src ID} & \colhead{R.A.} & \colhead{Decl.}  & \colhead{Offset} & \colhead{{\it V}} & \colhead{$\rm{\delta}V$} & \colhead{{\it B-V}} & \colhead{$\rm{\delta}\rm{(}B-V\rm{)}$} &  \colhead{{\it U-B}} & \colhead{$\rm{\delta}\rm{(}U-B\rm{)}$}  & \colhead{} \\
\colhead{} & \colhead{($\arcsec$)}  & \colhead{} &\multicolumn{2}{c}{(J2000.0)} & \colhead{($\arcsec$)} & \colhead{} & \colhead{} & \colhead{} & \colhead{} & \colhead{} & \colhead{} & \colhead{} \\[0.2cm]
\colhead{[1]} & \colhead{[2]} & \colhead{[3]} & \colhead{[4]} & \colhead{[5]} & \colhead{[6]} & \colhead{[7]} & \colhead{[8]} & \colhead{[9]} & \colhead{[10]} & \colhead{[11]}  & \colhead{[12]}  & \colhead{[13]} }
\startdata
6\_1   & 1.50    &   {\bf O\_6\_77228}   &  00 52 08.95  &  -72 38 02.9   &  0.58  & 15.03 & 0.02 & 0.14 & 0.03 &  \nodata  &  \nodata  &  c  \\ 
    &     &   Z\_2319498    &  00 52 09.07  &  -72 38 03.1   &  0.64  & 15.23 & 0.03 & -0.08 & 0.03 & -0.81 & 0.04 &                  \\ 
 
\hline 
6\_2  & 1.91    &   {\bf O\_7\_47103}   &  00 54 55.87  &  -72 45 10.7   &  0.40  & 15.01 & 0.01 & -0.02 & 0.01 &  \nodata  &  \nodata   & c  \\  
    &     &   Z\_2729974    &  00 54 55.91  &  -72 45 10.6    &  0.61  & 15.00 & 0.03 & -0.03 & 0.04 & -0.93 & 0.18 &  \\ 
 
\hline 
6\_3   & 1.50    &   {\bf O\_6\_153267}  &  00 52 42.20  &  -72 47 21.0   &  1.38  & 18.15 & 0.07 & -0.14 & 0.08 &  \nodata  &  \nodata  & n \\ 
    &     &   Z\_2399147    &  00 52 42.27  &  -72 47 20.9    &  1.30  & 17.81 & 0.06 & 0.04 & 0.08 & -0.52 & 0.07 &  \\[0.1cm]
 
    &     &   O\_6\_154606  &  00 52 42.24  &  -72 47 19.1   &  0.54  & 20.16 & 0.09 & 0.52 & 0.15 &  \nodata  &  \nodata  &   \\[0.1cm] 
 
    &     &   O\_6\_155950  &  00 52 42.40  &  -72 47 20.6   &  1.15  & 20.39 & 0.21 & 0.09 & 0.27 &  \nodata  &  \nodata  &  \\
\enddata
\tablenotetext{\dag}{Sources detected in two neighboring OGLE-II fields.}
\tablenotetext{\ddag}{The {\it I}\, band photometry in the MCPS catalog was replaced with that from the OGLE-II (see Zaritsky et al. (2002) for more details).}
\tablenotetext{o}{Just outside the error circle of the X-ray source there is another OGLE-II source (not listed in these tables), that it is most likely related to the detected MCPS source within the search radius.}
\tablenotetext{fc}{The finding chart of \chandra source 6\_18 includes also source 6\_15 as these sources are too near to each other.}
\tablenotetext{\ast}{Sources which have OGLE-II coverage (thus available on-line finding charts) but no counterpart(s) (either in OGLE-II or MCPS catalogs).}
\end{deluxetable}

\makeatletter
\def\jnl@aj{AJ}
\ifx\revtex@jnl\jnl@aj\let\tablebreak=\nl\fi
\makeatother
\begin{deluxetable}{ccccccccccccc}
\tablewidth{0pt}
\tabletypesize{\scriptsize}
\tablecolumns{13}
\tablecaption{Optical counterparts of X-ray sources in \chandra field 7\label{field7}}
\tablehead{\multicolumn{2}{c}{X-ray} & \multicolumn{10}{c}{Optical counterpart} & \colhead{Notes} \\
\colhead{Src} & \colhead{Rad.} & \colhead{Src ID} & \colhead{R.A.} & \colhead{Decl.}  & \colhead{Offset} & \colhead{{\it V}} & \colhead{$\rm{\delta}V$} & \colhead{{\it B-V}} & \colhead{$\rm{\delta}\rm{(}B-V\rm{)}$} &  \colhead{{\it U-B}} & \colhead{$\rm{\delta}\rm{(}U-B\rm{)}$}  & \colhead{} \\
\colhead{} & \colhead{($\arcsec$)}  & \colhead{} &\multicolumn{2}{c}{(J2000.0)} & \colhead{($\arcsec$)} & \colhead{} & \colhead{} & \colhead{} & \colhead{} & \colhead{} & \colhead{} & \colhead{} \\[0.2cm]
\colhead{[1]} & \colhead{[2]} & \colhead{[3]} & \colhead{[4]} & \colhead{[5]} & \colhead{[6]} & \colhead{[7]} & \colhead{[8]} & \colhead{[9]} & \colhead{[10]} & \colhead{[11]}  & \colhead{[12]}  & \colhead{[13]} }
\startdata
7\_1   & 1.50   &   {\bf O\_5\_65517}   &  00 49 03.34  &  -72 50 52.1  &  0.45  & 16.94 & 0.06 & 0.09 & 0.10 &  \nodata  &  \nodata  & c \\ 
    &    &   Z\_1916984    &  00 49 03.41  &  -72 50 52.3   &  0.35  & 16.78 & 0.04 & 0.08 & 0.06 & -0.84 & 0.05 &  \\[0.1cm] 
 
    &    &   O\_5\_65743   &  00 49 03.30  &  -72 50 53.7  &  1.20  & 18.86 & 0.23 & 0.15 & 0.38 &  \nodata  &  \nodata  &  \\ 
    &    &   (Z\_1916967)    &  00 49 03.40  &  -72 50 53.9  &  1.35  & 18.08 & 0.06 & 0.07 & 0.12 & -0.32 & 0.13 &  \\[0.1cm] 

    &    &   O\_5\_66768   &  00 49 03.24  &  -72 50 53.2  &  0.86  & 18.89 & 0.28 & -0.08 & 0.37 &  \nodata  &  \nodata  &  \\ 
 
\hline 
7\_2\tablenotemark{\ast\ast}   & 1.50   &   {\bf Z\_1976595}    &  00 49 32.05  &  -72 39 25.2   &  0.40  & 20.50 & 0.09 & 0.33 & 0.13 & 99.99 & 99.99 &  n \\ 
 
\hline 	 
7\_3   & 1.60   &   O\_5\_147467  &  00 50 00.38  &  -72 49 28.9  &  0.76  & 20.75 & 0.29 & 99.99 & 99.99 &  \nodata  &  \nodata  &  n, c \\	 
    &    &   O\_5\_149072  &  00 50 00.17  &  -72 49 28.8  &  0.19  & 20.78 & 0.42 & 99.99 & 99.99 &  \nodata  &  \nodata  &   \\
\enddata
\tablenotetext{\dag}{Sources detected in two neighboring OGLE-II fields.}
\tablenotetext{o}{Just outside the error circle of the X-ray source there is another OGLE-II source (not listed in these tables), that it is most likely related to the detected MCPS source within the search radius.}
\tablenotetext{\ast}{Sources with OGLE-II coverage (thus available on-line finding chart) but no counterpart(s) (either in OGLE-II or MCPS catalogs).}
\tablenotetext{\ast\ast}{Sources without OGLE-II coverage (thus no available on-line finding charts) but MCPS counterparts within their search radius.}
\tablecomments{Sources 7\_8 and 7\_14 do not have OGLE-II coverage (thus no finding charts), and any MCPS counterpart(s).}
\end{deluxetable}

\makeatletter
\def\jnl@aj{AJ}
\ifx\revtex@jnl\jnl@aj\let\tablebreak=\nl\fi
\makeatother
\begin{deluxetable}{ccccccc}
\tabletypesize{\scriptsize}
\tablecolumns{7}
\tablewidth{0pt}
\tablecaption{Total number of counterparts\label{tableresults}}
\tablehead{\colhead{} & \colhead{} & \colhead{} & \multicolumn{4}{c}{X-ray sources} \\
\colhead{\chandra} & \colhead{Overlap by} & \colhead{Catalog} & \multicolumn{4}{c}{with total number of counterparts}  \\
\colhead{field ID}  & \colhead{OGLE-II survey} & \colhead{used} & \colhead{0} & \colhead{1} & \colhead{2} & \colhead{$>2$} \\[0.2cm]
\colhead{[1]} & \colhead{[2]} & \colhead{[3]} & \colhead{[4]} & \colhead{[5]} & \colhead{[6]} & \colhead{[7]}}
\startdata
3   & partial    & O           & 2    & 3    & 0  & 1  \\
    &            &  Z          & 14  & 11  & 2  & 0  \\
    &            &  combined    &  14 & 10  & 2  & 1  \\
\hline
4   & full       & O          &  3   & 14  & 8   & 11  \\
    &            &  Z          & 14 & 16  & 3   & 3  \\
    &            &  combined    & 3    & 12  & 11 & 10  \\   
\hline
5   & partial    & O           & 1  & 3    & 3  & 4  \\
    &            &  Z          & 7  & 19  & 6  & 4  \\
    &            &  combined    &  6 & 13  & 9  & 8  \\     
\hline
6   & full       & O           & 5  & 9   & 8  & 6  \\
    &            &  Z          & 11& 13 & 3  & 1  \\
    &            &  combined    & 5   & 9   & 8  & 6  \\     
\hline
7   & partial    & O           & 4  &  6 & 4  & 8  \\
    &            &  Z          & 9  & 10  & 5  & 1  \\
    &            &  combined    & 5  & 8   & 3  & 10
\enddata
\end{deluxetable}

\makeatletter
\def\jnl@aj{AJ}
\ifx\revtex@jnl\jnl@aj\let\tablebreak=\nl\fi
\makeatother
\begin{deluxetable}{ccccccc}
\tablewidth{0pt}
\tabletypesize{\scriptsize}
\tablecolumns{7}
\tablecaption{Additional optical info\label{tableotherinfo}}
\tablehead{\colhead{X-ray} & \colhead{Type\tablenotemark{*}} & \colhead{Ref.} & \colhead{C/part of} & \colhead{Ref.} & \colhead{Other} & \colhead{Ref.} \\
\colhead{source} & \colhead{} & \colhead{} & \colhead{associated} & \colhead{} & \colhead{sources} & \colhead{} \\
\colhead{ID} & \colhead{} & \colhead{} & \colhead{X-ray source} & \colhead{} & \colhead{within $1.5\arcsec$} & \colhead{}  \\[0.2cm]
\colhead{[1]} & \colhead{[2]} & \colhead{[3]} & \colhead{[4]} & \colhead{[5]} & \colhead{[6]} & \colhead{[7]}}
\startdata
3\_1   & AGN                     & [DSM03]         & OGLE00571981-7225337 & [SPH03]   & [USNO-A2.0]0150-00625436                       & [SPH03] \\ 
           & variable               & [ZSW01]        &                                                  &           &                                               &         \\ 					   

\hline
3\_2  & Be-XBP                            & [ECG04]         & Z\_2806702   & [ECG04],[CEG05] & \nodata & \nodata \\
      & or bkg AGN?                  & [HE08]          &              &                 &        &       \\

\hline					   
3\_3   & Be-XBP    & [MFL03]    &[MA93]1020     & [MFL03],[HP04]                     & [M02]40790     & [M02] \\ 
       & Be?                            & [SPH03]         & [MA93]1020$\equiv$O\_8\_49531     & [SPH03]     &     &   \\
       & Be candidate      & [MPG02]           & O\_8\_49531$\equiv$                                   & [CEG05]   &                            &         \\  
       &                        &          & $\equiv$[MACHO]207.16432.1575          & [CEG05]                    &                            & \\
\enddata
\tablenotetext{*}{Be? H${\alpha}$ emission-line object as possible counterpart, no optical spectrum is available in the literature in this case, HMXB?: HMXB candidate, P: pulsar, Be/X?: very promising candidate for Be-XRBs, Be-XBP: Be-XRB pulsar}
\tablerefs{\tiny{Azzopardi \& Vigneau (1982) [AV82], Bica \etal (1995) [BS95], Buckley \etal (2001) [BCS01], Cioni \etal (2000) [CI00], Clark \etal (1978) [CDL78], Coe \etal (2005a) [CEG05], Coe \etal (2005b) [CNM05], Coe \& Orosz (2000) [CO00], Corbet \etal (1998) [CML98], Corbet \etal (2002) [CMM02], Cowley \etal (1997) [CSM97], DENIS Consortium (2003) [DENIS03], DENIS Consortium (2005) [DENIS05], Dobrzycki \etal (2003a) [DMS03], Dobrzycki \etal (2003b) [DSM03], Edge \& Coe (2003) [EC03], Edge \etal (2004) [ECG04], Evans \etal (2004) [EHI04], Fabrycky, D. (2005) [F05], Garmany, Conti \& Massey (1987) [GCM87], Haberl \etal (2000) [HFP00],  Haberl \& Pietsch (2004) [HP04], Haberl \etal (2004) [HPS04], Haberl \& Sasaki (2000) [HS00], Haberl \& Eger (2008) [HE08], Haberl \etal (2008) [HEP08], Israel \etal (1995) [ISA95], Ita \etal (2004) [ITM04], Macomb \etal (2003) [MFL03], Marshall \etal (1998) [MLS98], Massey 2002 [M02], Massey \& Duffy (2001) [MD01], Mennickent \etal (2002) [MPG02], Meyssonnier \& Azzopardi (1993) [MA93], Monet \etal (2003) [MLC03], Murphy \& Bessell (2000) [MB00], Oey \etal (2004) [OKP04], Paczynski, B. (2001) [PAC01], Sasaki \etal (2000) [SHP00], Sasaki \etal (2003) [SPH03], Schmidtke \& Cowley (2005,2006) [SC05,SC06], Shtykovskiy \& Gilfanov (2005) [SG05], Soszynski \etal (2002) [SZU02], Stevens, Coe \& Buckley (1999) [SCB99], STScI 2001 [GSC2.2], Tucholke, de Boer \& Seitter (1996) [TBS96], Wyrzykowski \etal (2004) [WUK04], Yokogawa \etal (2003) [YIT03], Zacharias \etal (2004) [ZA04], Zebrun \etal (2001) [ZSW01]}}
\end{deluxetable}

\makeatletter
\def\jnl@aj{AJ}
\ifx\revtex@jnl\jnl@aj\let\tablebreak=\nl\fi
\makeatother
\begin{deluxetable}{cccccc}
\tabletypesize{\scriptsize}
\tablecolumns{6}
\tablewidth{0pt}
\tablecaption{Chance associations between the \chandra sources and the OGLE-II stars\label{chanceOGLE}}
\tablehead{\colhead{} &  \colhead{} &  \multicolumn{4}{c}{Chance coincidence probability} \\
\colhead{Magnitude} & \colhead{Color} & \multicolumn{2}{c}{single matches} & \multicolumn{2}{c}{multiple matches} \\
\colhead{range} & \colhead{range} & \colhead{r=1.5$\arcsec$} & \colhead{r=2.5$\arcsec$} & \colhead{r=1.5$\arcsec$} & \colhead{r=2.5$\arcsec$}  \\[0.2cm]
\colhead{[1]} & \colhead{[2]} & \colhead{[3]} & \colhead{[4]} & \colhead{[5]} & \colhead{[6]}} 
\startdata
$M_{{V}_{o}}\leq-4.5$ &  all                 &$0.01\pm0.16$  & $0.02\pm0.14$   & $0.01\pm0.16$  & $0.02\pm0.12$ \\[0.15cm] 
                                        & $(B-V)_{o}\leq -0.31$\tablenotemark{\dag} & \nodata               & \nodata                 & \nodata               & \nodata  \\[0.15cm]
                                        & $(B-V)_{o}> -0.31$     & $0.01\pm0.15$  & $0.02\pm0.14$  & $0.01\pm0.15$  & $0.02\pm0.12$  \\[0.2cm]
							                					
\hline
\\							                					
$-4.5<M_{{V}_{o}}\leq-0.25$ &  all                 & $0.47\pm0.35$  & $0.82\pm0.17$   & $0.24\pm0.23$  & $0.65\pm0.24$ \\[0.15cm]

                                                    & $(B-V)_{o}\leq -0.11$ & $0.19\pm0.30$  & $0.70\pm0.28$  & $0.19\pm0.32$  & $0.52\pm0.33$  \\[0.15cm]
                                                    & $(B-V)_{o}> -0.11$ & $0.31\pm0.34$  & $0.81(>0.61)$ & $0.24\pm0.28$  & $0.67\pm0.31$  \\[0.2cm]

\hline
\\
\enddata
\tablecomments{These are normalized chance coincidence probabilities
  (OGLE-II matches per \chandra source), thus not additive.} 
\tablenotetext{\dag}{In the range $M_{{V}_{o}}\leq-4.5$ and $(B-V)_{o}\leq -0.31$ there are not stars, thus the chance coincidence probability is not estimated there.}
\end{deluxetable}

\makeatletter
\def\jnl@aj{AJ}
\ifx\revtex@jnl\jnl@aj\let\tablebreak=\nl\fi
\makeatother
\begin{deluxetable}{cccccc}
\tabletypesize{\scriptsize}
\tablecolumns{6}
\tablewidth{0pt}
\tablecaption{Chance associations between the \chandra sources and the MCPS stars\label{chanceMCPS}}
\tablehead{\colhead{} &  \colhead{} &  \multicolumn{4}{c}{Chance coincidence probability} \\
\colhead{Magnitude} & \colhead{Color} & \multicolumn{2}{c}{single matches} & \multicolumn{2}{c}{multiple matches} \\
\colhead{range} & \colhead{range} & \colhead{r=1.5$\arcsec$} & \colhead{r=2.5$\arcsec$} & \colhead{r=1.5$\arcsec$} & \colhead{r=2.5$\arcsec$}  \\[0.2cm]
\colhead{[1]} & \colhead{[2]} & \colhead{[3]} & \colhead{[4]} & \colhead{[5]} & \colhead{[6]}} 
\startdata
$M_{{V}_{o}}\leq-4.5$ &  all                 & $0.01\pm0.15$ & $0.05\pm0.16$   & $0.01\pm0.15$ & $0.04\pm0.13$  \\[0.15cm] 
                                        & $(B-V)_{o}\leq -0.31$\tablenotemark{\dag} & \nodata                & \nodata               & \nodata               & \nodata \\[0.15cm]
                                        & $(B-V)_{o}> -0.31$     & $0.01\pm0.15$  & $0.05\pm0.15$  & $0.01\pm0.15$ & $0.04\pm0.13$ \\[0.2cm]
							                					
\hline
\\							                					
$-4.5<M_{{V}_{o}}\leq-0.25$ &  all                 & $0.72(>0.41)$  & $0.89(>0.78)$   & $0.25\pm0.17$  & $0.57\pm0.17$ \\[0.15cm]

                                                    & $(B-V)_{o}\leq -0.11$ & $0.14\pm0.26$  & $0.56\pm0.35$  & $0.11\pm0.19$ & $0.39\pm0.30$ \\[0.15cm]
                                                    & $(B-V)_{o}> -0.11$     & $0.67\pm0.32$  & $0.89(>0.76)$  & $0.33\pm0.24$ & $0.59\pm0.19$ \\[0.2cm]

\hline
\\
\enddata
\tablecomments{These are normalized chance coincidence probabilities
  (MCPS matches per \chandra source), thus not additive.} 
\tablenotetext{\dag}{In the range $M_{{V}_{o}}\leq-4.5$ and $(B-V)_{o}\leq -0.31$ there are not stars, thus the chance coincidence probability is not estimated there.}
\end{deluxetable}

\makeatletter
\def\jnl@aj{AJ}
\ifx\revtex@jnl\jnl@aj\let\tablebreak=\nl\fi
\makeatother
\ptlandscape 
\begin{deluxetable}{cccccccccccccccc}
\tablewidth{0pt}
\tabletypesize{\tiny}
\tablecolumns{16}
\rotate
\tablecaption{Summary of optical and X-ray properties for bright sources\label{propertiestable}}
\tablehead{\colhead{X-ray} & \multicolumn{8}{c}{Optical properties} & \multicolumn{3}{c}{X-ray properties} & \colhead{$\xi$} & \colhead{[MA93]} & \multicolumn{2}{c}{Classification\tablenotemark{\ddag}}  \\
\colhead{src. ID} & \colhead{src. ID} & \colhead{$V$} & \colhead{$\delta V$} & \colhead{$BV$} & \colhead{$\delta(BV)$} & \colhead{$UB$} & \colhead{$\delta(UB)$} & \colhead{Off.} & \colhead{spectrum\tablenotemark{*}} & \colhead{pulsations} & \colhead{${\rm F_{x}^{obs}}$\tablenotemark{\dag}} & \colhead{} & \colhead{} & \colhead{this work} & \colhead{previous} \\[0.2cm]
\colhead{[1]} & \colhead{[2]} & \colhead{[3]} & \colhead{[4]} & \colhead{[5]} & \colhead{[6]} & \colhead{[7]}  & \colhead{[8]}  & \colhead{[9]} & \colhead{[10]} & \colhead{[11]} & \colhead{[12]} & \colhead{[13]} & \colhead{[14]} & \colhead{[15]} & \colhead{[16]}}
\startdata
 3\_1 &    O\_7\_267132 & 19.76 & 0.30 & 99.99 & 99.99 &  \nodata & \nodata & 0.13 & soft (S) & no$^{(1)}$ & $28.89_{-1.62}^{+1.52}$  & $14.89_{-0.31}^{+0.31}$ & \nodata          & unclass. & AGN\\[0.12cm]
          &    O\_7\_267163 & 19.54 & 0.33 & 99.99 & 99.99 & \nodata & \nodata & 0.54 &               &                     &                                                & $14.67_{-0.34}^{+0.33}$ &                         &                       & (type I)?   \\[0.12cm]
\hline
 3\_2 &    Z\_2804383   & 20.30 & 0.07 & -0.38 &  0.14 & -0.33 &  0.16 & 2.01 &   hard (S) & SXP34.1$^{(2)}$ & $23.53_{-1.38}^{+1.51}$  & $15.21_{-0.09}^{+0.10}$ & \nodata         & Be-XBP & Be-XBP\\[0.12cm]
          &    Z\_2804939   & 20.30 & 0.08 &  1.17 &  0.20 & -1.85 &  0.22 & 0.28 &                   &                       &                                                & $15.21_{-0.10}^{+0.11}$ &                        &                                                                  & or     \\[0.12cm]

      &    Z\_2806702\tablenotemark{\bullet}   &  16.78 & 0.03 & -0.12 & 0.04 & -0.77 & 0.06 & 4.27  &        &        &        & $11.69_{-0.07}^{+0.08}$  &         &        &  AGN?  \\[0.12cm]   

\hline
 3\_3 &    O\_8\_49531  & 16.01 & 0.02 & -0.02 &  0.04 & -1.05 &  0.17 & 0.14 &   hard (S) & SXP565$^{(3)}$ & $10.96_{-1.01}^{+1.00}$  & $10.09_{-0.10}^{+0.10}$ &  1020 ($0.4\arcsec$) & Be-XBP &          Be-XBP\\[0.12cm]
\hline
 3\_4 &    O\_8\_56601  & 18.19 & 0.02 &  0.37 &  0.04 & -0.60 &  0.06 & 0.54 &   soft (S)   & \nodata         & $8.68_{-0.96}^{+0.80}$   & $12.01_{-0.12}^{+0.10}$ &  \nodata         & unclass.  &    \nodata        \\[0.12cm]
 \hline
3\_7 &    Z\_3075967   & 14.71 & 0.03 & -0.07 &  0.03 & -1.04 &  0.04 & 1.30 & hard? (HR) & no$^{(4)}$ & $3.10_{-0.54}^{+0.56}$   & $7.42_{-0.19}^{+0.20}$  &  \nodata         &   HMXB   & HMXB? \\
\enddata
\end{deluxetable}

\makeatletter
\def\jnl@aj{AJ}
\ifx\revtex@jnl\jnl@aj\let\tablebreak=\nl\fi
\makeatother
\ptlandscape 
\begin{deluxetable}{cccccccccccccccc}
\addtocounter{table}{-1}
\tablewidth{0pt}
\tabletypesize{\tiny}
\tablecolumns{16}
\rotate
\tablecaption{Summary of optical and X-ray properties of bright sources\label{propertiestable}}
\tablehead{\colhead{X-ray} & \multicolumn{8}{c}{Optical properties} & \multicolumn{3}{c}{X-ray properties} & \colhead{$\xi$} & \colhead{[MA93]} & \multicolumn{2}{c}{Classification\tablenotemark{\ddag}}  \\
\colhead{src. ID} & \colhead{src. ID} & \colhead{$V$} & \colhead{$\delta V$} & \colhead{$BV$} & \colhead{$\delta(BV)$} & \colhead{$UB$} & \colhead{$\delta(UB)$} & \colhead{Off.} & \colhead{spectrum\tablenotemark{*}} & \colhead{pulsations} & \colhead{${\rm F_{x}^{obs}}$\tablenotemark{\dag}} & \colhead{} & \colhead{} & \colhead{this work} & \colhead{previous} \\[0.2cm]
\colhead{[1]} & \colhead{[2]} & \colhead{[3]} & \colhead{[4]} & \colhead{[5]} & \colhead{[6]} & \colhead{[7]}  & \colhead{[8]}  & \colhead{[9]} & \colhead{[10]} & \colhead{[11]} & \colhead{[12]} & \colhead{[13]} & \colhead{[14]} & \colhead{[15]} & \colhead{[16]}}
\startdata
4\_19 &    O\_5\_262679 & 19.41 & 0.09 & -0.26 & 0.10 & -0.24 & 0.11 & 1.40 & hard? (HR) & \nodata &  $1.02_{-0.32}^{+0.30}$   & $10.70_{-0.35}^{+0.33}$  &  \nodata  & new HMXB & \nodata \\[0.12cm]
      &    O\_5\_261182\tablenotemark{\bullet} & 18.16 & 0.03 & -0.08 & 0.04 & -0.61 & 0.05 & 2.27 &            &         &                           & $9.45_{-0.34}^{+0.32}$   &           &             &         \\
      & $\equiv$Z\_2153646   &       &      &       &       &       &       &       &           &         &         &         &         &         &           \\[0.12cm]
\hline
4\_21 &    O\_5\_256084 & 17.93 & 0.06 &  0.88 &  0.15 &  \nodata & \nodata & 2.59 & \nodata & \nodata  & $4.72_{-0.64}^{+0.67}$   & $11.10_{-0.16}^{+0.16}$ & \nodata  &   unclass.  & SNR\tablenotemark{\otimes}  \\[0.12cm]
           &    O\_5\_256315 & 18.63 & 0.04 &  0.76 &  0.09 & -0.81 &  0.12 & 1.39 &                      &                    &                                                & $11.80_{-0.15}^{+0.16}$ &                        &                       &          \\[0.12cm]
           &    O\_5\_256826 & 18.33 & 0.04 & -0.18 &  0.05 & -0.62 &  0.06 & 1.75 &                      &                    &                                               & $11.50_{-0.15}^{+0.16}$ &                         &                       &           \\[0.12cm]
\hline
4\_22 &    O\_5\_90840  & 18.40 & 0.03 &  0.68 &  0.06 & 99.99 & 99.99 & 3.20 & \nodata & \nodata & $1.01_{-0.29}^{+0.40}$   & $9.90_{-0.31}^{+0.42}$  & \nodata         &    new HMXB? & \nodata      \\[0.12cm]
           &    O\_5\_91270  & 18.26 & 0.03 & -0.07 &  0.03 & -0.41  &  0.08  & 2.32 &                &                &                                               & $9.76_{-0.31}^{+0.42}$  &                        &                          &             \\
\enddata
\end{deluxetable}

\makeatletter
\def\jnl@aj{AJ}
\ifx\revtex@jnl\jnl@aj\let\tablebreak=\nl\fi
\makeatother
\ptlandscape 
\begin{deluxetable}{cccccccccccccccc}
\addtocounter{table}{-1}
\tablewidth{0pt}
\tabletypesize{\tiny}
\tablecolumns{16}
\rotate
\tablecaption{Summary of optical and X-ray properties for bright sources\label{propertiestable}}
\tablehead{\colhead{X-ray} & \multicolumn{8}{c}{Optical properties} & \multicolumn{3}{c}{X-ray properties} & \colhead{$\xi$} & \colhead{[MA93]} & \multicolumn{2}{c}{Classification\tablenotemark{\ddag}}  \\
\colhead{src. ID} & \colhead{src. ID} & \colhead{$V$} & \colhead{$\delta V$} & \colhead{$BV$} & \colhead{$\delta(BV)$} & \colhead{$UB$} & \colhead{$\delta(UB)$} & \colhead{Off.} & \colhead{spectrum\tablenotemark{*}} & \colhead{pulsations} & \colhead{${\rm F_{x}^{obs}}$\tablenotemark{\dag}} & \colhead{} & \colhead{} & \colhead{this work} & \colhead{previous} \\[0.2cm]
\colhead{[1]} & \colhead{[2]} & \colhead{[3]} & \colhead{[4]} & \colhead{[5]} & \colhead{[6]} & \colhead{[7]}  & \colhead{[8]}  & \colhead{[9]} & \colhead{[10]} & \colhead{[11]} & \colhead{[12]} & \colhead{[13]} & \colhead{[14]} & \colhead{[15]} & \colhead{[16]}}
\startdata
 6\_3 &    O\_6\_153267 & 18.15 & 0.07 & -0.14 &  0.08 & -0.52 &  0.07 & 1.38 & soft? (HR) & \nodata &  $4.23_{-0.65}^{+0.58}$   & $11.19_{-0.18}^{+0.16}$ & \nodata                     &    unclass.  & \nodata     \\[0.12cm]
\hline
 6\_4 &    O\_6\_147662 & 14.42 & 0.05 & -0.10 &  0.05 & -0.97 &  0.04 & 0.27 & hard? (HR) & \nodata & $2.85_{-0.53}^{+0.53}$   & $7.03_{-0.21}^{+0.21}$  & 618 ($1.1\arcsec$) &       Be-XRB       &      Be-XRB\\[0.12cm]
\hline
 6\_8 &    O\_6\_306524 & 17.60 & 0.01 &  1.12 &  0.06 &  0.74 &  0.11 & 2.02 & soft? (HR) & \nodata & $1.60_{-0.42}^{+0.40}$   & $9.59_{-0.28}^{+0.27}$  & \nodata                     &  unclass.  & \nodata     \\[0.12cm]
\hline
6\_11 &    Z\_2657679   & 17.07 & 0.03 &  0.36 &  0.05 &  0.03 &  0.05 & 2.92 &  hard? (HR) & \nodata & $1.33_{-0.35}^{+0.41}$   & $8.86_{-0.29}^{+0.33}$  & \nodata   & new HMXB  & \nodata   \\
      & $\equiv$O\_6\_320020 & 17.10 & 0.02 & 0.26 & 0.03 & & & & & & & & & &\\
      & $\equiv$O\_7\_57270  & 17.14 & 0.02 & 0.25 & 0.04 & & & & & & & & & &\\[0.12cm]
\hline
6\_20 &    O\_6\_311169 & 13.71 & 0.14 &  0.39 &  0.19 & -0.67 &  0.12 & 0.62 & \nodata & \nodata & $0.75_{-0.29}^{+0.26}$   & $4.87_{-0.44}^{+0.40}$  & 739 ($0.8\arcsec$) & unclass.   &       B[e]\\[0.12cm]
\hline
6\_23 &    O\_6\_158108 & 17.10 & 0.05 &  1.42 &  0.15 &  1.92 &  0.16 & 0.56 & \nodata & \nodata & $0.48_{-0.19}^{+0.21}$   & $7.77_{-0.44}^{+0.49}$  &  \nodata                     &  unclass.  & \nodata   \\
\enddata
\end{deluxetable}

\makeatletter
\def\jnl@aj{AJ}
\ifx\revtex@jnl\jnl@aj\let\tablebreak=\nl\fi
\makeatother
\begin{deluxetable}{cccccccccccc}
\tablewidth{0pt}
\tabletypesize{\scriptsize}
\tablecolumns{12}
\tablecaption{Bright optical counterparts ($M_{{V}_{o}}\leq-0.25$) of X-ray sources within the
$2\sigma$ search radius (no match(es) found in the $1\sigma$ search radius)\label{BrightNor}}
\tablehead{\multicolumn{2}{c}{X-ray} & \multicolumn{10}{c}{Optical counterpart} \\
\colhead{Src} & \colhead{Rad.\tablenotemark{\star}} & \colhead{Src ID} & \colhead{R.A.} & \colhead{Decl.} & \colhead{Offset} & \colhead{{\it V}} & \colhead{$\rm{\delta}V$} & \colhead{{\it B-V}} & \colhead{$\rm{\delta}\rm{(}B-V\rm{)}$} &  \colhead{{\it U-B}} & \colhead{$\rm{\delta}\rm{(}U-B\rm{)}$} \\
\colhead{} & \colhead{($\arcsec$)}  & \colhead{} & \multicolumn{2}{c}{(J2000.0)} & \colhead{($\arcsec$)} & \colhead{} & \colhead{} & \colhead{} & \colhead{} & \colhead{} & \colhead{} \\[0.2cm]
\colhead{[1]} & \colhead{[2]} & \colhead{[3]} & \colhead{[4]} & \colhead{[5]} & \colhead{[6]} & \colhead{[7]} & \colhead{[8]} & \colhead{[9]} & \colhead{[10]} & \colhead{[11]}  & \colhead{[12]} }
\startdata
3\_10 &  3.00 & Z\_2838059\tablenotemark{\dag} & 00 55 41.83  & -72 18 10.9 & 1.93 & 14.18 & 0.04 &  1.46 &  0.26 &  1.28 &  0.26 \\ 
\hline
3\_24 &  3.00 & Z\_2989515\tablenotemark{\dag} & 00 56 46.83  & -72 22 50.9 & 2.23 & 18.55 & 0.09 &  0.62 &  0.10 &  0.09 &  0.08 \\ 
\hline 
3\_25 &  3.00 & Z\_2924823\tablenotemark{\dag} & 00 56 18.97  & -72 15 09.8 & 2.08 & 17.90 & 0.03 &  1.16 &  0.04 &  0.91 &  0.12 \\ 
\hline
3\_27 &  3.00 & Z\_2894870\tablenotemark{\dag} & 00 56 06.19  & -72 20 47.0 & 2.13 & 16.57 & 0.03 &  1.70 &  0.05 &  1.83 &  0.12 \\ 
\hline
\hline
5\_35 &  3.00 & Z\_2438540 & 00 52 58.84  & -72 26 05.3 & 2.58 & 18.84 & 0.05 &  0.00 &  0.06 & -0.48 &  0.08 \\ 
\hline 
5\_37\tablenotemark{\ddag} & 3.38  & O\_6\_324148 & 00 53 58.77  & -72 31 14.3  & 2.25 & 16.94 &  0.04 &  1.09 &  0.08 & \nodata & \nodata \\
                           &       & Z\_2582218   & 00 53 58.82  & -72 31 14.3  & 2.07 & 16.90 &  0.03 &  1.11 &  0.04 &  0.74 &  0.07 \\ 
\hline 
\hline 
6\_27\tablenotemark{\ddag} & 3.81  & O\_6\_227467 & 00 53 33.75  & -72 48 17.9  & 2.77 & 18.30 &  0.03 &  1.02 &  0.04 & \nodata & \nodata \\
\enddata
\tablenotetext{\star}{This is the $2\sigma$ search radius given in arcseconds.}
\tablenotetext{\dag}{Single counterpart within the $2\sigma$ search radius.}
\tablenotetext{\ddag}{For this source there is a finding chart.}
\end{deluxetable}

\makeatletter
\def\jnl@aj{AJ}
\ifx\revtex@jnl\jnl@aj\let\tablebreak=\nl\fi
\makeatother
\begin{deluxetable}{ccccccccc}
\tablewidth{0pt}
\tabletypesize{\scriptsize}
\tablecolumns{9}
\tablecaption{Summary of optical and X-ray properties of sources
presented in Table \ref{BrightNor}\label{properties2r}}
\tablehead{\colhead{X-ray} & \multicolumn{3}{c}{Optical properties} & \multicolumn{3}{c}{X-ray properties} & \colhead{$\xi$} & \colhead{Classification}  \\
\colhead{Src ID} & \colhead{$M_{{V}_{o}}$} &
\colhead{$(B-V)_{\rm{o}}$} & \colhead{Offset ($\arcsec$)} &
\colhead{spectrum} & \colhead{pulsations} &  \colhead{${\rm F_{x}^{obs}}$\tablenotemark{\dag}} & \colhead{} & \colhead{this work \& previous} \\[0.2cm]
\colhead{[1]} & \colhead{[2]} & \colhead{[3]} & \colhead{[4]} & \colhead{[5]} & \colhead{[6]} & \colhead{[7]}  & \colhead{[8]}  & \colhead{[9]}}
\startdata
3\_10  & -5.01  & 1.37  & 1.93 &  hard? (HR)   & \nodata   & $1.50_{-0.37}^{+0.39}$ & $6.10_{-0.28}^{+0.28}$ & unclassified \\
\hline
3\_24  & -0.64  & 0.53  & 2.23 & \nodata       & \nodata   & $0.39_{-0.16}^{+0.26}$ & $9.00_{-0.45}^{+0.73}$ & unclassified \\
\hline
3\_25  & -1.29  & 1.07  & 2.08 & \nodata       & \nodata   & $0.44_{-0.18}^{+0.24}$ & $8.48_{-0.46}^{+0.60}$ & unclassified \\
\hline
3\_27  & -2.62  & 1.61  & 2.13 & \nodata       & \nodata   & $0.37_{-0.17}^{+0.25}$ & $6.97_{-0.51}^{+0.72}$ & unclassified \\
\hline
\hline					   
5\_35  & -0.35  & -0.09 & 2.58 & \nodata       & \nodata   & $0.27_{-0.15}^{+0.20}$ & $8.91_{-0.59}^{+0.78}$ & unclassified \\
\hline 
5\_37\tablenotemark{\ddag}  & -2.25  & 1.00  & 2.25 & \nodata       & \nodata   & $0.34_{-0.17}^{+0.25}$ & $7.23_{-0.56}^{+0.80}$ & unclassified \\
                            & -2.29  & 1.02  & 2.07 & \nodata       &           &                        &                        &              \\
\hline
\hline
6\_27\tablenotemark{\ddag}  & -0.89  & 0.93  & 2.77 & \nodata       & \nodata   & $0.30_{-0.16}^{+0.22}$ & $8.48_{-0.58}^{+0.77}$  & unclassified \\
\enddata
\tablenotetext{\dag}{Absorbed flux in the $2.0-10.0$ keV energy band calculated from net counts, in units of $10^{-14}$\funit (Zezas \etal 2007, in preparation).}
\end{deluxetable}

\clearpage

\clearpage

\clearpage
\begin{center} 
\includegraphics[width=.45\textwidth]{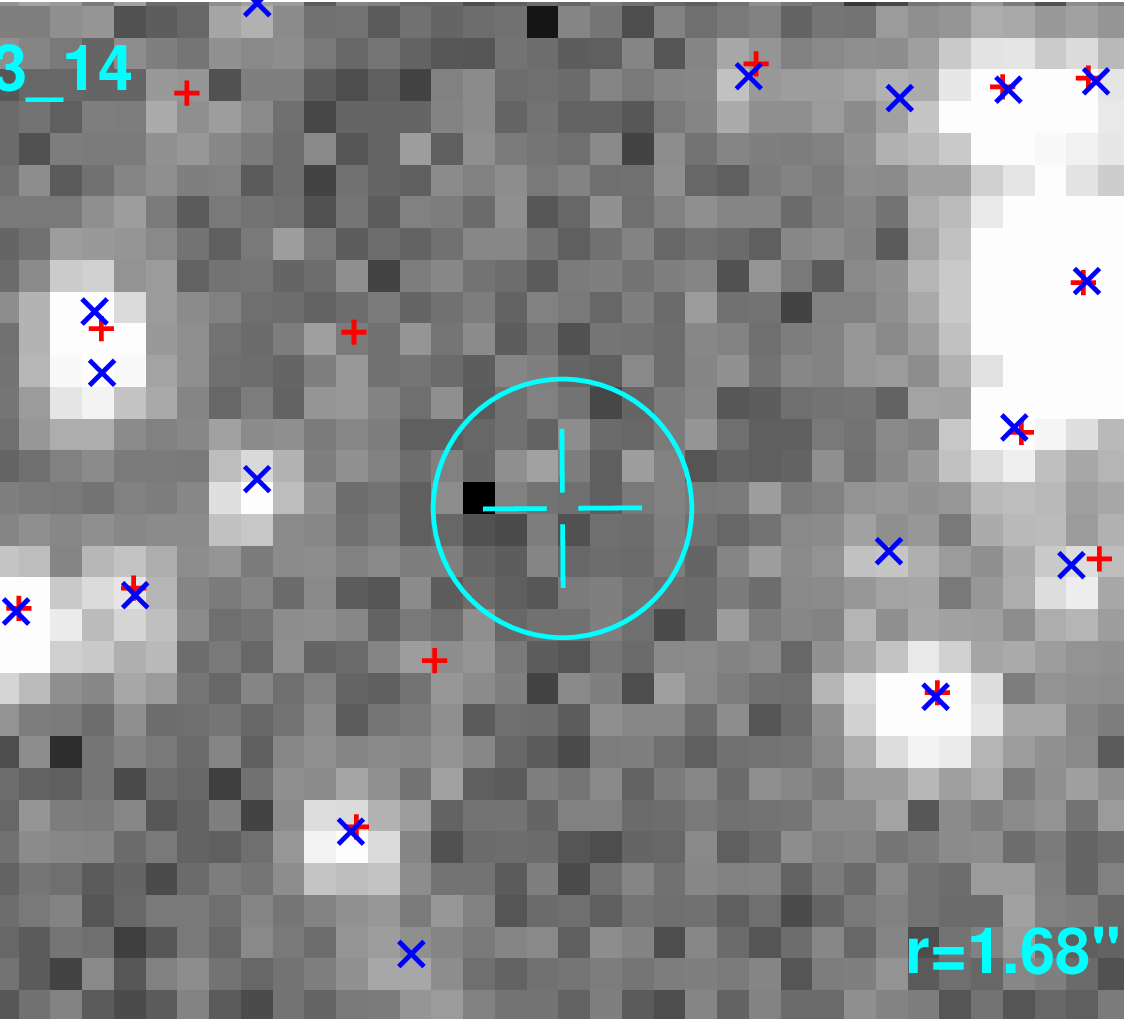} 
\includegraphics[width=.45\textwidth]{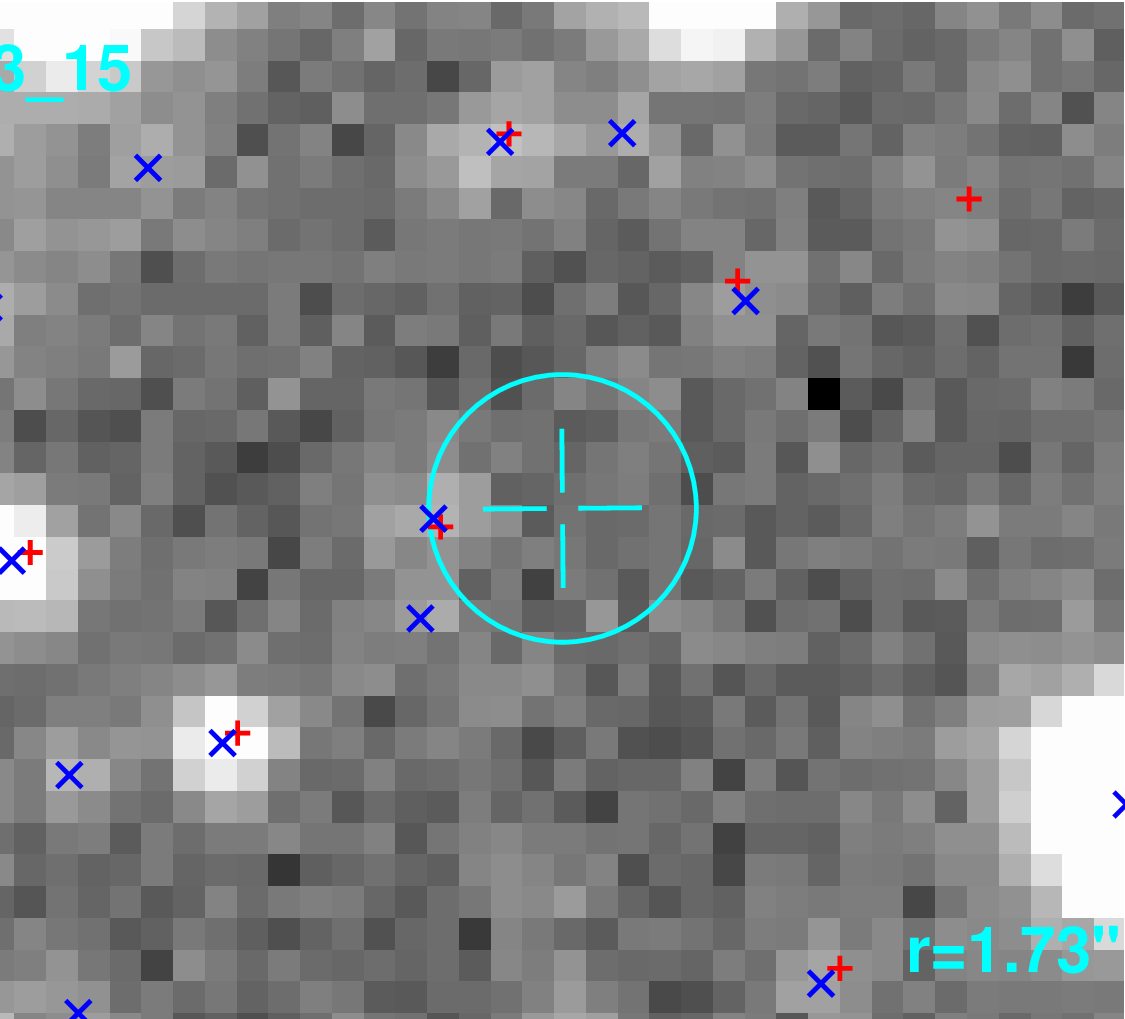}
\end{center} 
\clearpage

\clearpage
\begin{center} 
\includegraphics[width=.45\textwidth]{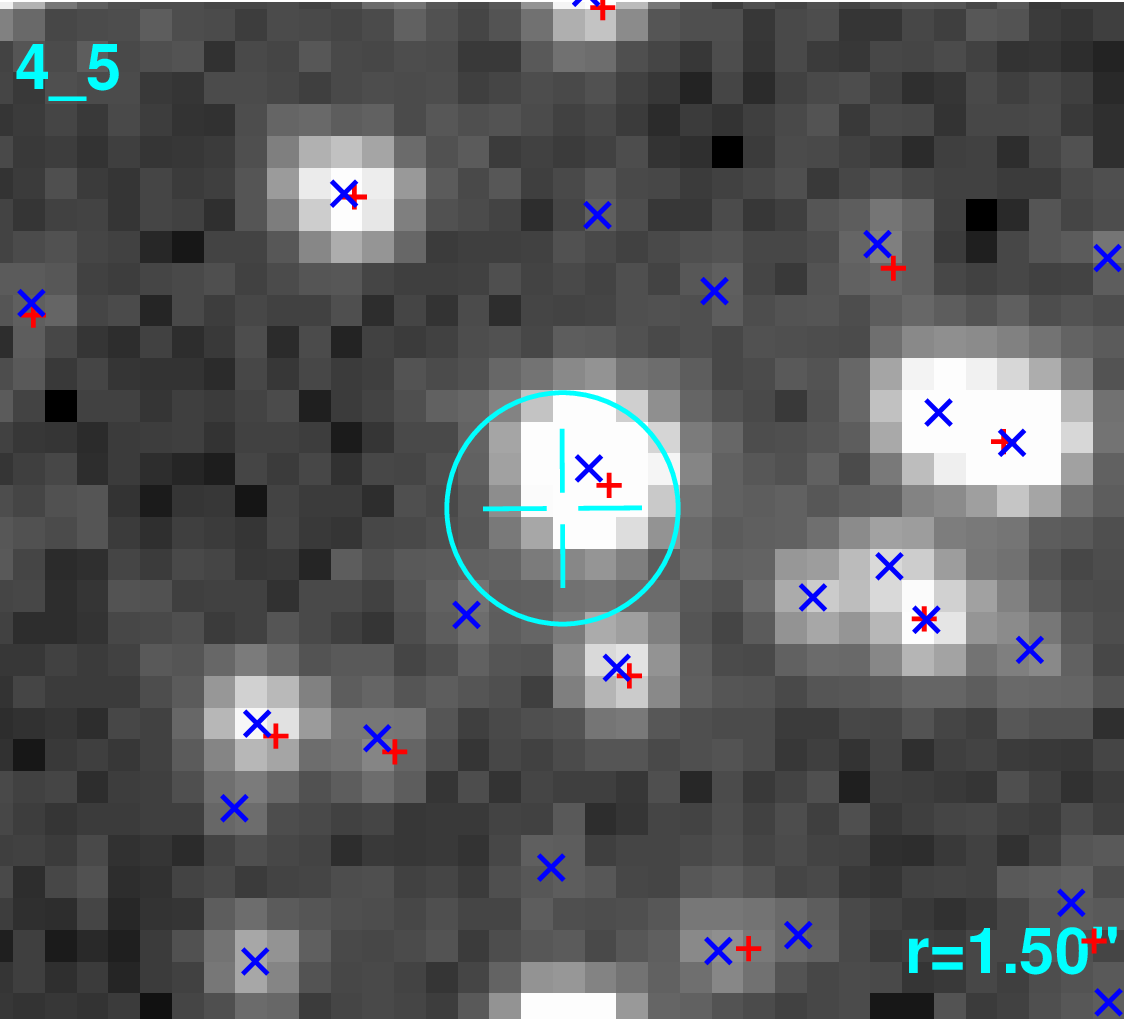} 
\includegraphics[width=.45\textwidth]{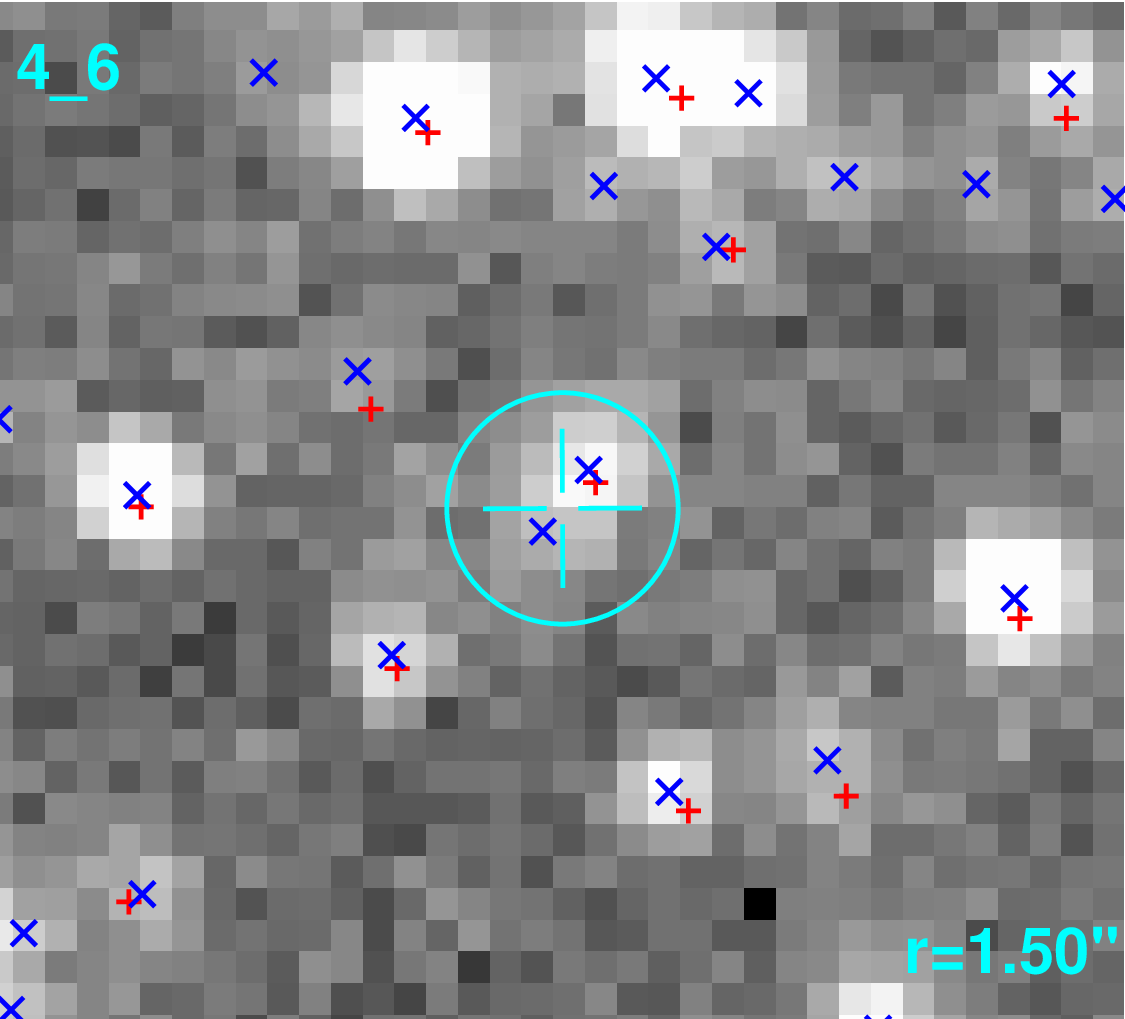} 
\includegraphics[width=.45\textwidth]{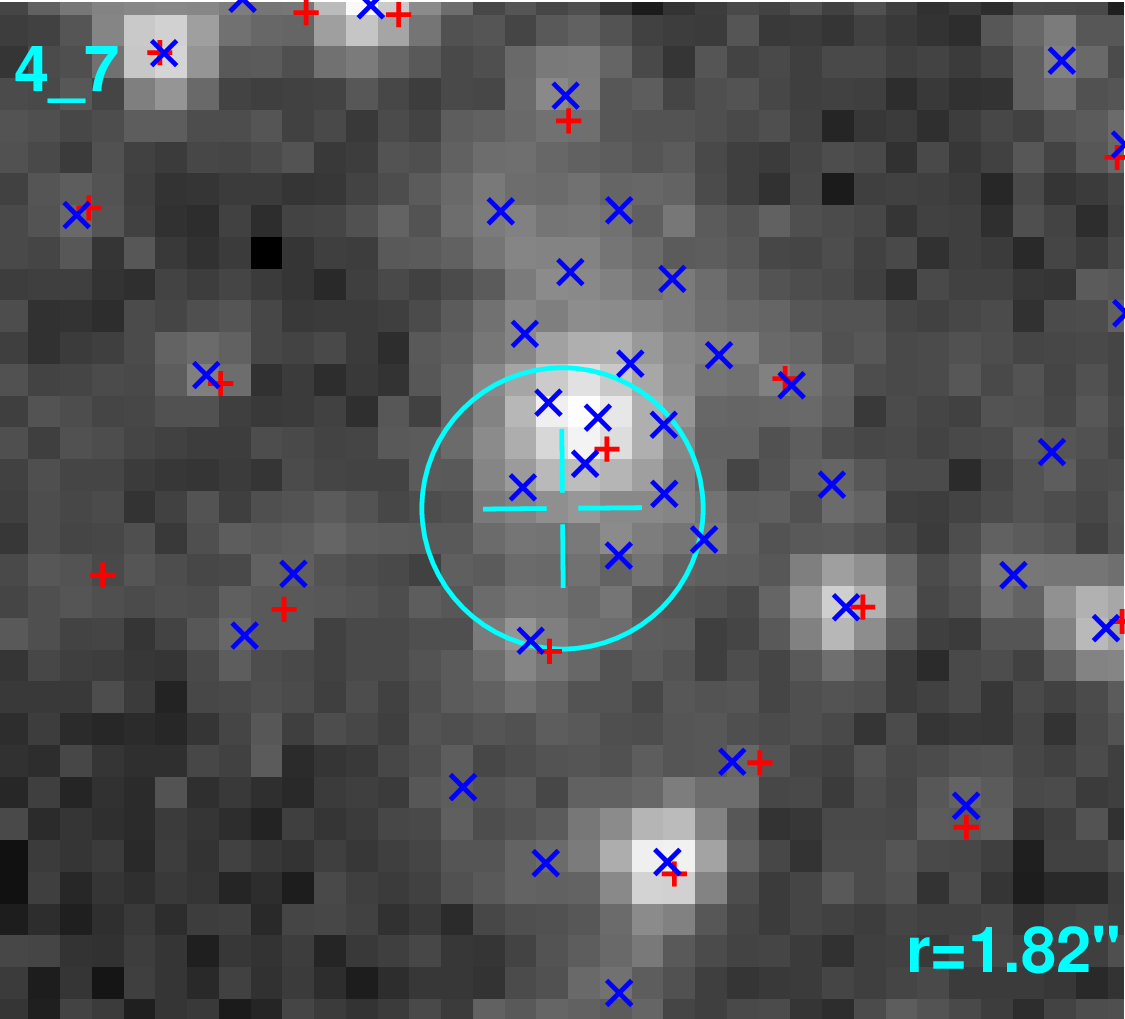} 
\includegraphics[width=.45\textwidth]{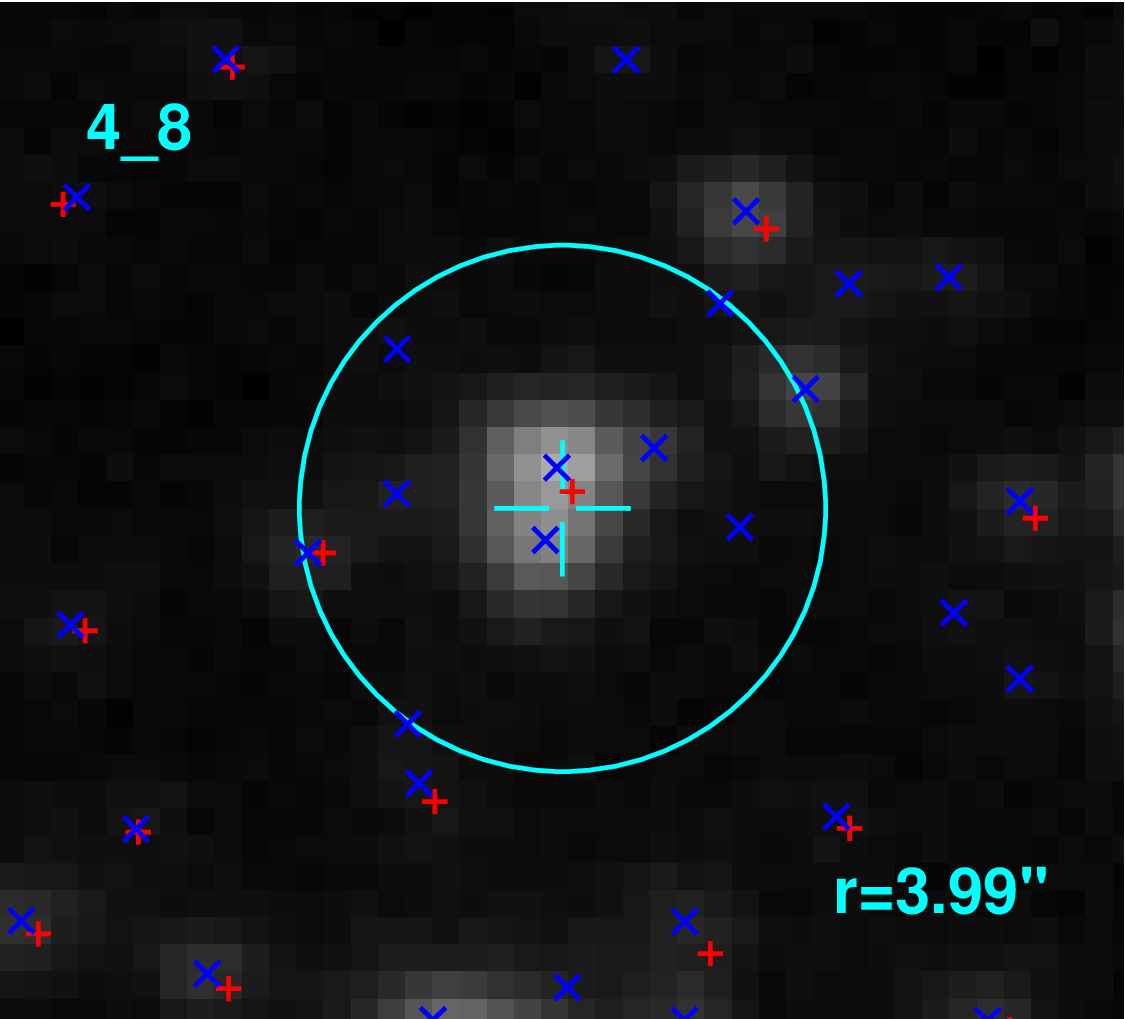} 
\end{center} 
\clearpage
\begin{center} 
\includegraphics[width=.45\textwidth]{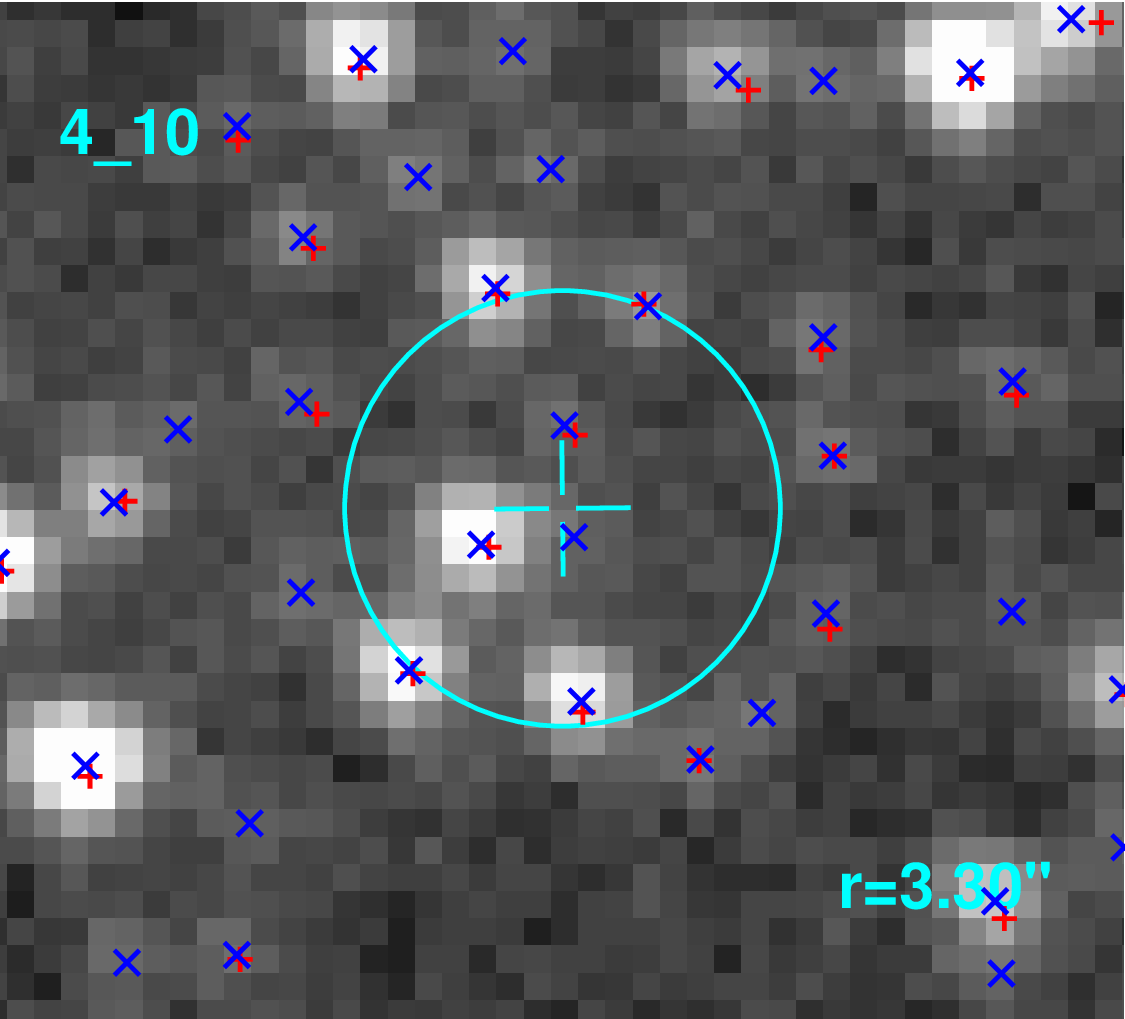} 
\includegraphics[width=.45\textwidth]{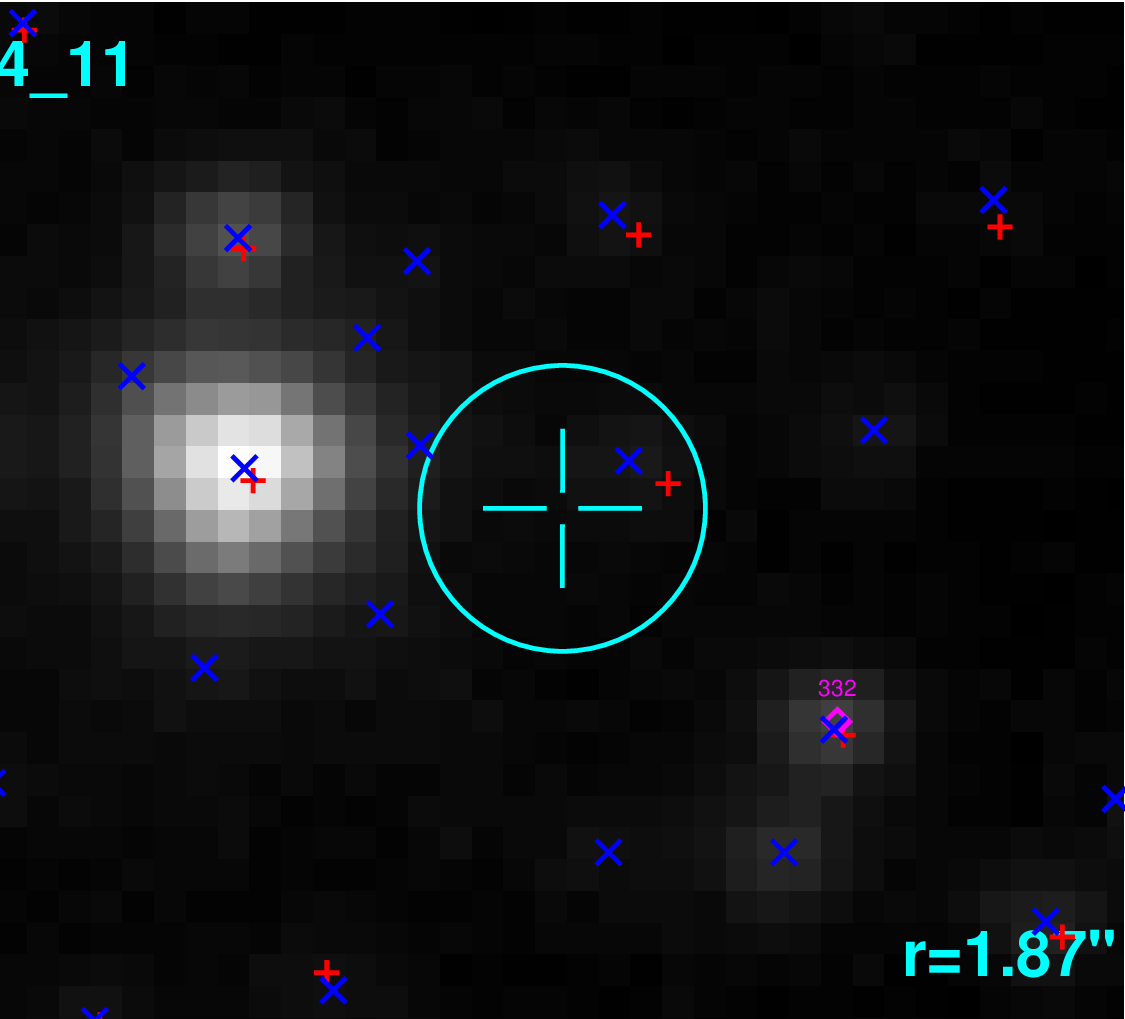} 
\includegraphics[width=.45\textwidth]{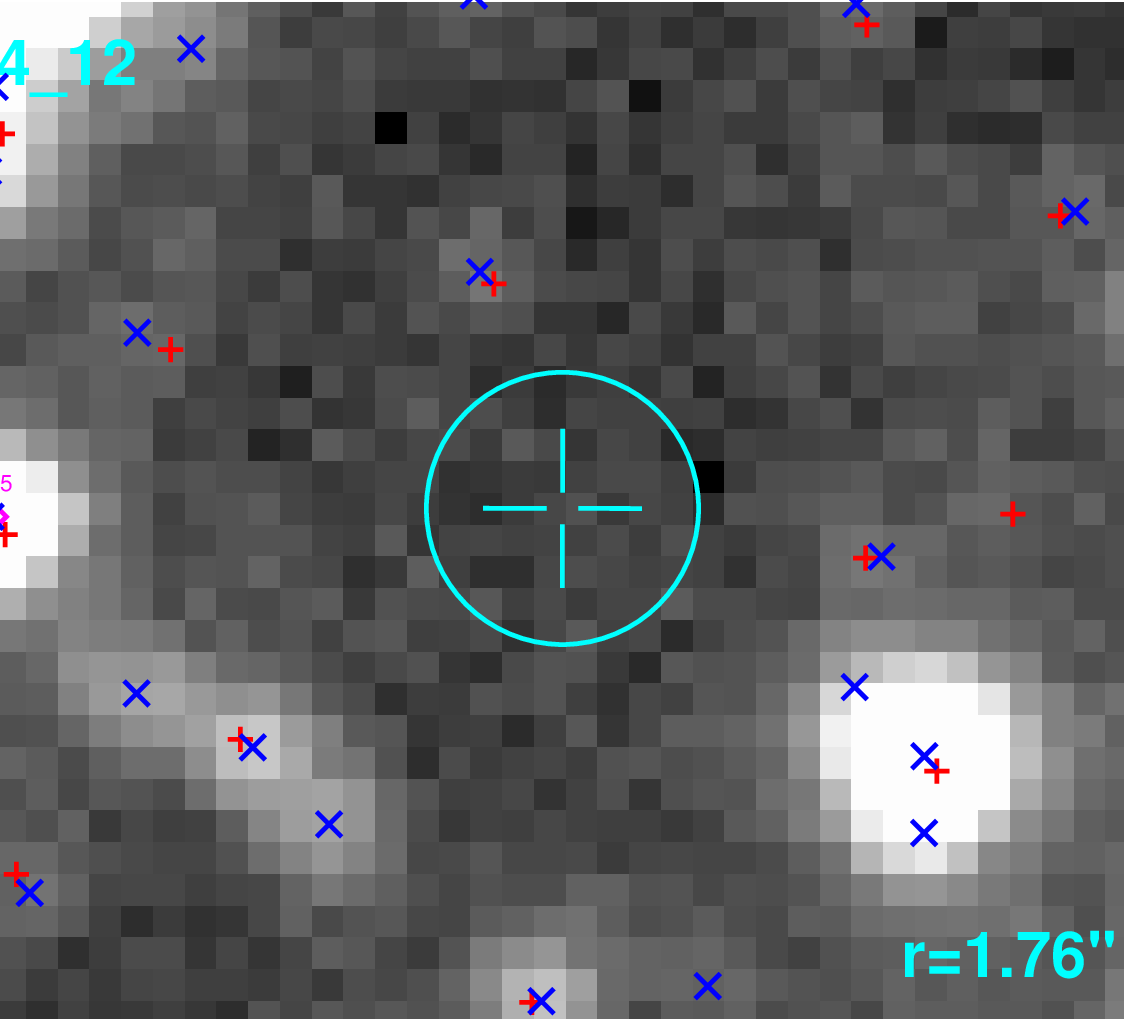} 
\includegraphics[width=.45\textwidth]{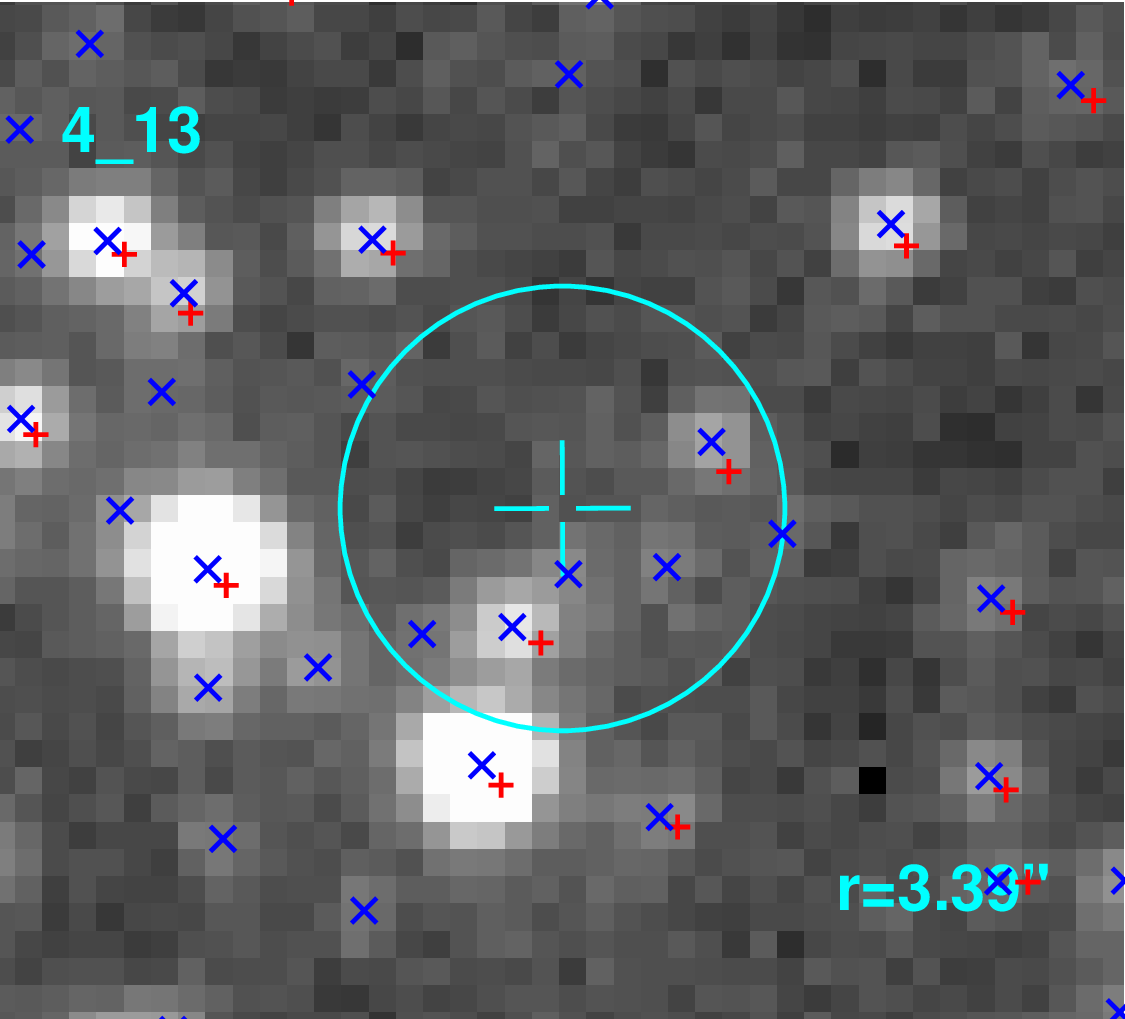} 
\end{center} 
\clearpage
\begin{center} 
\includegraphics[width=.45\textwidth]{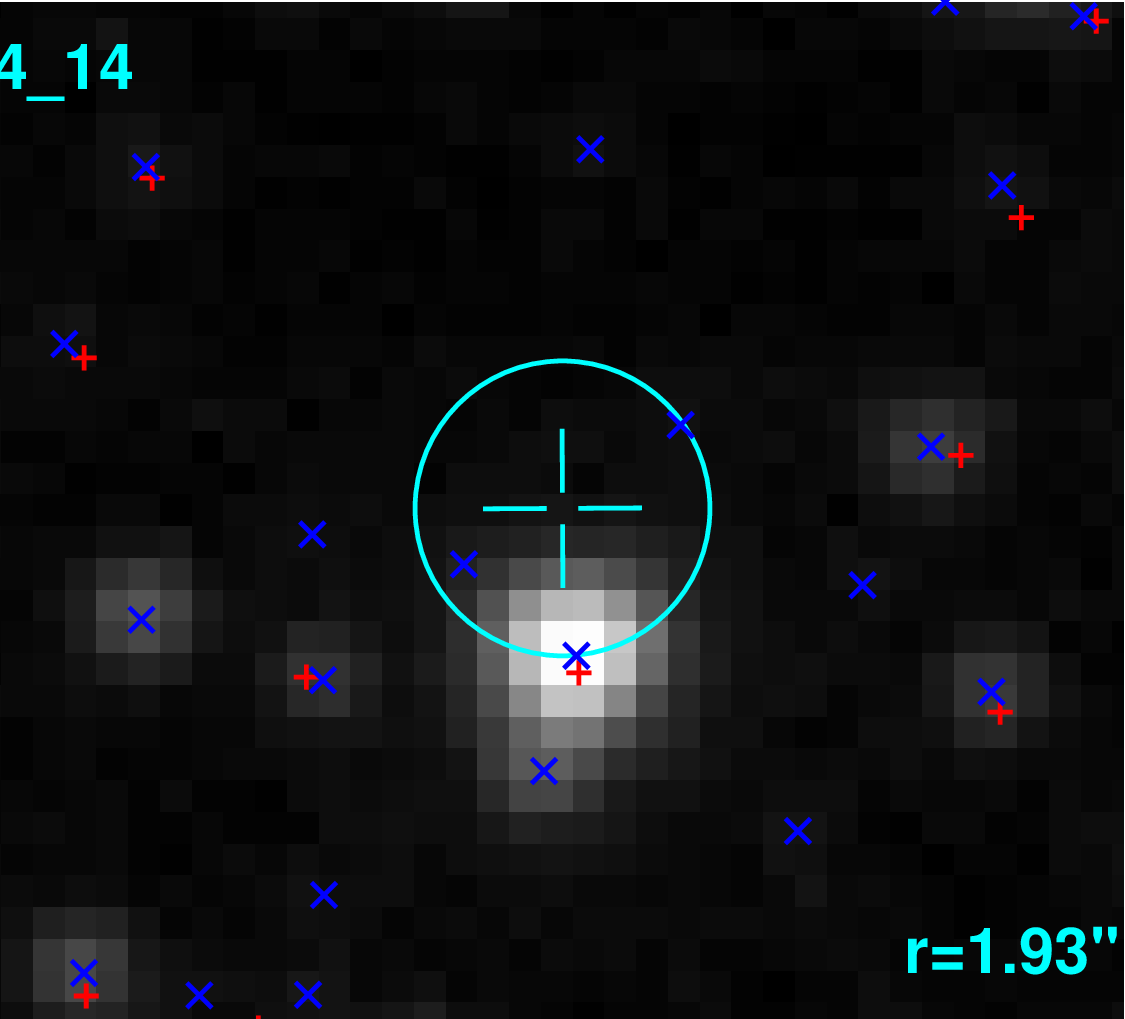} 
\includegraphics[width=.45\textwidth]{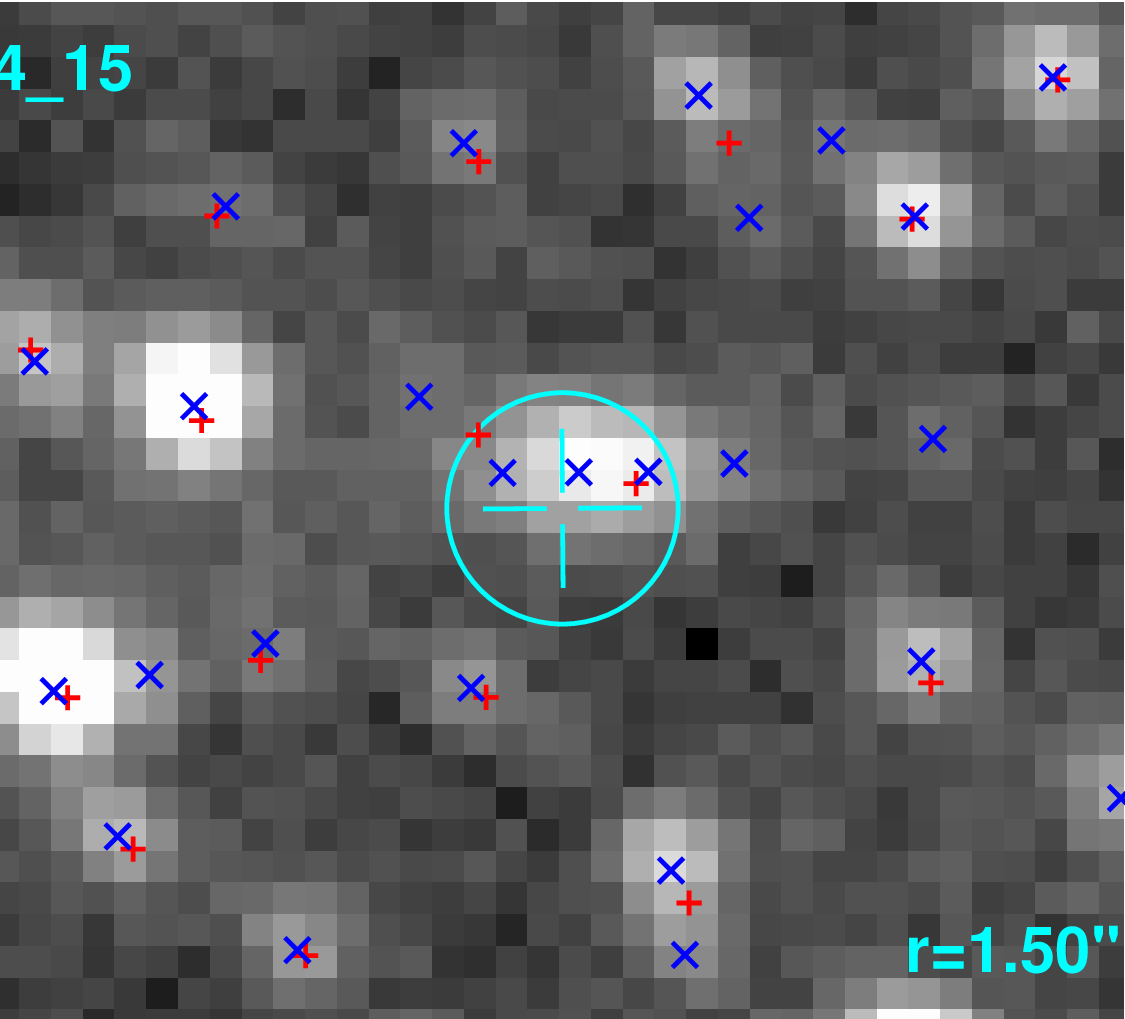} 
\includegraphics[width=.45\textwidth]{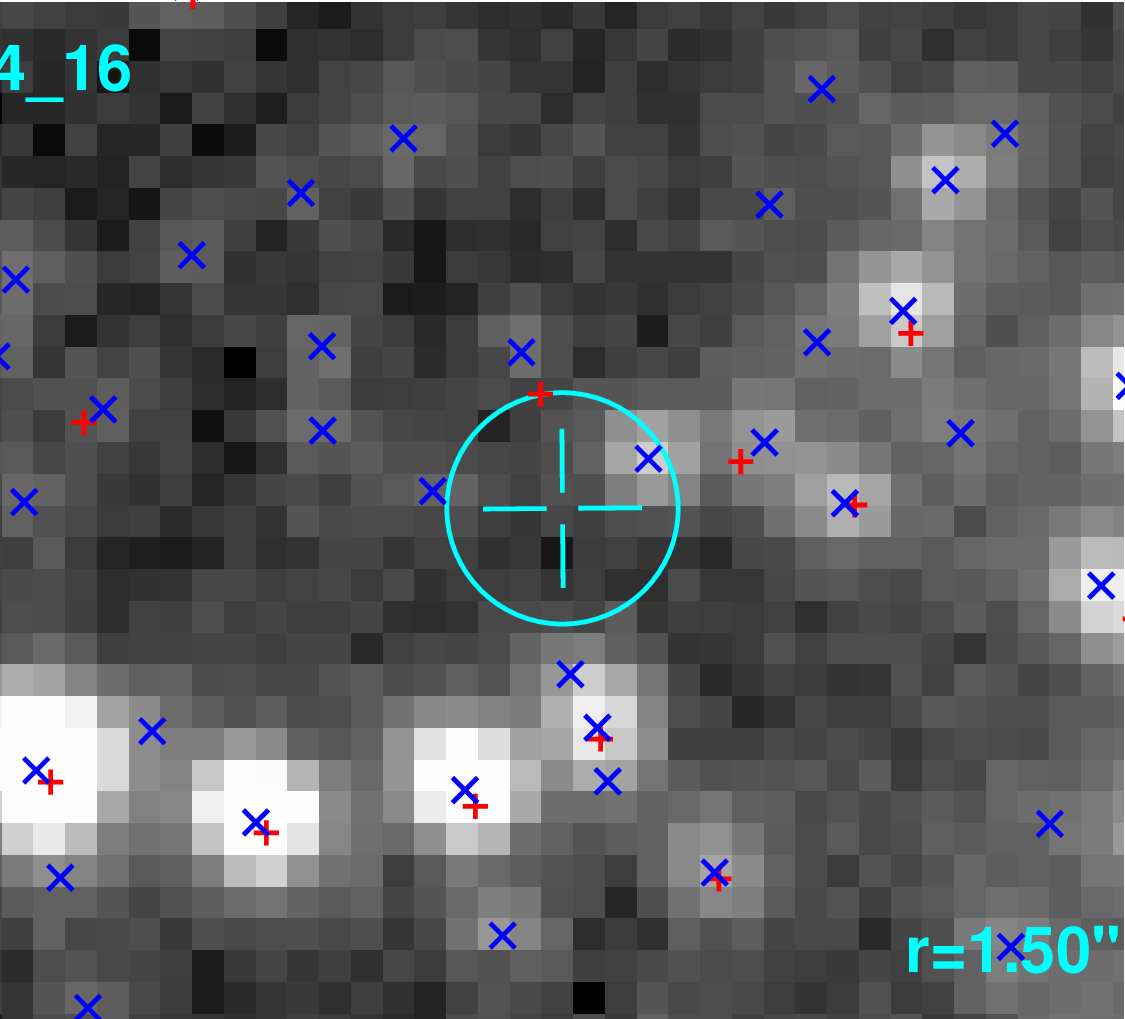} 
\includegraphics[width=.45\textwidth]{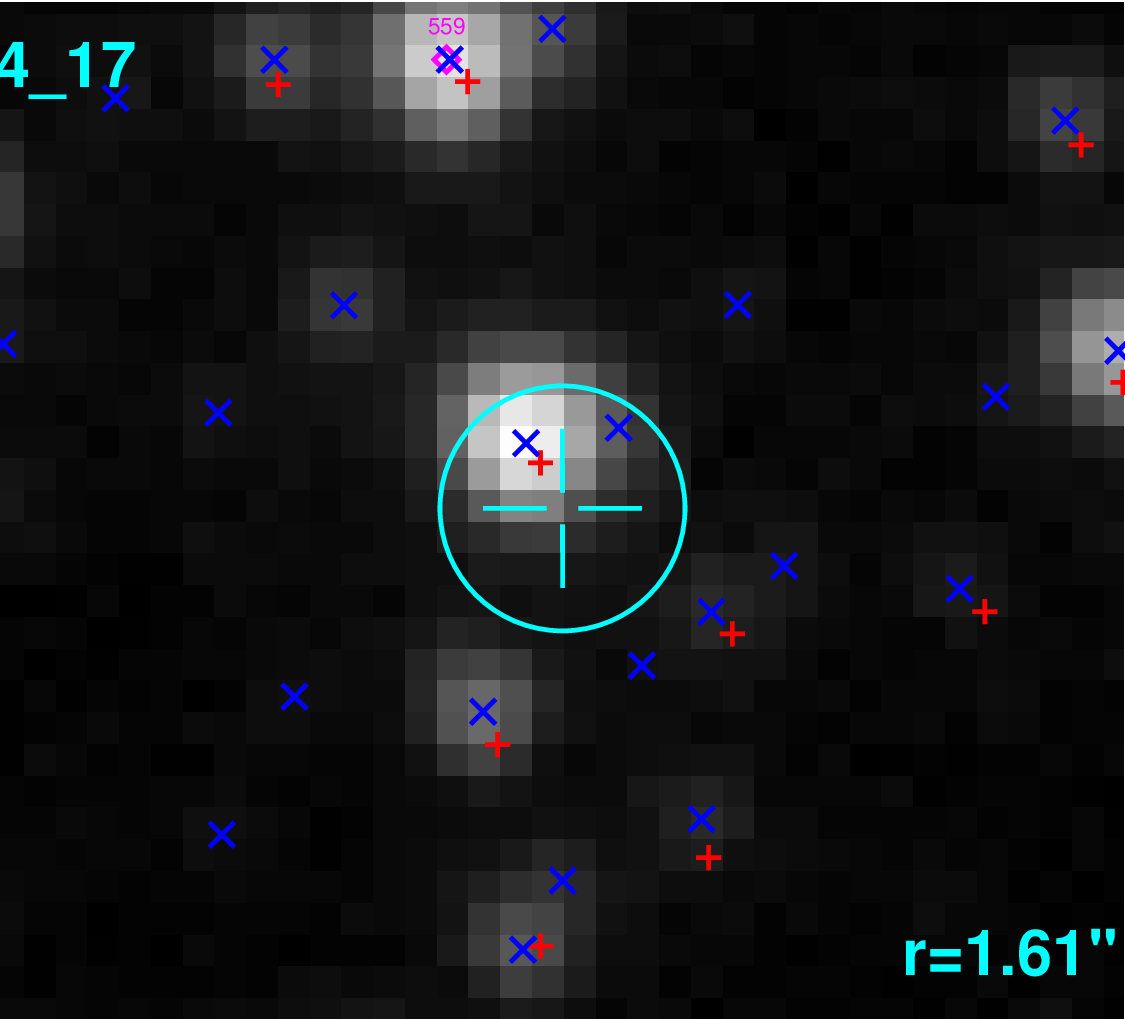} 
\end{center} 
\clearpage
\begin{center} 
\includegraphics[width=.45\textwidth]{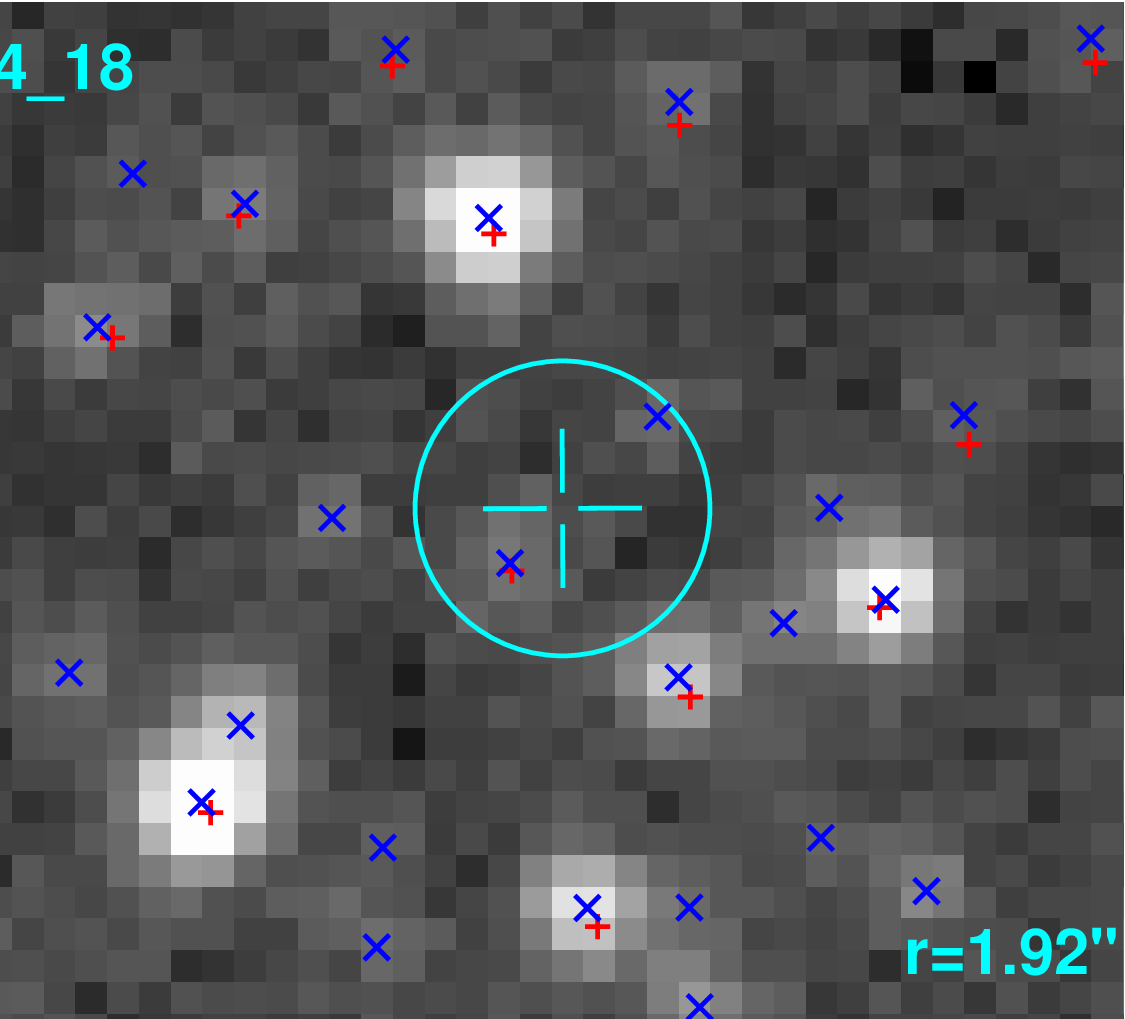} 
\includegraphics[width=.45\textwidth]{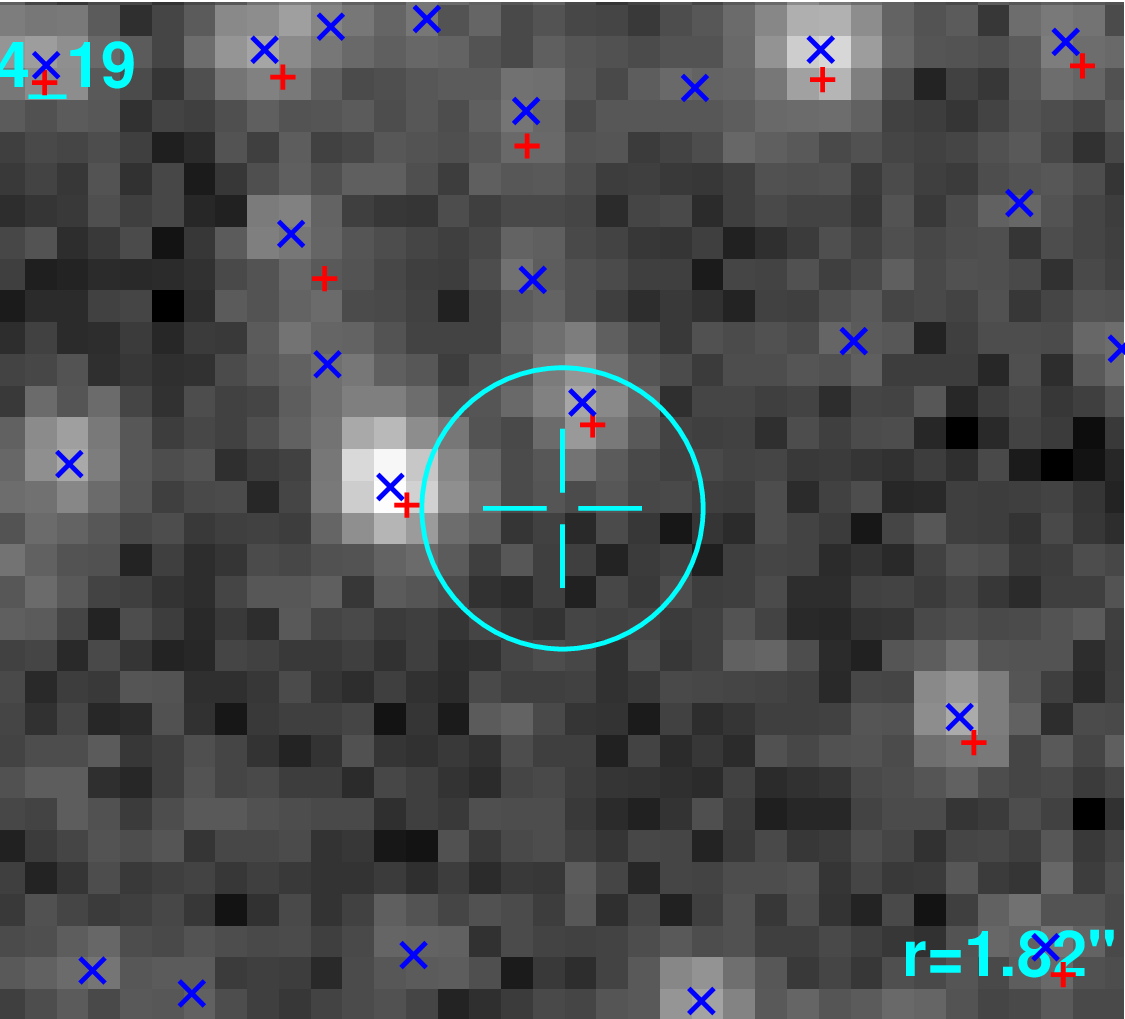} 
\includegraphics[width=.45\textwidth]{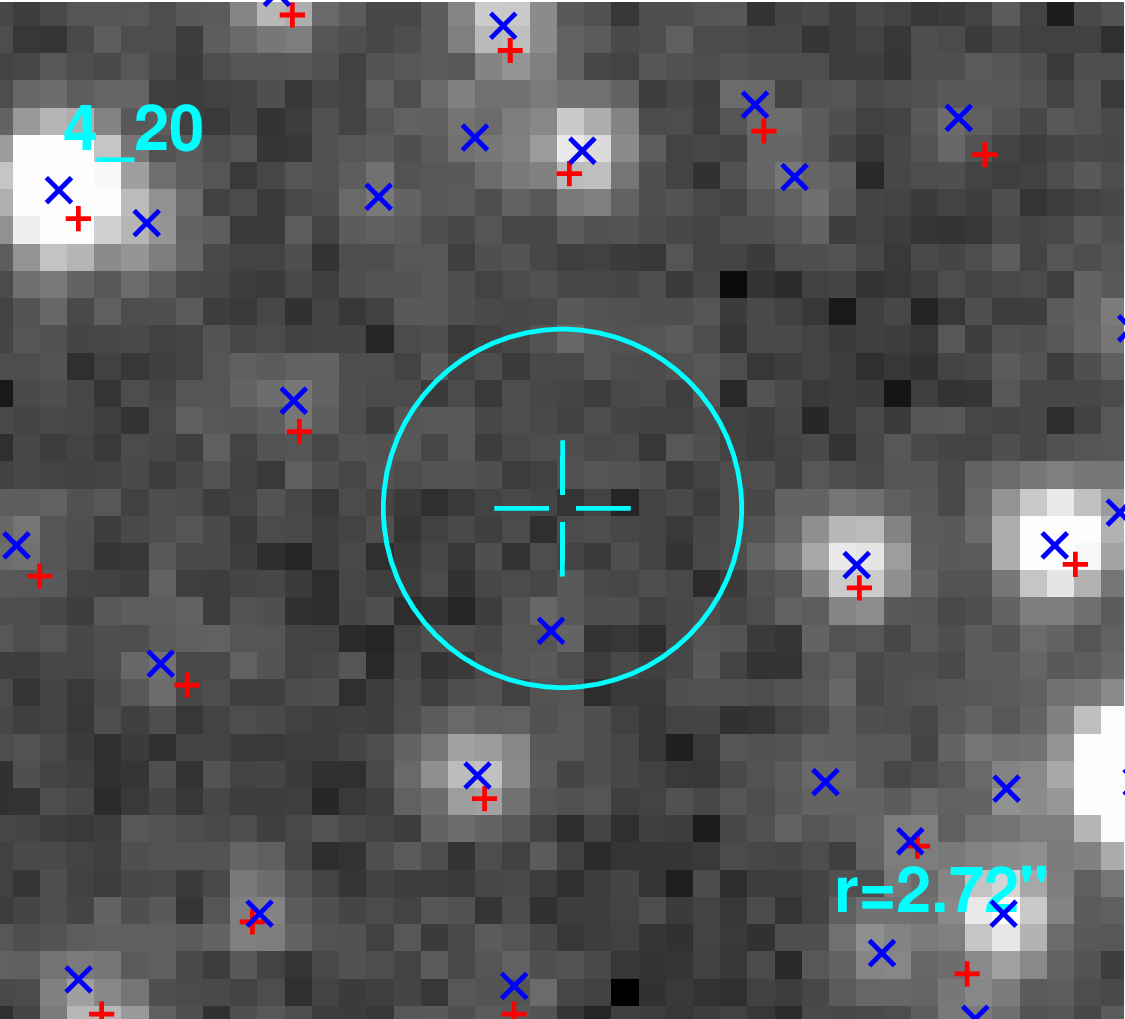} 
\includegraphics[width=.45\textwidth]{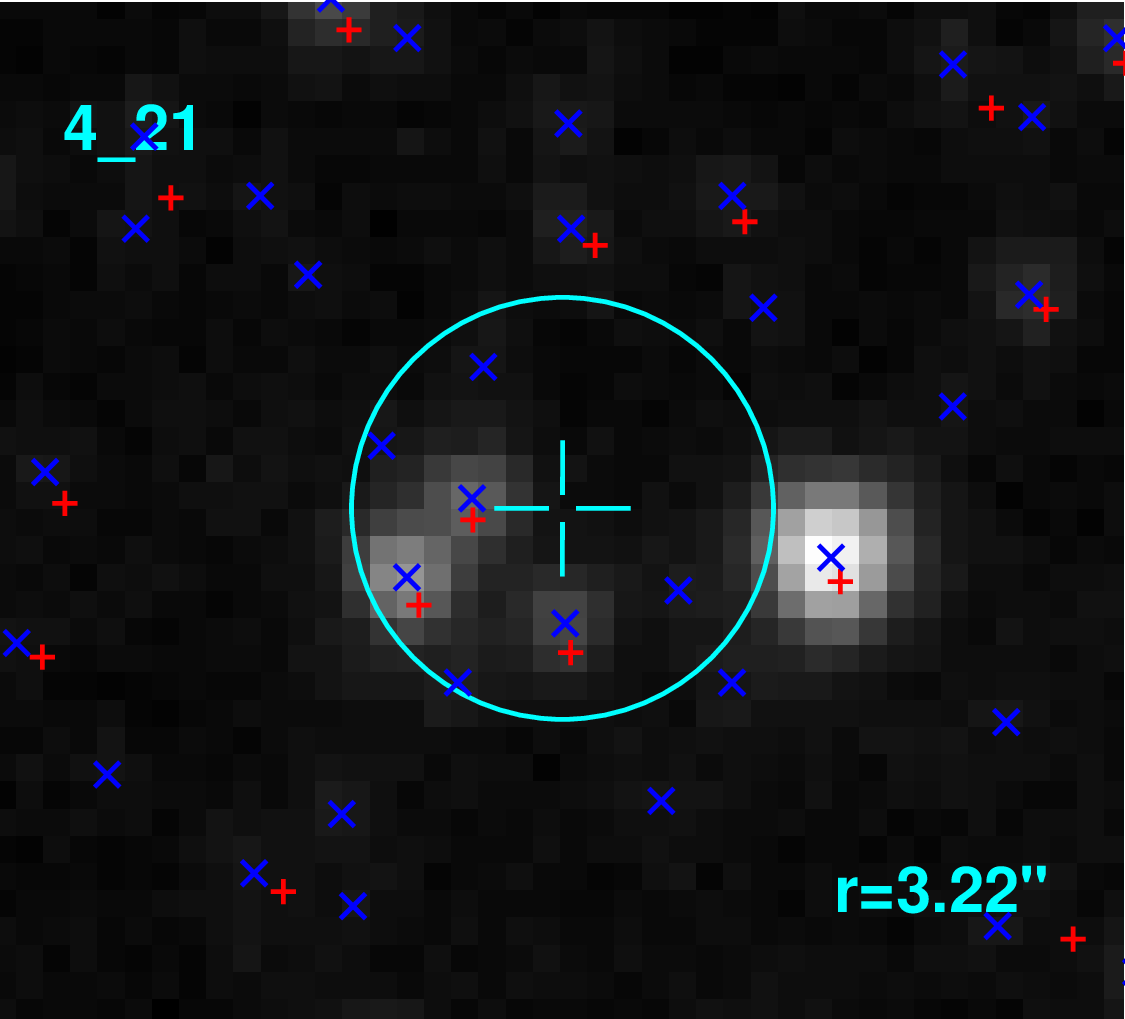} 
\end{center} 
\clearpage
\begin{center} 
\includegraphics[width=.45\textwidth]{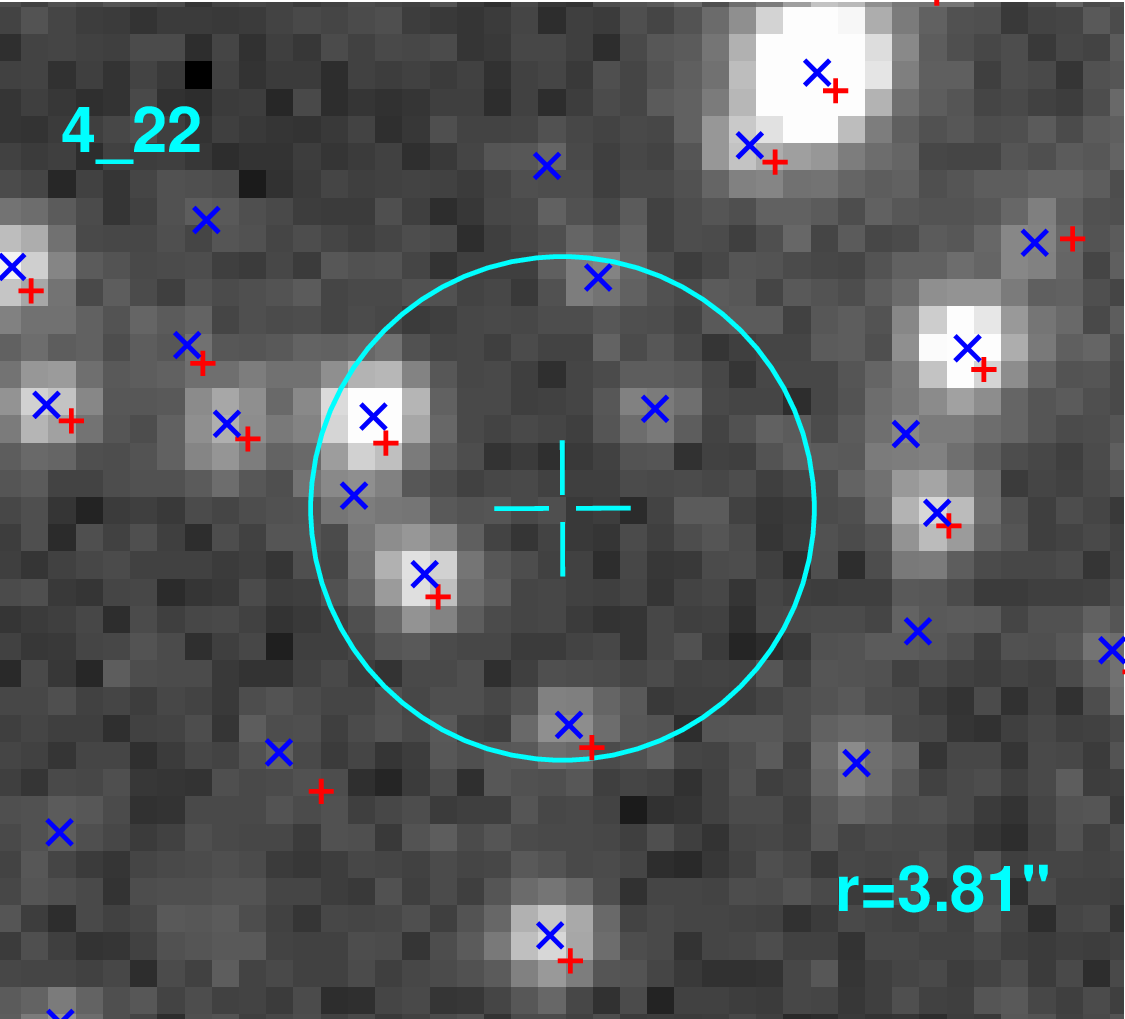} 
\includegraphics[width=.45\textwidth]{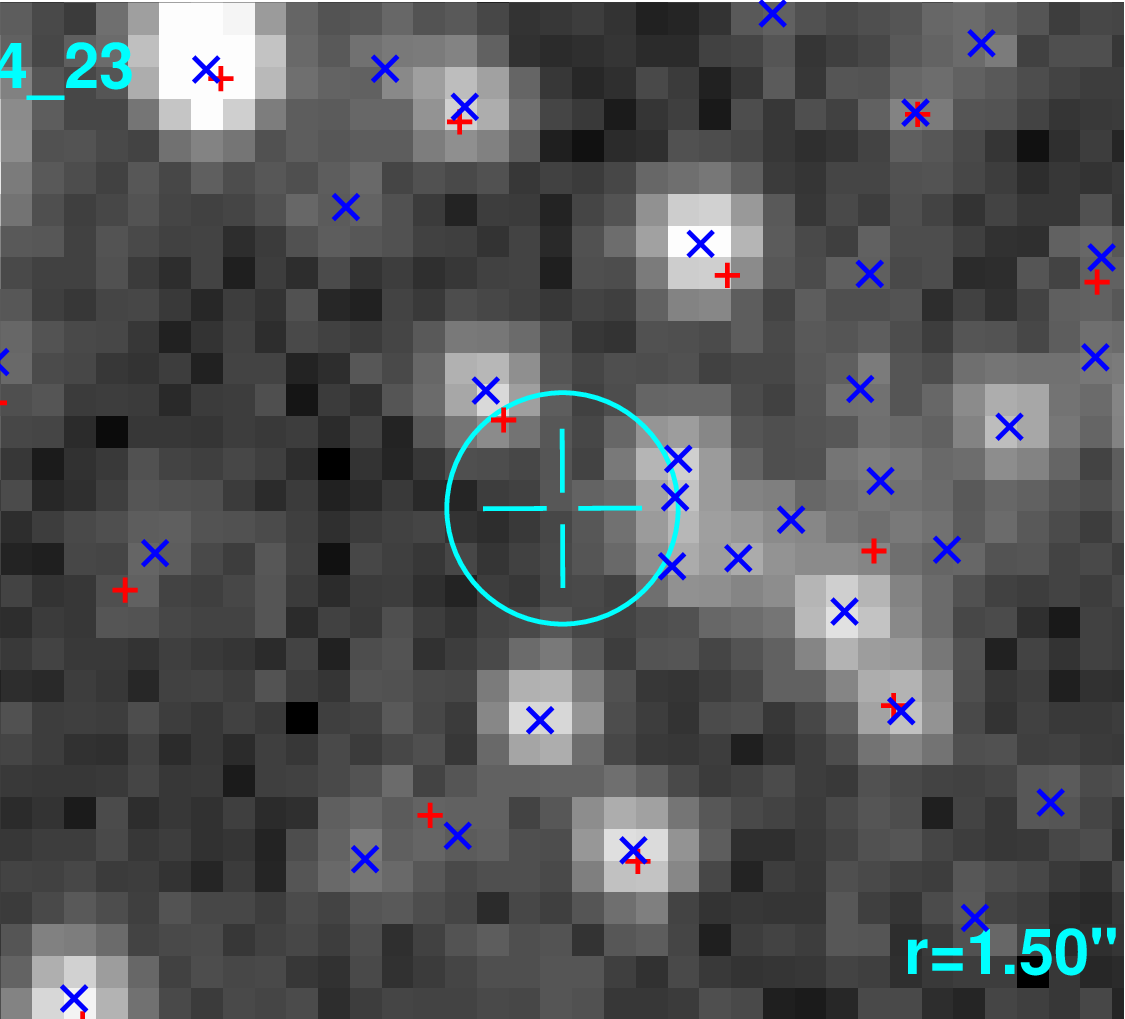} 
\includegraphics[width=.45\textwidth]{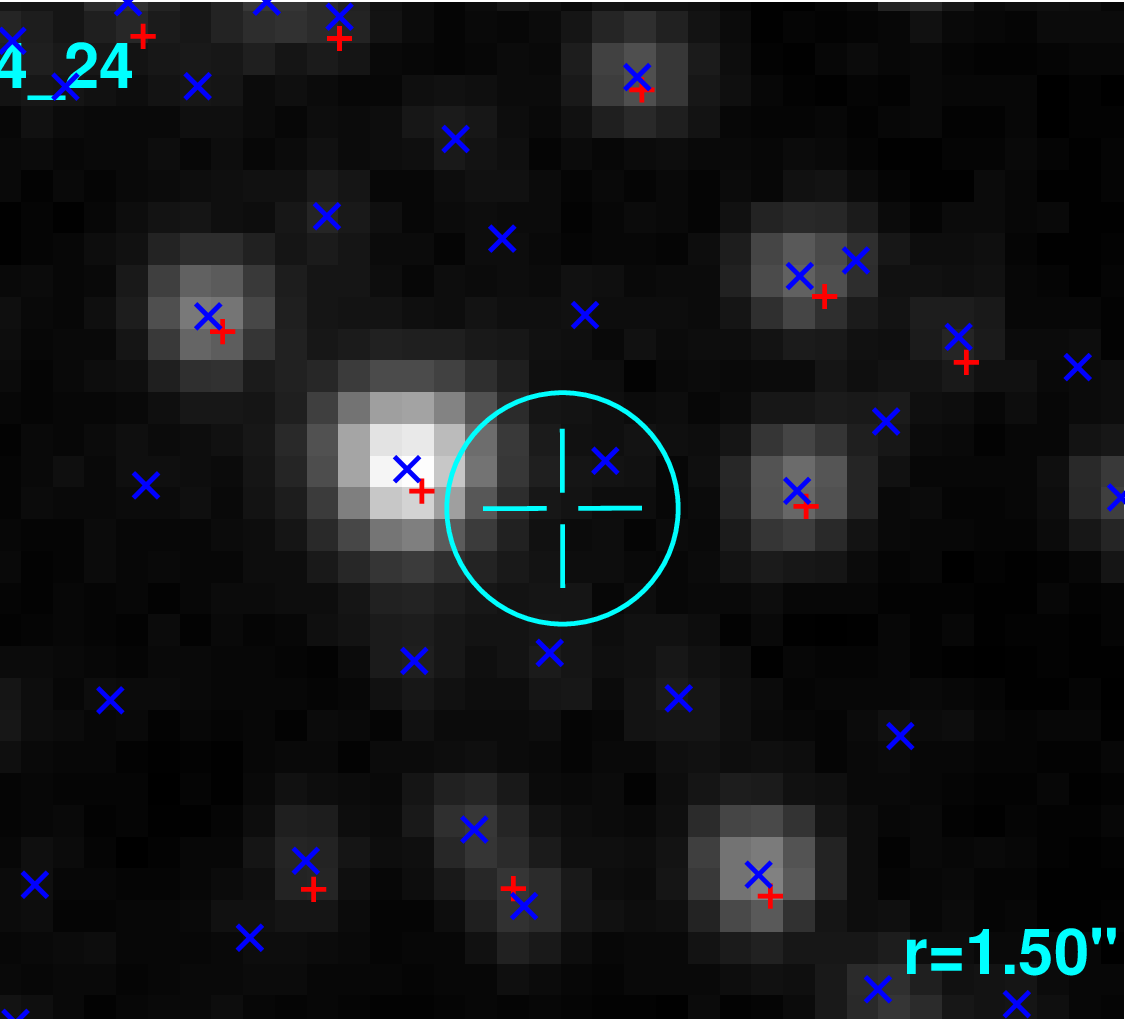} 
\includegraphics[width=.45\textwidth]{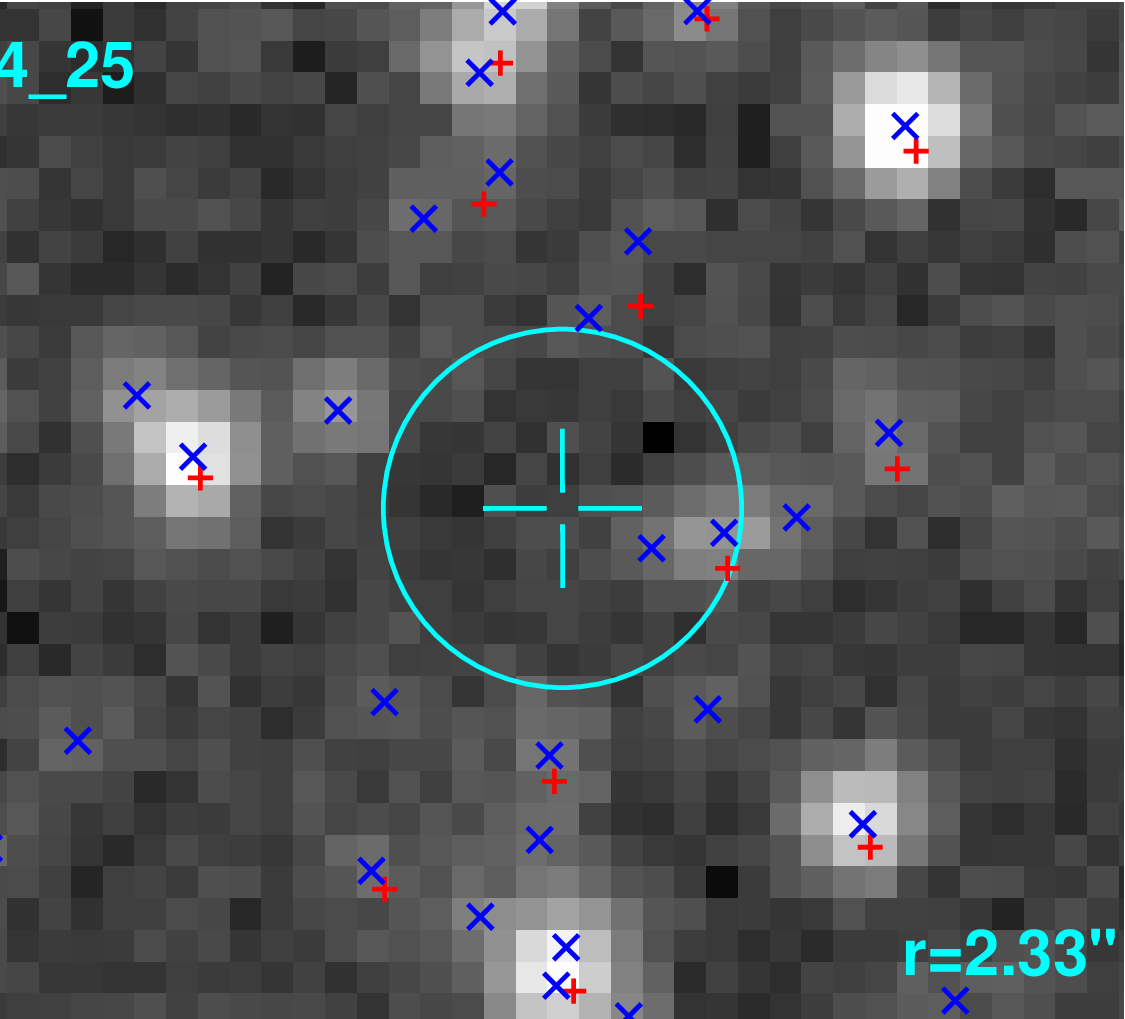} 
\end{center} 
\clearpage
\begin{center} 
\includegraphics[width=.45\textwidth]{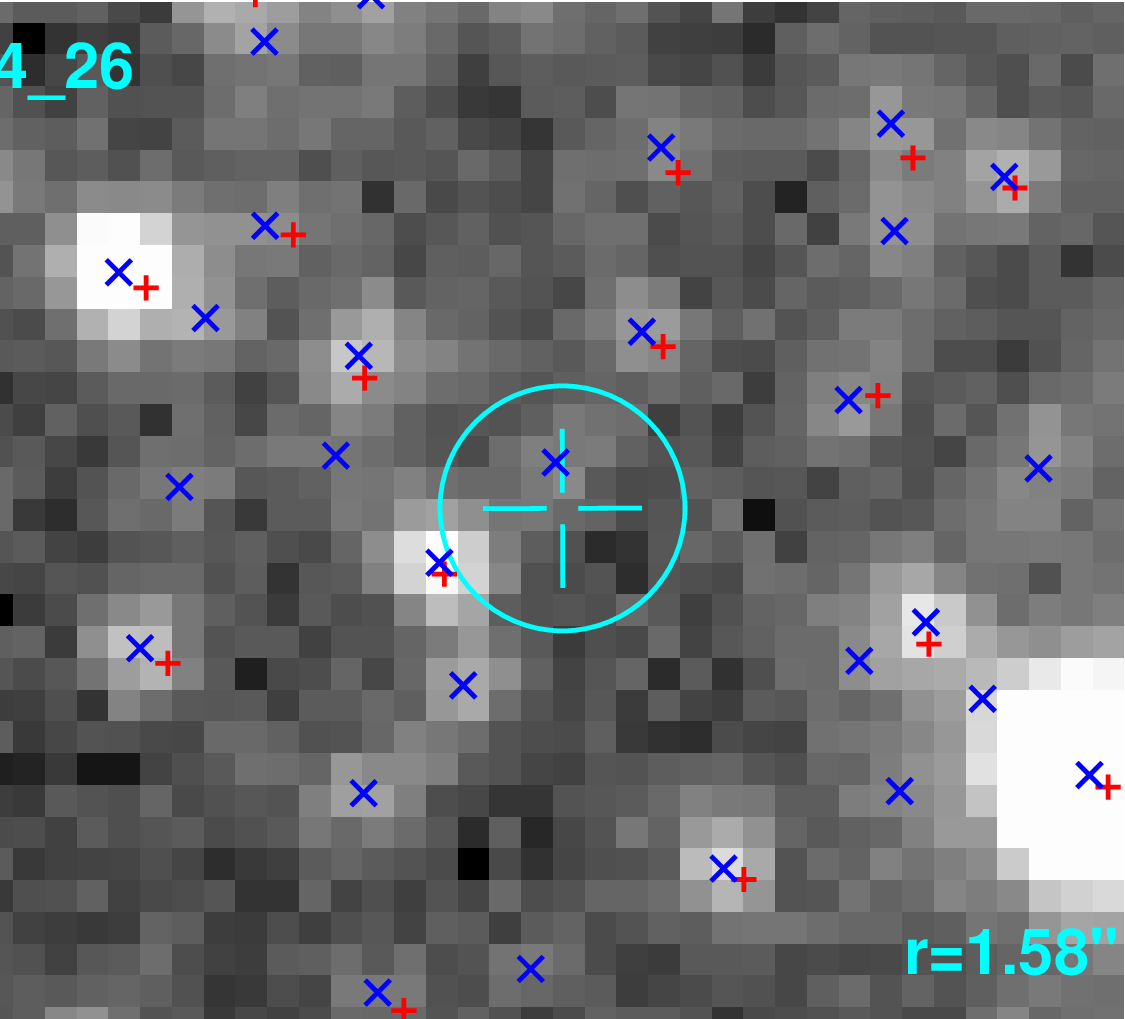} 
\includegraphics[width=.45\textwidth]{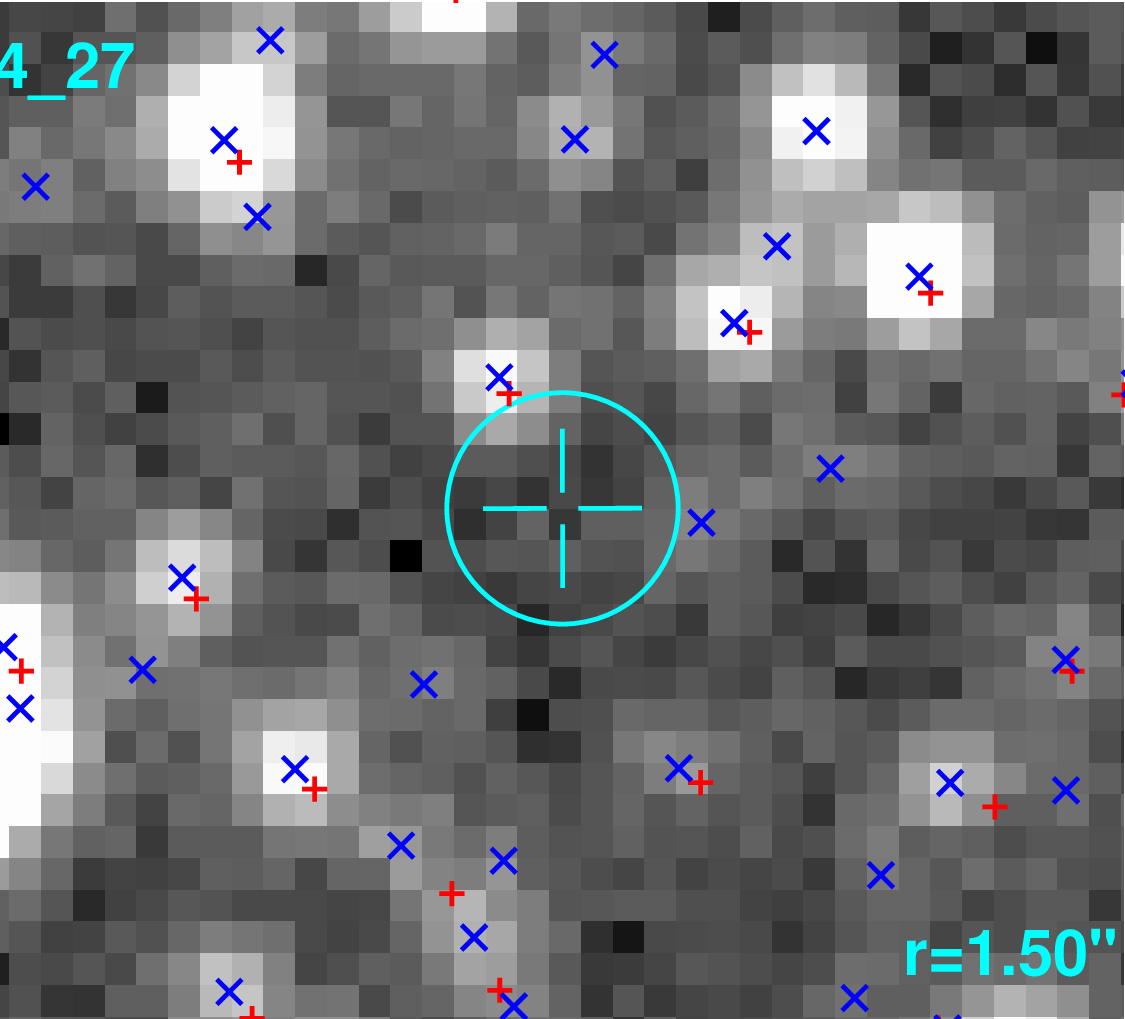} 
\includegraphics[width=.45\textwidth]{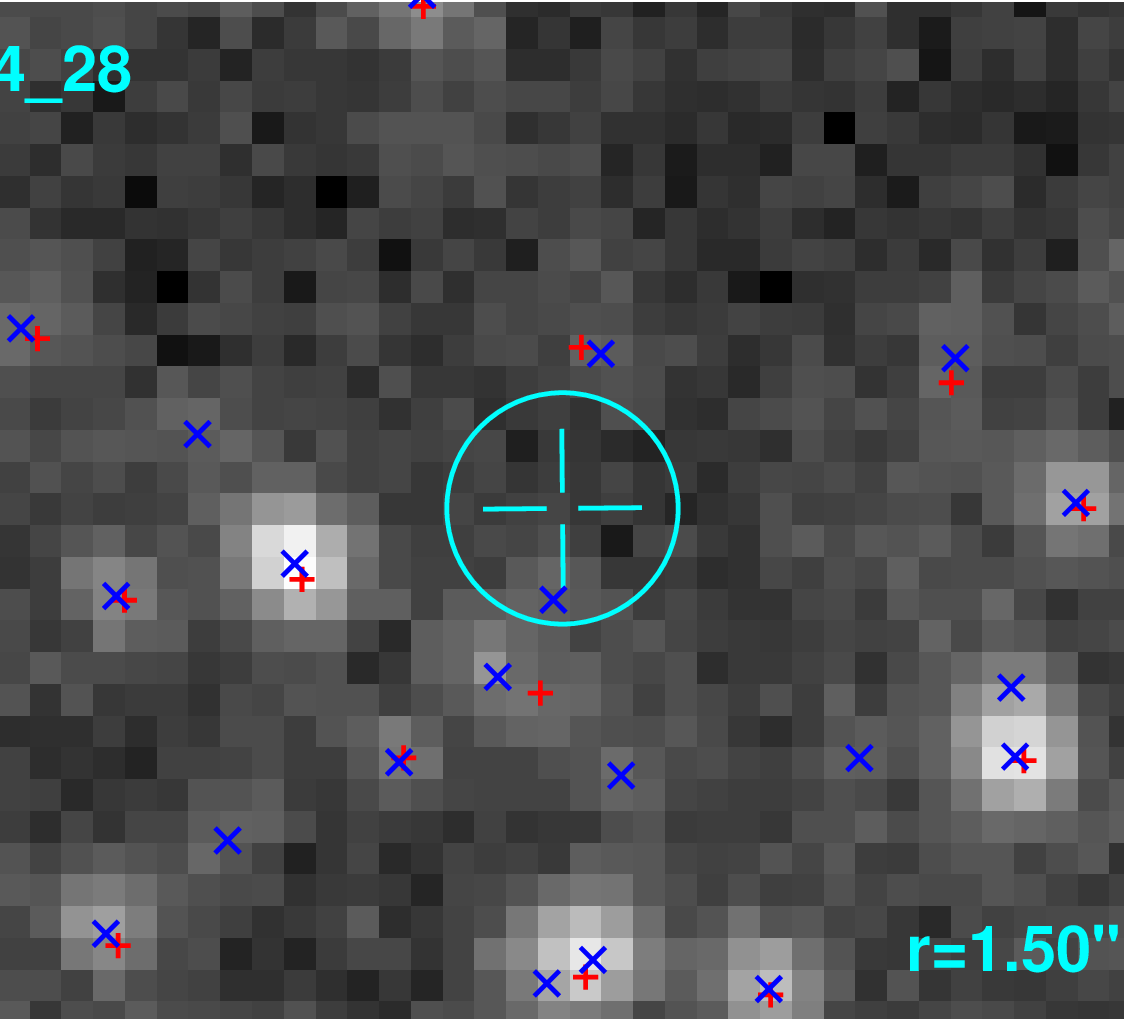} 
\includegraphics[width=.45\textwidth]{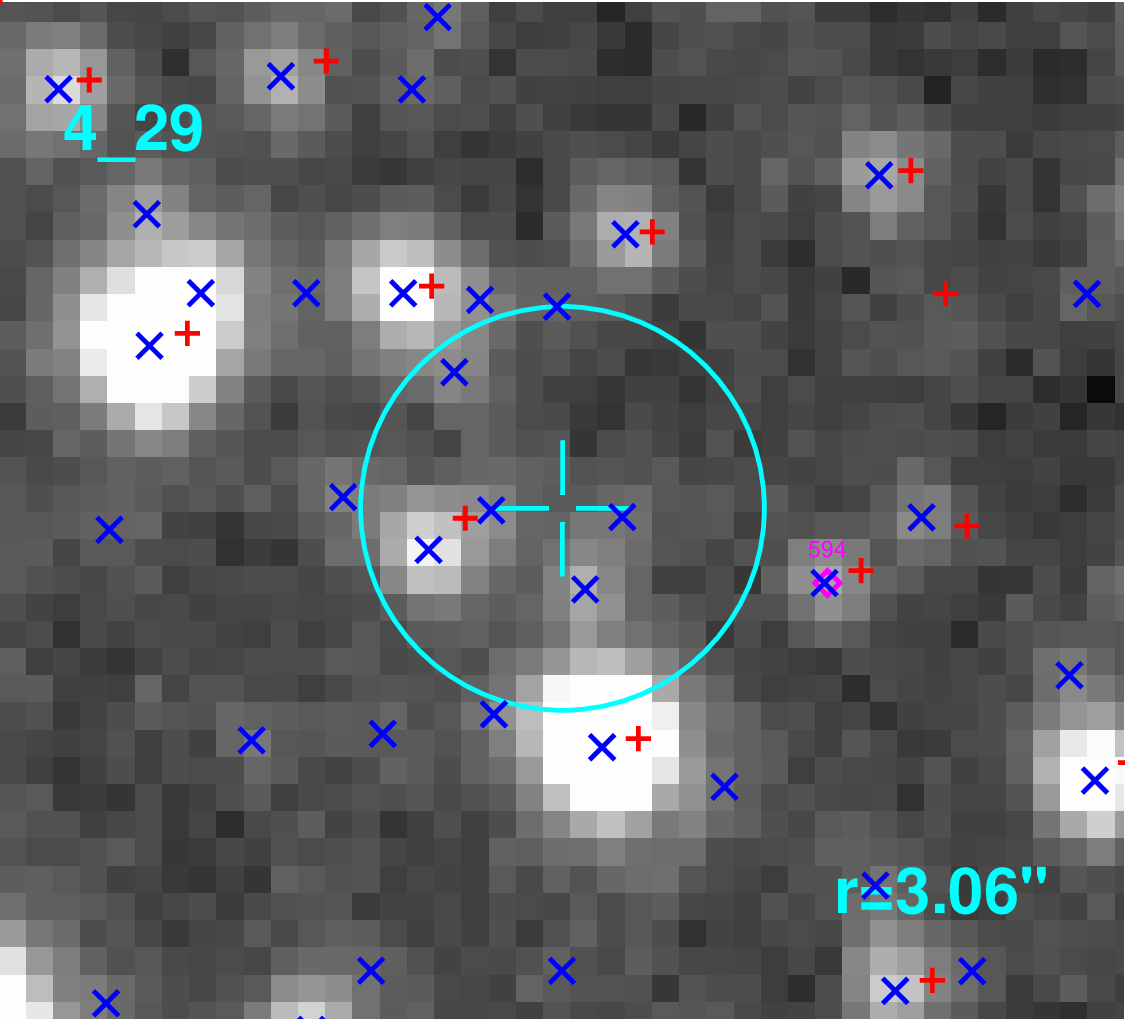} 
\end{center} 
\clearpage
\begin{center} 
\includegraphics[width=.45\textwidth]{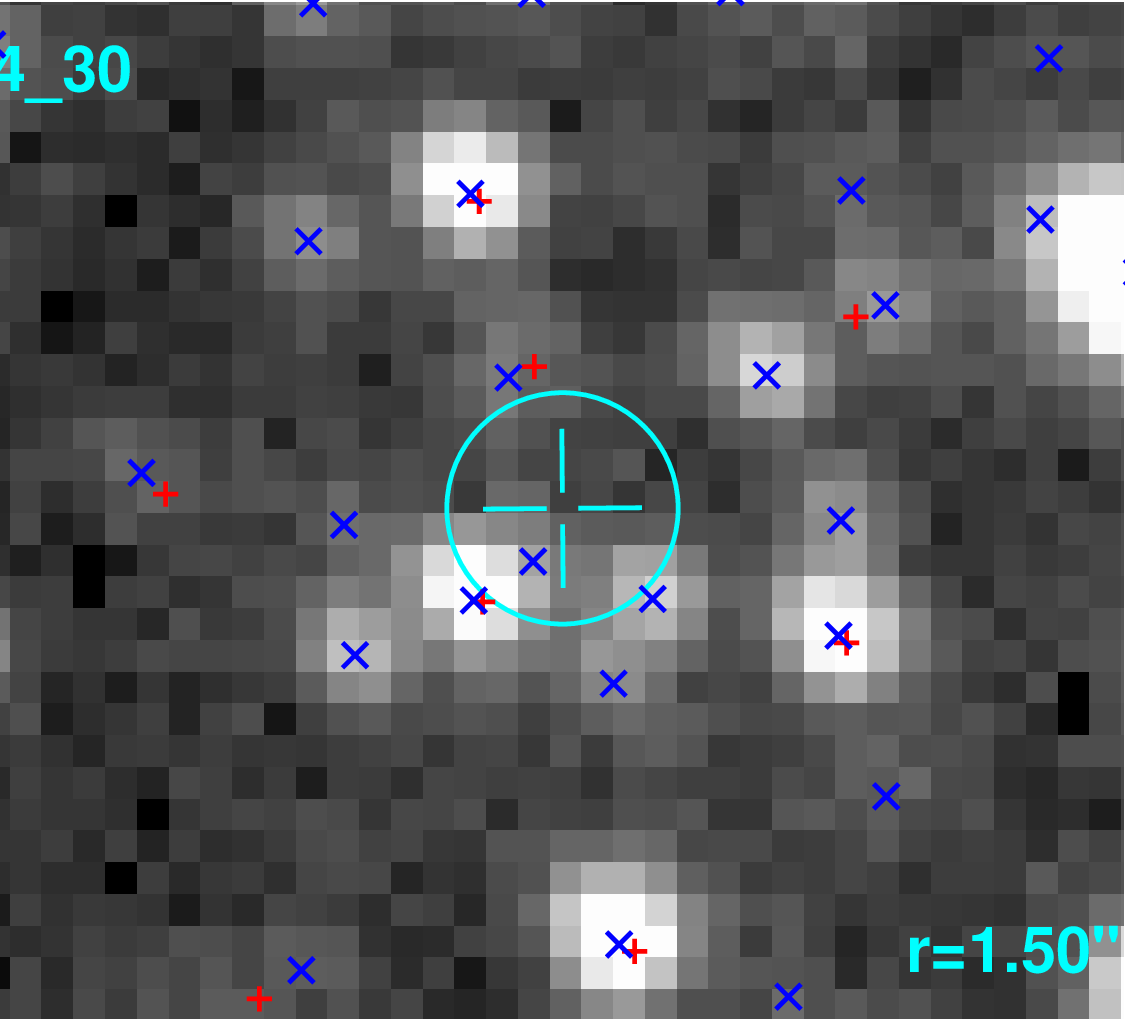} 
\includegraphics[width=.45\textwidth]{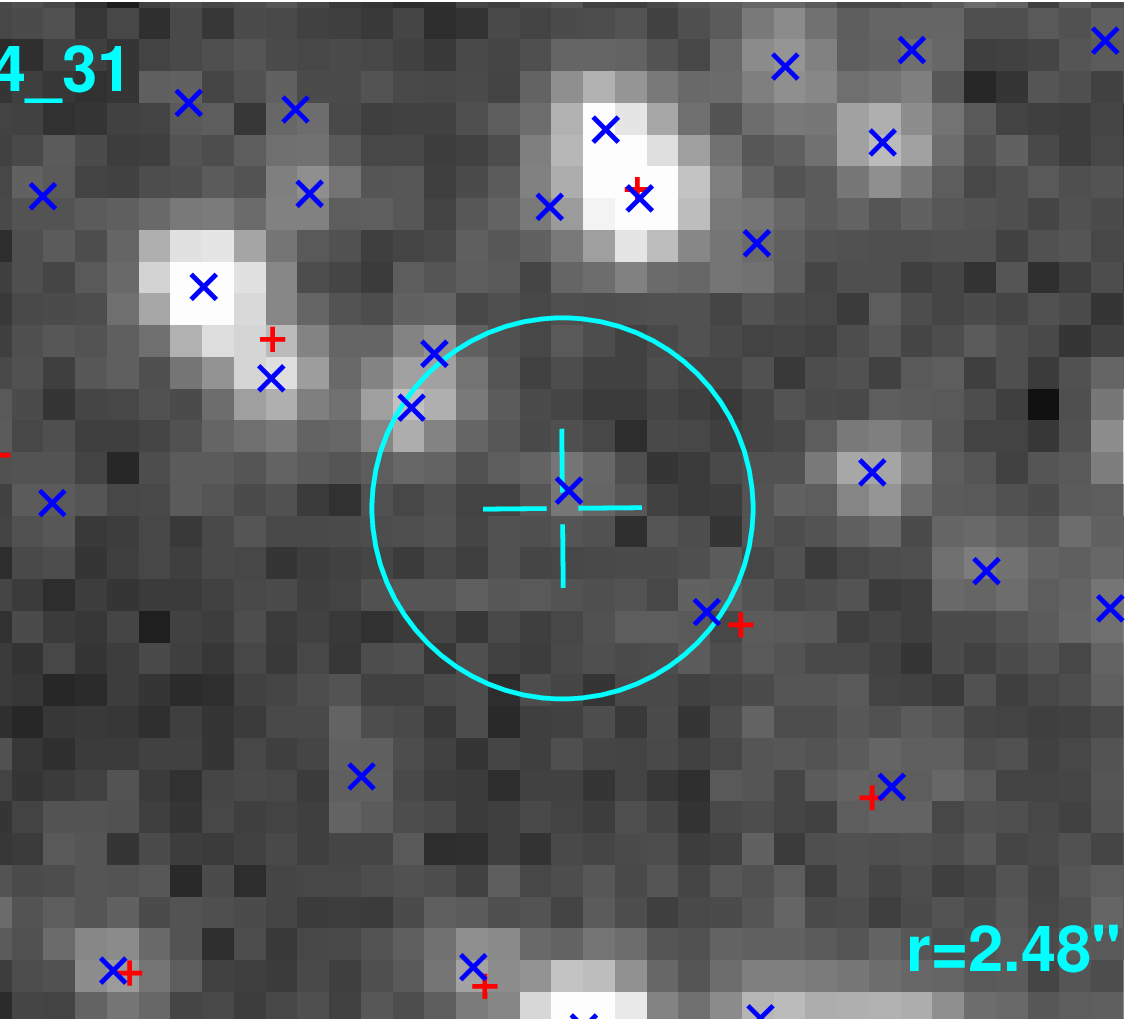} 
\includegraphics[width=.45\textwidth]{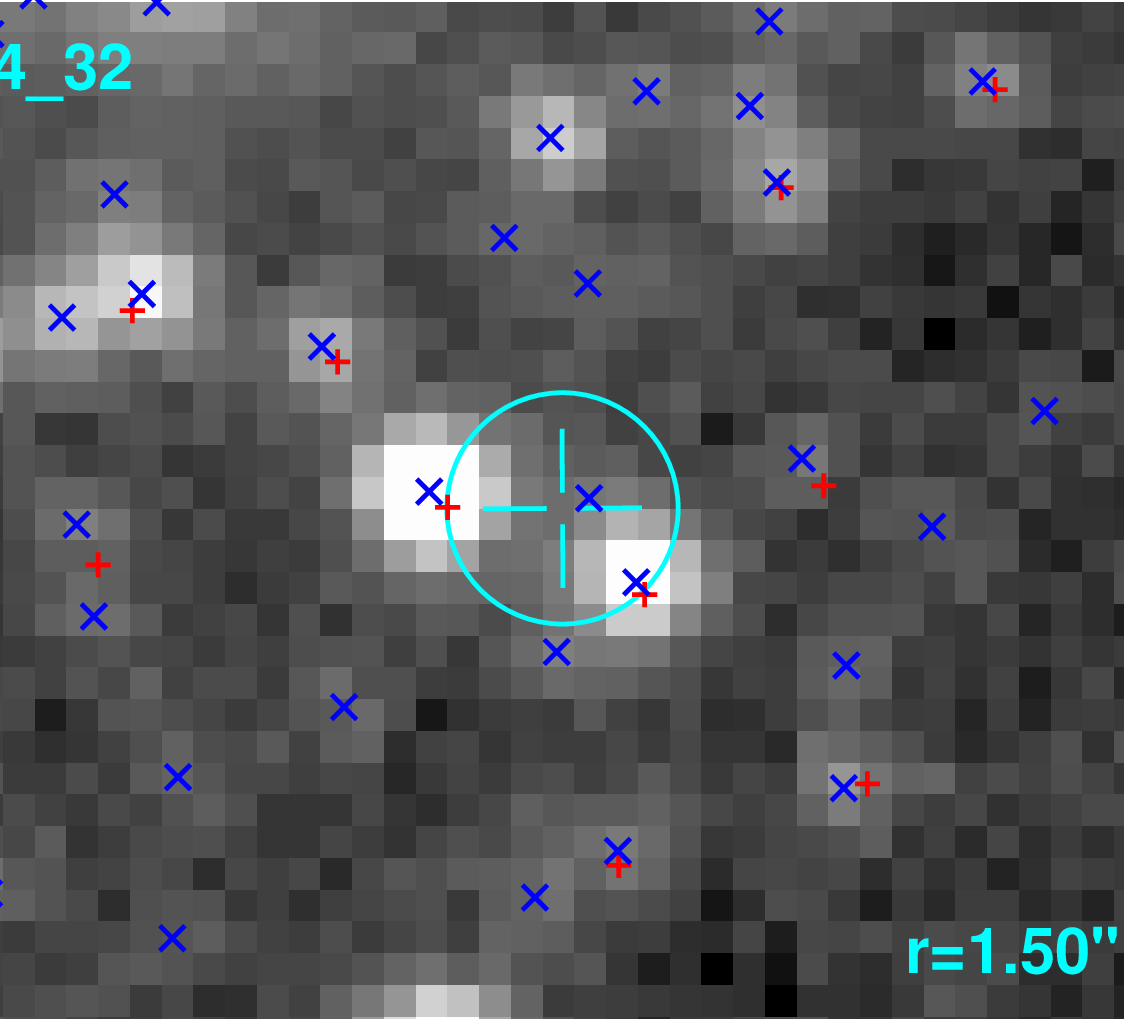} 
\includegraphics[width=.45\textwidth]{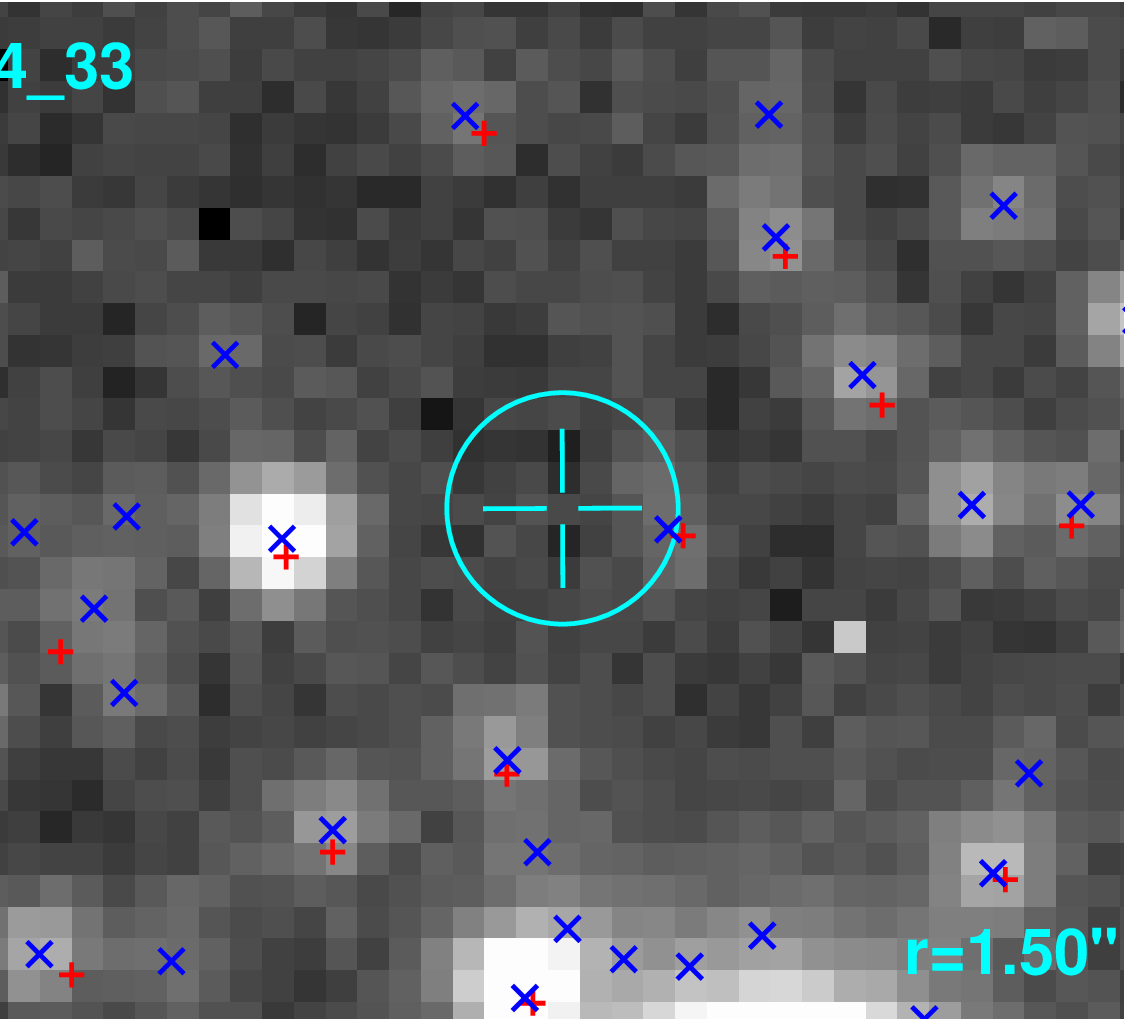} 
\end{center} 
\clearpage
\begin{center} 
\includegraphics[width=.45\textwidth]{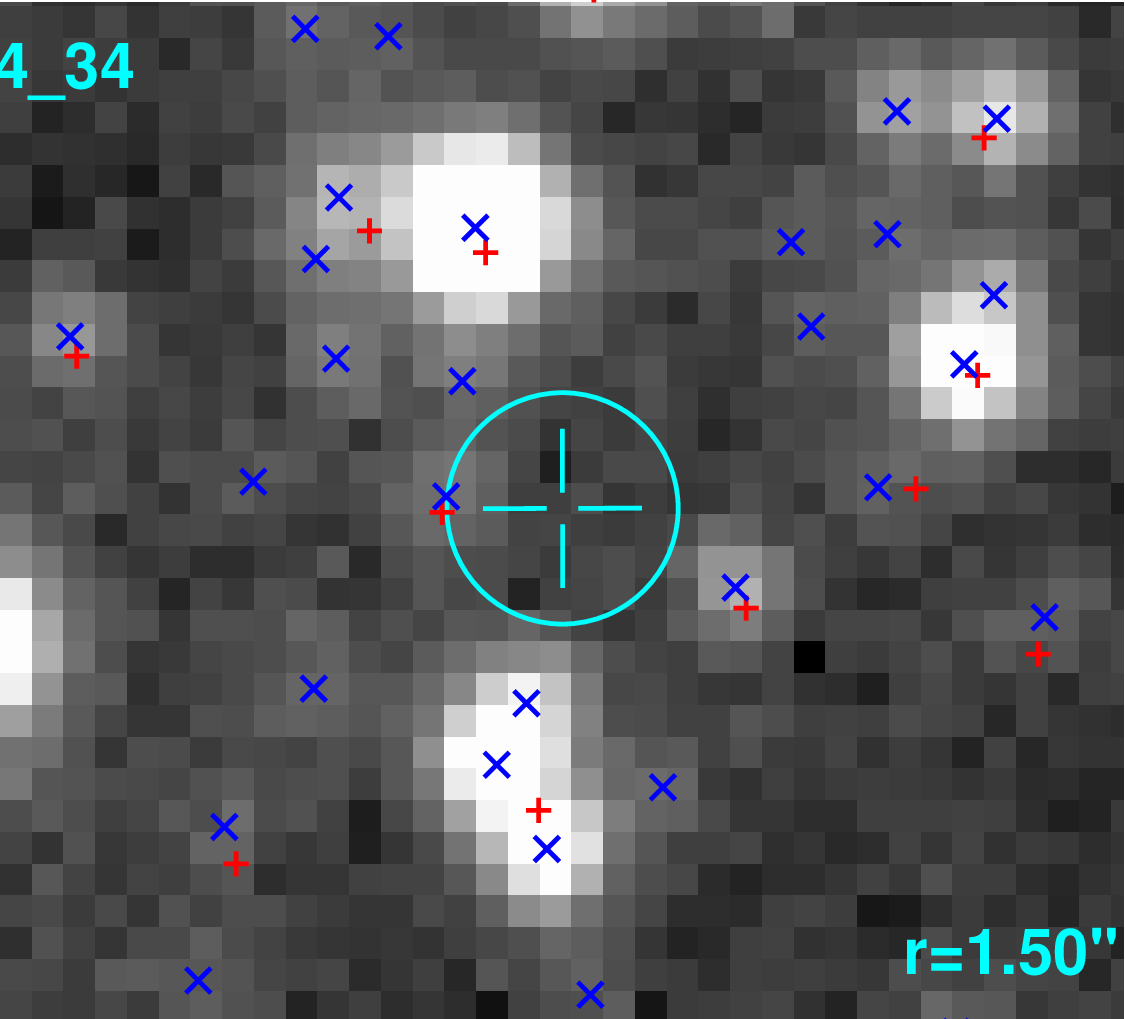} 
\includegraphics[width=.45\textwidth]{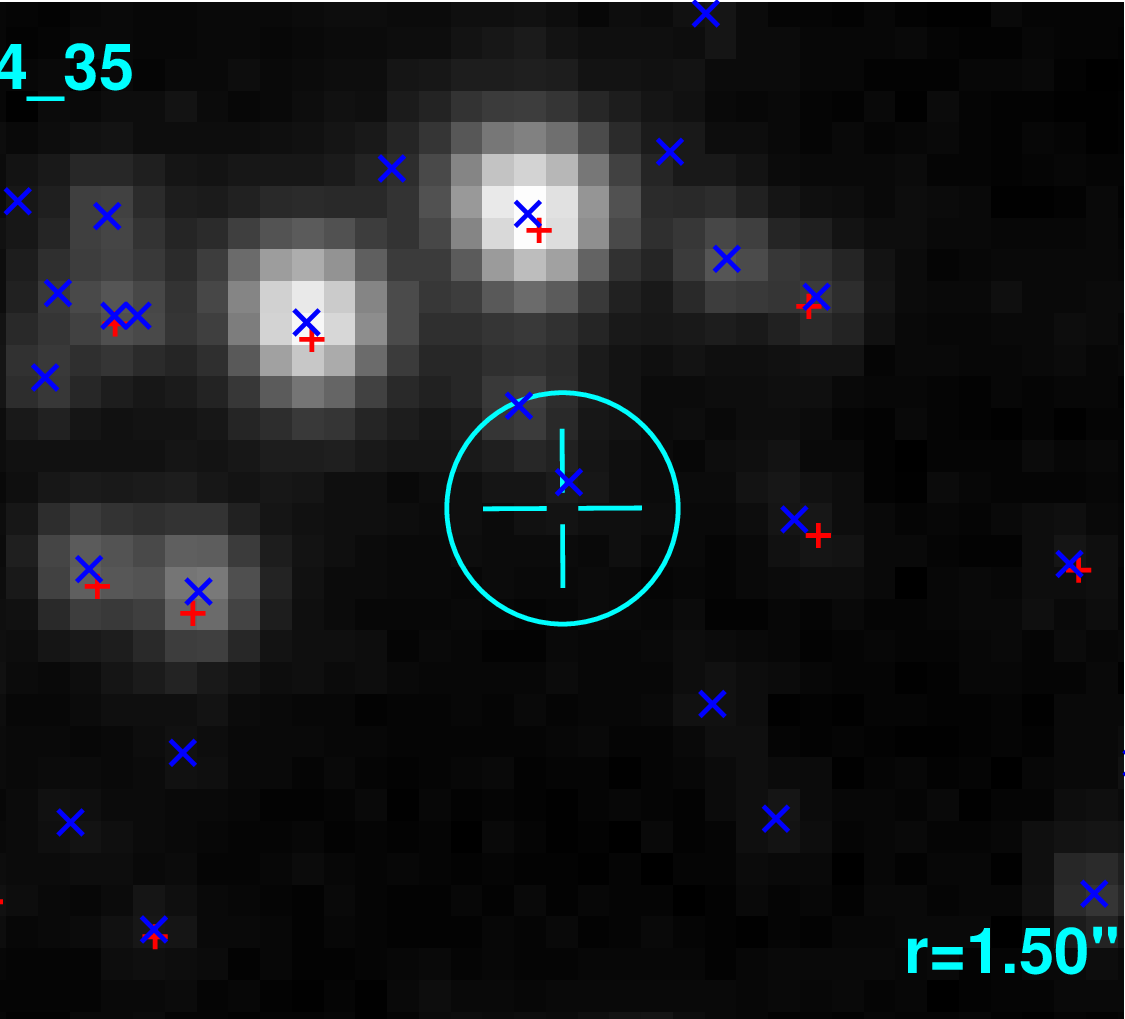} 
\includegraphics[width=.45\textwidth]{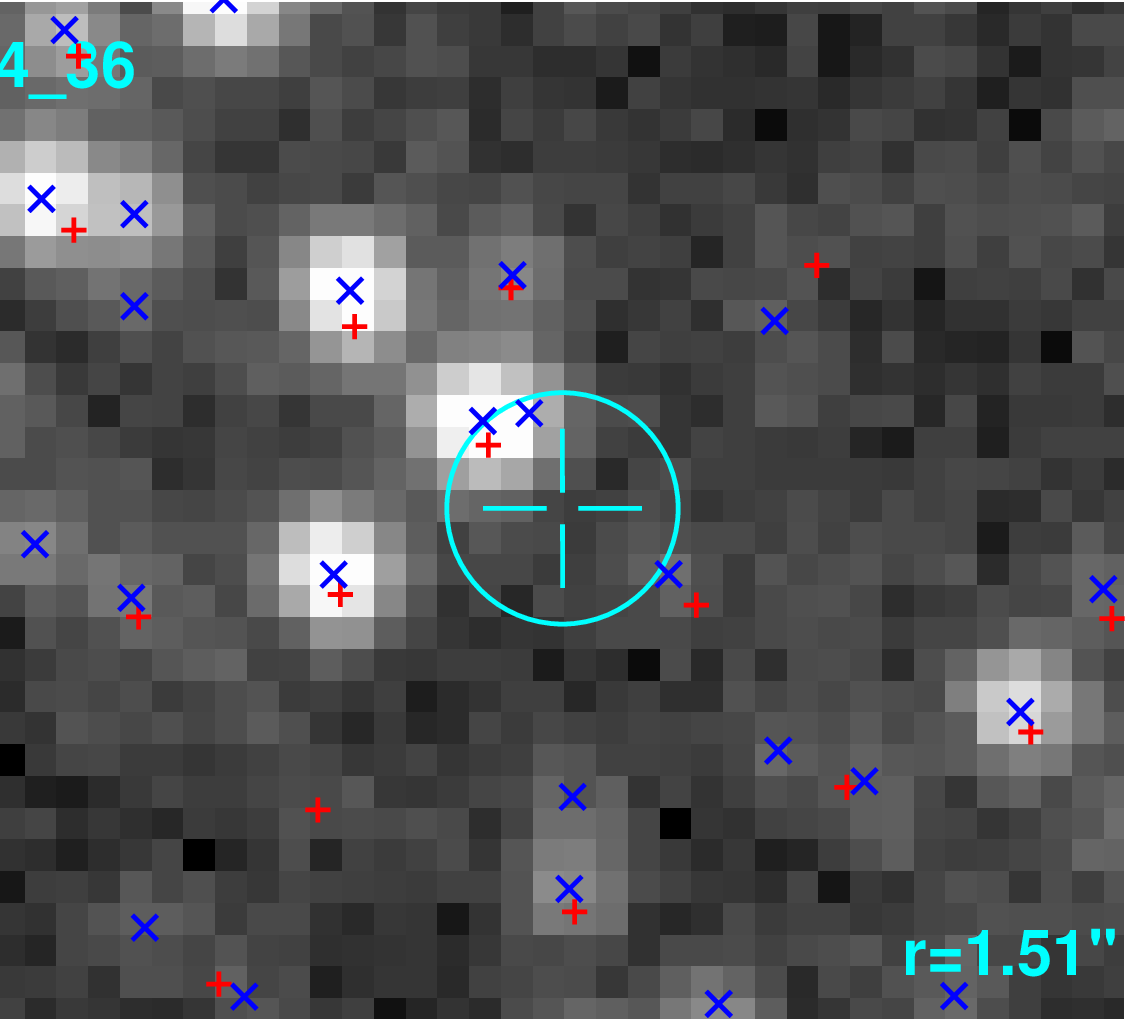} 
\includegraphics[width=.45\textwidth]{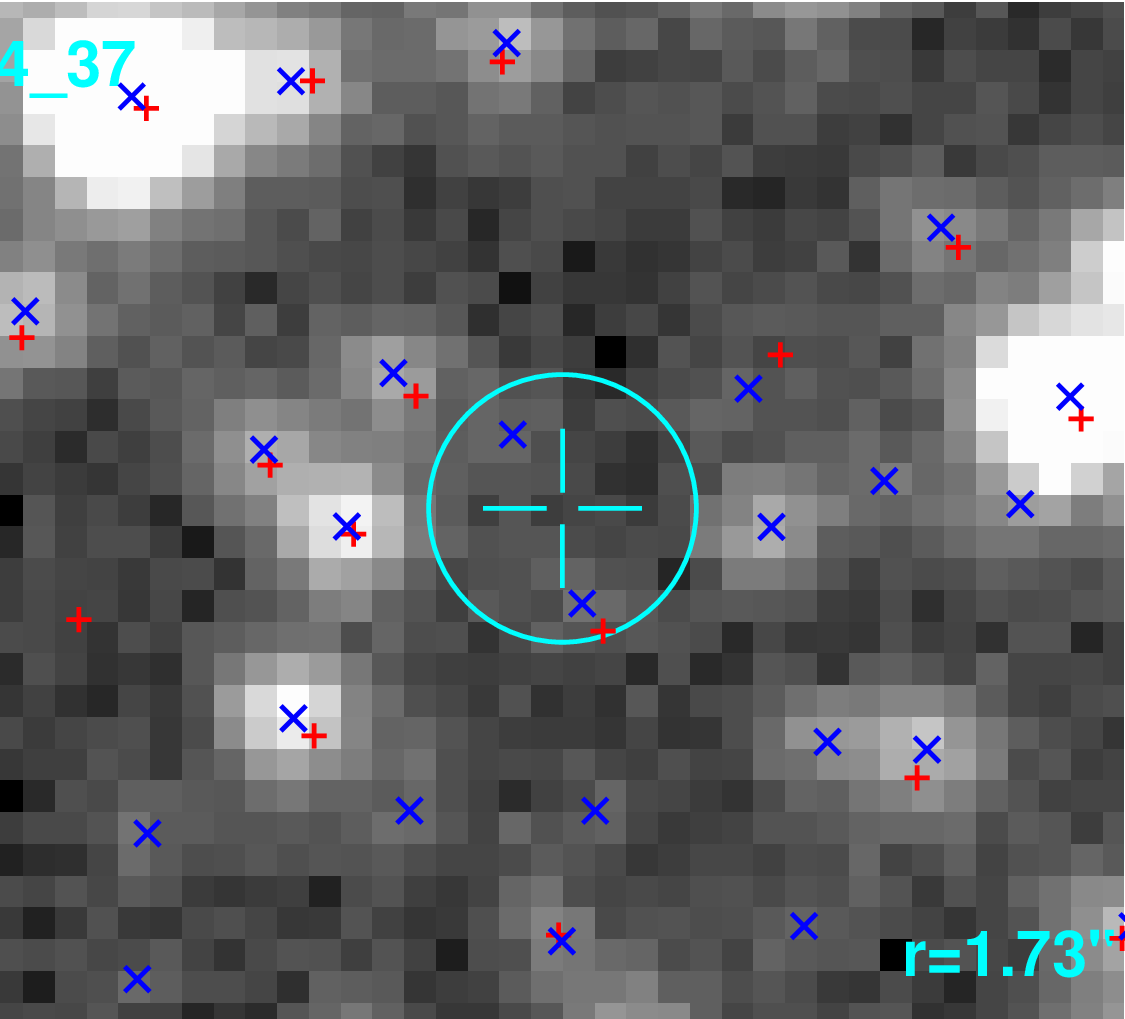} 
\end{center} 
\clearpage

\clearpage
\begin{center} 
\includegraphics[width=.45\textwidth]{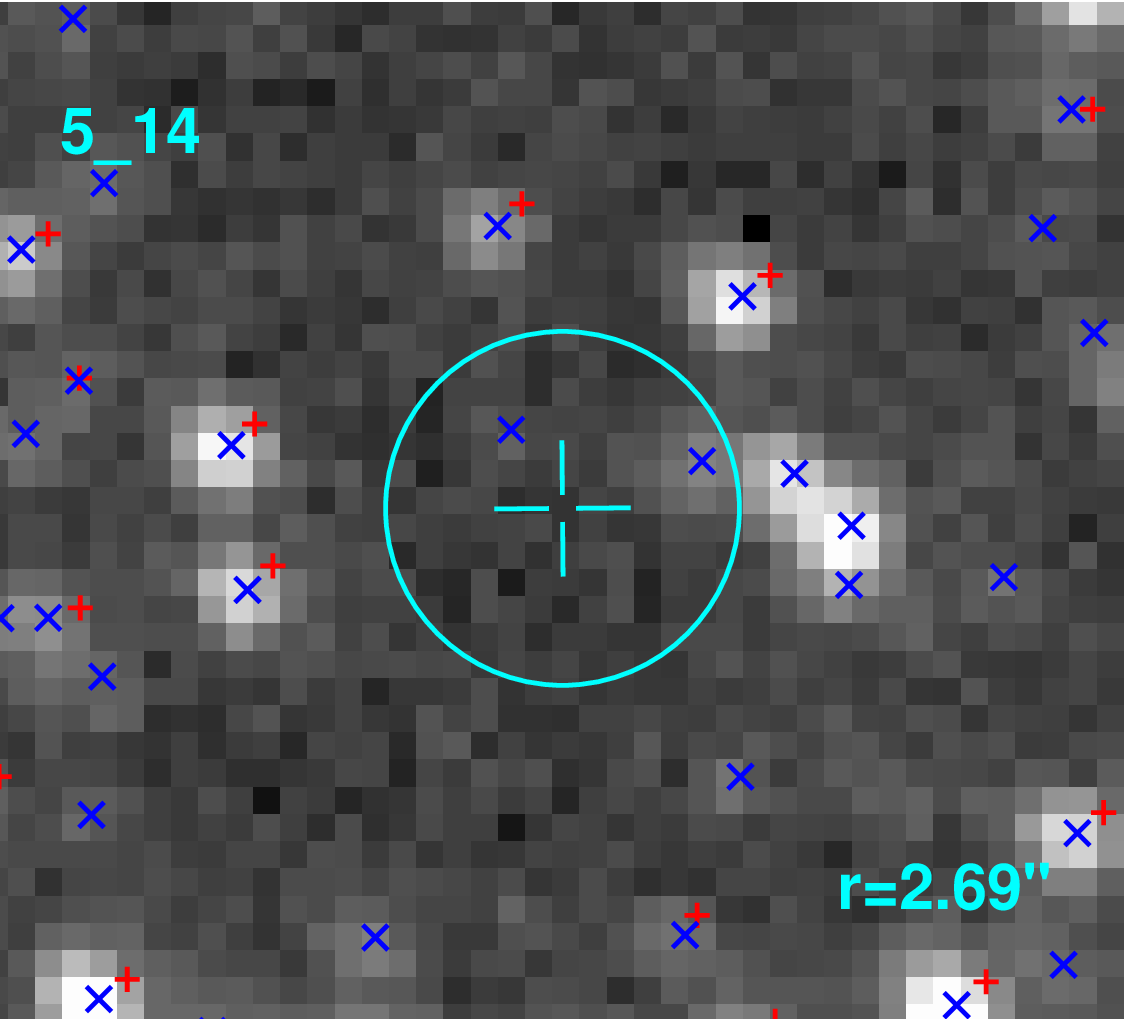} 
\includegraphics[width=.45\textwidth]{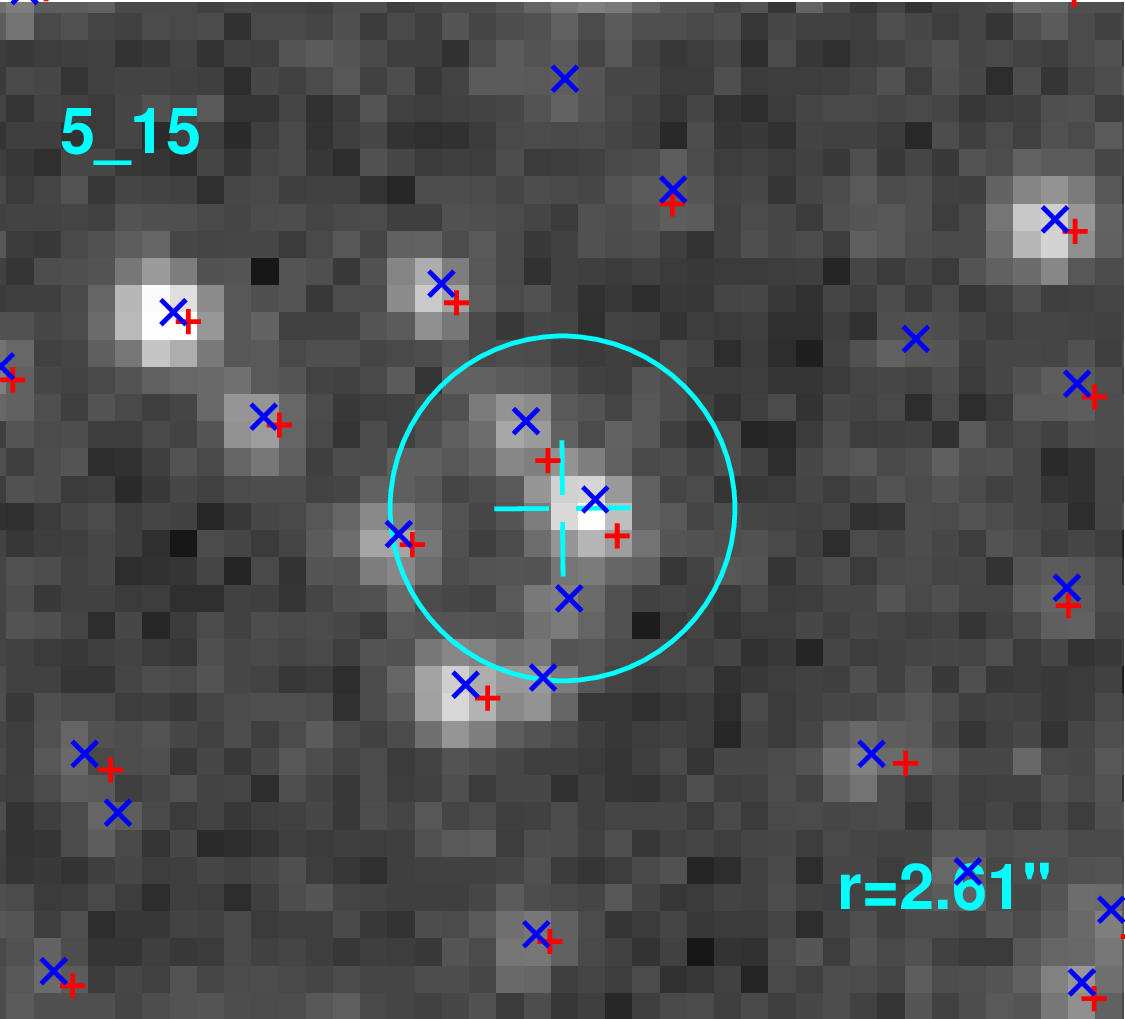} 
\includegraphics[width=.45\textwidth]{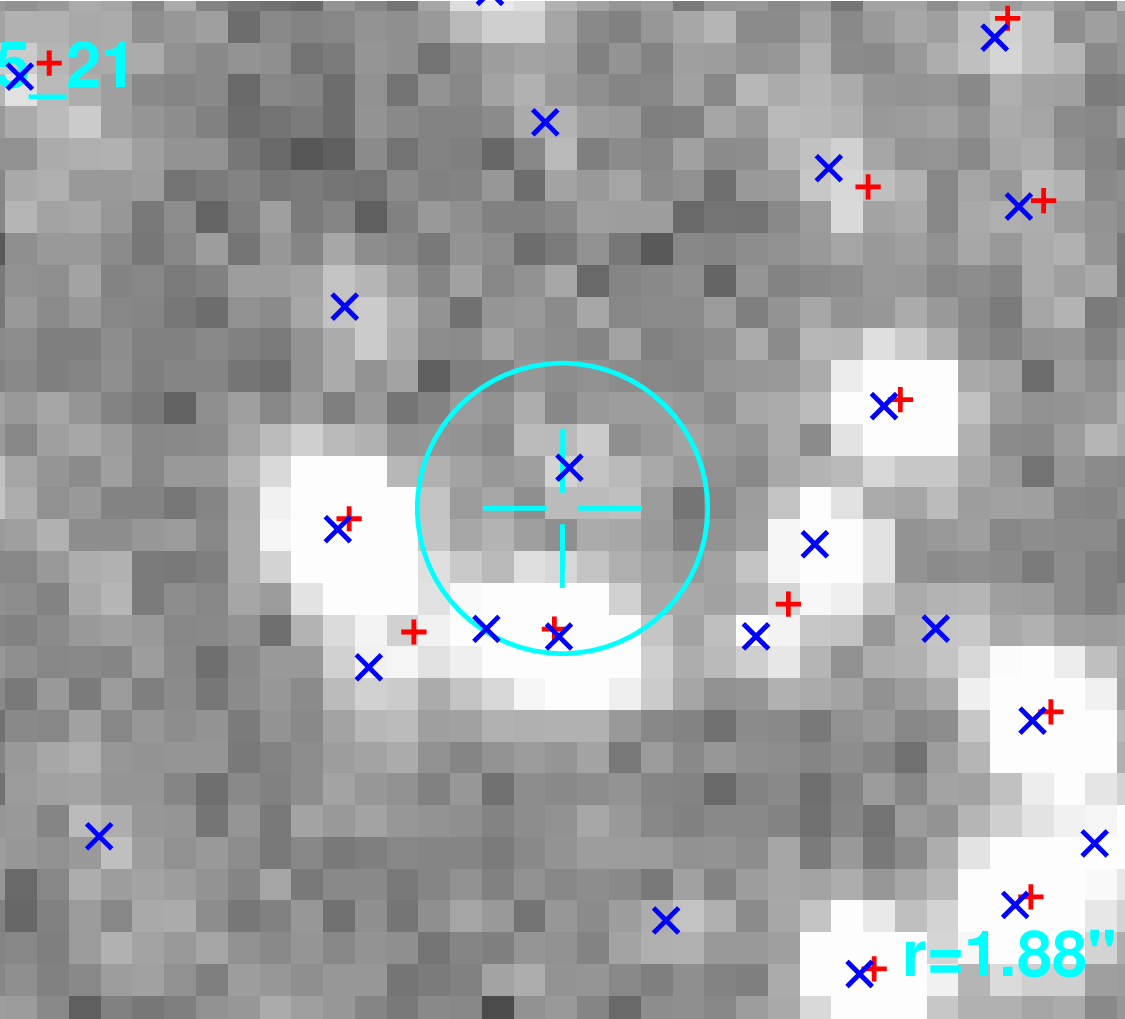} 
\includegraphics[width=.45\textwidth]{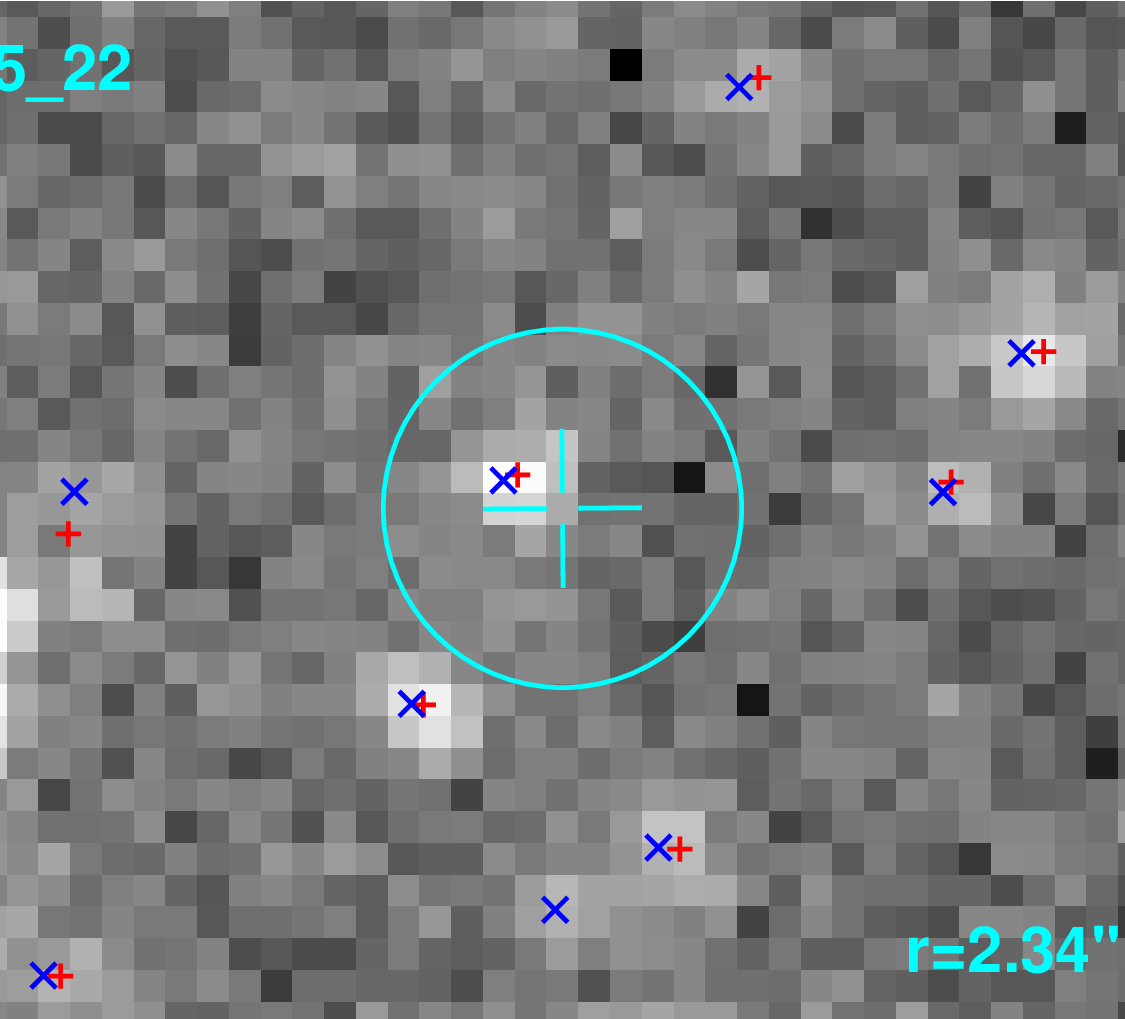} 
\end{center} 
\clearpage
\begin{center} 
\includegraphics[width=.45\textwidth]{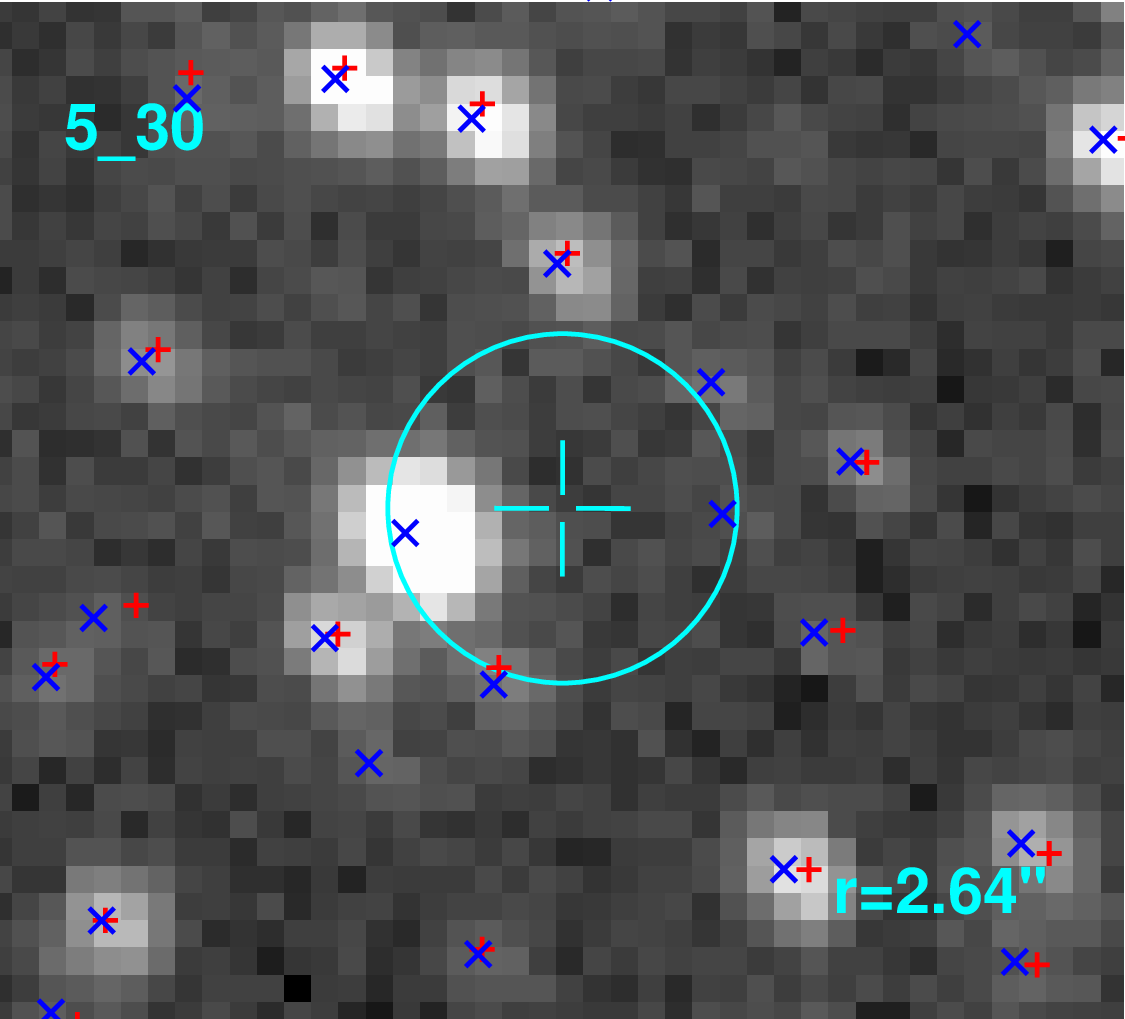} 
\includegraphics[width=.45\textwidth]{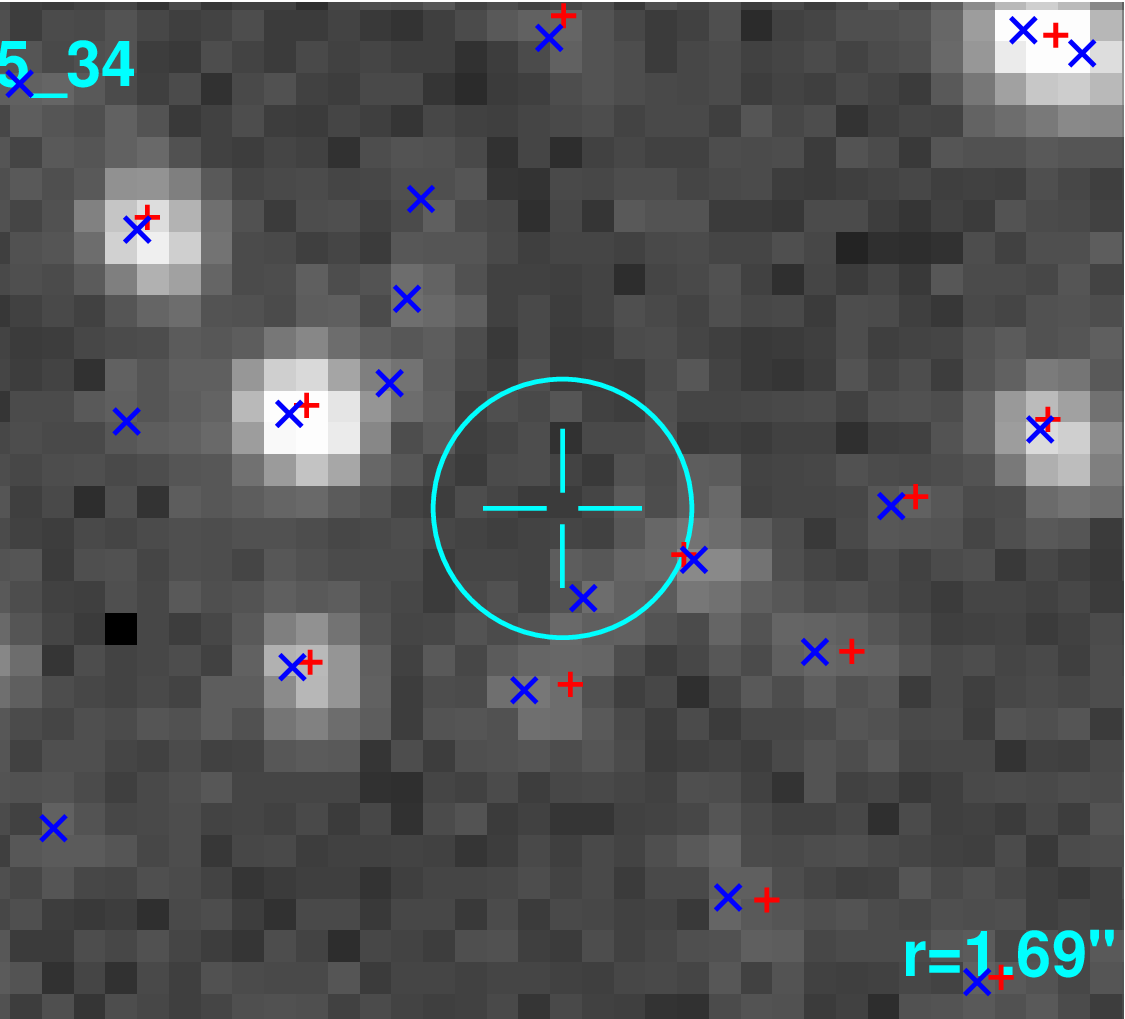} 
\includegraphics[width=.45\textwidth]{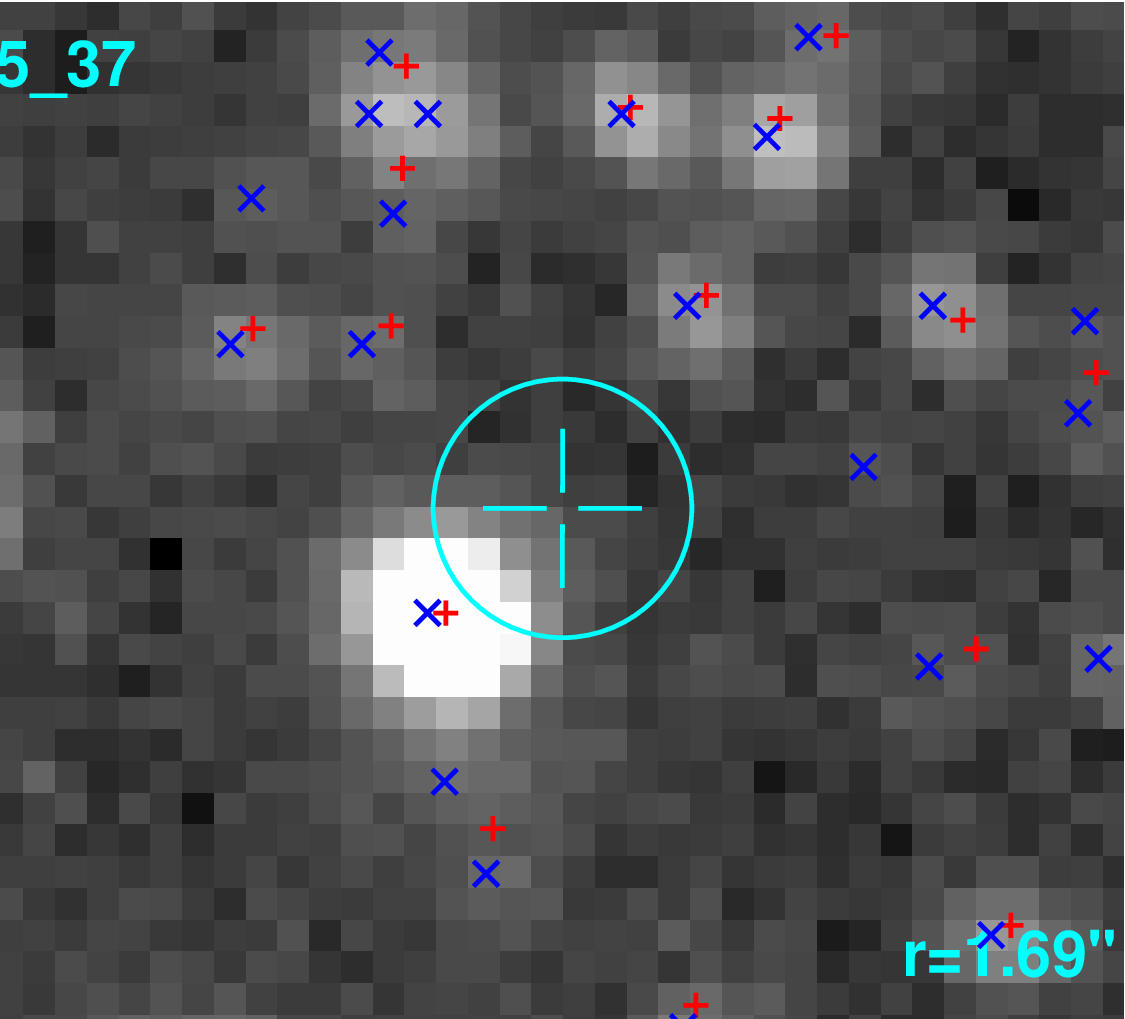} 
\end{center} 
\clearpage

\clearpage
\begin{center} 
\includegraphics[width=.45\textwidth]{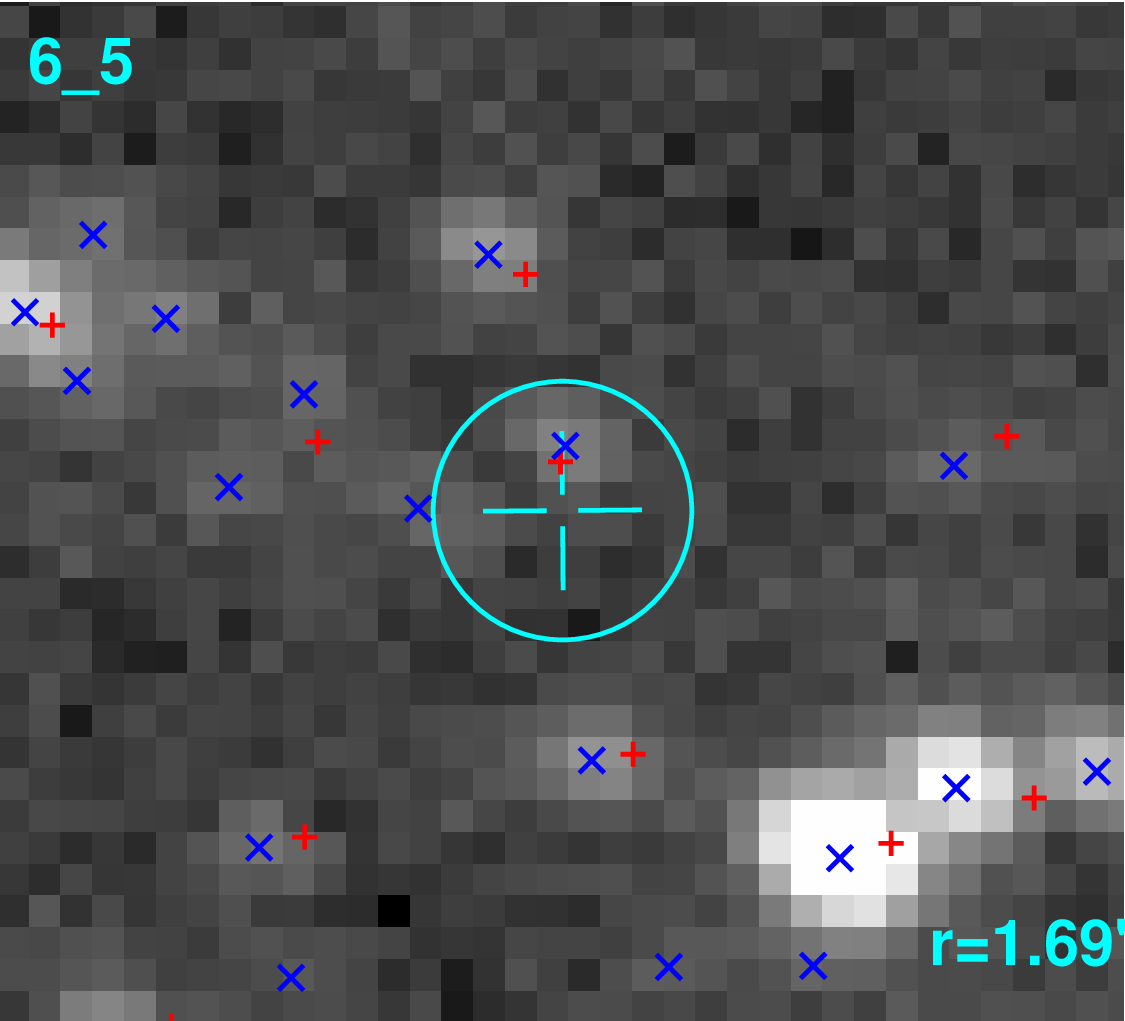} 
\includegraphics[width=.45\textwidth]{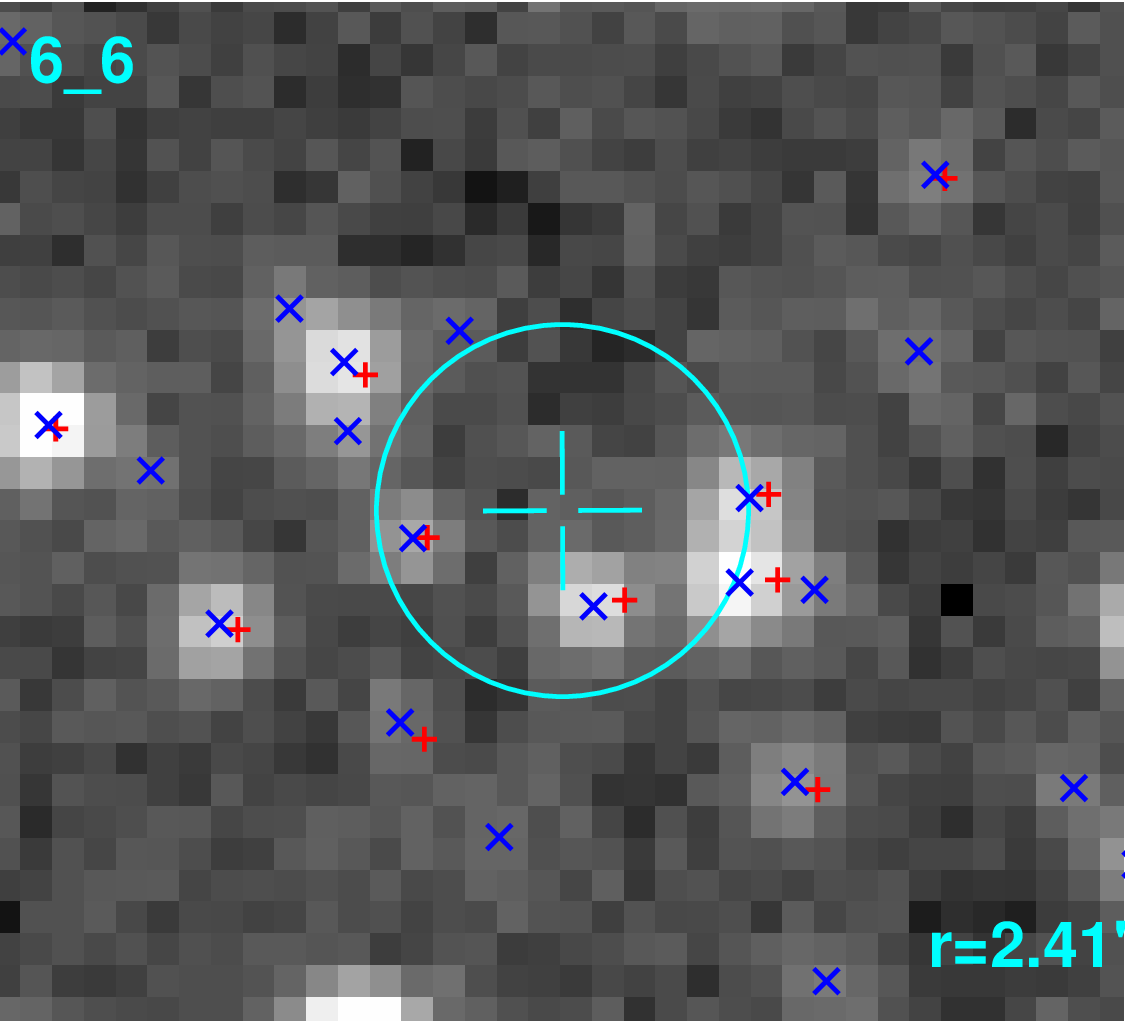} 
\includegraphics[width=.45\textwidth]{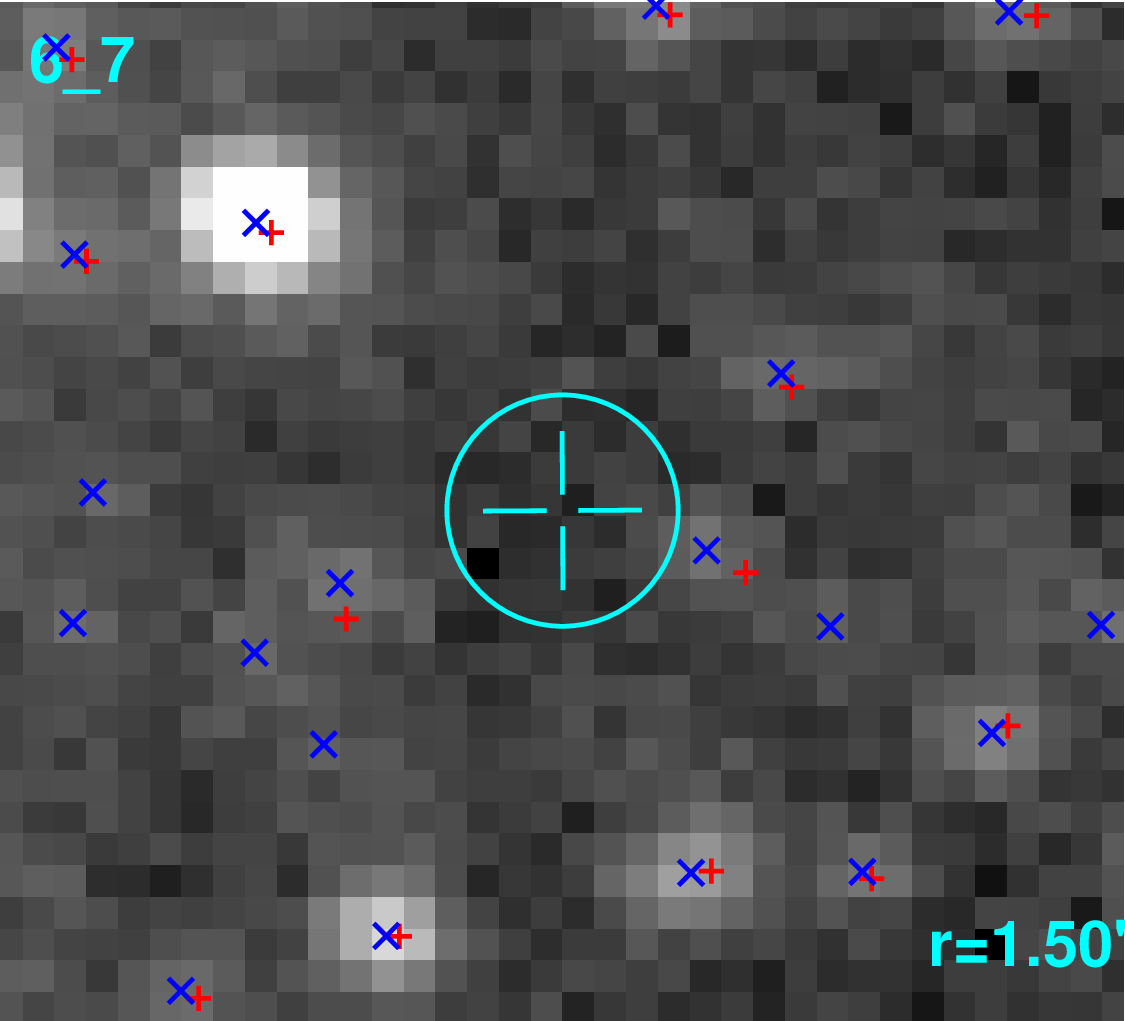} 
\includegraphics[width=.45\textwidth]{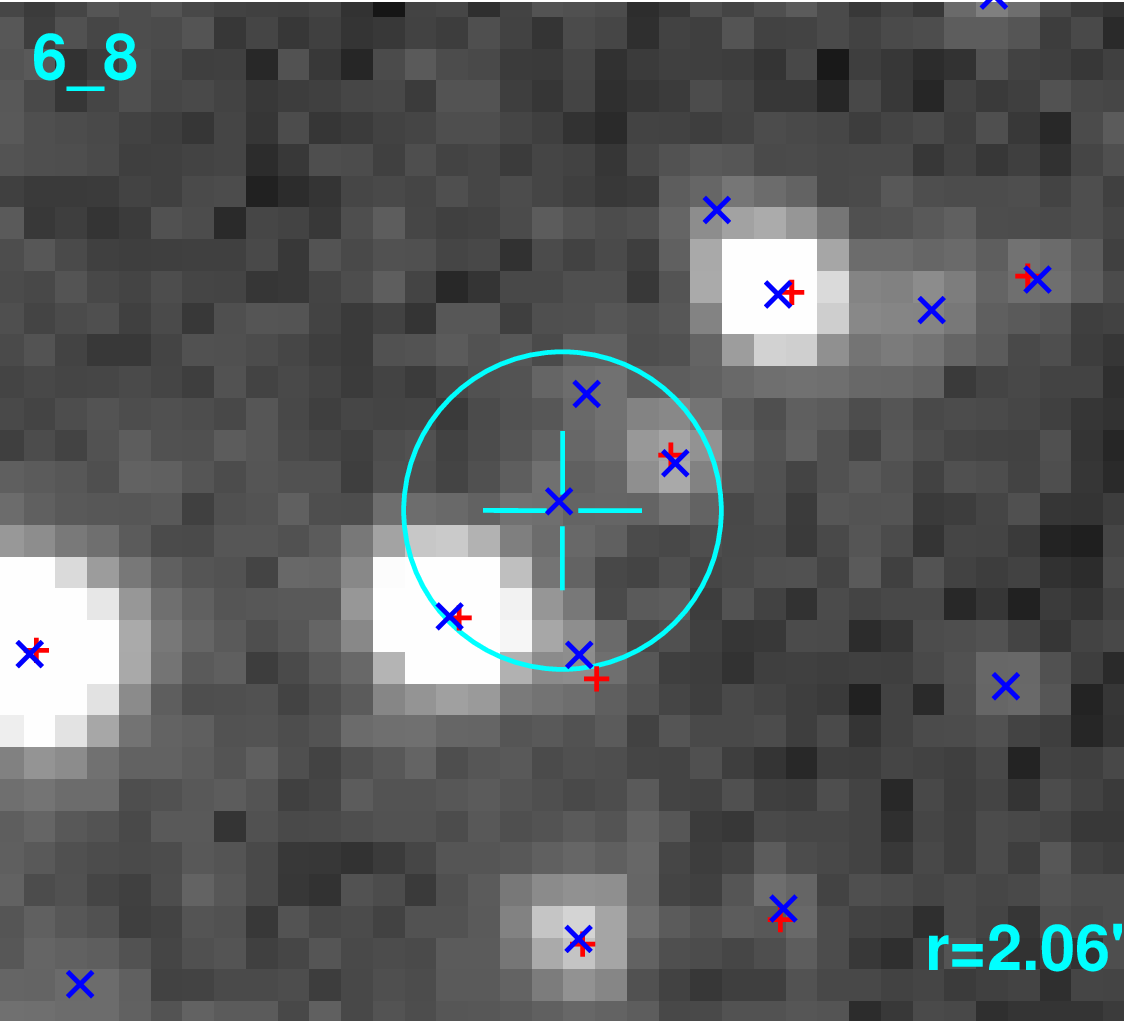} 
\end{center} 
\clearpage
\begin{center} 
\includegraphics[width=.45\textwidth]{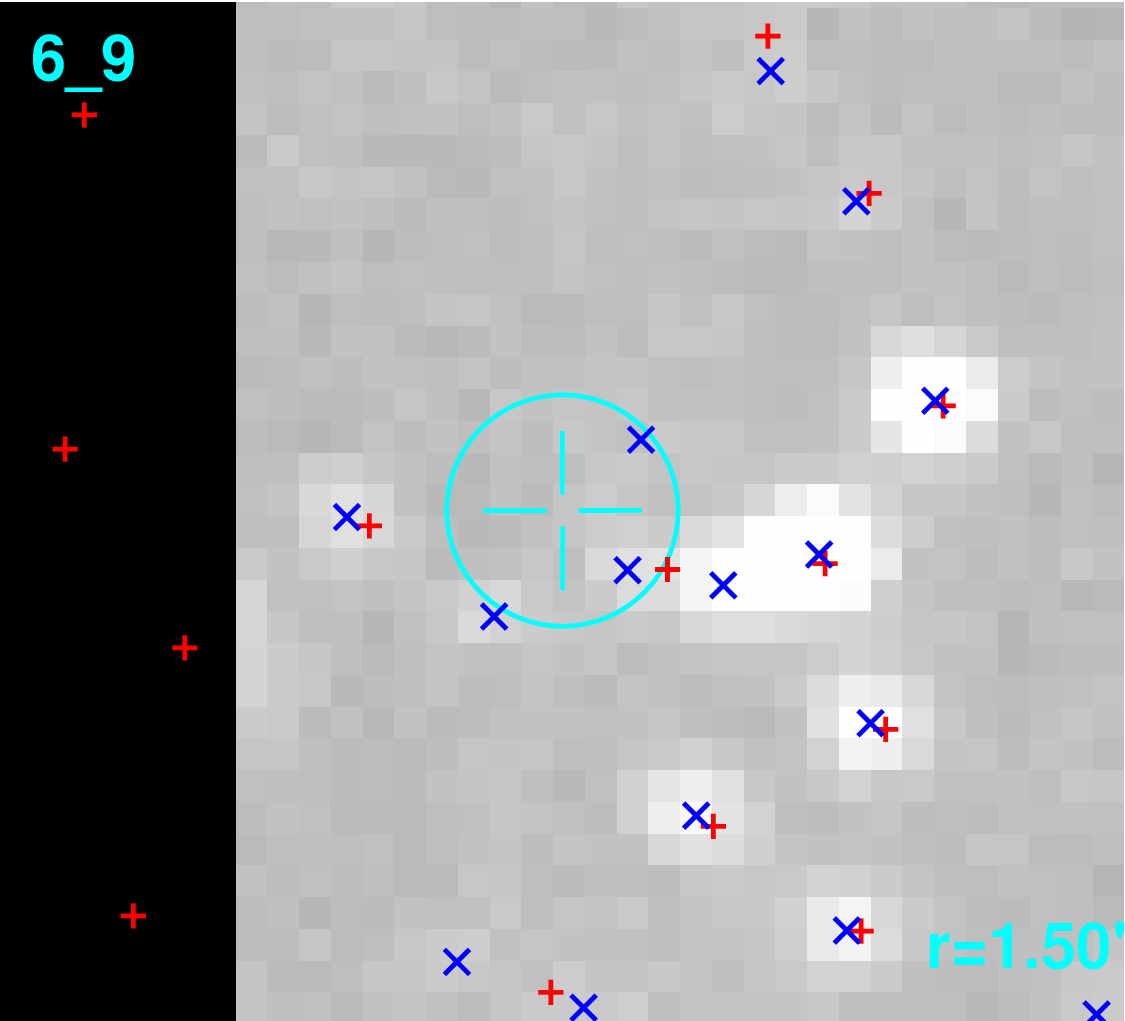} 
\includegraphics[width=.45\textwidth]{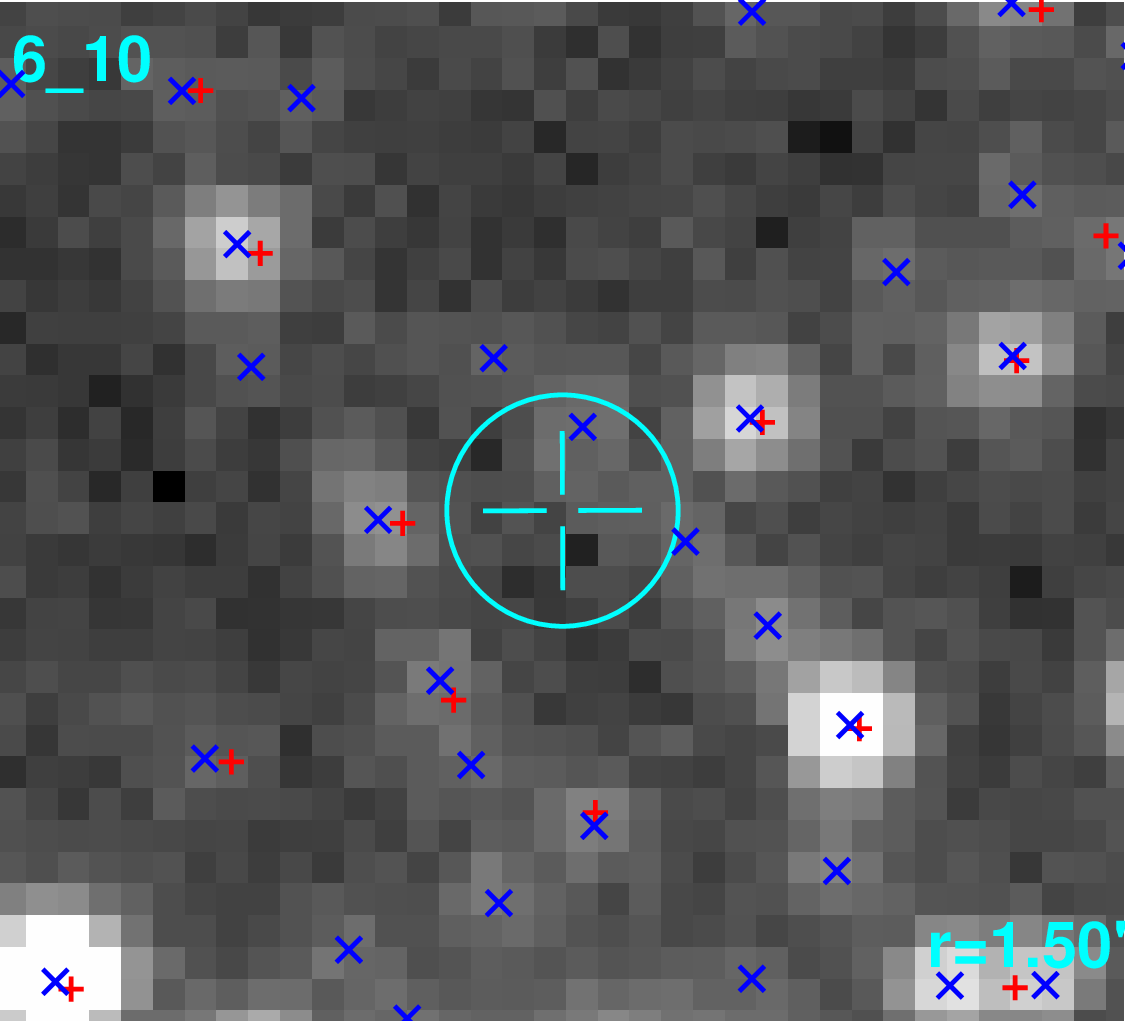} 
\includegraphics[width=.45\textwidth]{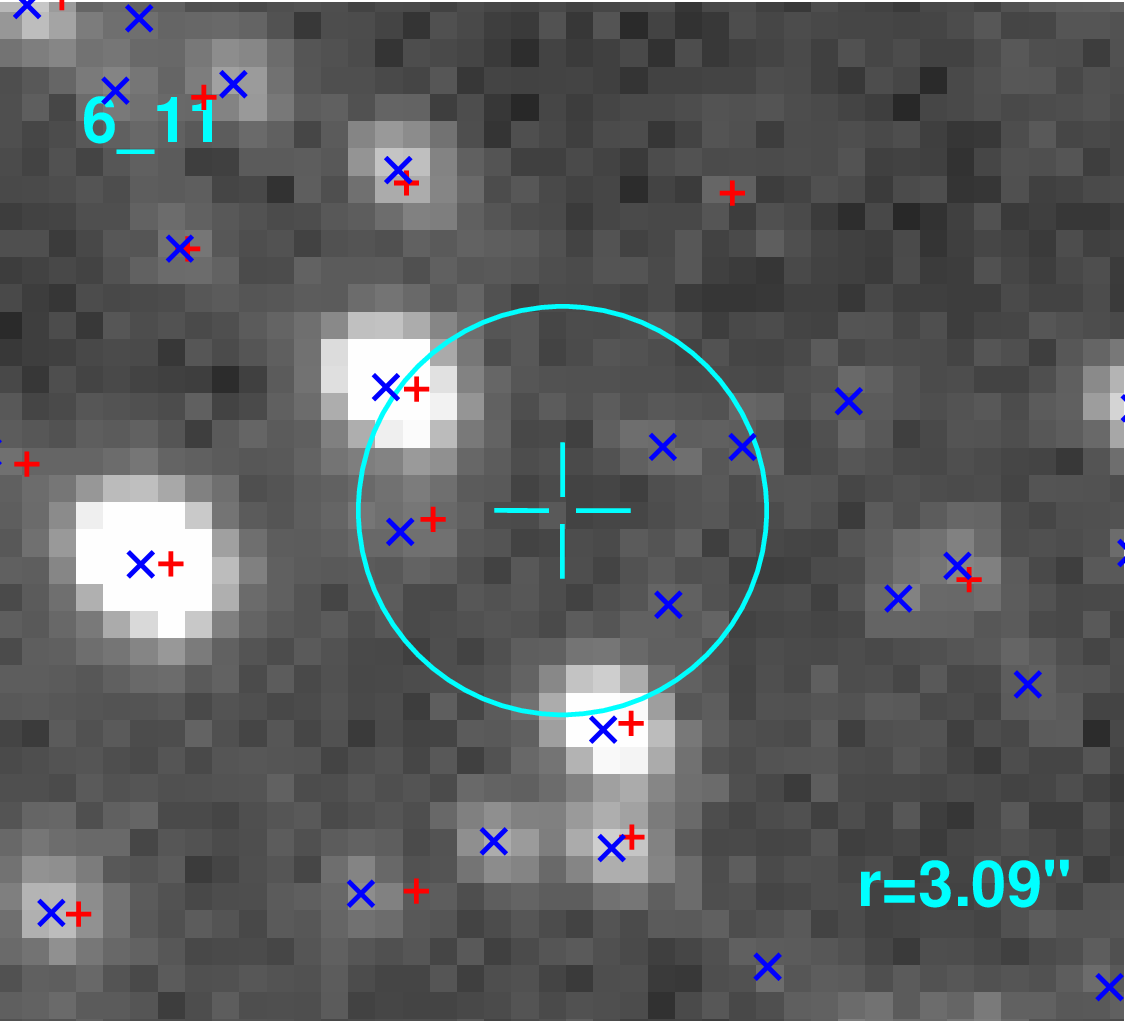} 
\includegraphics[width=.45\textwidth]{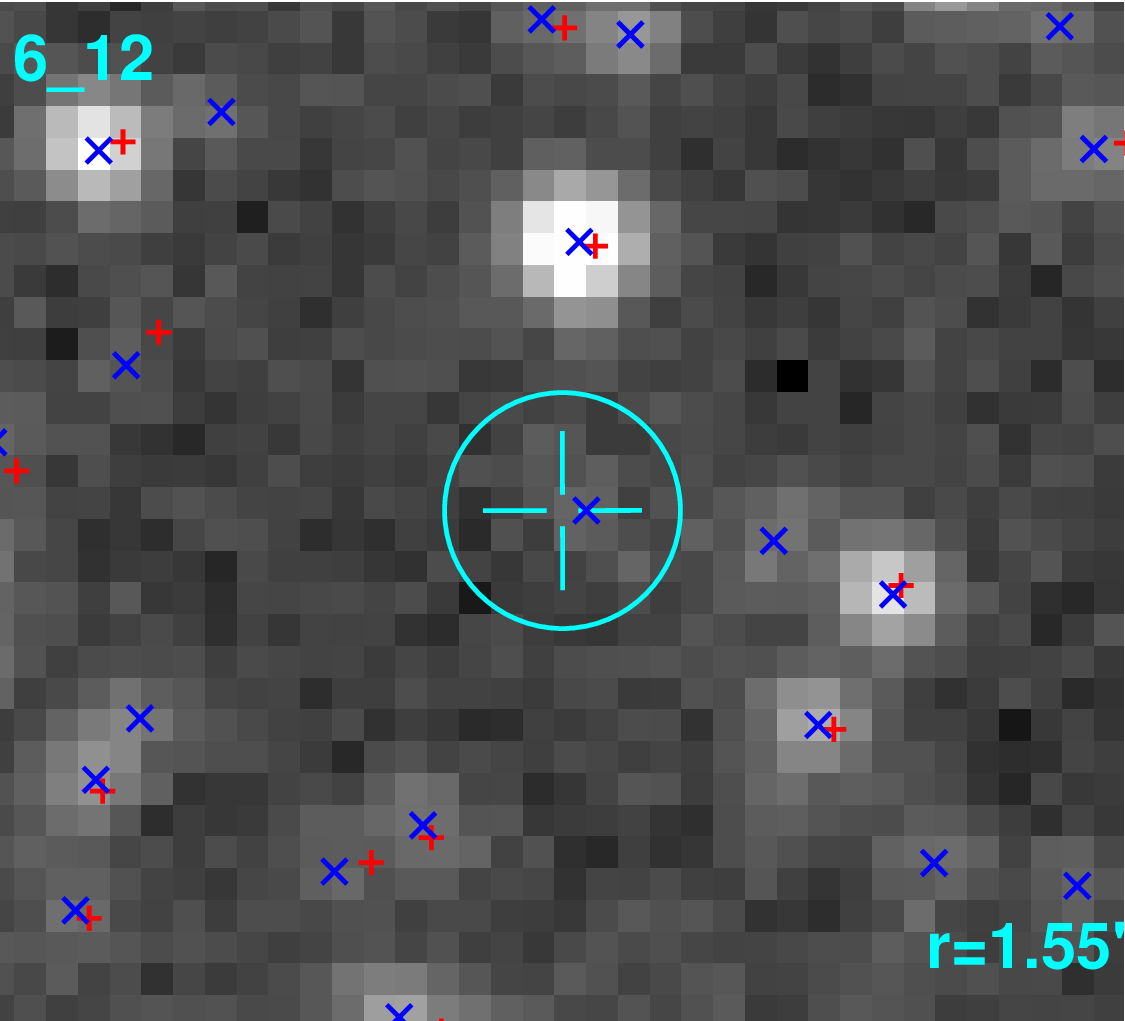} 
\end{center} 
\clearpage
\begin{center} 
\includegraphics[width=.45\textwidth]{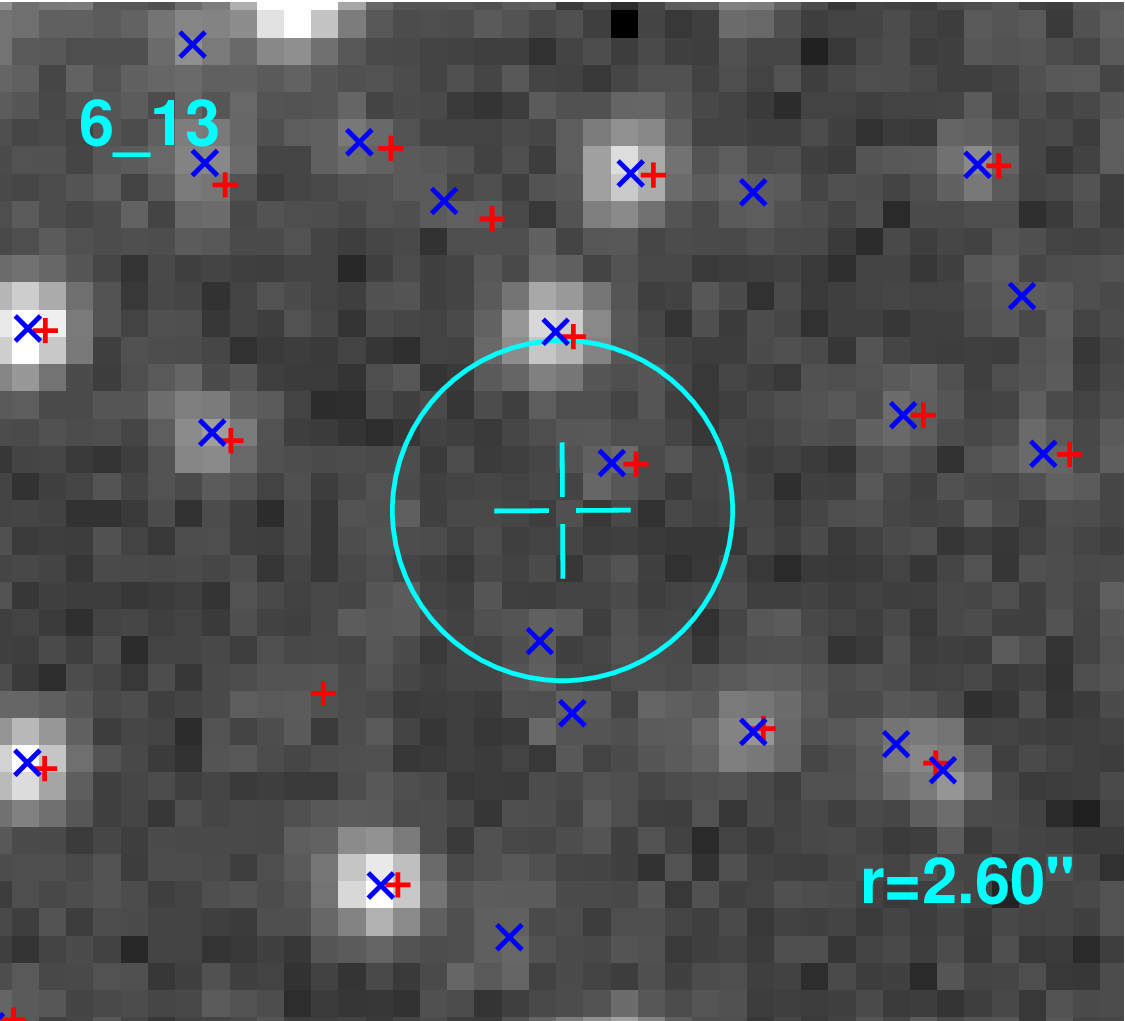} 
\includegraphics[width=.45\textwidth]{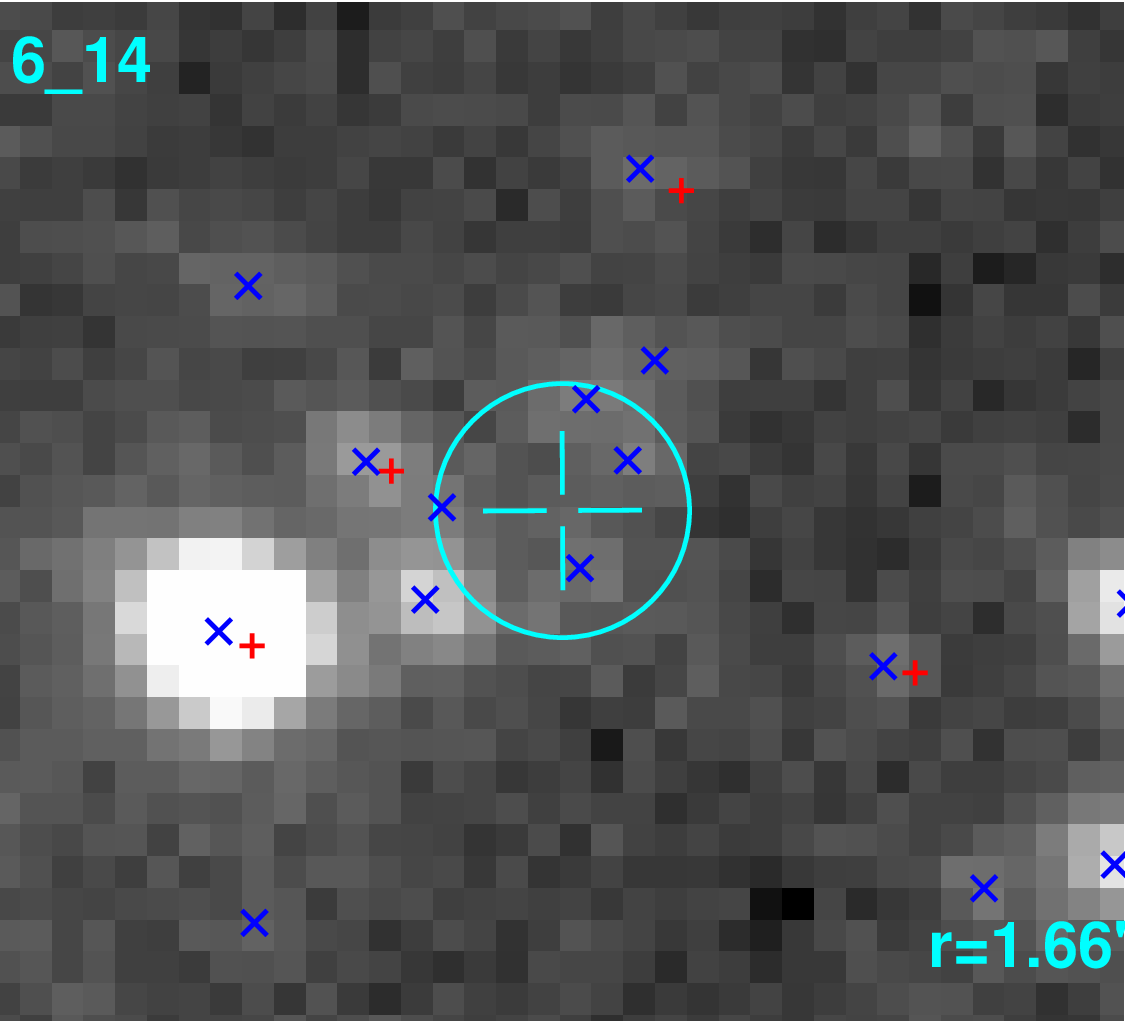} 
\includegraphics[width=.45\textwidth]{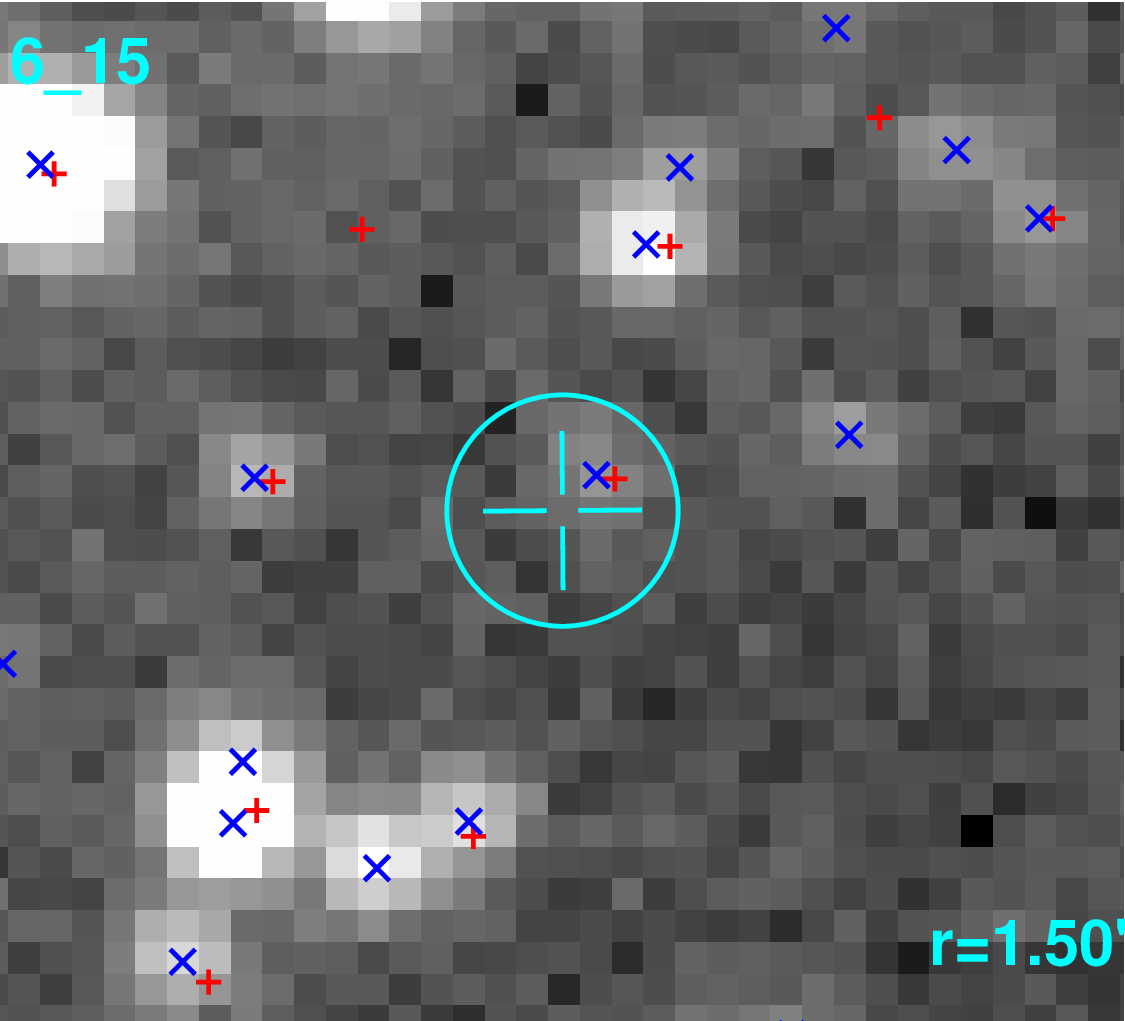} 
\includegraphics[width=.45\textwidth]{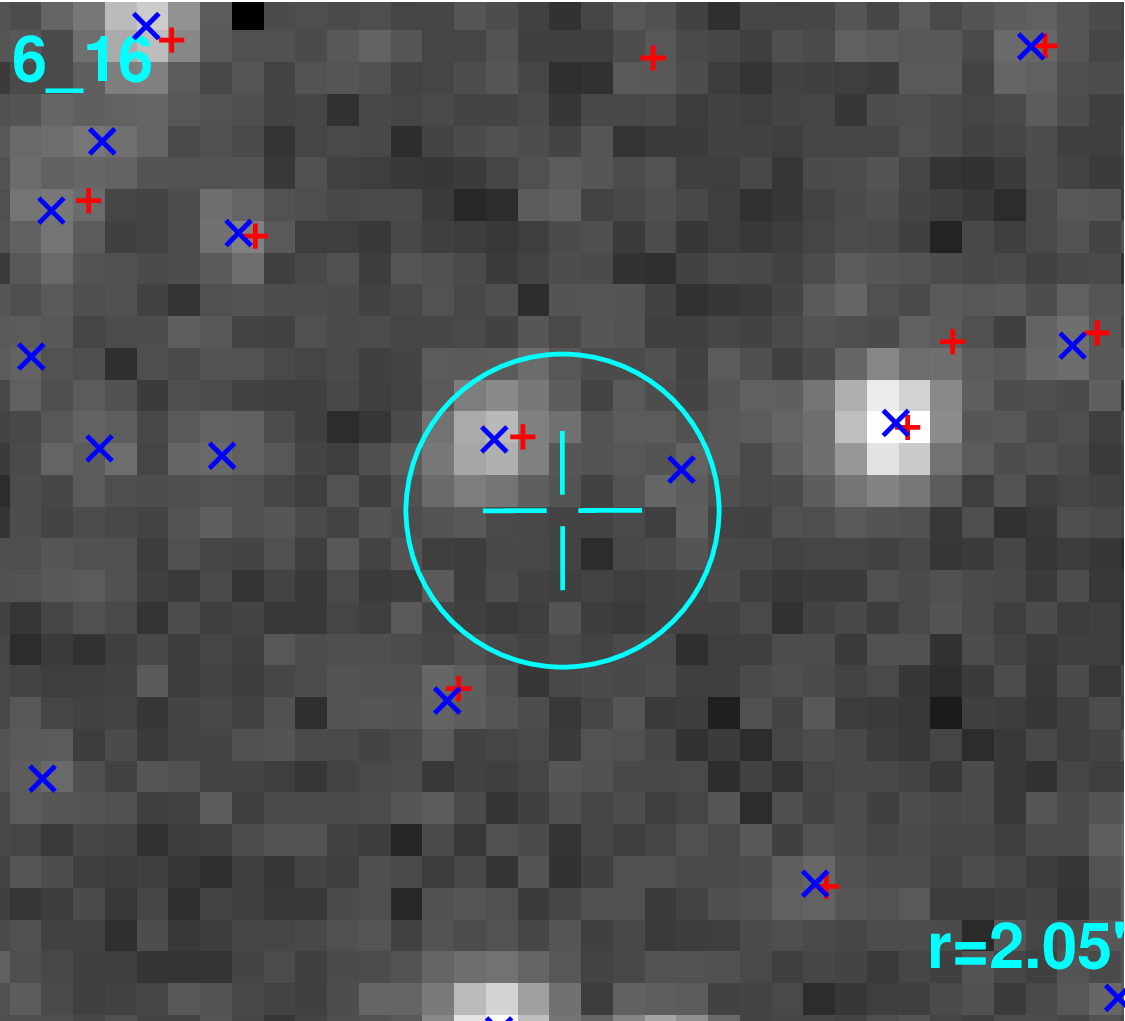} 
\end{center} 
\clearpage
\begin{center} 
\includegraphics[width=.45\textwidth]{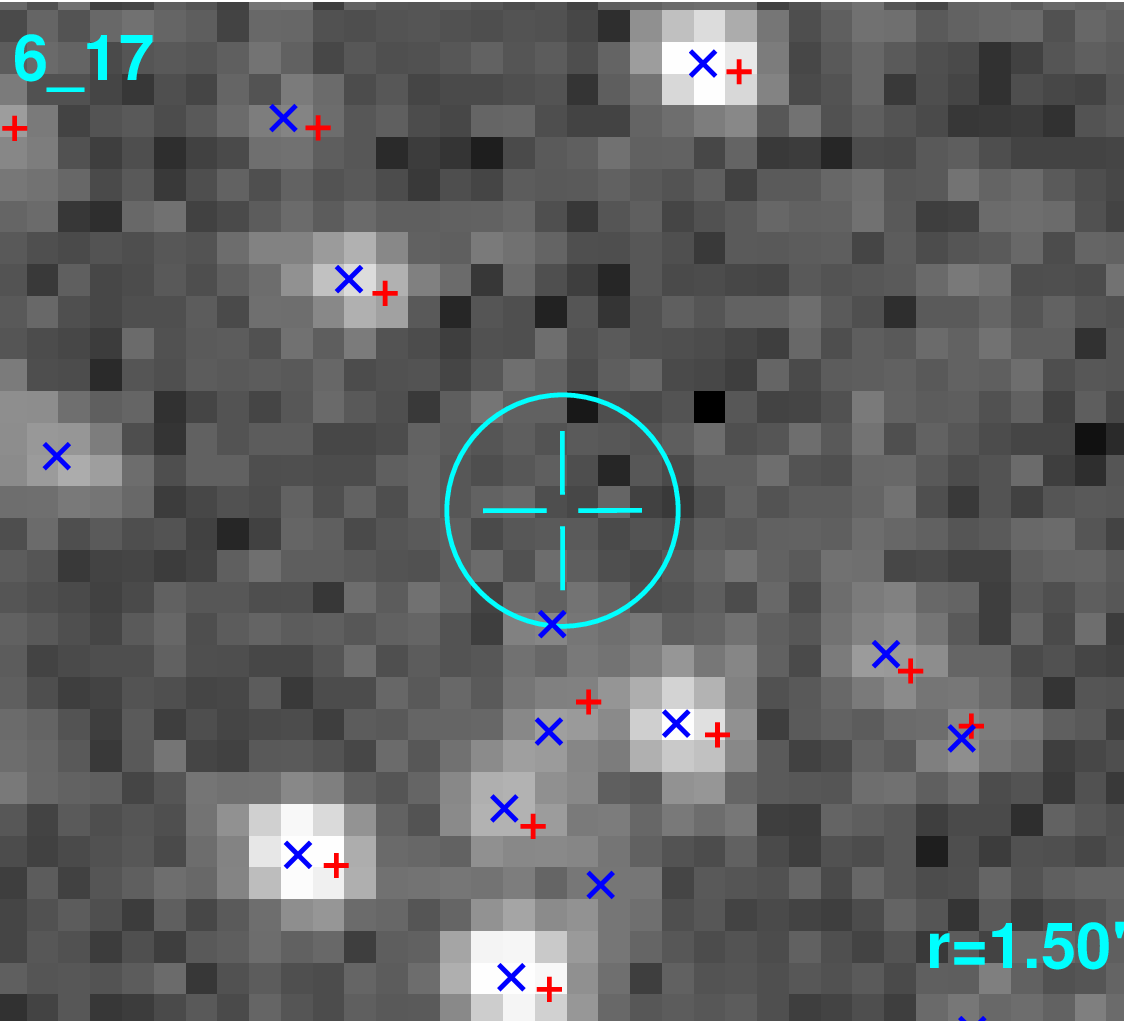} 
\includegraphics[width=.45\textwidth]{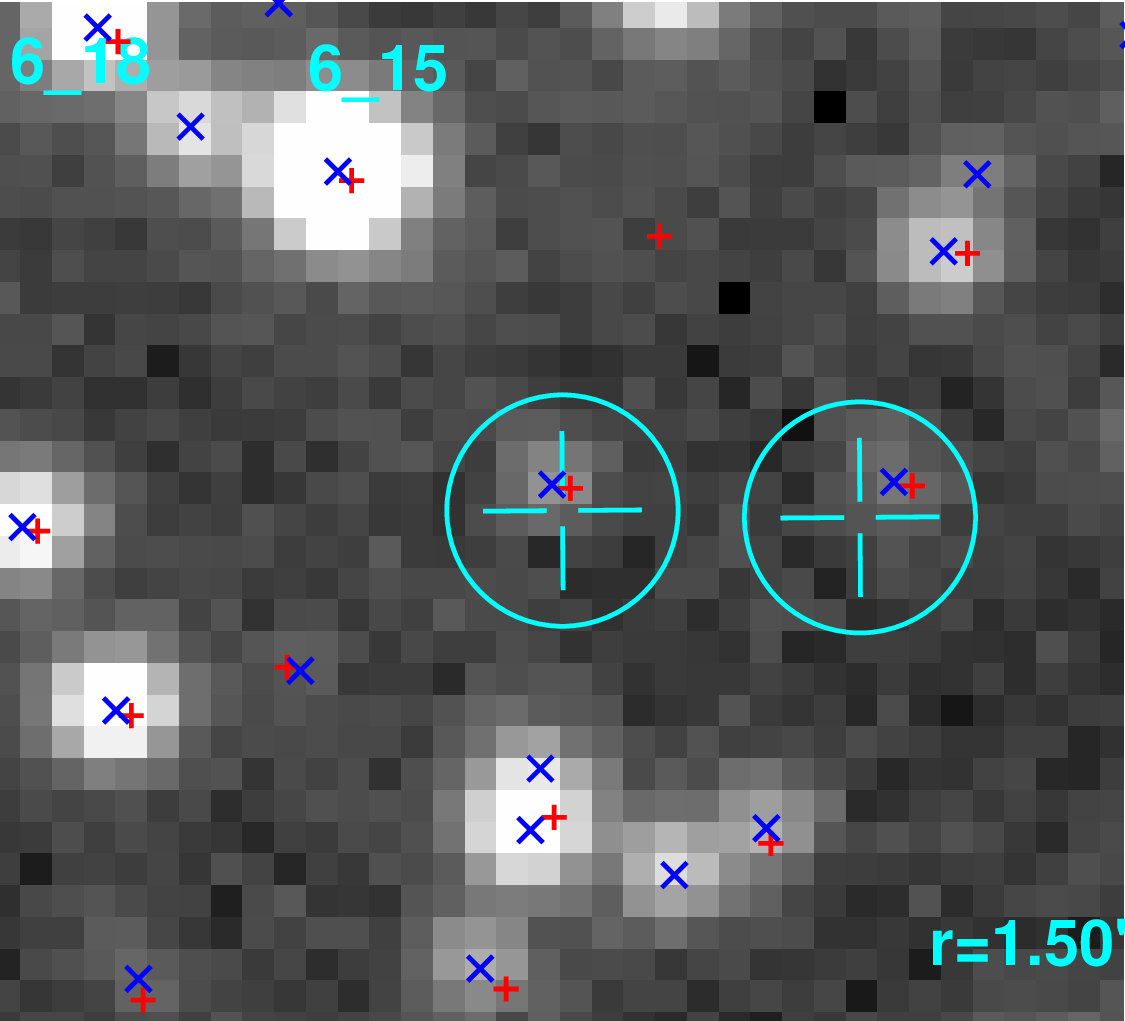} 
\includegraphics[width=.45\textwidth]{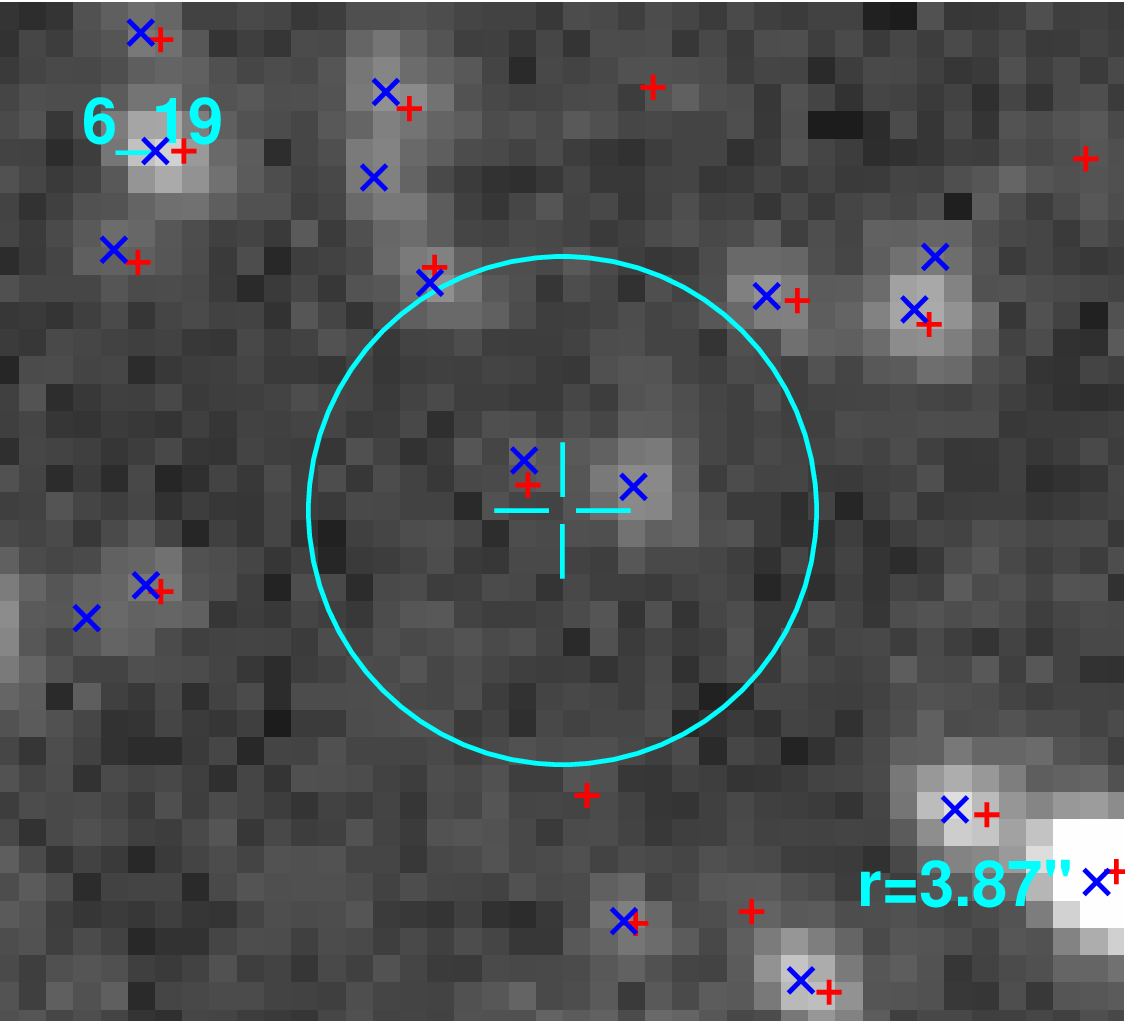} 
\includegraphics[width=.45\textwidth]{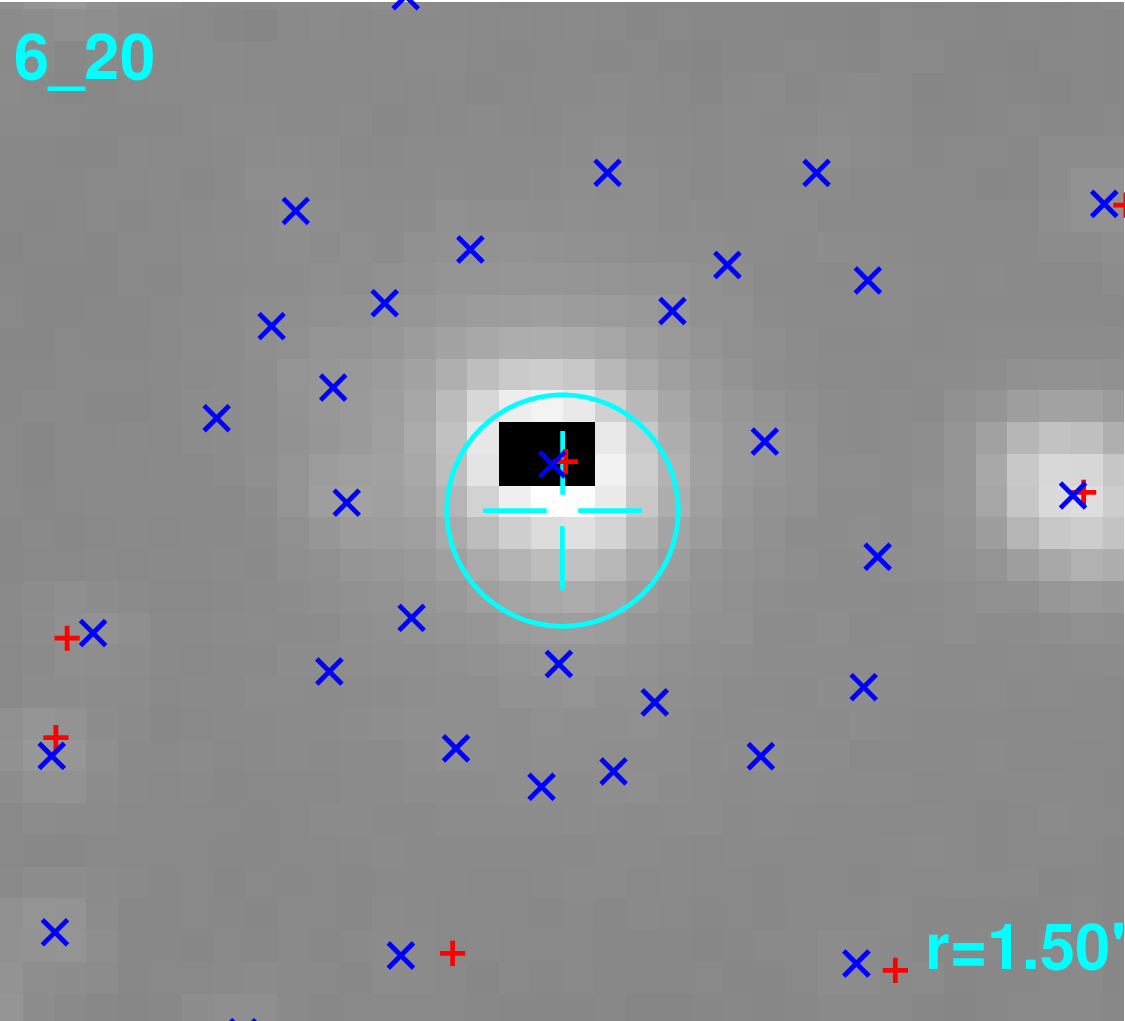} 
\end{center} 
\clearpage
\begin{center} 
\includegraphics[width=.45\textwidth]{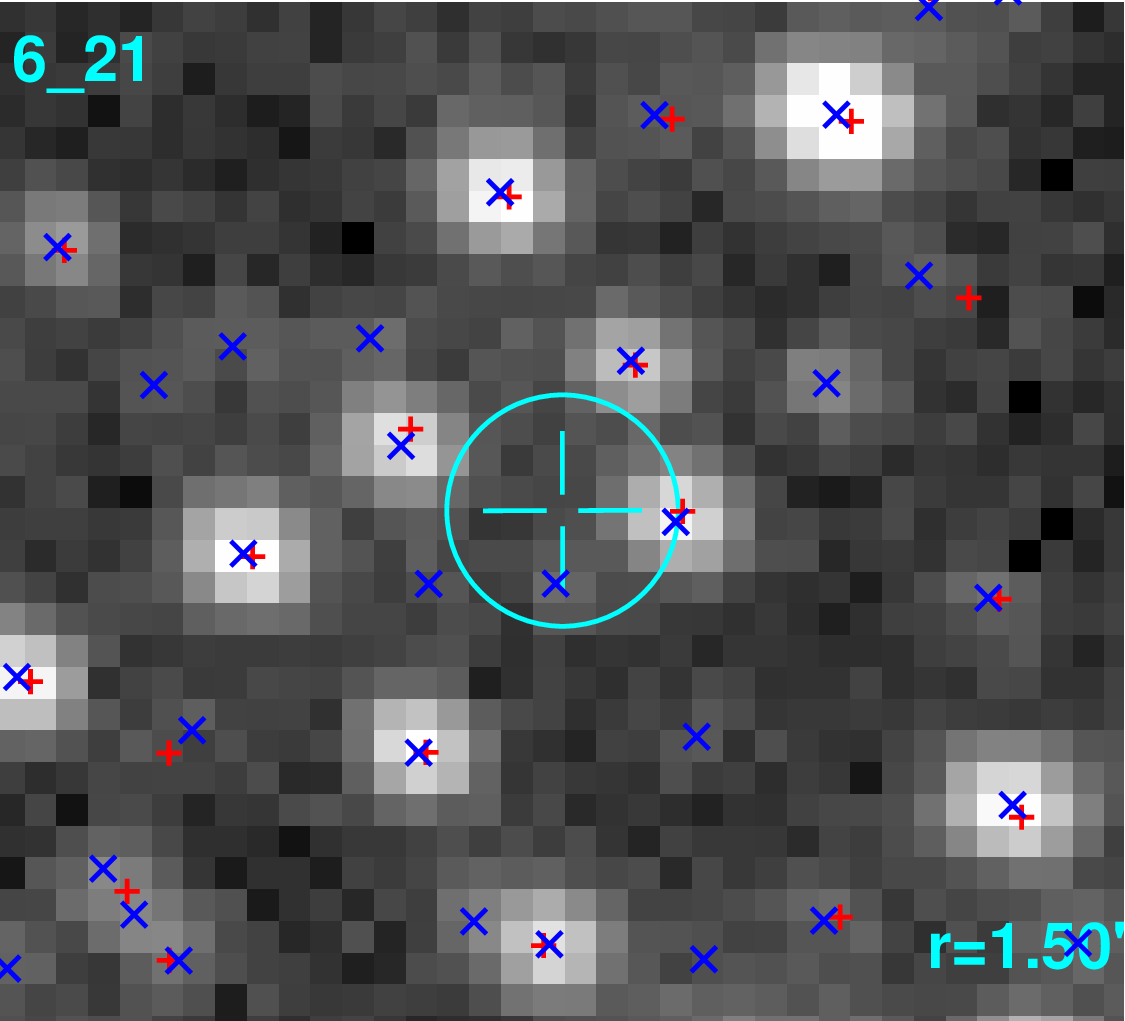} 
\includegraphics[width=.45\textwidth]{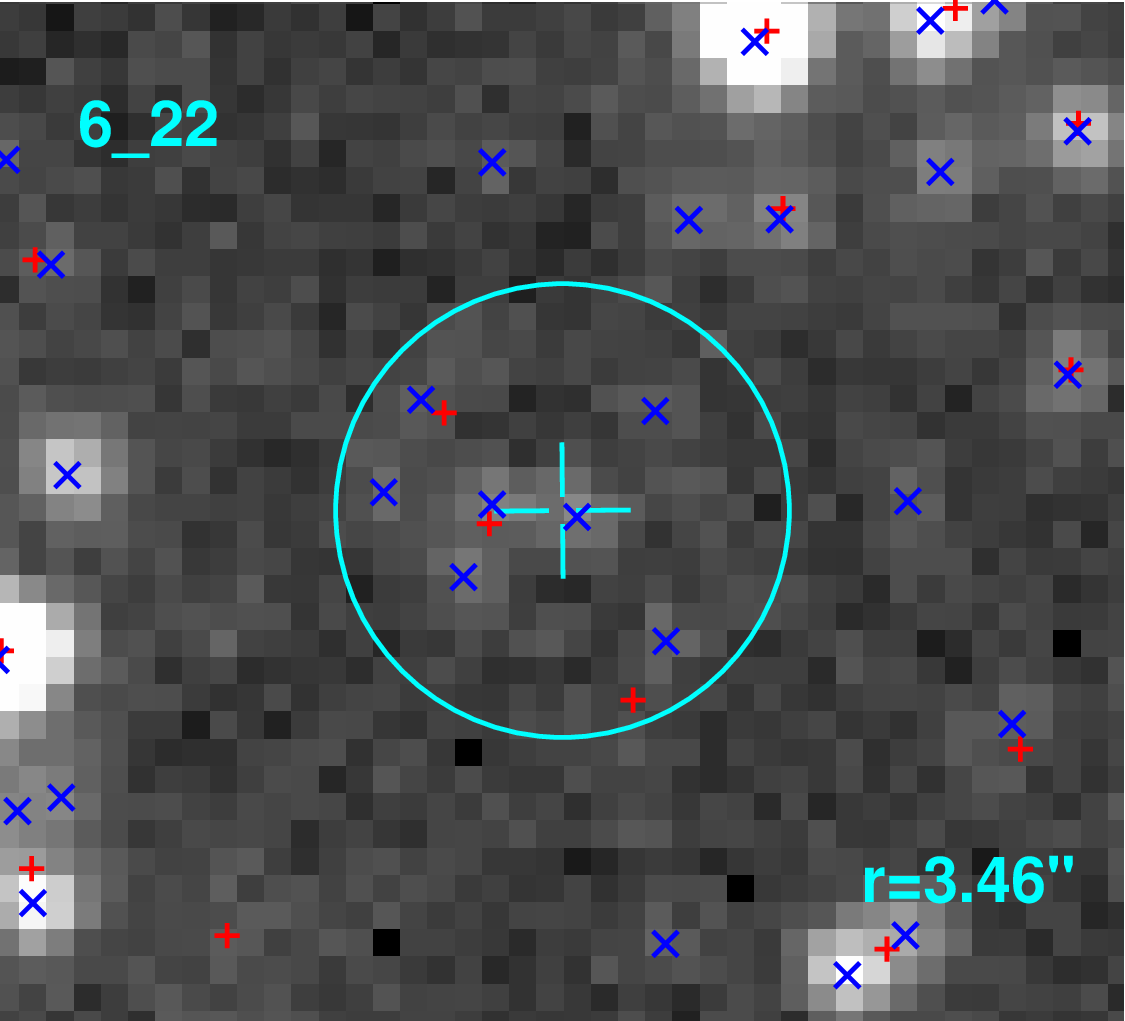} 
\includegraphics[width=.45\textwidth]{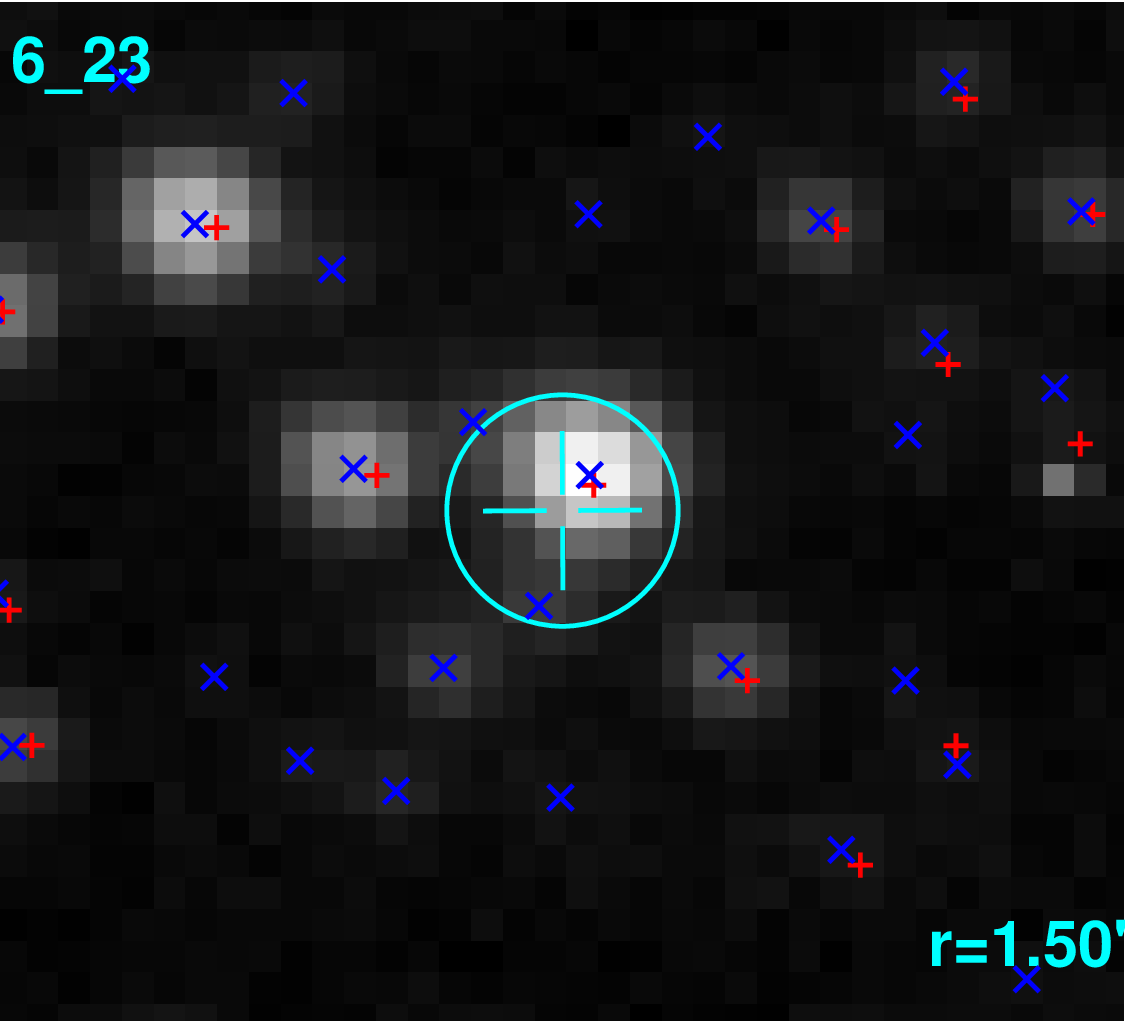} 
\includegraphics[width=.45\textwidth]{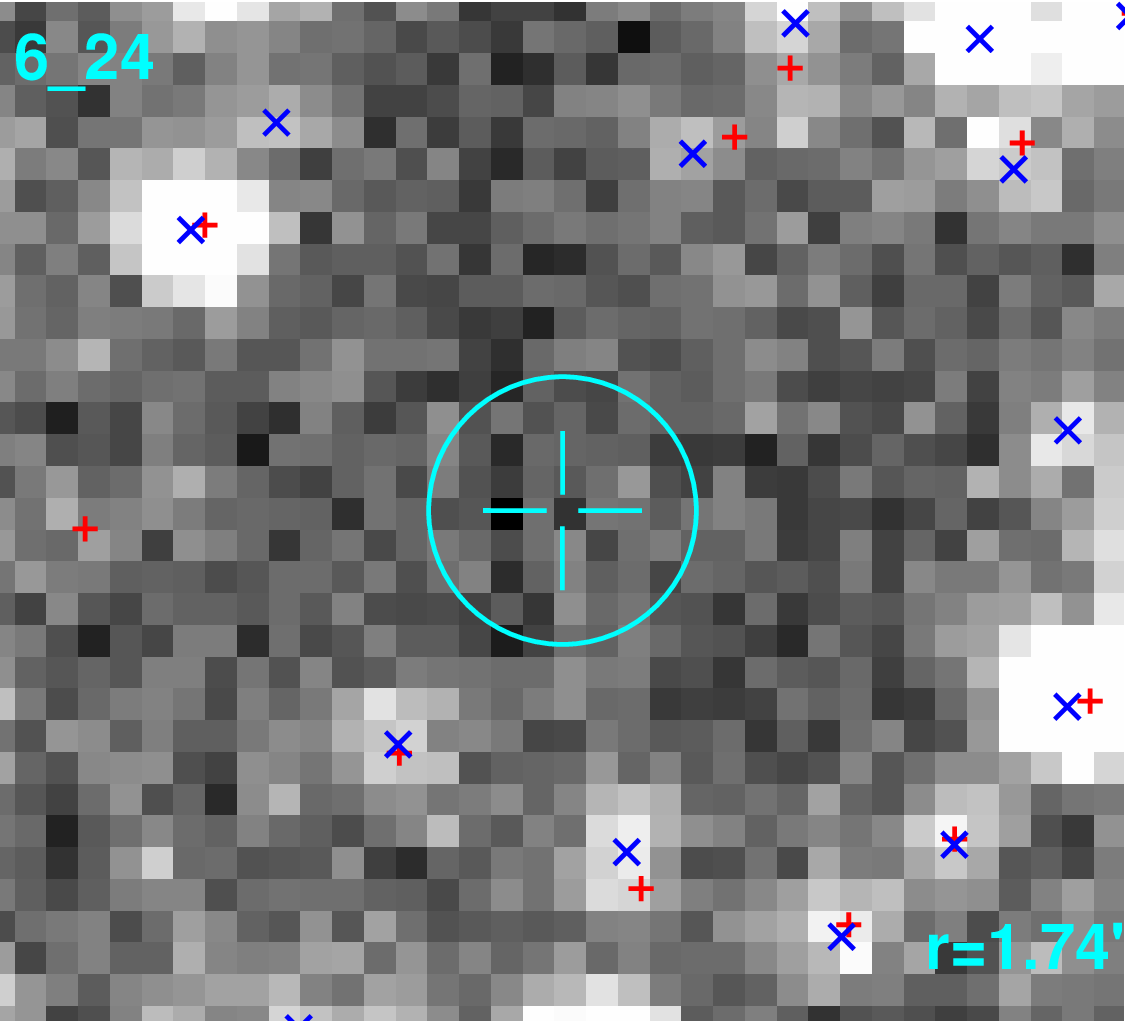} 
\end{center} 
\clearpage
\begin{center} 
\includegraphics[width=.45\textwidth]{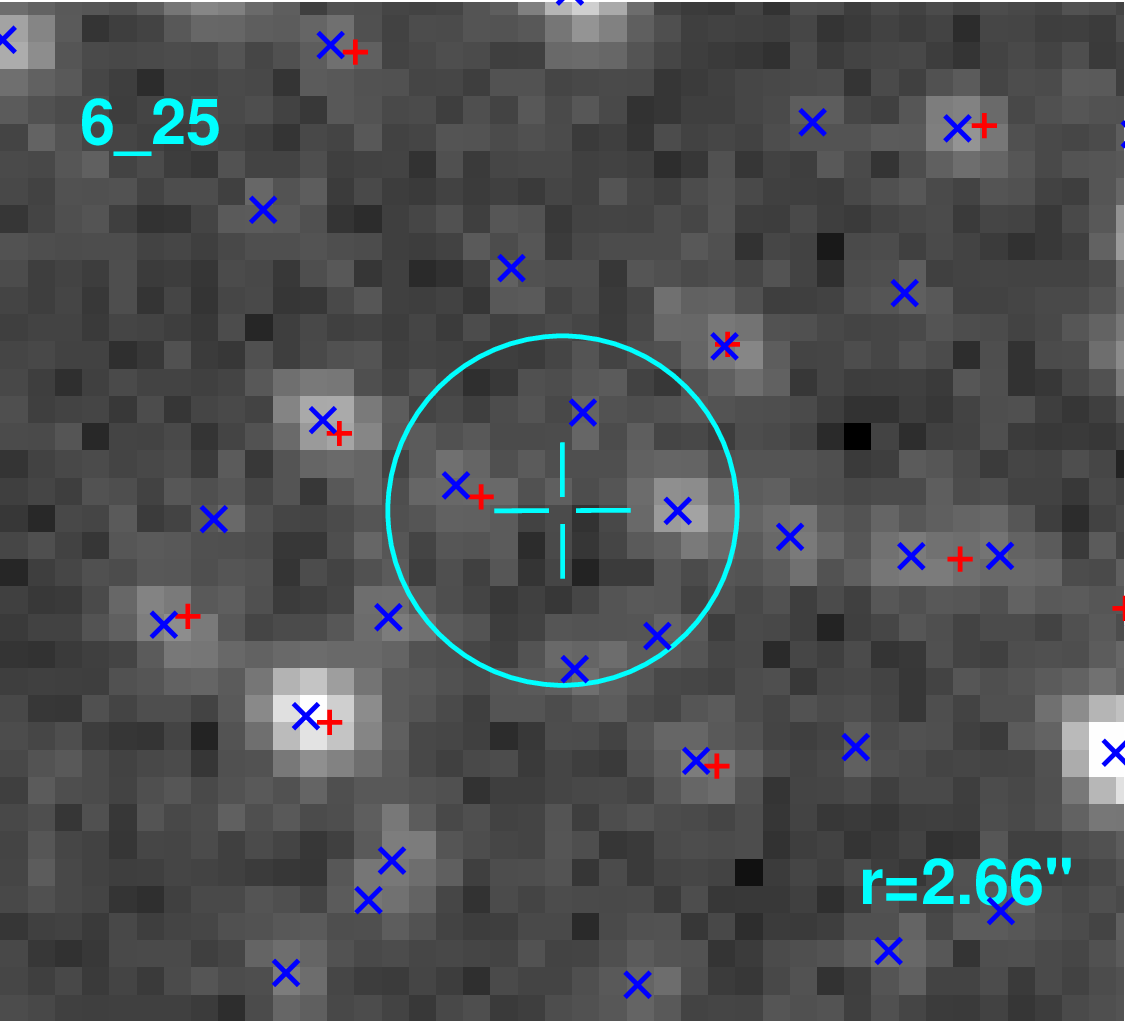} 
\includegraphics[width=.45\textwidth]{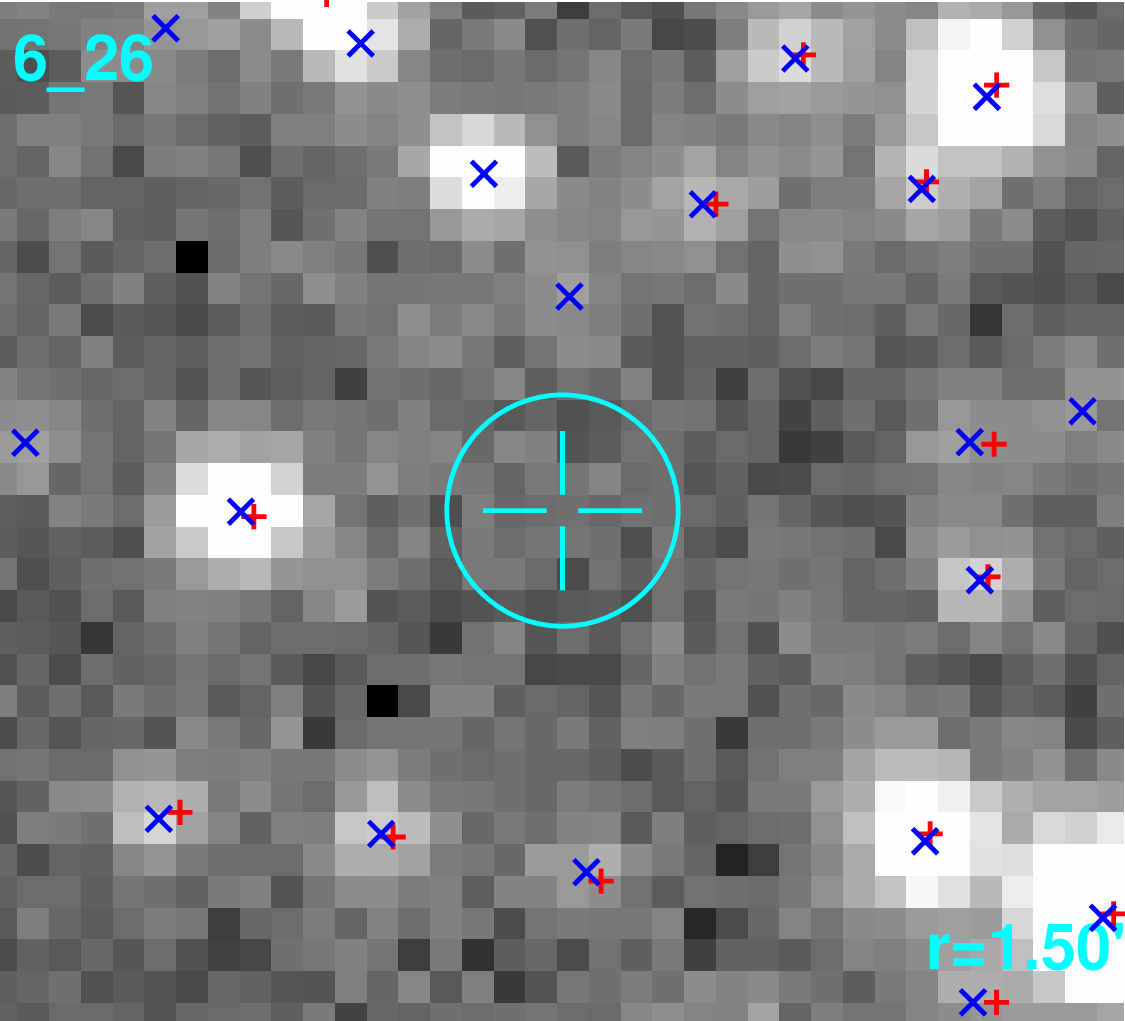} 
\includegraphics[width=.45\textwidth]{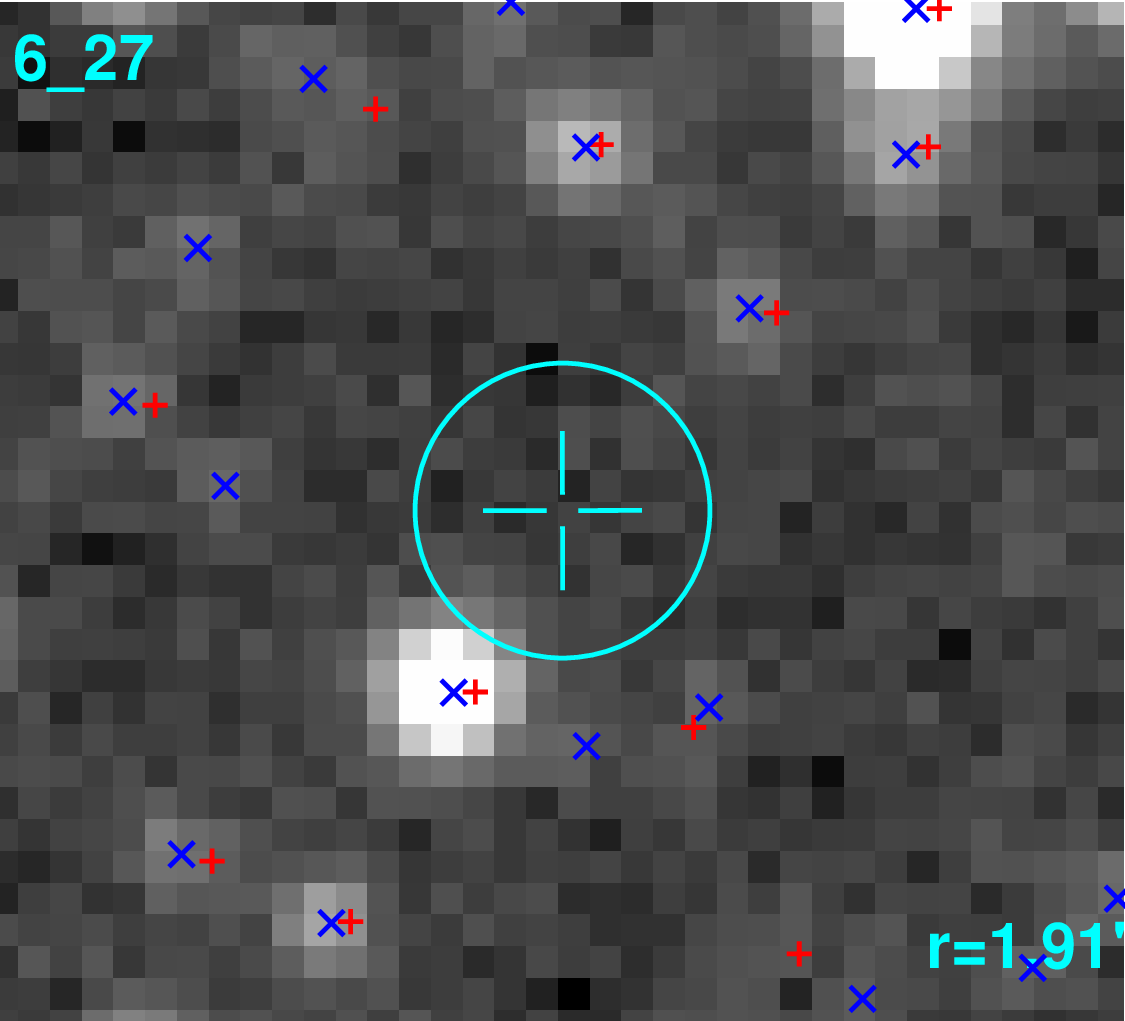} 
\includegraphics[width=.45\textwidth]{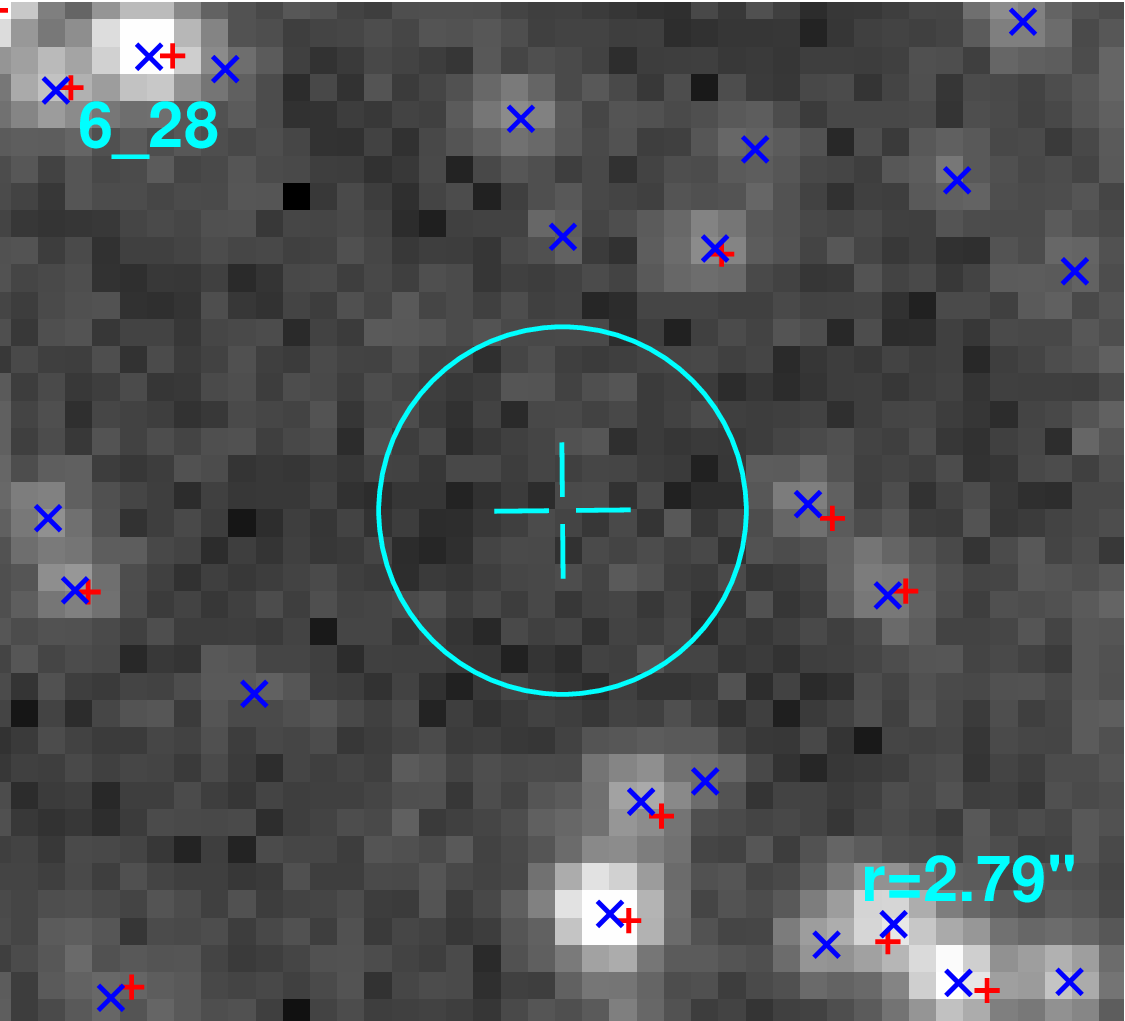} 
\end{center} 
\clearpage

\clearpage
\begin{center} 
\includegraphics[width=.45\textwidth]{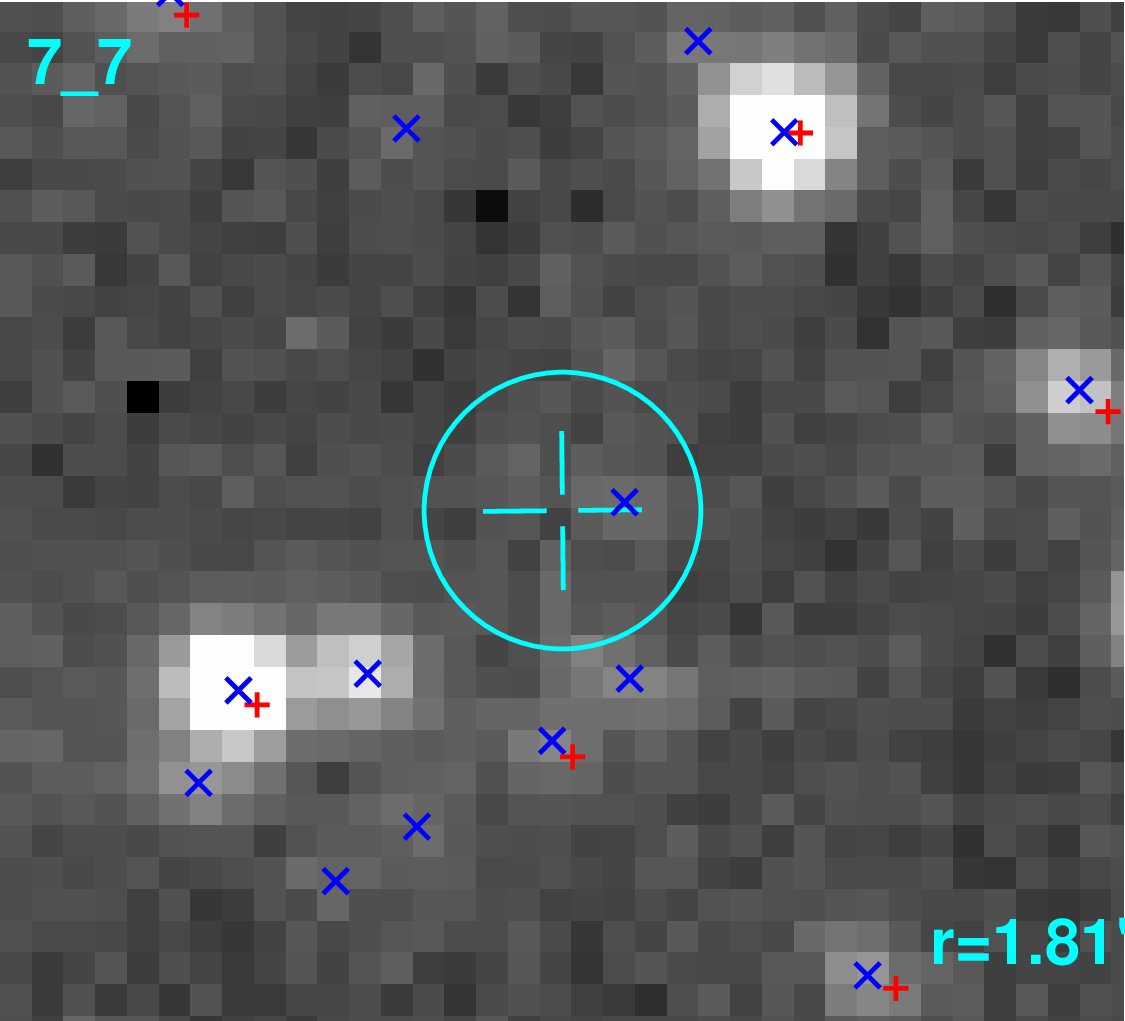} 
\includegraphics[width=.45\textwidth]{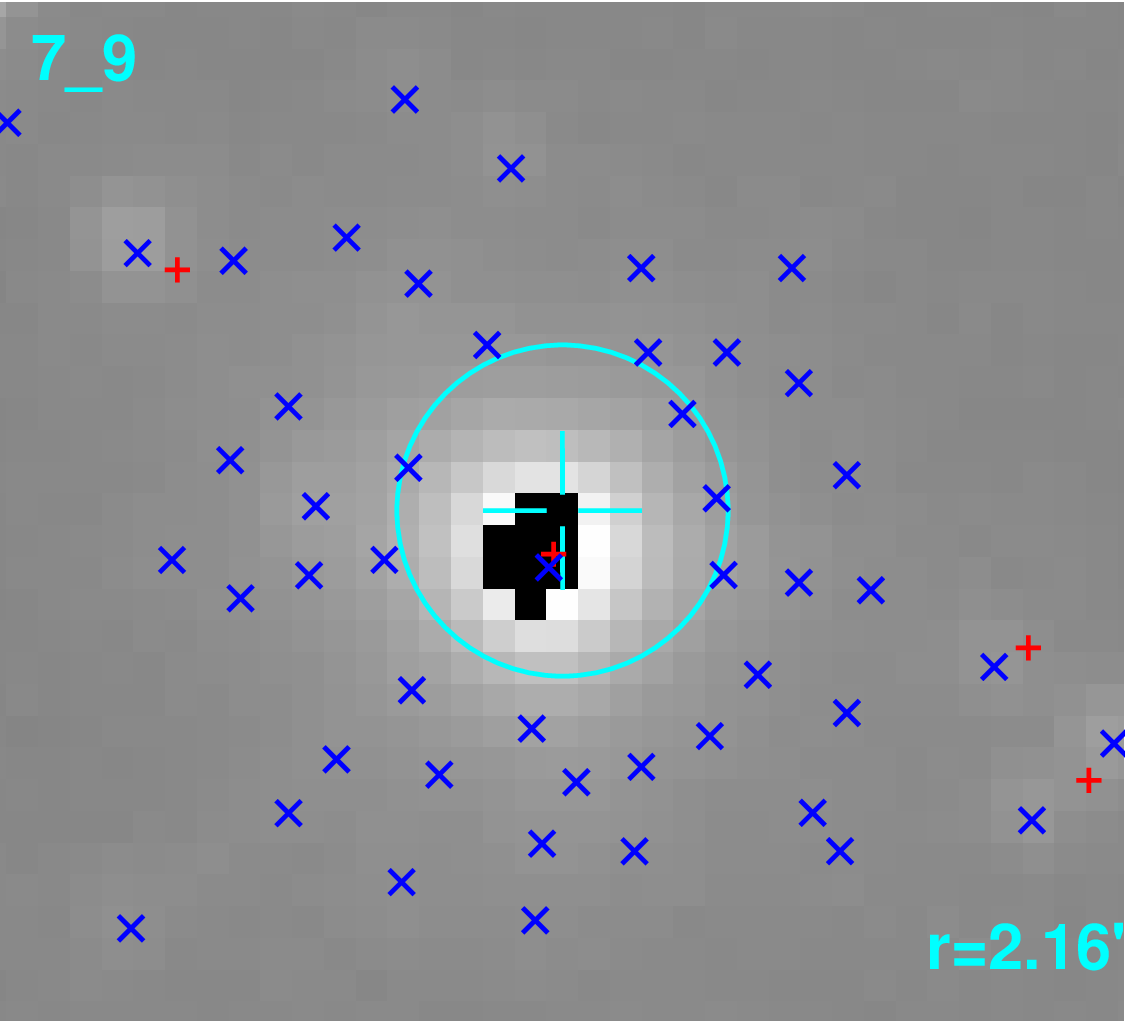} 
\includegraphics[width=.45\textwidth]{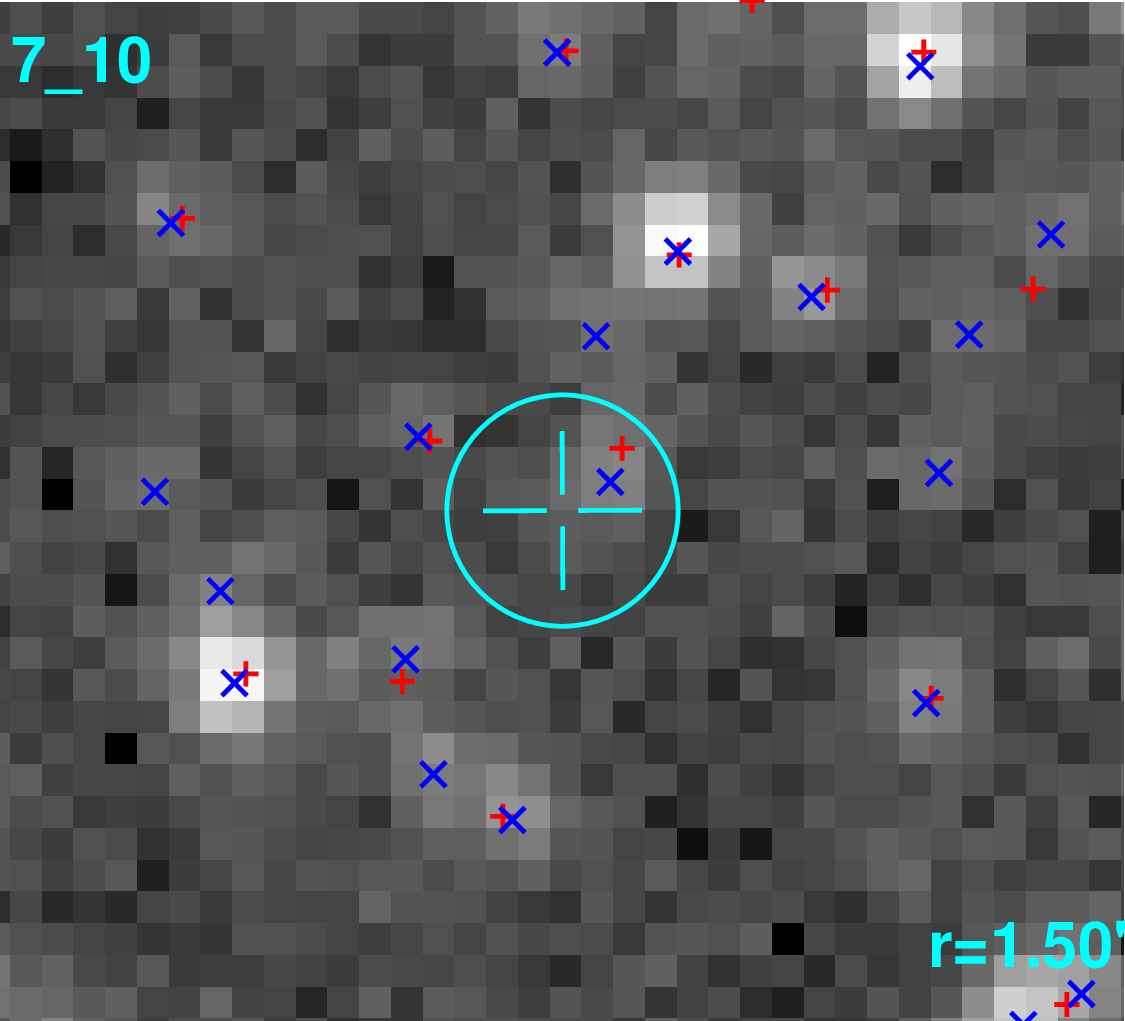} 
\includegraphics[width=.45\textwidth]{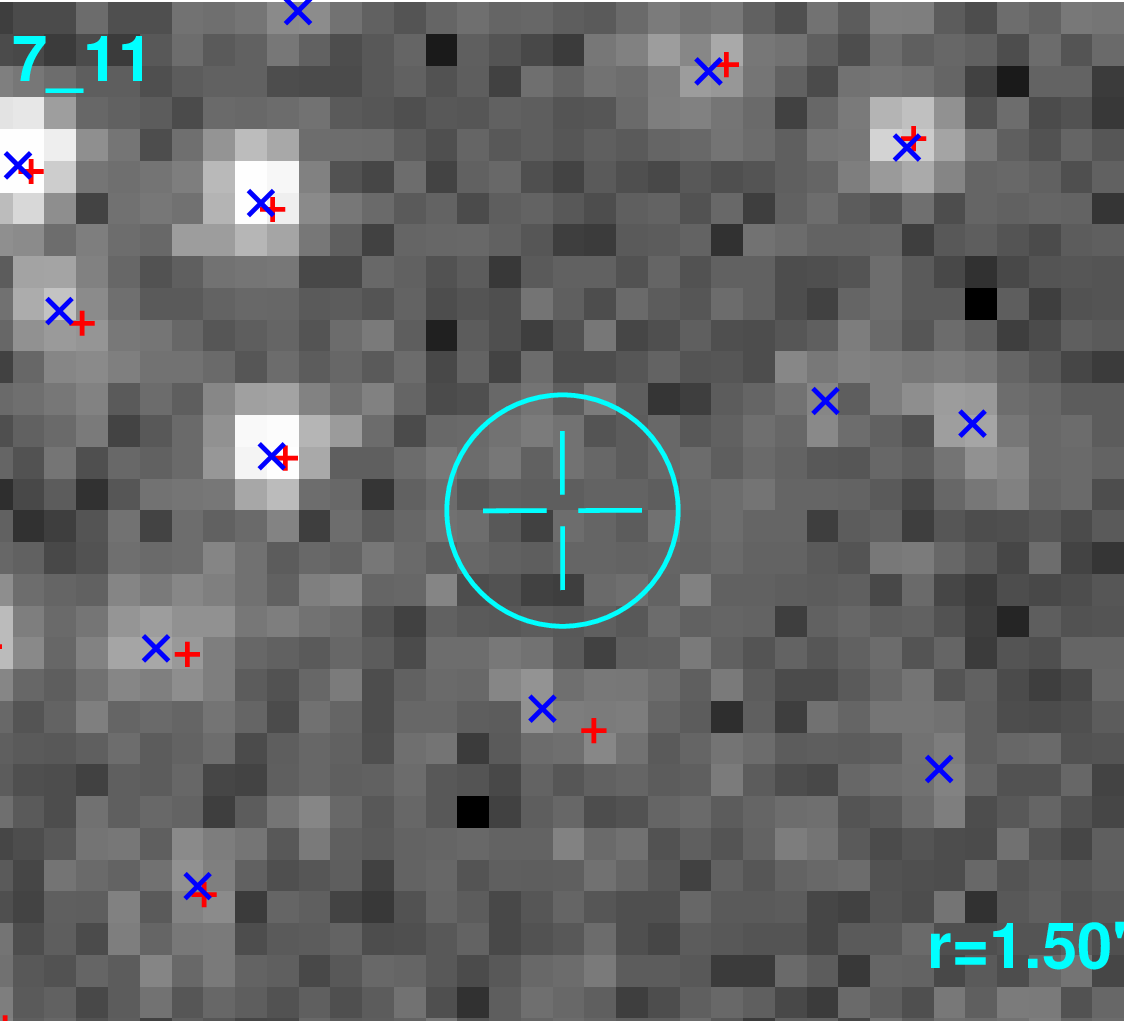} 
\end{center} 
\clearpage
\begin{center} 
\includegraphics[width=.45\textwidth]{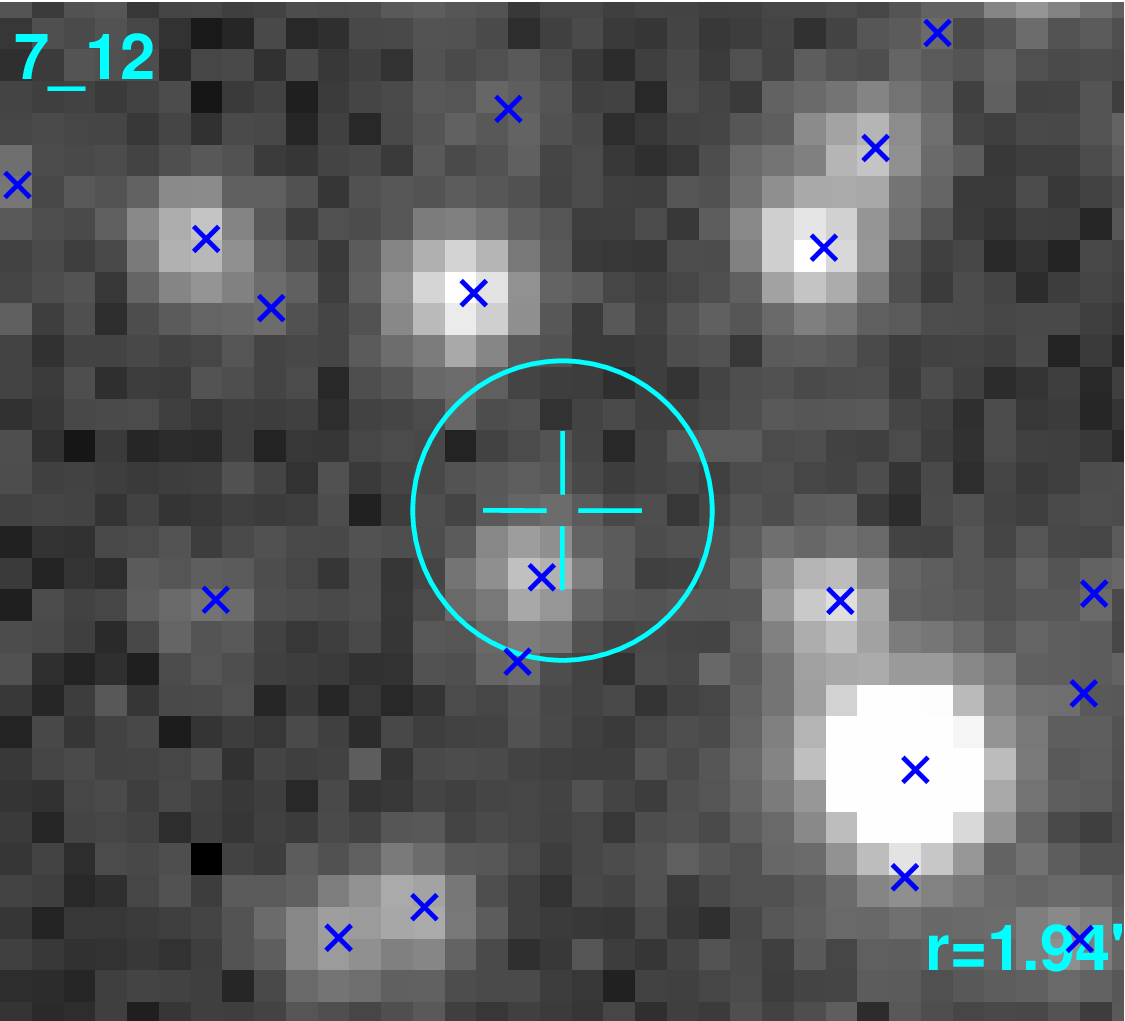} 
\includegraphics[width=.45\textwidth]{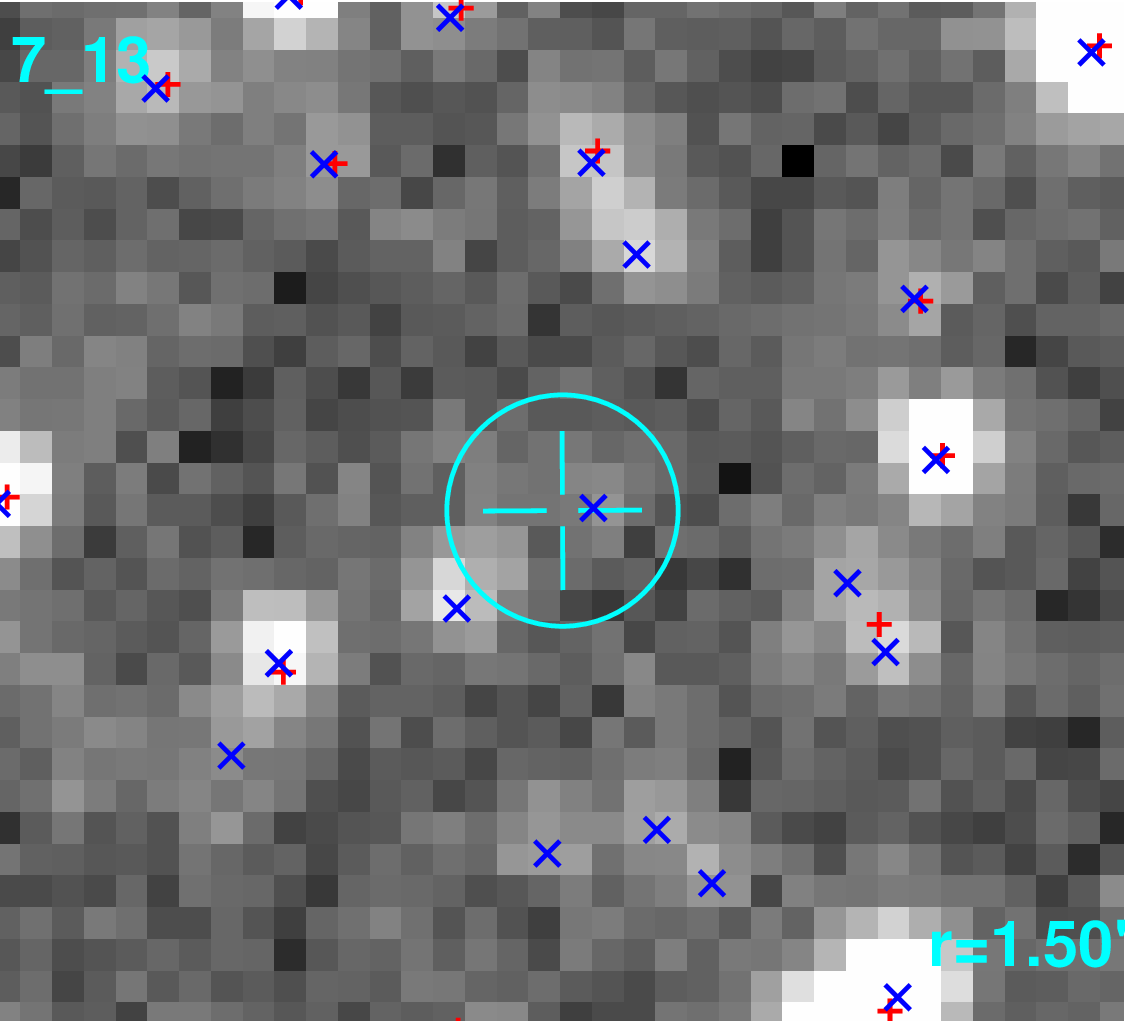} 
\includegraphics[width=.45\textwidth]{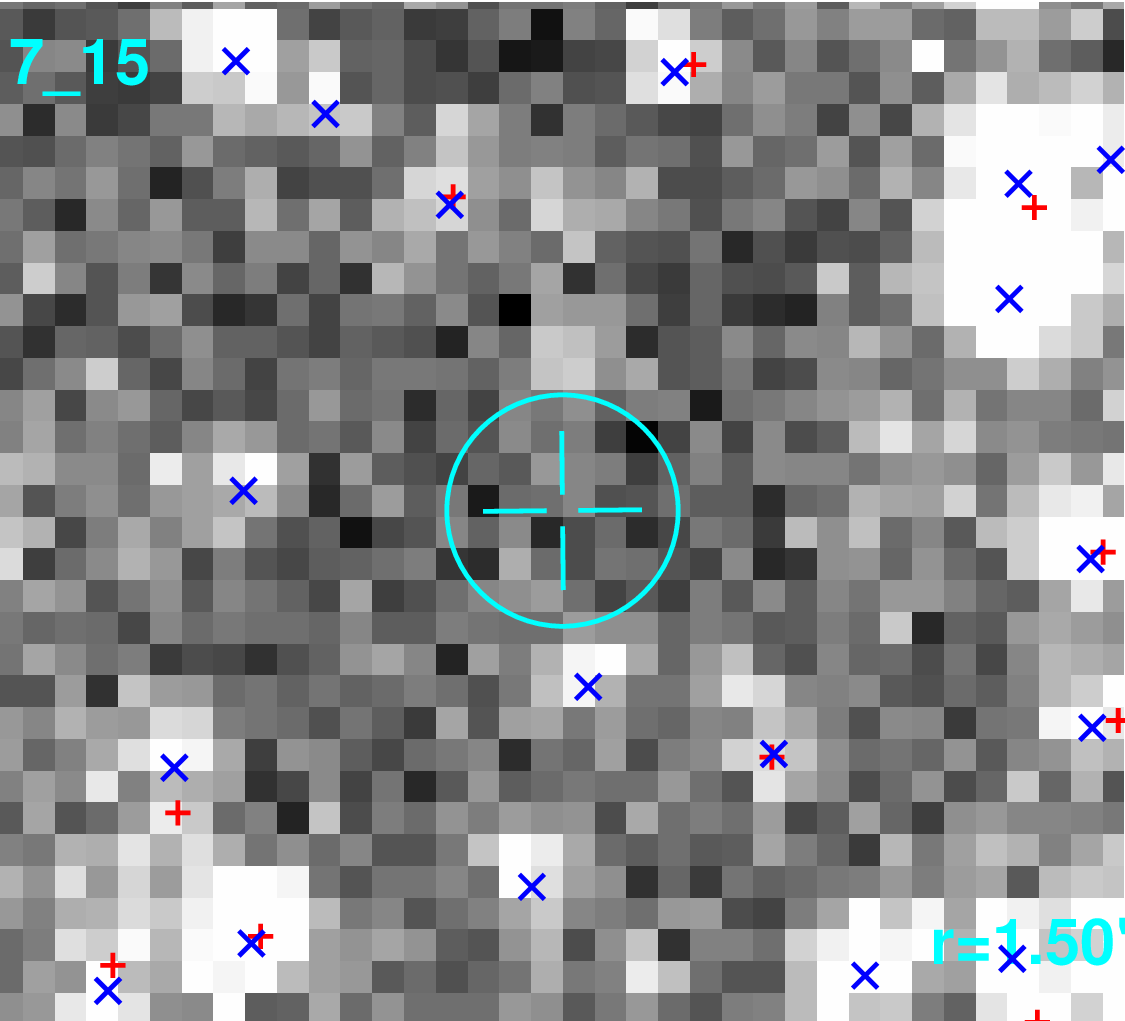} 
\includegraphics[width=.45\textwidth]{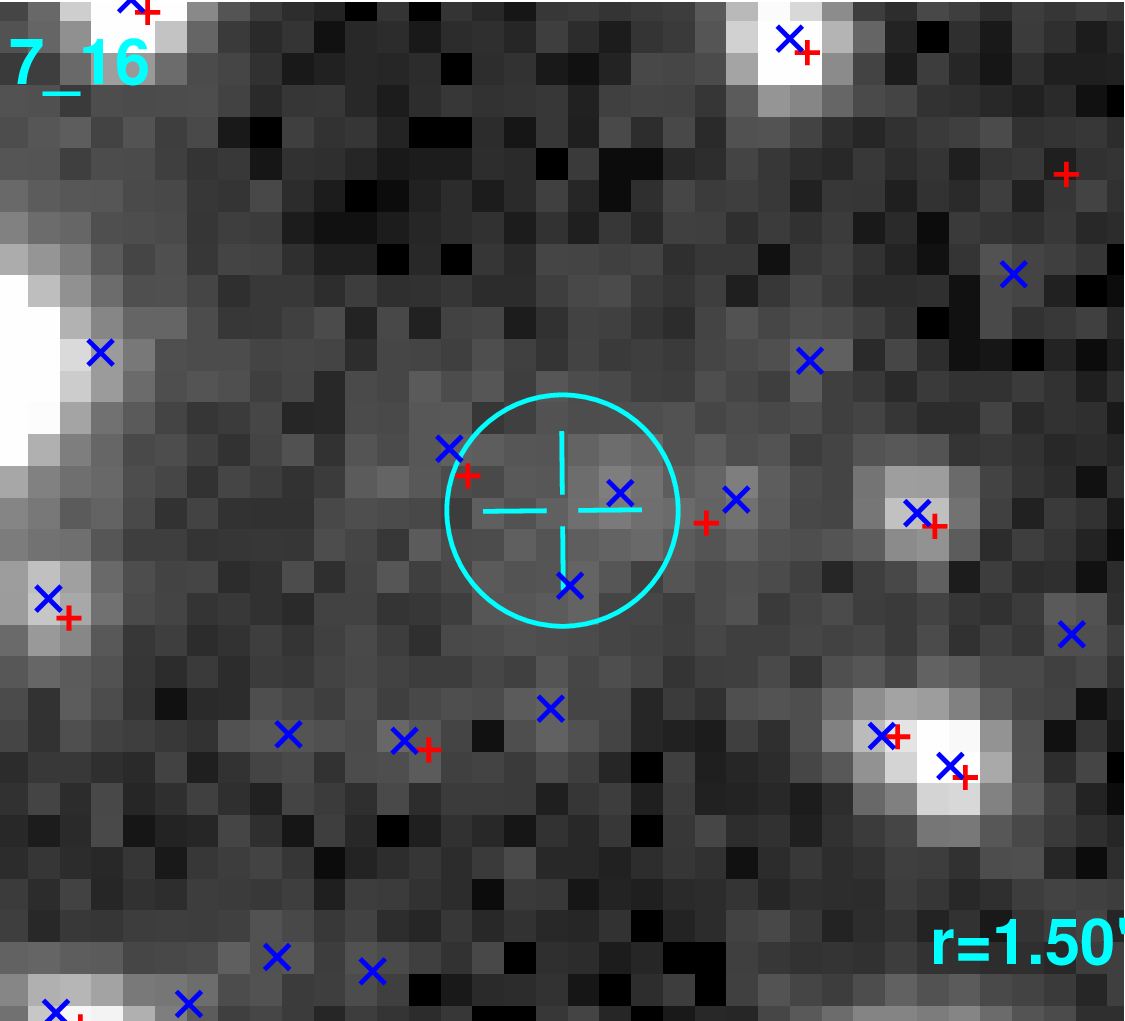} 
\end{center} 
\clearpage
\begin{center} 
\includegraphics[width=.45\textwidth]{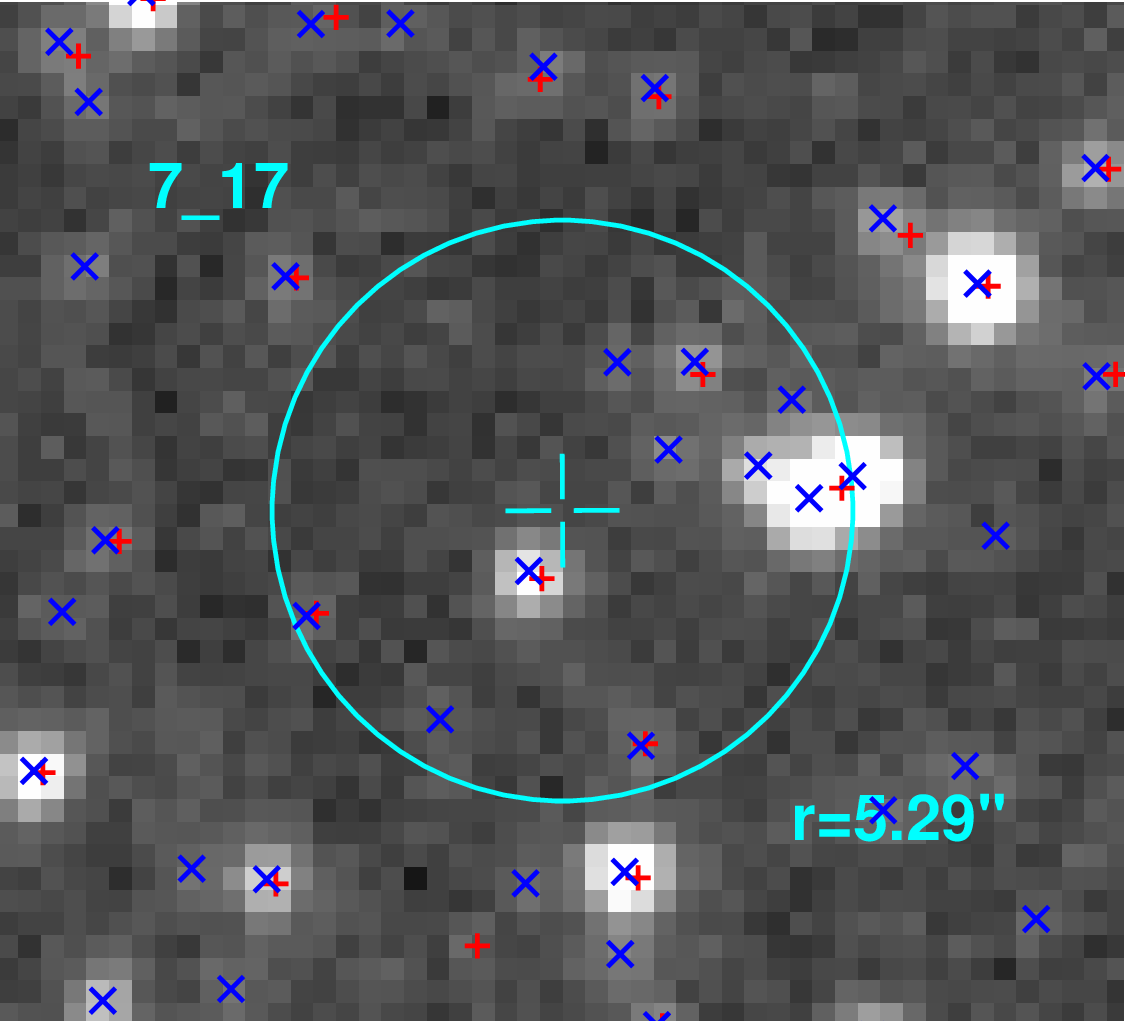} 
\includegraphics[width=.45\textwidth]{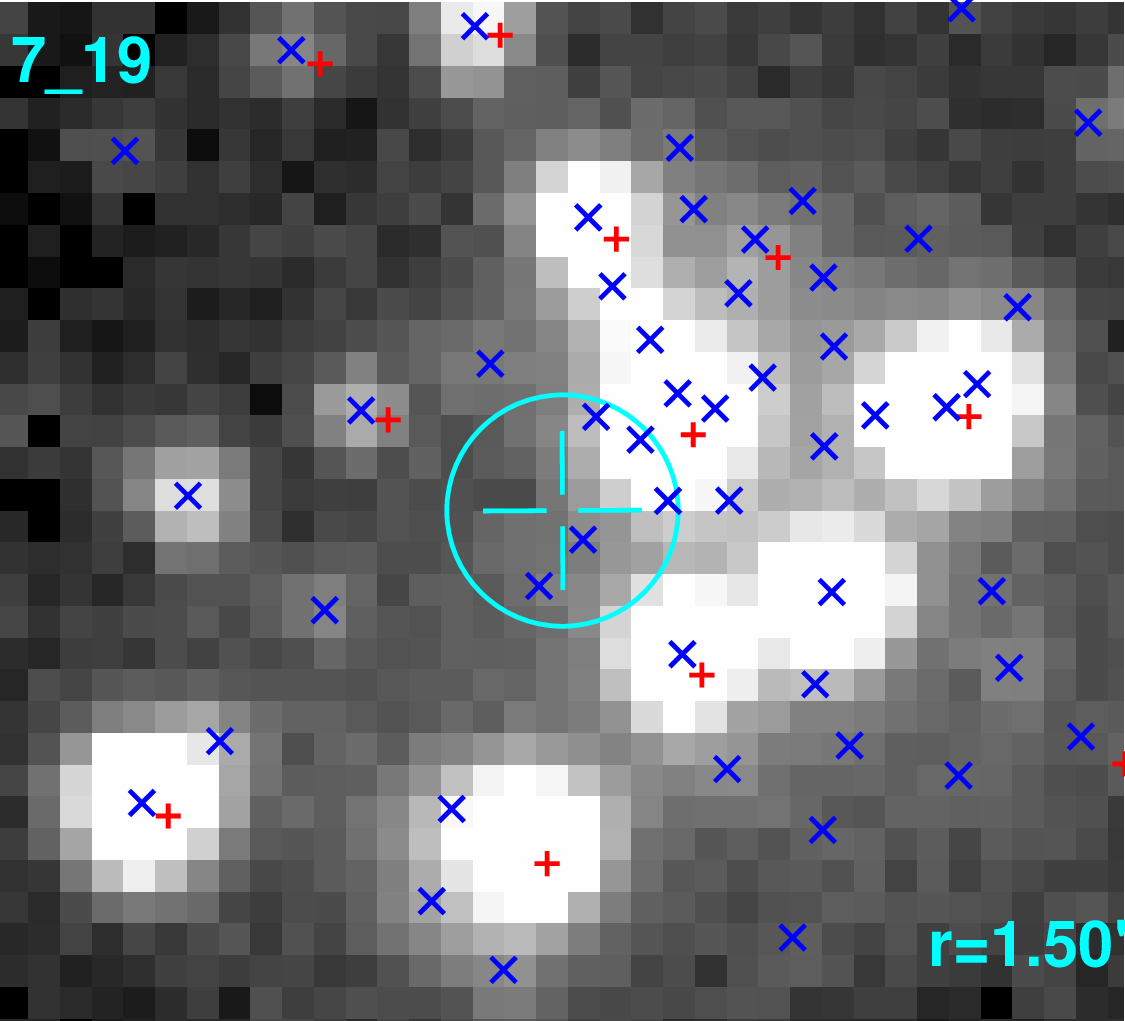} 
\includegraphics[width=.45\textwidth]{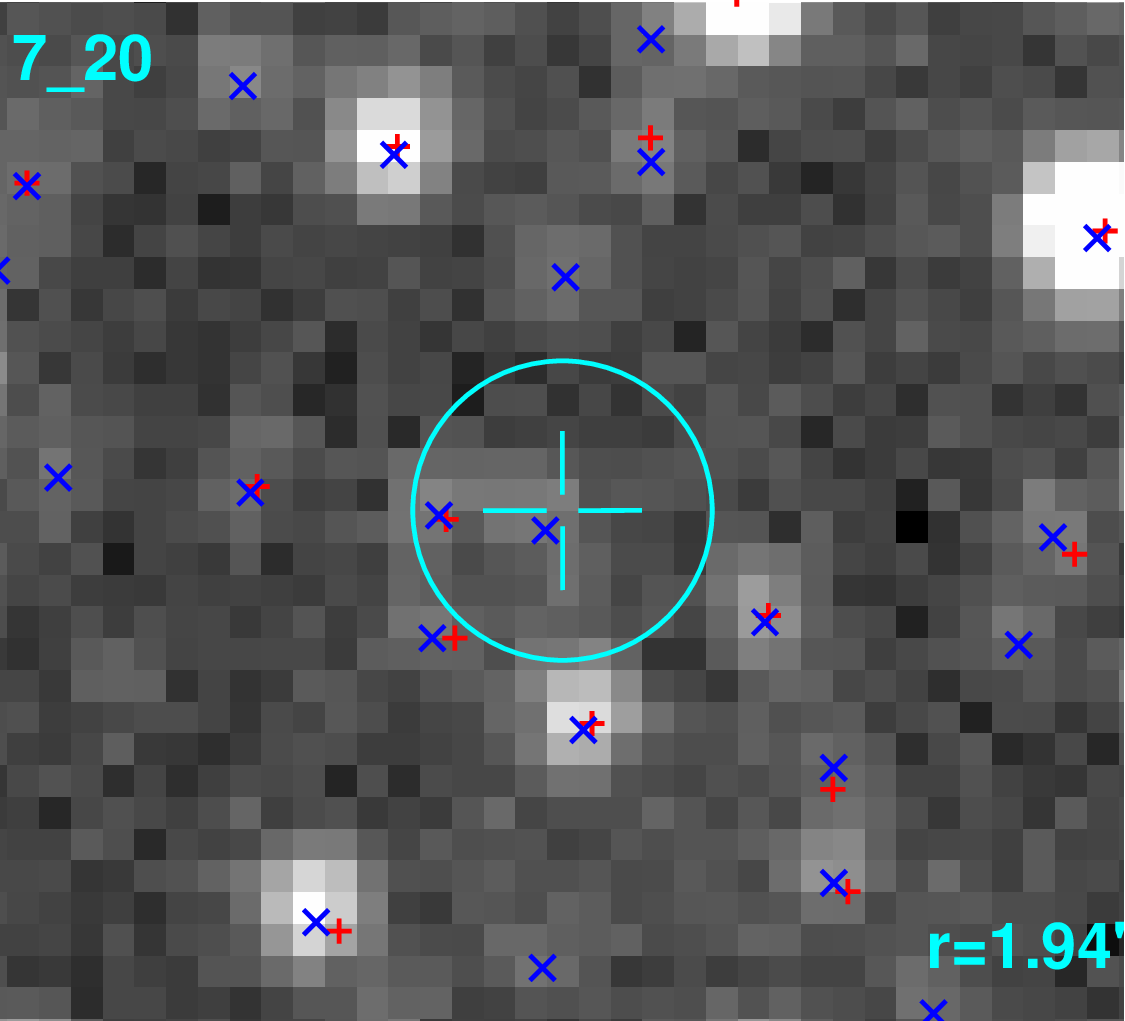} 
\includegraphics[width=.45\textwidth]{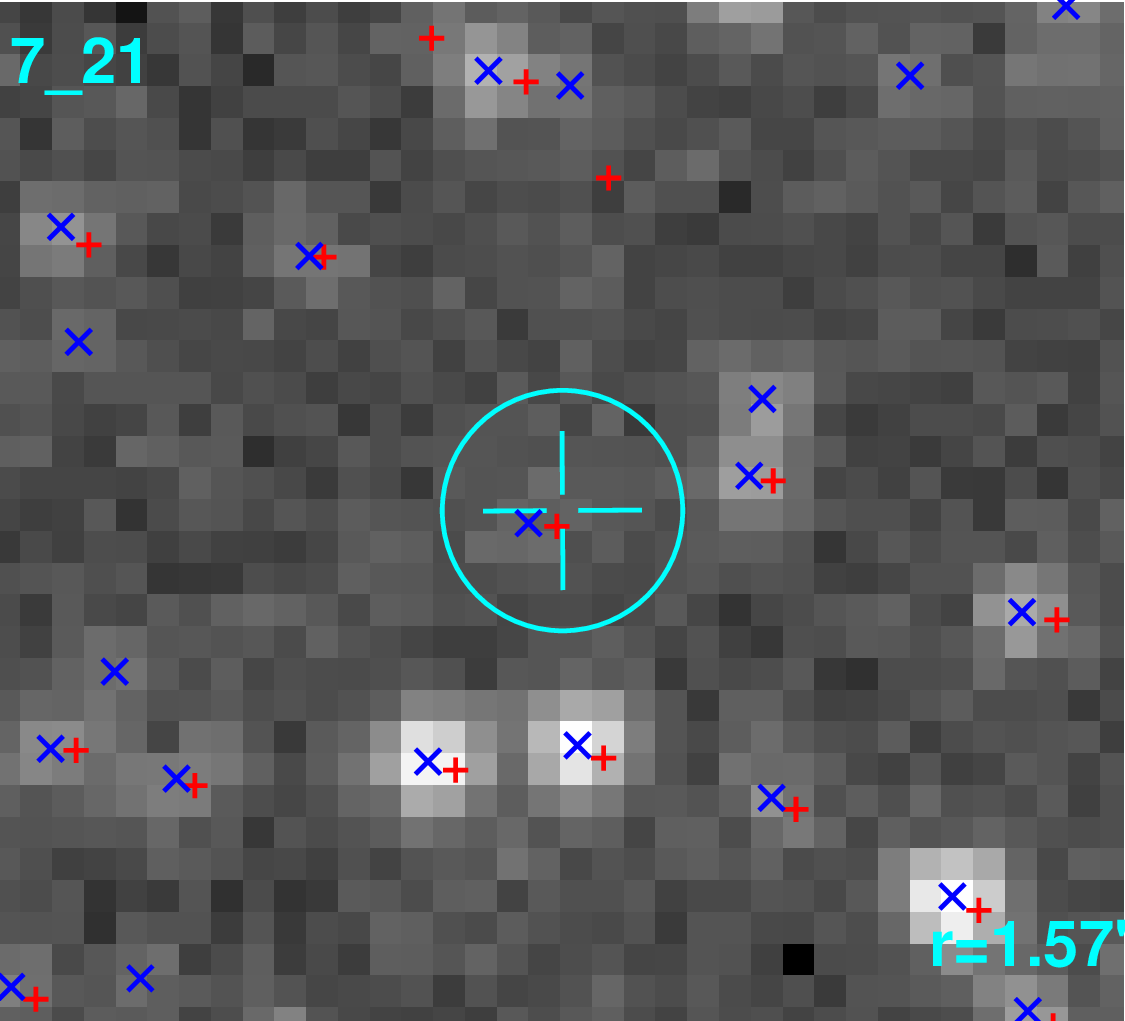} 
\end{center} 
\clearpage
\begin{center} 
\includegraphics[width=.45\textwidth]{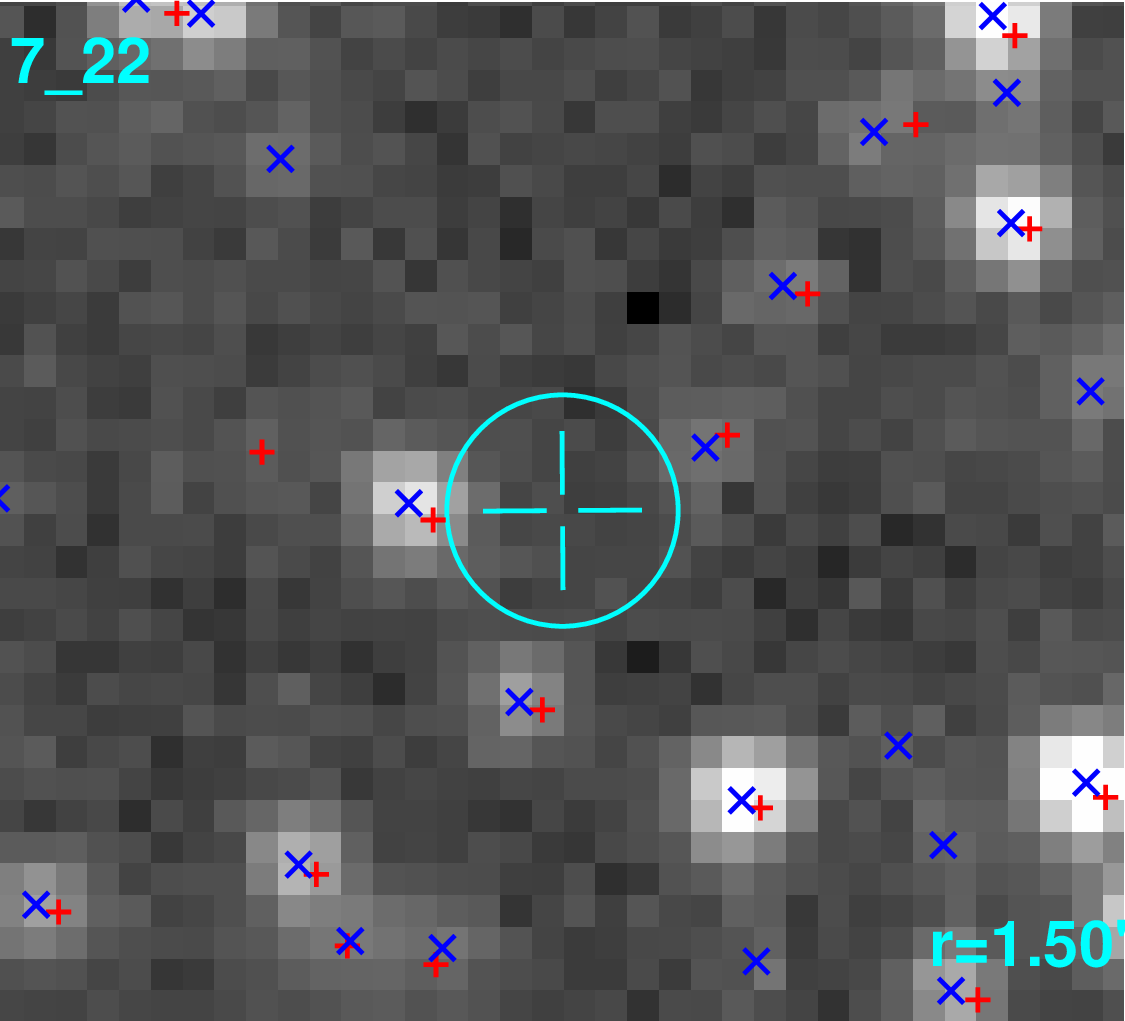} 
\includegraphics[width=.45\textwidth]{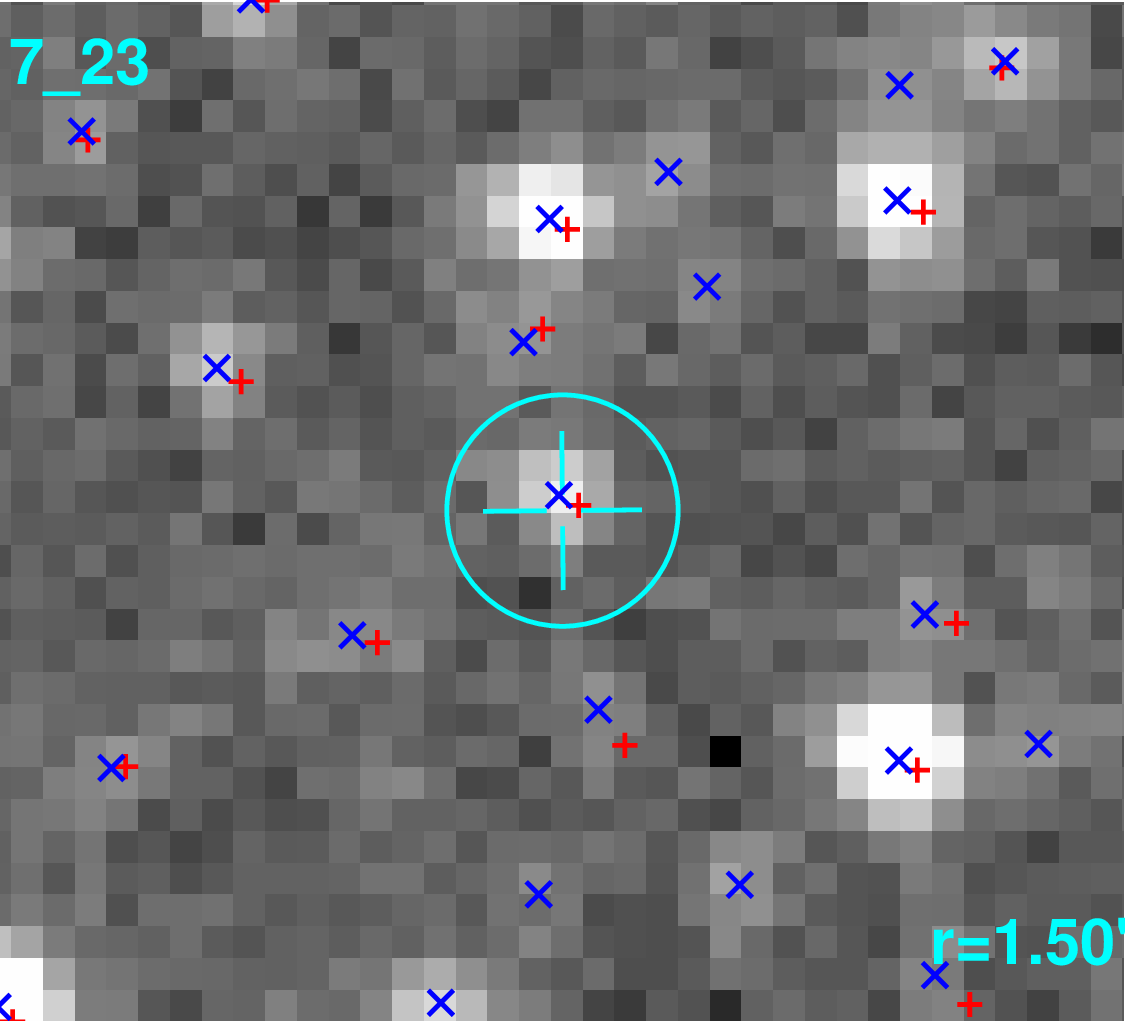} 
\includegraphics[width=.45\textwidth]{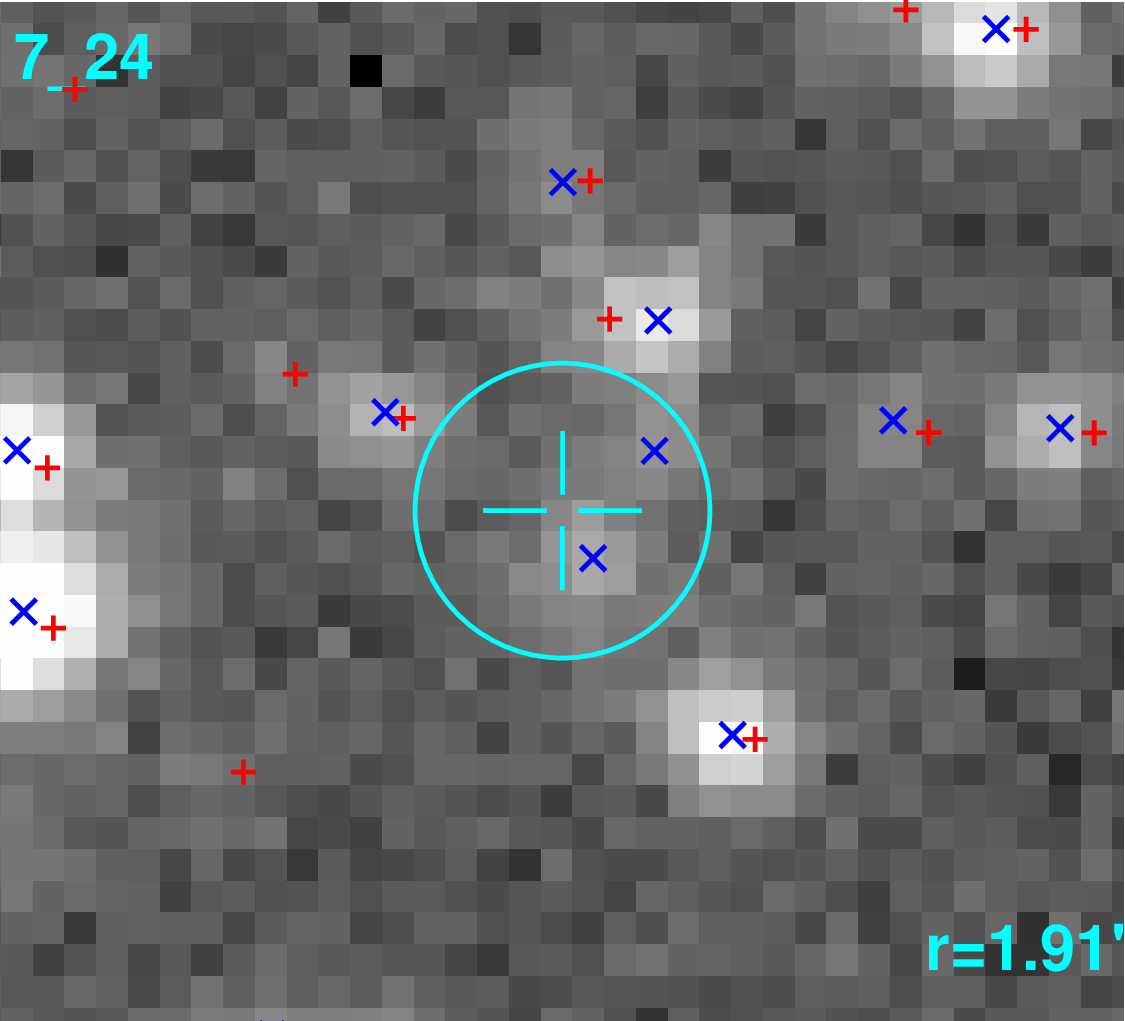} 
\includegraphics[width=.45\textwidth]{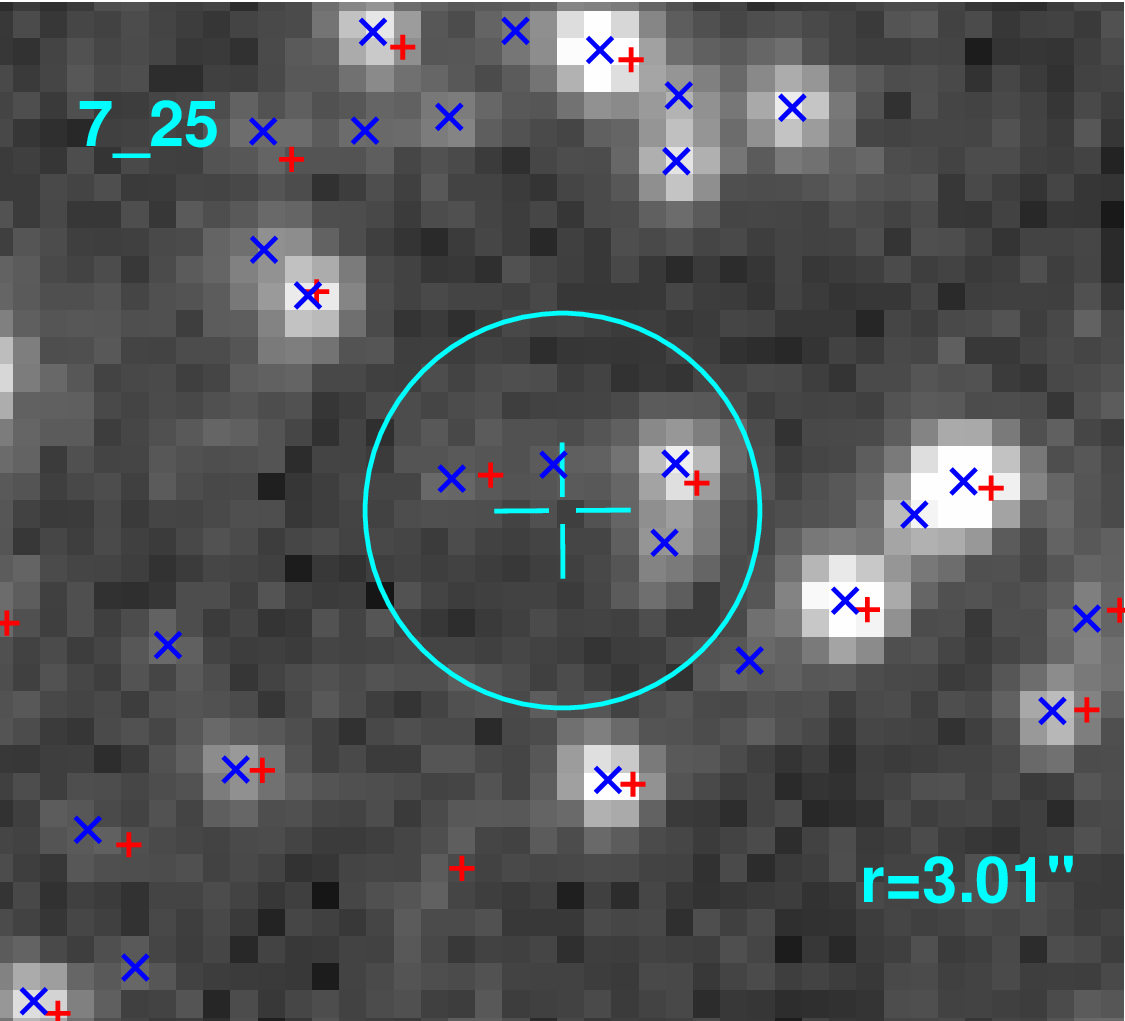} 
\end{center} 
\clearpage
\begin{center} 
\includegraphics[width=.45\textwidth]{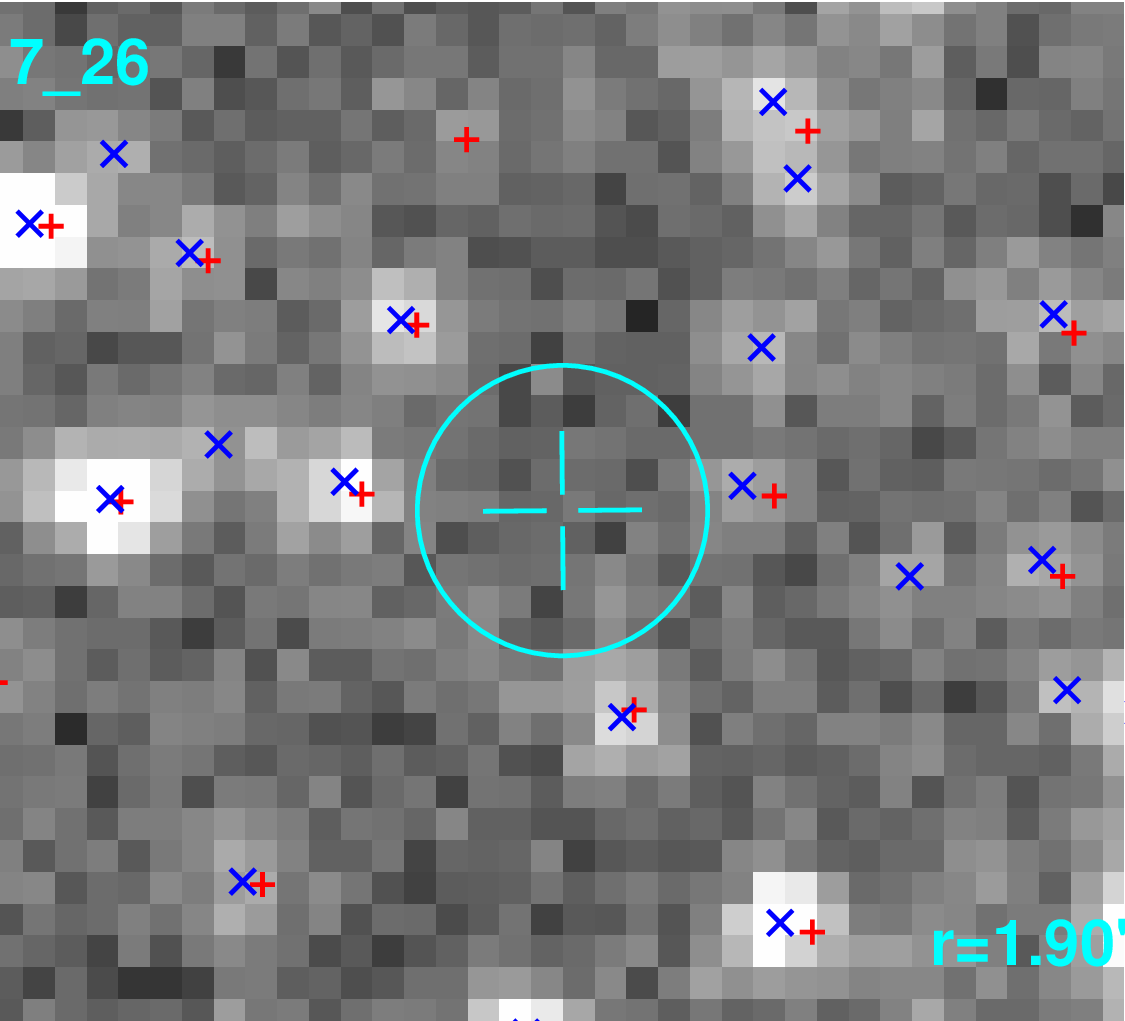} 
\includegraphics[width=.45\textwidth]{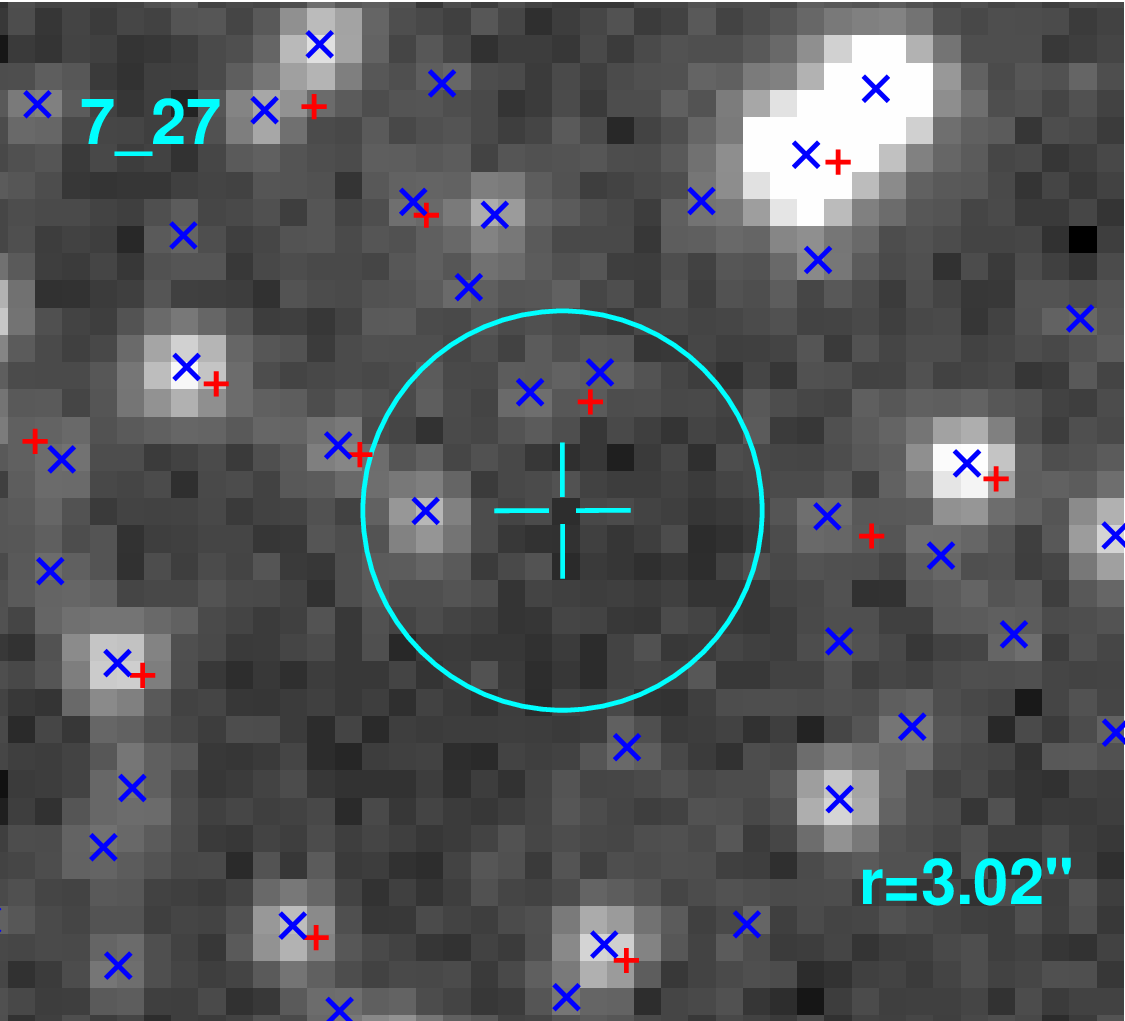} 
\end{center} 
\clearpage

\end{document}